\documentclass[fleqn,usenatbib]{mnras}

\usepackage{newtxtext,newtxmath}

\usepackage[T1]{fontenc}
\usepackage{ae,aecompl}

\usepackage{graphicx}	
\usepackage{amsmath}	
\usepackage{enumerate}   
\usepackage{color}
\usepackage{ulem}
\usepackage{multirow}   
\usepackage{xspace}
\usepackage{tabularx}
\usepackage{lipsum} 

\usepackage{amssymb}	
\defcitealias{keszthelyi2019}{\nobreak{Paper~I}} 
\newcommand{\PaperI}{\citetalias{keszthelyi2019}\xspace} 

\defcitealias{keszthelyi2020}{\nobreak{Paper~II}} 
\newcommand{\PaperII}{\citetalias{keszthelyi2020}\xspace} 

\defcitealias{keszthelyi2021}{\nobreak{Paper~III}} 
\newcommand{\PaperIII}{\citetalias{keszthelyi2021}\xspace} 

\title[The effects of surface fossil magnetic fields on massive star evolution: 
IV.]{The effects of surface fossil magnetic fields on massive star evolution: 
IV. Grids of models at Solar, LMC, and SMC metallicities}
\author[Z. Keszthelyi et al.]{
Z. Keszthelyi$^{1,2}$\thanks{E-mail: zsolt.keszthelyi@nao.ac.jp},
A. de Koter$^{1,3}$,
Y. G\"otberg$^{4}$\thanks{Hubble Fellow},
G. Meynet$^{5}$,
S.A. Brands$^{1}$,
V. Petit$^{6}$,   \newauthor
M. Carrington$^{7}$, 
A. David-Uraz$^{8,9}$, 
S.T. Geen$^{1}$, 
C. Georgy$^{5}$,
R. Hirschi$^{10,11}$, 
J. Puls$^{12}$,  \newauthor
K.J. Ramalatswa$^{13,14}$,
M.E. Shultz$^{6}$, 
A. ud-Doula$^{15}$
\\  
$^{1}$Anton Pannekoek Institute for Astronomy, University of Amsterdam, Science Park 904, 1098 XH, Amsterdam, The Netherlands \\
$^{2}$Center for Computational Astrophysics, Division of Science, National Astronomical Observatory of Japan, 2-21-1, Osawa, Mitaka, Tokyo 181-8588, Japan \\
$^{3}$Institute of Astronomy, KU Leuven, Celestijnenlaan 200D, 3001 Leuven, Belgium \\
$^{4}$The observatories of the Carnegie institution for science, 813 Santa Barbara Street, Pasadena, CA 91101, USA \\
$^{5}$Geneva Observatory, University of Geneva, Maillettes 51, 1290 Sauverny, Switzerland \\
$^{6}$Dept. of Physics and Astronomy, Bartol Research Institute, University of Delaware, 217 Sharp Lab, Newark, DE 19716, USA \\
$^{7}$Department of Physics and Space Science, Royal Military College of Canada, PO Box 1700, Station Forces, Kingston, ON K7K 0C6, Canada \\ 
$^{8}$Department of Physics and Astronomy, Howard University, Washington, DC 20059, USA \\ 
$^{9}$Center for Research and Exploration in Space Science and Technology, and X-ray Astrophysics Laboratory, NASA/GSFC, Greenbelt, MD 20771, USA \\ 
$^{10}$Astrophysics Group, Keele University, Keele, Staffordshire ST5 5BG, UK\\
$^{11}$Institute for Physics and Mathematics of the Universe (WPI), University of Tokyo, 5-1-5 Kashiwanoha, Kashiwa 277-8583, Japan\\
$^{12}$LMU M\"unchen, Universit\"atssternwarte, Scheinerstr. 1, 81679 M\"unchen, Germany \\ 
$^{13}$Department of Astronomy, University of Cape Town, Private Bag X3, Rondebosch 7701, South Africa \\ 
$^{14}$South African Astronomical Observatory, PO Box 9, Observatory, 7935, South Africa \\ 
$^{15}$Dept. of Physics, Penn State Scranton, 120 Ridge View Drive, Dunmore, PA 18512, USA \\ 
}
\date{Accepted 2022 September 9. Received 2022 August 25; in original form 2022 February 25
}

\pubyear{2022}

\begin{document}
\label{firstpage}
\pagerange{\pageref{firstpage}--\pageref{lastpage}}
\maketitle

\begin{abstract}
Magnetic fields can drastically change predictions of evolutionary models of massive stars via mass-loss quenching, magnetic braking, and efficient angular momentum transport, which we aim to quantify in this work.
We use the \textsc{mesa} software instrument to compute an extensive main-sequence grid of stellar structure and evolution models, as well as isochrones, accounting for the effects attributed to a surface fossil magnetic field. 
The grid is densely populated in initial mass (3-60~M$_\odot$), surface equatorial magnetic field strength \nobreak{(0-50~kG)}, and metallicity (representative of the Solar neighbourhood and the Magellanic Clouds).
We use two magnetic braking and two chemical mixing schemes and compare the model predictions for slowly-rotating, nitrogen-enriched ("Group~2") stars with observations in the Large Magellanic Cloud. We quantify a range of initial field strengths that allow for producing Group~2 stars and find that typical values (up to a few kG) lead to solutions. 
Between the subgrids, we find notable departures in surface abundances and evolutionary paths. In our magnetic models, chemical mixing is always less efficient compared to non-magnetic models due to the rapid spin-down. We identify that quasi-chemically homogeneous main sequence evolution by efficient mixing could be prevented by fossil magnetic fields.
We recommend comparing this grid of evolutionary models with spectropolarimetric and spectroscopic observations with the goals of i) revisiting the derived stellar parameters of known magnetic stars, and ii) observationally constraining the uncertain magnetic braking and chemical mixing schemes.

\end{abstract}

 

\section{Introduction} \label{sec:intro}

Magnetism is ubiquitously present in the Universe, from the scale of sub-atomic particles up to the scale of galaxy clusters \citep[e.g.,][]{neronov2010}. For example, magnetic fields play a vital role in regulating star formation as molecular clouds collapse \citep[e.g.,][]{commercon2011,mackey2011,crutcher2012,Hennebelle2013,Koertgen2020,Seifried2020a}, and in the formation and physics of neutron stars \citep[e.g.,][]{reisenegger2009,beloborodov2009,takiwaki2009,takiwaki2011,mosta2015,kuroda2020,aloy2021,reboul2021, masada2022}. In the phase between star formation and stellar end products, massive star evolution remains uncertain in part due to the incomplete understanding of stellar magnetic fields.

%
Spectropolarimetric surveys revealed that a subset (about 10\%) of hot ($\gtrsim$~10 kK), massive (8 - 60 M$_\odot$) and intermediate-mass (2~-~8~M$_\odot$) stars in the Galaxy with spectral types O, B, and A host large-scale, globally-organised surface magnetic fields \citep{morel2015,fossati2016,wade2016,alecian2016,grunhut2017,shultz2018,sikora2019,petit2019}. Surface magnetic fields are detected both in chemically peculiar Bp/Ap stars, as well as in O, B, and A stars without observed chemical peculiarities \citep[e.g.,][]{donati2009,henrichs2013,neiner2015,grunhut2017,shultz2018,sikora2019}. In addition, all known Of?p stars (showing variable emission in the C\textsc{iii}$~\lambda$4647-4650-4652 complex, of comparable strength at its maximum to emission in the N\textsc{iii}$~\lambda$4634-4640-4642 complex, \citealt{walborn1972}) in the Galaxy are observed to be magnetic\footnote{However, not all magnetic O-type stars belong to this class \citep[][]{donati2002,petit2013,petit2017,grunhut2017}.}.
Alongside spectropolarimetric observations, several properties may be used to identify magnetic candidates from photometric and spectroscopic studies, including multi-wavelength diagnostics \citep[e.g.,][]{babel1997,cohen2003,marcolino2012,rivinius2013,naze2014,oksala2015,bram2018,walborn2015,leto2021}. Most recently, TESS photometric data is being used to identify candidate magnetic stars based on characteristic light-curve variations and subsequently observe them via spectropolarimetry \citep[][]{mobster1,mobster2,mobster3,mobster4,mobster5}. 

The observed surface magnetic fields of hot, massive and intermediate-mass stars do not show any apparent correlation with stellar and rotational parameters unlike in lower-mass ($<$2~M$_\odot$), cool stars ($<$10 kK), where magnetism due to surface convection and differential rotation ubiquitously produce dynamo activity \citep{donati2009,neiner2015}. Consequently, the organised, large-scale magnetic fields of hot stars are expected to be of fossil origin \citep{cowling1945,spitzer1958,mestel1967, mestel2003,moss2003,mestel2010,ferrario2015,braithwaite2004,braithwaite2017}. 
The exact origin of observed magnetic fields remains debated. In about 10\% of intermediate-mass Herbig Be/Ae stars (which 10\% is thought to be the precursors of main sequence Bp/Ap stars), large-scale surface magnetic fields are already observed on the pre-main sequence \citep{stepien2000,alecian2009, alecian2013,villebrun2019,lavail2020}, which may be acquired from the star-forming disk or generated via a dynamo action inside the star during a fully convective pre-main sequence phase \citep[e.g.][]{moss2003,braithwaite2012}.
In addition to magnetic fields possibly remaining from the star formation or pre-main sequence phases, stellar mergers could also amplify seed magnetic fields to a strength sufficient to be detectable \citep{ferrario2009,wickramasinghe2014,schneider2016,schneider2019,schneider2020}, suggesting that there may exist multiple channels to generate globally-organised, large-scale fossil magnetic fields. Merger events of compact remnants have also been proposed to explain strongly magnetised white dwarfs and neutron stars \citep[e.g.,][]{tout2008,giacomazzo2015,ferrario2020,caiazzo2021,shultz2021m}. 

The nature of fossil fields is fundamentally different from contemporaneously generated dynamo fields by a mechanical source (such as convection or differential rotation). Fossil field evolution is purely dissipative with no active field generation counteracting its slow dissipation \citep[][]{braithwaite2017}. In stellar layers where large-scale fossil fields spread through, it is expected that solid-body rotation will develop \citep[e.g.][]{mestel1999}. In those stellar layers\footnote{However, dynamo-generated fields and fossil fields may co-exist in some stellar layers, for example, at the core-envelope interface \citep[][]{feat2009}. Whether such an interaction could lead to a more rapid dissipation of the fossil field remains an open question.} the mechanical source of differential rotation is absent, and consequently small-scale dynamo fields in radiative stellar layers cannot be induced \citep{spruit2004}.  
The Tayler instability \citep[][]{tayler73,goldstein2019}, for example, cannot develop if the radial rotation profile is completely flat, which means that the Tayler-Spruit (or "$\Omega$-type") dynamo cannot be induced in the presence of a fossil field \citep[e.g.,][]{spruit2004}. In fact, while this type of dynamo mechanism in radiative stellar layers was proposed by \cite{spruit2002}, there remains ongoing debate about the necessary electromotive force to operate the dynamo cycle \citep{fuller2019}. The simulations of \cite{zahn2007} suggest that this dynamo cycle does not operate.
Despite the contradictory numerical results and the lack of direct observational evidence, dynamos in radiative stellar layers are commonly accounted for in evolutionary models \citep[e.g.,][]{spruit2002,maeder2003,maeder2004,maeder2005,Maeder2009a,heger2005,yoon2006,denissenkov2007, potter2012b,quentin2018,fuller2019,fuller2019b,takahashi2021}. 
We emphasise that these implementations are not suitable (at least directly) to model stars that are known to host fossil fields. 

The time evolution of fossil magnetic fields also remains an unresolved problem. Observed samples of magnetic A-type stars and compact remnants are consistent with the magnetic flux being conserved over time \citep[e.g.,][]{landstreet2007,landstreet2008,wickramasinghe2005,neiner2017,martin2018,sikora2019}, whereas other observational evidence (including that for OB stars) suggests magnetic flux decay \citep[e.g.,][]{fossati2016,shultz2019b}. Fossil magnetic fields are expected to evolve only by Ohmic dissipation \citep{wright1969,spruit2004,duez2010,braithwaite2017}, which has a longer timescale than the main sequence nuclear timescale \citep{cowling1945,spitzer1958}. However, Ohmic dissipation remains riddled with uncertainties depending on the exact value of magnetic diffusivity \citep[e.g.,][]{charbonneau2001} and the geometry of the magnetic field since more complex fields dissipate faster. 

Despite the uncertainties regarding the origin and evolution of fossil magnetic fields, it is now well established that they lead to various changes in stellar structure and evolution \citep[e.g.,][]{mestel1989,mestel1999,duez2010,macdonald2019,jermyn2020}. Two main surface effects, mass-loss quenching and magnetic braking (discussed in detail below), have been shown to drastically modify evolutionary model predictions \citep[e.g.,][]{meynet2011,keszthelyi2017a,keszthelyi2019,keszthelyi2020,keszthelyi2021}. For instance, heavy stellar-mass black holes and pair-instability supernovae could be formed from magnetic progenitors even at solar metallicity \citep[][]{petit2017,georgy2017}. 

Thus far, surface magnetic fields have only been detected in Galactic stars. Currently, high-resolution spectropolarimeters used for stellar magnetometry are employed on 4m-class telescopes, which limits observations to bright nearby stars. Using low-resolution spectropolarimetry, \citet{bagnulo2017,bagnulo2020} searched for strong magnetic fields in the Magellanic Clouds through the Zeeman effect, which did not lead to definite detections in any of the targets. While high-resolution spectropolarimetry remains largely limited to a Galactic environment, very extensive spectroscopic campaigns in the Magellanic Clouds -- in addition to the identified Of?p stars  \citep[e.g.,][]{walborn2015,munoz2020} -- suggest that the nature of some stars may be explained by invoking surface magnetic fields \citep[][]{hunter2008,brott2011,riverogonzalez2012, potter2012b,grin2017,dufton2018,dufton2020,varsha2021}.
Observations of known magnetic stars in the Galaxy are often compared to evolutionary models that do not include fossil field effects (e.g., those of \citealt{brott2011,ekstroem2012,chieffi2013,choi2016,costa2021, grasha2021}), possibly making inferences of stellar parameters rather uncertain. In turn, this can largely impact the derived ages of individual stars and bias isochrone fitting of stellar clusters. 
Therefore, there is a need for stellar evolution models (and model grids), which take into account mass-loss quenching and magnetic braking (although see \citealt[][]{potter2012b} for the latter), thereby affecting detailed evolutionary model predictions and population synthesis studies in both Galactic and extragalactic environments.

The motivation of this study is to help to resolve these issues by presenting and studying an extensive grid of stellar structure and evolution models with metallicities typical of environments in the Solar neighbourhood\footnote{Since we follow the elemental abundance determinations given by \cite{prz2008,prz2013}, \cite{nieva2012} and \cite{asplund2009}, we refer to this set of models as Solar and not Galactic.}, Large Magellanic Cloud (LMC), and Small Magellanic Cloud (SMC) that include the effects of surface fossil magnetic fields. The model computations are open source and the entire library of models is available to the community via Zenodo at \url{https://doi.org/10.5281/zenodo.7069766}.

%
%
%
%
This paper is part of a series in which we aim to explore the effects of surface fossil magnetic fields on massive star evolution. 
In the first paper of the series \citep[][hereafter \PaperI]{keszthelyi2019}, we used the Geneva stellar evolution code \citep{eggenberger2008,ekstroem2012,georgy2013,groh2019,murphy2021} and discussed the mutual impact of magnetic mass-loss quenching, magnetic braking, and field evolution on a typical massive star of initially 15~M$_\odot$ at solar metallicity. We studied both solid-body and differentially-rotating models and evaluated their key evolutionary characteristics, showing that strong surface nitrogen enrichment is expected for magnetic models with differential rotation.
In the second paper \citep[][hereafter \PaperII]{keszthelyi2020}, we elaborated on the implementation of massive star magnetic braking in the \textsc{mesa} software instrument \citep{paxton2011,paxton2013,paxton2015,paxton2018,paxton2019} and detailed the magnetic and rotational evolution of the models by performing a parameter test in initial mass, magnetic field strength, and rotational velocity space with 35 models. Then, 72 tailored models were compared with a sample of observed magnetic B-type stars from \cite{shultz2018,shultz2019a,shultz2019b}. A key finding of \PaperII is that magnetic stars could originate from ZAMS progenitors with a variety of parameter combinations.
In \PaperIII (\citealt[][]{keszthelyi2021}) we focused on the scenario that some magnetic stars may originate from rapidly-rotating progenitors at the ZAMS, and specifically applied it to the case of the magnetic early B-type star $\tau$ Sco. We found that for this star the simultaneous nitrogen enrichment and slow rotation poses a significant challenge for single-star evolution.


The paper is organised as follows. In Section~\ref{sec:two}, we detail the assumptions and input physics used in the models. In Section~\ref{sec:three}, we present and scrutinise the stellar structure and evolution models from our computations. In Section~\ref{sec:four}, we discuss implications and future work. Finally, we conclude our findings in Section~\ref{sec:concl}.

%
%

\begin{figure}
\includegraphics[width=0.49\textwidth]{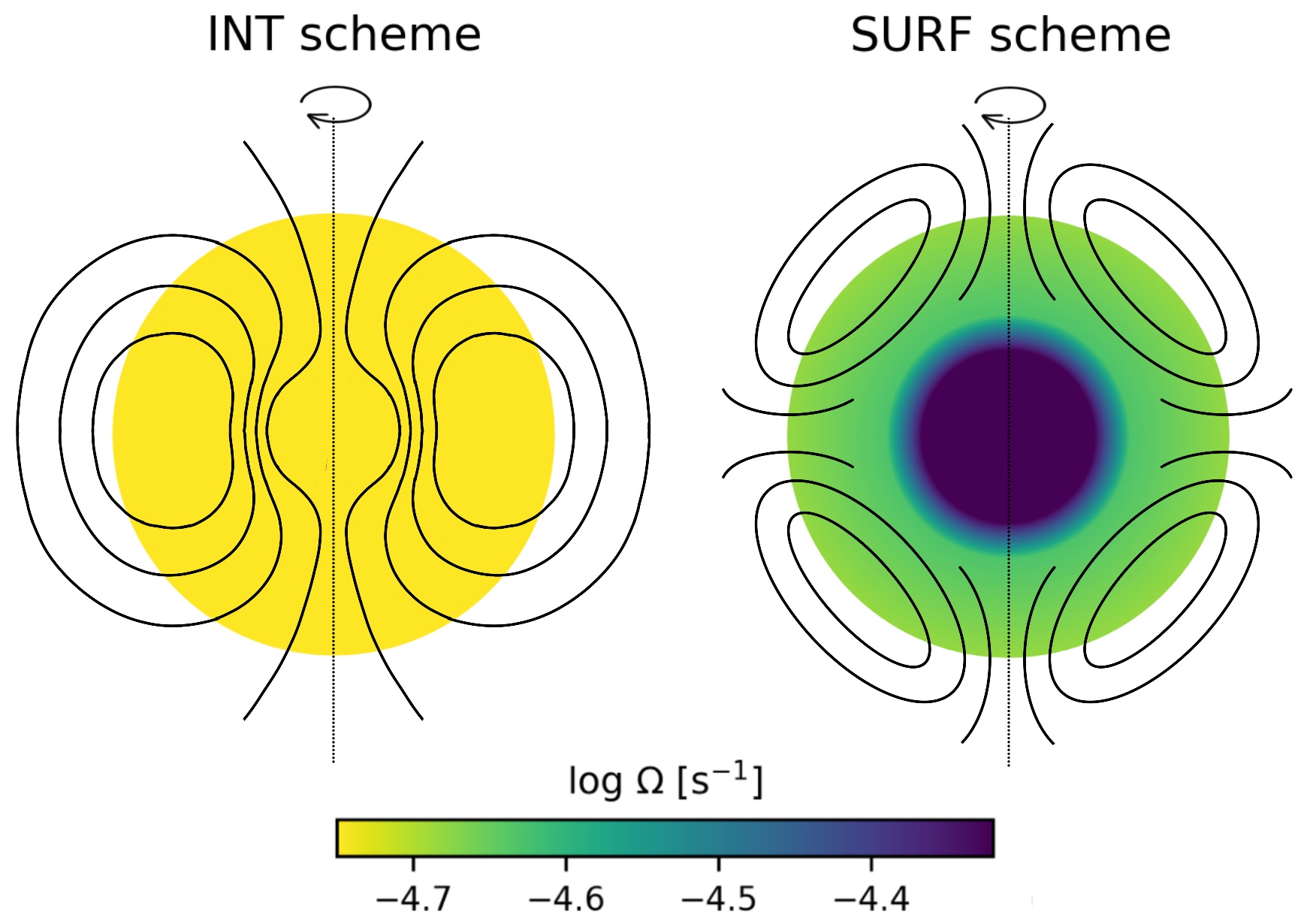}
\centering 
\caption{Schematic illustration of the magnetic field geometries (solid lines) considered in this study. These geometries are not directly implemented in our 1D models, instead they are used to constrain the corresponding scaling relations. Colours indicate the logarithmic angular velocity; the rotation axis is shown with dotted lines. \textit{Left:} INT scheme, representative of a dipolar field configuration leading to internal magnetic braking, spinning down all layers of the star. As a phenomenological picture, we assume that convective expulsion (see Appendix~\ref{sec:convex}) would not allow the field to relax in the stellar core; however, we assume that the braking can propagate uniformly (including the core). The angular rotation is uniform throughout the star, enforced by a high diffusivity attributed to the dipole field. \textit{Right:} SURF scheme, representative of a more complex field geometry leading to surface magnetic braking. The magnetic energy can be stored in higher-order spherical harmonics, and twisted field lines only penetrate to some extent of the envelope which we assume to be 20 per cent of the mass fraction. In those layers efficient angular momentum transport is present but we assume that core-envelope coupling is not achieved. Magnetic braking is only applied to those upper layers (see Section~\ref{sec:magbraking}). To be able to incorporate the effects of such fields into 1D models, we assume that a quadrupole scaling may be used for calculating the Alfv\'en radius.}\label{fig:intsurf}
\end{figure}

\section{Modelling assumptions and setup}\label{sec:two}

\subsection{General strategy}

In this work, we follow the general strategy of adopting suitable parametric prescriptions to model the effects of fossil magnetic fields, similar to the approaches presented in \PaperI, \PaperII, \PaperIII and references therein. 
While one-dimensional magnetohydrodynamic (MHD) approaches are possible, they have mostly been developed for dynamo models \citep[e.g.,][]{potter2012b,feiden2012,feiden2013,quentin2018,takahashi2021} and as such are not directly applicable to model fossil fields of intermediate-mass and massive stars. In particular, magnetic transport equations have been developed and used previously in the context of dynamo-generated fields \citep[e.g.,][]{spruit2002,maeder2003,maeder2004,maeder2005,heger2005,yoon2006,potter2012b,kissin2018,quentin2018,fuller2019,fuller2019b,takahashi2021}. Although the characteristics (for example, the scale and time evolution) of dynamo models are incompatible with those of fossil fields (see Section~\ref{sec:intro}), the transport equations may follow similar implementations. Here, we opt to artificially increase the diffusivity instead of testing "magnetic" transport equations (Section~\ref{sec:amtr}). Such equations would introduce more free parameters and further assumptions regarding the geometry, structure, and radial dependence of the magnetic field. Clearly, further research and observational verification is required before an appropriate one-dimensional magnetic transport process could be reliably incorporated to model stellar evolution with fossil magnetic fields (however, see \citealt[][]{duez2010,schneider2020}). \citet{duez2010} and \citet{duez2010b} presented a  comprehensive approach applicable for fossil fields, showing however that the impact on hydrodynamic equilibrium and energy transport are modest even for strong magnetic fields. 
To this extent, it is indeed appropriate to use parametric prescriptions and focus on the major, measurable effects\footnote{See e.g., \cite{driessen2019b} for mass-loss quenching, and e.g. \cite{town2010,oksala2012,song2021} for magnetic braking.} that fossil magnetic fields have, namely changing the mass loss (Section \ref{sec:mdot}) and rotation (Section \ref{sec:rot}) of the star, affecting chemical mixing and angular momentum transport. 

One of the major modelling challenges is that the geometry and alignment of the magnetic field play a significant role in the corresponding physical description. It has been demonstrated that a seed magnetic field can relax into a stable axisymmetric (around the magnetic axis) configuration if the magnetic flux is centrally concentrated, or into a non-axisymmetric (around the magnetic axis) configuration otherwise \citep{braithwaite2004,braithwaite2006,braithwaite2008}. In both cases, the latitudinal averaging is inappropriate to model the magnetic field in 1D\footnote{For example, in the case of an axisymmetric dipole geometry, both poloidal and toroidal components must exist in the stellar interior \citep{braithwaite2004}. The poloidal field strength and orientation relative to the normal to the surface varies over latitudes. The toroidal field, confined by closed poloidal lines, has zero strength along the polar rotation axis and reaches its maximum along the equatorial plane (see, e.g., Figure 4 of \citealt{braithwaite2008}). A latitudinal averaging instead assigns a mean value to the poloidal and toroidal field components along a radius.}. 

For simplicity, we assume that the field is aligned with the rotation axis of the star since appropriate scaling relations for oblique fields (tilted with respect to the rotation axis) are still in development. However, the obliquity angles inferred from observations appear to follow a random distribution, which suggests that, apart from a few possible exceptions, massive stars generally possess magnetic fields that are inclined with respect to the rotation axis \citep[e.g.,][]{khalack2003,shultz2019a,sikora2019}. Recent work from \cite{ud2020} suggests that oblique rotation leads to decreasing the efficiency of magnetic braking. This effect could be incorporated in our models via a suitable scaling factor in future studies. However, the efficiency of magnetic braking, in the evolutionary context, is also largely dependent on magnetic field evolution, which still needs to be better constrained (e.g., \PaperIII).

In fact, magnetic field evolution is closely tied to the question of the field geometry and, in this regard, new insights are gained from extensive monitoring campaigns, which can reveal the surface properties of magnetic fields. Although a purely dipolar field geometry generally matches observations \citep[][]{grunhut2017,shultz2018}, modest deviations from pure dipolar geometries are now identified \citep[][]{leto2018,das2020,daviduraz2021}. In other cases, contributions from higher-order harmonics are also identified (e.g., \citealt{shultz2018,kochukhov2019}); however, the dipole is the strongest component, which consequently drives the main physical effects. In a few cases, observations have also identified stars with uniquely complex magnetic fields, which cannot be described with a dominant dipole component \citep[e.g., $\tau$ Sco,][]{donati2006,kochukhov2016,shultz2018}. In these cases, most of the magnetic energy is stored in higher-order spherical harmonics, although a weak dipole contribution may still be present. 

Since the structure of large-scale magnetic fields in stellar interiors is still considerably uncertain, we aim to test two limiting cases - illustrated in Figure \ref{fig:intsurf} - to apply depth-dependent magnetic braking (c.f. \PaperII). In the models with internal magnetic braking (INT, described in detail below), we assume that the magnetic field is dipolar. We further assume that the field is present in the entire stellar envelope and is able to achieve core-envelope coupling (however, only as a phenomenological picture, we assume that it is excluded from the stellar core). In nature, the stellar envelope must also have a closed toroidal field for a stable configuration \citep[][]{wright1969,wright1973, braithwaite2004,braithwaite2006,akgun2013}, which we cannot directly take into account in our 1D models. The magnetic field is curl-free above the stellar surface; any toroidal field diffuses to a poloidal structure. 
We also introduce a set of models with surface magnetic braking (SURF, see also below) as a limiting scenario to contrast the INT models with. The complex magnetic field geometries clearly cannot be translated to our 1D models, therefore we make various simplifying assumptions (discussed in detail throughout Section \ref{sec:two}). We reiterate that the field geometry cannot be directly included in our models, only via the scaling relations. The limitations related to the 1D parametrisation can likely only be resolved once 2D stellar evolution modelling becomes feasible \citep{espinosa2011,lovekin2020,reese2021}. 

Angular momentum is always lost from the outermost layers of the star. We use the INT/SURF schemes to control the propagation of this loss to the stellar interiors along with efficient angular momentum transport. In the INT models, all stellar layers efficiently transport and lose specific angular momentum. Motivated by the simulations of \cite{braithwaite2008}, we assume that in the SURF models the magnetic-field driven angular momentum transport and rotational braking only affect the upper 20\% of stellar mass. Although even for complex surface magnetic fields there may be weak dipole contributions, we neglect here any possible magnetic angular momentum transport in the deep stellar interiors so that we are able to test a limiting case in which radial differential rotation may develop in regions of the star where the magnetic flux is assumed to be negligible. Furthermore, to generalise from the range of possible field configurations deduced from observations, we assume that at least the Alfv\'en radius (see section \ref{sec:alf}) has to be smaller in the SURF models than in the INT models for a given surface field strength. For this reason, we use a quadrupole scaling in the SURF models to obtain the Alfv\'en radius. This results in less efficient magnetic braking for complex fields compared to dipole-dominated geometries. The field geometry defines the wind flow and mass flux from the stellar surface. We evaluate in Appendix \ref{sec:quad} how an actual quadrupole field geometry would affect mass-loss quenching; however, since this effect only concerns the highest-mass models, for simplicity we adopt a formalism where only the Alfv\'en radius is changed in the SURF models compared to the INT models (see Section~\ref{sec:mdot}).

\subsection{General model setup}

We use the software instrument Modules for Experiments in Stellar Astrophysics \textsc{mesa} release 15140 \citep[][]{paxton2011,paxton2013,paxton2015,paxton2018,paxton2019} and Software Development Kit (\textsc{SDK}) version 20.12.1 \citep[][]{sdk}, and carry out our computations on the Dutch supercomputers Cartesius\footnote{\url{https://userinfo.surfsara.nl/systems/cartesius}} and Snellius\footnote{\url{https://userinfo.surfsara.nl/systems/snellius}}. 
In this study, we consider main sequence models with initial masses from 3 to 60~M$_\odot$ (Table~\ref{tab:t1}). This choice is made since there are still considerable uncertainties in stellar evolution, as well as in magnetic field evolution on the post-main sequence. The computations begin by relaxing the initial model and we consider the Zero Age Main Sequence (ZAMS) to begin at the time when the initial abundance of core hydrogen has decreased by 0.3 percent. The endpoint of the models is the Terminal Age Main Sequence (TAMS), which we consider at the time when the core hydrogen mass fraction drops below $10^{-5}$. 

The \textsc{mesa} microphysics are summarised in Appendix~\ref{sec:micro}. 
To calculate the nuclear reaction rates, we use \textsc{mesa}'s default "basic.net" option.
To model convective mixing, we assume a mixing length parameter of $\alpha_{\rm MLT} = 1.8$ \citep[e.g.,][]{canuto1992,canuto1996} and the Henyey formalism \citep{henyey1965} in \textsc{mesa}. Common values of $\alpha_{\rm MLT}$ range from 1.5 to 2.0, mostly based on solar and solar-type star calibrations \citep[e.g.,][]{bonaca2012}. For simplicity, we assume the same value in all models, although some studies predict mass dependence \citep[e.g.,][]{yildiz2006}. Semiconvective and thermohaline mixing are not used\footnote{Neither of these mixing mechanisms is expected to significantly change main sequence models. It was shown by \cite{charbonnel2007} that a strong magnetic field could inhibit thermohaline mixing in descendants of Ap stars (see also, e.g., \citealt{denissenkov2009}). On the other hand, \cite{harrington2019} finds that thermohaline mixing could be enhanced by an aligned magnetic field.}. Convective boundaries are determined via the Ledoux criterion. Core overshooting is applied in the exponential scheme with parameters $f_{\rm ov} = 0.015$ and  $f_{0} = 0.005$, which roughly corresponds to a step overshoot parameter $\alpha_{\rm ov}$ of 10 percent of the local pressure scale height \citep[][see further discussion in Appendix~\ref{sec:convex}]{ekstroem2012}. $\alpha_{\rm ov}$ may be mass dependent \citep{castro2014}. Commonly used values of $\alpha_{\rm ov}$ range from 0.1 to 0.335, \citep[][]{schaller1992,brott2011}. Exponential overshooting at non-burning convective regions is adopted with $f_{\rm ov} = 0.0010$ and  $f_{0} = 0.0005$. \textsc{mesa}'s MLT++ scheme is not applied (see e.g., \citealt[][]{pon2021}). 

We employ high spatial and temporal resolution by setting \texttt{mesh\_delta\_coeff = 1.} and \texttt{time\_delta\_coeff = 1.},
in addition to setting \texttt{varcontrol\_target = 1.d-4}. This results in an average of 2000-3000 zones for our stellar structure models, and the evolutionary models consist of a hundred to a few thousand structure models (each corresponding to one time step), mostly depending on the initial mass.

We consider one initial rotation rate\footnote{The actual initial rotation rate of stars in general remains an open question. Spectroscopic studies have focused on large samples to obtain the distribution of projected rotational velocities in the Galaxy \citep[][]{howarth1997,huang2010,simon2014,simon2016,simon2017,holgado2022} and in the Magellanic Clouds \citep[][]{martayan2006,martayan2007,ramireza2013,ramireza2015,dufton2019,dufton2020}. The findings indicate a Gaussian distribution of $v \sin i$, with different peak values depending on physical (spectral type, mass, etc.) and observational characteristics (sample size, magnitude limit, etc.). The typical peaks of the distributions are around 100~km\,s$^{-1}$. Considering that this value needs to be corrected for the (usually unknown) inclination angles and that it reflects on the current rotation after a given star or population have evolved away from the ZAMS, it is generally assumed that the canonical initial rotational velocities of massive stars are of the order of 300~km\,s$^{-1}$. Although for Galactic O-stars the IACOB survey \citep{simon2014,simon2016,simon2017,holgado2022} has shown somewhat lower values than previous studies ($\approx$ 100-200~km\,s$^{-1}$), which could be consistent with lower initial rotational velocities. Our chosen input parameter for the initial rotation rate of $\Omega/\Omega_{\rm crit} = 0.5$ reflects closely on the canonical value around 300~km\,s$^{-1}$, identified in the sample of \cite{howarth1997}.} in all models by relaxing an initial ratio of $\Omega/\Omega_{\rm crit} = 0.5$. In our models with solar metallicity, this approximately corresponds to an initial equatorial surface rotational velocity (\texttt{surf\_avg\_vrot} in \textsc{mesa}) of 300 to 370 km\,s$^{-1}$ in the initial stellar mass range from 3 to 60~M$_\odot$. The critical angular velocity is adopted as defined in \textsc{mesa} \texttt{star\_utils.f90}: 
\begin{equation}\label{eq:crit1}
 \Omega_{\rm crit} = \sqrt{(1 - \Gamma) \frac{G m}{r^3}} \, , 
\end{equation}
\noindent where the Eddington parameter is $\Gamma = L_{\rm rad}/L_{\rm Edd}$, G is the gravitational constant, and $m$ and $r$ are the mass and radius taken at the photosphere. This definition only plays a role in setting the initial rotational velocity\footnote{In fact, in \textsc{mesa} the Eddington luminosity is calculated from the total opacity. Since the precise definition of $\Omega_{\rm crit}$ is significantly more complex (see \citealt[][]{puls2008} for further discussion), we stress here that $\Omega_{\rm crit}$ is only used as an input option to set the rotational velocity.}. The initial model relaxes from solid-body rotation to a new configuration constrained by angular momentum distribution and transport.

In \PaperII we found that a lower rotation rate (initially $\Omega/\Omega_{\rm crit} = 0.2$) leads to smaller differences between models with and without magnetic fields. This is simply because magnetic braking is less efficient when the rotation is slow\footnote{In \PaperI, we also demonstrated that for a typical non-rotating 15~M$_\odot$ model at solar metallicity, mass-loss quenching is modest. As shown by \cite{petit2017}, who computed non-rotating solar metallicity models between 40 and 80~M$_\odot$, the evolutionary impact of magnetic mass-loss quenching becomes significant at higher masses.}. 
In \PaperII, we showed that at least a few magnetic stars were best matched with models that had an initial rotation rate of $\Omega/\Omega_{\rm crit} = 0.8$. If the initial rotation rate was higher than considered in this study, it would i) alter the early evolution of the models on the Hertzsprung-Russell diagram (HRD, as shown in Figure~3 of \PaperII), and ii) impact the quantitative predictions regarding the surface chemical enrichment. It is worth noting that the most rapidly rotating (presumably young) magnetic stars  have present-day surface rotational velocities of about 300 km~s$^{-1}$, which is close to  50\% of the critical rotation defined in our Equation~\ref{eq:crit1} \citep[e.g.,][]{oksala2010,grunhut2012,shultz2019b,song2021}. Nevertheless it remains unknown what initial rotation rates could characterise the entire sample of magnetic massive stars. 

Finally, given the supporting observational evidence by some studies \citep[][]{kochukhov2006,wickramasinghe2005,neiner2017,martin2018,sikora2019}, we assume magnetic flux conservation (Alfv\'en's theorem, \citealt[][]{alfven1942}) such that the surface magnetic field strength is obtained from:
\begin{equation}\label{eq:fieldevol}
    B_{\rm eq} =  B_{\rm eq, ini} \left(\frac{R_{\rm \star, ini}}{R_{\star}} \right)^{2}
\end{equation}
with $B_{\rm eq, ini}$ being the initial surface equatorial magnetic field strength (which is assumed in the models at the ZAMS), while $R_{\rm \star, ini}$ and $R_\star$ are the initial and current stellar radii. For further discussions on magnetic field evolution, we refer the reader to \PaperI, \PaperII, \PaperIII and references therein. We adopt a large range of initial equatorial magnetic field strengths (from 0 to 50 kG; \citealt[e.g.,][]{donati2009,shultz2019b}.)

The \texttt{run\_star\_extras} file (available as part of a full reproduction package on Zenodo at \url{https://doi.org/10.5281/zenodo.7069766}) is used to modify the wind, torque, angular momentum transport, and chemical element transport routines. The parameters varied in the models are summarised in Tables~\ref{tab:t1}, \ref{tab:met}, and \ref{tab:t2}, and are discussed below.

%
%
%
\begin{table*}
\caption{Comprehensive grid of 8,748 \textsc{mesa} evolutionary models including the effects of surface fossil magnetic fields, covering a large parameter space in initial mass and magnetic field strength (see below) for 3 metallicities (Table \ref{tab:met}) and for 2 braking schemes and 2 mixing schemes (Table \ref{tab:t2}).}
\begin{tabular}{l|l}   
\hline \hline
$M_{\rm \star, ini}$ [M$_\odot$] & 3, 4, 5, 6, 7, 8, 9, 10, 11, 12, 13, 14, 15, 16, 17, 18, 19, 20, 21, 22, 23, 24, 25, 30, 40, 50, 60 \\
\hline
$B_{\rm eq, ini}$ [kG] & 0, 0.25, 0.50, 0.75, 1.0, 1.5, 2.0, 2.5, 3.0, 3.5, 4.0, 4.5, 5.0, 5.5, 6.0, 6.5, 7.0, 7.5, 8.0, 8.5, 9.0, 9.5, 10, 15, 20, 30, 50  \\ 
\hline
\end{tabular} \label{tab:t1}
\end{table*}

%
%
%
%

%
%
%
\begin{table*}
\caption{Key initial elemental abundances (in mass fractions and compared to the adopted solar value) included in our models.}
\begin{tabular}{ccccccccccccc}   
\hline \hline
 &X$_{\rm ini}$ & $\frac{\mathrm{X_{ini}}}{\mathrm{X_{\odot, ini}}}$  & 
 Y$_{\rm ini}$ & $\frac{\mathrm{Y_{ini}}}{\mathrm{Y_{\odot, ini}}}$ & Z$_{\rm ini}$  & $\frac{\mathrm{Z_{ini}}}{\mathrm{Z_{\odot, ini}}}$ &  C$_{\rm ini}$ & $\frac{\mathrm{C_{ini}}}{\mathrm{C_{\odot, ini}}}$ &  N$_{\rm ini}$ & $\frac{\mathrm{N_{ini}}}{\mathrm{N_{\odot, ini}}}$ & O$_{\rm ini}$ &  $\frac{\mathrm{O_{ini}}}{\mathrm{O_{\odot, ini}}}$ \\
\hline \hline
Solar & 0.72000 & 1 & 0.26600 & 1 &  1.40000 $\cdot$10$^{-2}$ & 1 & 1.84720  $\cdot$10$^{-3}$  & 1&  6.21528  $\cdot$10$^{-4}$  & 1&  6.62907   $\cdot$10$^{-3}$ & 1 \\
\hline
LMC & 0.73685 & 1.02 & 0.25671 & 0.97 &  6.43605 $\cdot$10$^{-3}$ & 0.46 &  9.25898 $\cdot$10$^{-4}$ & 0.50 &  1.45717 $\cdot$10$^{-4}$ & 0.24 &  2.96143 $\cdot$10$^{-3}$ & 0.45\\
\hline
SMC & 0.74840 & 1.04 &  0.24900 & 0.94 &  2.60758 $\cdot$10$^{-3}$ & 0.19 & 2.15433  $\cdot$10$^{-4}$ & 0.12 & 6.76488 $\cdot$10$^{-5}$ & 0.11 &  1.34354  $\cdot$10$^{-3}$ & 0.20 \\
\hline
\end{tabular} \label{tab:met}
\end{table*}

\subsection{Metallicity}

We compute models for 3 different metallicities (the key elements are summarised in Table~\ref{tab:met} -- the full list of abundances is available via the model files shared on Zenodo) that are representative of the Solar neighbourhood, LMC, and SMC. For the Solar composition, we assume the hydrogen and helium mass fractions along with a metallicity of $Z=0.014$ \citep{asplund2009}. The metal fractions are adopted following the works of \cite{prz2008,prz2013} and \cite{nieva2012}, which updated some elements compared to the \cite{asplund2005,asplund2009} abundances, considering B-type stars in the solar neighbourhood. The baseline values for the Magellanic Clouds are still subject to ongoing investigations \citep[e.g.,][]{dufton2020,bouret2021}.
For the helium and metal abundances in the LMC and SMC, we adopt the mean values listed in Tables 5 and 6 of \cite{dopita2019}. These mean values are a result of 9 separate investigations using different approaches, which include atmospheric determinations of hot stars, supernova remnants, and H~\,\textsc{ii} regions. 
In all 3 metallicities, we use the \cite{lodders2003} isotopic ratios. The metallicity adopted for the chemical composition is fully consistent with the metallicity used for the opacity tables.

\subsection{Alfv\'en radius}\label{sec:alf}

The Alfv\'en radius characterises a critical distance at which the magnetic energy density and the gas kinetic energy density are equal. Alternatively, it can also be cast as the inverse square of the Alfv\'enic Mach number. Its definition plays an important role in both mass-loss quenching (Equations \ref{eq:fb1}-\ref{eq:fb2}) and magnetic braking (Equations \ref{eq:br}-\ref{eq:br2}). For a dipolar field configuration, \cite{ud2009} use a numerical fitting for a quartic equation to obtain: 
\begin{equation}\label{eq:alf1}
    \frac{R_{\rm A}}{R_\star} \approx 1 + (\eta_\star + 0.25)^{1/4} - (0.25)^{1/4} \, ,
\end{equation}
with $\eta_\star$ the equatorial magnetic confinement parameter, defined as: 
\begin{equation}
    \eta_\star = \frac{B_{\rm eq}^2 R_\star^2}{\dot{M}_{B=0} v_\infty} \, ,
\end{equation} 
\noindent with $B_{\rm eq}$ the equatorial magnetic field strength, $\dot{M}_{B=0}$ the mass-loss rate in absence of a magnetic field, and $v_\infty$ the terminal velocity\footnote{Also calculated for in absence of a magnetic field.} \citep[][]{ud2009}. Observations typically reconstruct a polar field strength $B_{\rm p}$ from the line-of-sight disc-integrated (so called longitudinal) magnetic field strength \citep{donati2009}. The equatorial field strength is exactly one half of the polar field strength. 
We use Equation~\ref{eq:alf1} to obtain the Alfv\'en radius in the INT models, which we assume to be characterised by a predominantly dipolar field configuration (Figure \ref{fig:intsurf}). In the SURF models we assume the field to be more complex, in which case the definition of the Alfv\'en radius is non trivial. For the sake of simplicity we assume that the Alfv\'en radius  takes the form of a scaling appropriate for a quadrupole field geometry, such that:  
\begin{equation}\label{eq:alf2}
    \frac{R_{\rm A}}{R_\star} \approx 1 + (\eta_\star + 0.25)^{1/6} - (0.25)^{1/6} \, ,
\end{equation}
following the parametrisation in Equation 9 of \cite{ud2008}. This ensures that for a given field strength $R_{\rm A}$ is less in the SURF case than in the INT case, leading to less efficient magnetic braking.

\subsection{Stellar winds}\label{sec:mdot}

\subsubsection{Mass-loss schemes and terminal velocities}

The models include mass loss. Even though this is modest for the lower-mass stars (and the driving mechanism is not unambiguously identified as for more massive stars), it can impact their rotational evolution given the longer nuclear timescale. For this reason, we apply commonly used mass-loss rates of hot massive stars also to lower mass main sequence stars in our grid. While the higher-mass stars typically reach the TAMS at $T_{\rm eff} > 20$~kK, we describe here the detailed treatment implemented in our \textsc{mesa} extension for completeness and to aid further studies focusing on complementing this work with post-main sequence models.
For massive stars with $T_{\rm eff} > 10$~kK, the mass loss is powered by radiative line driving \citep[e.g.,][]{lucy1970,cak1975,puls2008}. In this regime, we apply the rates derived by \cite{vink2000,vink2001}, decreased by a factor of 2 for all models in the 3 - 60 M$_\odot$ range for consistency. The choice to reduce the nominal mass-loss rates is motivated by the growing evidence both from observations, suggesting that mass-loss rates are lower when accounting for wind clumping \citep[e.g.,][]{bouret2005,fullerton2006, trundle2005,dealmeida2019,brands2022}, and from new modelling approaches \citep[e.g.,][]{muijres2012,krticka2014,krticka2017,krticka2021,sundqvist2019,bjorklund2021}. When using these rates, we apply metallicity-dependent winds with a scaling of $\dot{M} \sim Z^{0.85}$ \citep[e.g.,][]{vink2001,mokiem2007}.

Similarly to \cite{keszthelyi2017b} and \PaperII, we implement the partitioning in effective temperature related to the bi-stability jump at 20 kK in agreement with observational and new theoretical works \citep{prinja1990,prinja1998,lamers1995,petrov2016}, rather than adopting it at 25-27 kK as in evolutionary models of e.g., \citet{brott2011}, \citet{ekstroem2012}, and \citet{choi2016}. This is further supported by measurements of projected rotational velocities that suggest a lack of bi-stability braking \citep{crowther2006,vink2010,keszthelyi2017b,gagnier2019a,gagnier2019b,krticka2021,vink2021} at least until about 20 kK \citep[][]{howarth1997,huang2010}.
Although we adopt an increase in mass-loss rates at the bi-stability jump, we note that this prediction still lacks empirical evidence in typical B-type supergiants \citep{crowther2006,markova2008,rubio2022} and is challenged by new numerical simulations \citep[][]{sundqvist2019,driessen2019,bjorklund2021}.

Below approximately 10 kK the nature of wind-driving remains poorly understood. We opt to use the rates of \cite{vanloon2005} for all models in this domain, which
only concerns a few lower-mass models in the present grid.
New modelling approaches have confirmed that the \textit{second} bi-stability jump due to Fe\,\textsc{iii} recombining to Fe\,\textsc{ii} is expected at $T_{\rm eff} \sim $~9\,kK \citep{petrov2016} in contrast with earlier indications of $\sim$~12.5\,kK \citep[][]{vink1999,vink2000}, and implementations in evolutionary models of $\sim$~17-15\,kK \citep[][]{ekstroem2012,brott2011}. Therefore we avoid the use of the second bi-stability jump that is typically included in other grids of models (for further details see, e.g., Figure~3 of \citealt[][]{keszthelyi2017b}). 
If the effective temperature is higher than 10 kK and the surface hydrogen mass fraction becomes less than 0.4, we apply the Wolf-Rayet rates of \cite{nugis2002}. This concerns some of our most massive models with efficient mixing.

In agreement with the partitionings in effective temperature, we estimate the terminal wind velocity $v_{\infty}$ via:
\begin{equation}\label{eq:eq1}
v_{\infty} = f_{\infty} \cdot \, v_{\rm esc} = f_{\infty} \, \sqrt{\frac{2 G M_\star }{R_\star} \left( 1 - \Gamma_e \right) } \, , 
\end{equation}
\noindent where $G$, $M_\star$, $R_\star$, and $\Gamma_e$ are respectively the gravitational constant, the stellar mass, the stellar radius, and the Eddington parameter for pure electron scattering. The terminal wind velocity is obtained from the escape velocity as a simple step function by adopting $f_{\infty} = 2.6, 1.3, 0.7$ at $T_{\rm eff} > 20$~kK, $ 20\, \mathrm{kK} > T_{\rm eff} > 10$~kK, and $T_{\rm eff} < 10$~kK, respectively \citep[][]{lamers1995,vink2000,kudritzki2000}. The typical terminal velocities at the ZAMS range from 800 to 3000 km\,s$^{-1}$ for models with initial masses from 3 to 60~M$_\odot$, respectively.
We calculate the rotational enhancement on the mass-loss rates\footnote{See also the recent study of \cite{brink2021}.} as described by \cite{maeder2000}. This requires defining the difference of the force multiplier parameters $\alpha'$ ($=\alpha - \delta$, that is, the exponent related to the line-strength distribution function minus the exponent quantifying the change in ionisation balance), which we adopt as a simple step function with values of 0.6, 0.5, 0.4, corresponding to the above-mentioned effective temperature ranges \citep[see][]{pauldrach1986,lamers1995,puls2000}.
The alternative calculation of rotational enhancement built into \textsc{mesa} is not used (see \PaperII Section 3.9 for details). 

%
%
\subsubsection{Magnetic mass-loss quenching}

The overall field configuration that extends into the wind outflow is governed by the competition between the kinetic energy of the wind and the magnetic energy of the field. The ionised stellar wind material is forced to flow along magnetic field lines. However, as the wind kinetic energy density has a shallower decline than the magnetic energy density, the field loops can only confine wind material up to a certain radius. Within closed field loops, material becomes trapped and eventually falls back onto the surface (unless centrifugally supported).
To account for the global, time-averaged effect of the magnetosphere, the mass-loss rates are systematically reduced. Following the works of \cite{ud2008,ud2009}, the mass-loss quenching parameter $f_{\rm B}$ is defined as:
\begin{equation}\label{eq:fb1}
f_{\rm B} = \, \frac{\dot{M}}{\dot{M}_{B=0}} = \, 1 - \sqrt{1 - \frac{1}{R_{\rm c}}}  \quad \mathrm{if} \quad R_{\rm A} < R_{\rm K} 
\end{equation}
and  
\begin{equation}\label{eq:fb2}
f_{\rm B} = \, \frac{\dot{M}}{\dot{M}_{B=0}} \, = 2 - \sqrt{1 - \frac{1}{R_{\rm c}}} - \sqrt{1 - \frac{0.5}{R_{\rm K}}}    \quad \mathrm{if} \quad R_{\rm K}  < R_{\rm A} 
\end{equation}
\noindent where $R_{\rm A} $, $R_{\rm K} $, and $R_{\rm c} $ are the Alfv\'en radius, the Kepler co-rotation radius, and the closure radius in units of the stellar radius, respectively (see \citealt{petit2017}, \PaperI, \PaperII, \PaperIII, and references therein). 
The closure radius, defining the distance from the stellar surface to the last closed magnetic loop, is approximated as $R_{\rm c}~\sim~ R_\star~+~0.7~(R_{\rm A}~-~R_\star~)$, see \citet{ud2008}.
$\dot{M}$ is the mass-loss rate that a non-rotating magnetic star would have. $\dot{M}$ is further scaled by the rotational enhancement $f_{\rm rot}$ (specified in Section 3.9 of \PaperII) such that the effective mass-loss rate is obtained from $\dot{M}_{\rm eff} = f_{\rm B} \cdot f_{\rm rot} \cdot \dot{M}_{B=0}$. The magnetic mass-loss quenching parameter (equivalent to the escaping wind fraction\footnote{Calculated for different R$_{\rm A}$ in the INT and SURF schemes; however, see Appendix~\ref{sec:quad} for an actual quadrupole geometry.}) can take values between 0 and 1, depending on the magnetic field strength. A strong magnetic field (with a strength of tens of kG) may lead to only a few percent of the wind material actually escaping the star (\citealt[][]{petit2017,georgy2017}, \PaperI).
Let us also note that the conditions in the above equations are equivalent to distinguishing between dynamical magnetospheres (if $R_{\rm A}  < R_{\rm K}  $) and centrifugal magnetospheres (if $R_{\rm K}  < R_{\rm A} $), a classification introduced by \cite{petit2013}. 

The use of Equation~\ref{eq:fb2} is a refinement compared to previous implementations. For situations when the Alfv\'en radius is larger than the Kepler co-rotation radius (centrifugal magnetospheres), the magnetosphere is expected to be less efficient at quenching wind mass-loss compared to dynamical magnetospheres \citep{ud2008,ud2009}. This is because material injected by the wind into the centrifugal magnetosphere is not returned to the stellar surface by gravity, but is instead ejected away from the star once the critical centrifugal breakout density is exceeded \citep{shultz2020,owocki2020}. This can lead to substantially larger values of $f_{\rm B}$ for a rotating as compared to a non-rotating star, however in practice rapid spin-down means that centrifugal magnetospheres are relatively short-lived and the incorporation of this modification to the mass-quenching prescription does not have a strong effect on evolution. It is generally expected that the evolution proceeds from centrifugal to dynamical magnetospheres (\citealt[][]{shultz2019b}, \PaperI, \PaperII).

In this approach, the magnetic mass-loss quenching parameter is an average quantity. MHD simulations of non-rotating magnetospheres predict up- and down-flows of material varying on short dynamical timescales \citep{ud2002}, which can manifest as stochastic variability in magnetospheric emission lines \citep{ud2013}, however time-averaged models provide a good reproduction of emission line properties \citep[e.g.][]{2012MNRAS.423L..21S,owocki2016,erba2021} and these short-term, stochastic variations can therefore be confidently neglected over evolutionary timescales. In the case of rapid rotators, 2D MHD simulations led to the expectation that breakout events would be similarly stochastic, leading to emptying of the centrifugal magnetosphere and large-scale magnetospheric reorganisation \citep{ud2006,ud2008}. However, no indication of large-scale changes has been observed \citep{townsend2013, shultz2020}, leading \cite{shultz2020} and \cite{owocki2020} to infer that breakout events are characterised by small spatial scales and occur more or less continuously, such that the centrifugal magnetosphere is maintained nearly continuously at the breakout density. As a result, it is therefore appropriate to treat magnetospheric mass-drainage via breakout as an effectively continuous process over evolutionary timescales and apply Equations \ref{eq:fb1} and \ref{eq:fb2}.

%
%
\begin{table}
\caption{Magnetic braking and chemical mixing schemes}
\centering
\begin{tabular}{ll}   
\hline \hline
\textbf{INT}:& solid-body rotation, braking the rotation of the entire star \\
\textbf{SURF}:& differential rotation, braking the rotation of the surface \\
\textbf{Mix1}:& $D_{\rm chem} $ defined by Equation~\ref{eq:chem1} with $f_{\rm c } = 0.033, f_{\mu} = 0.1$ \\ 
\textbf{Mix2}:& $D_{\rm chem} $ defined by Equation~\ref{eq:chem2} with $f_{\rm c } = 1, f_{\mu} = 1$ \\ 
\hline
\end{tabular} \label{tab:t2}
\end{table}

\subsection{Angular momentum transport and loss}\label{sec:rot}

Magnetic fields are much more efficient at transporting angular momentum than purely hydrodynamic processes such as meridional currents and shear instabilities \citep[e.g.,][]{mestel1999,spruit1999,spruit2002,kulsrud2005,braithwaite2017}. The rotation of the star leads to Maxwell stresses, which result in losing angular momentum from the star.

In \PaperI, \textsc{genec} models were used, where magnetic braking is adopted as a boundary condition to internal angular momentum transport, directly affecting the uppermost layer of the stellar models. In \PaperII, two kinds of models were introduced to account for the uncertainty regarding how deeply fossil magnetic fields are anchored in massive stars. In the INT models, magnetic braking was applied to the entire star, decreasing uniformly the specific angular momentum in all layers. In the SURF models, magnetic braking was set to remove specific angular momentum from a very near-surface reservoir. 
In \PaperIII, \textsc{genec} models were contrasted with a \textsc{mesa} implementation where magnetic braking was applied to most of the stellar envelope. 
In \textsc{genec}, two configurations were used to model internal angular momentum transport: one with only hydrodynamic instabilities correctly accounted for via an advecto-diffusive equation (allowing for shears to develop in deeper layers), and one with a purely diffusive equation, in which solid-body rotation was established. In \textsc{mesa}, we relied on the nominal hydrodynamic transport processes since they are used in a purely diffusive assumption, leading to nearly solid-body rotation on the magnetic braking timescale. 
Here, we make some further refinements and adjustments compared to these approaches, particularly accounting (indirectly) for the field geometry as depicted in Figure~\ref{fig:intsurf}.

\subsubsection{Magnetic braking}\label{sec:magbraking}

Stellar rotation bends and twists magnetic field lines in the azimuthal direction. Magnetic field lines can transport and store angular momentum, and the associated Maxwell stresses are very efficient at transferring angular momentum to the surrounding plasma. Once the angular momentum is imparted from the field to the gas, the wind material carries it away, leading to a spin-down of the star. This process is commonly referred to as (wind) magnetic braking.

In a pioneering series of works, analytical and numerical MHD simulations were developed, confirming that the \citet{weber1967} model (see also, \citealt{parker1958,mestel1968}) leads to an appropriate scaling relation also for massive stars \citep{ud2002,ud2008,ud2009,owocki2004,townsend2005}. Following the work of \cite{ud2009}, the total -- wind and magnetic field induced -- loss of angular momentum can be expressed via:
\begin{equation}\label{eq:br}
\frac{\mathrm{d}J_{\rm B}}{\mathrm{d}t}  = \frac{2}{3} \dot{M}_{B=0} \,  \Omega_\star R_{A}^2 \, ,
\end{equation}
\noindent with $\mathrm{d}J_{\rm B}/\mathrm{d}t$ the rate of angular momentum loss from the system, $\Omega_\star$ the surface angular velocity, and $R_{A}$ the Alfv\'en radius (defined in Equations~\ref{eq:alf1} and \ref{eq:alf2}). As this equation accounts for the gas and field driven angular momentum loss \citep{ud2009}, it yields the angular momentum loss resulting purely from mass loss when $B_{\rm surf} = 0$. As specified in \PaperII, we have adjusted the angular momentum lost via mass loss to avoid double counting. In Equation~\ref{eq:br}, the numerical term 2/3 arises from integrating over latitudes. We note that this equation is not applicable when the effective mass-loss rate, as introduced above, is exactly zero (this situation does not happen in our models). In the strong confinement limit, when $f_{\rm B} \rightarrow 0$, the effective mass-loss rate can become very small. In this case, a strong magnetic braking can still be achieved since the Maxwell stresses driving the angular momentum transport are independent of the plasma flow. As long as there is wind material at a radial distance larger than the last closed magnetic field line, i.e., the star is not surrounded by vacuum, the field can impart angular momentum to the plasma.
In \PaperII and \PaperIII, Equation \ref{eq:br} was implemented into \textsc{mesa} via changing the specific angular momentum in given layers of the star, such that a summation over mass yields the total rate of angular momentum loss as defined in Equation \ref{eq:br}. It is coded as: 
\begin{equation}\label{eq:br2}
    \frac{\mathrm{d}J_{\rm B}}{\mathrm{d}t} = \sum_{k=1}^{k=x} \frac{\mathrm{d j_{\rm B}}}{\mathrm{d} t} = - \frac{\mathrm{d} J_{\rm B}}{J_{\rm INT/SURF}} \sum_{k=1}^{k=x}  \frac{\mathrm{d}j}{\mathrm{d}t} \, , 
\end{equation}
\noindent where $\mathrm{d}j_{\rm B}/\mathrm{d} t$ is the rate of specific angular momentum change (dubbed as "\texttt{extra\_jdot}" in \textsc{mesa}). The negative sign is added to reduce the reservoir (i.e., to account for loss), $\mathrm{d} J_{\rm B} = (\mathrm{d}J_{\rm B}/ \mathrm{d}t) \cdot \mathrm{d}t$ is the total angular momentum lost per time $\mathrm{d}t$, $J_{\rm INT/SURF}$ is the angular momentum reservoir of the entire star (INT) or of defined layers in the stellar envelope (SURF; see Figure~\ref{fig:intsurf} and discussion below), $j$ is the specific angular momentum of a layer (called "\texttt{j\_rot}" in \textsc{mesa}), d$t$ is one timestep in the computation\footnote{We use a timestep control, specified in \PaperII, which prevents the star model from fully exhausting specific angular momentum in any layer.}, $k$ is an index running through all layers, and $x$ is the index of the last layer where magnetic braking is applied. Therefore, Equation~\ref{eq:br2} indicates how to distribute the total angular momentum lost per unit time ($\mathrm{d}J_{\rm B}/\mathrm{d}t$ given by Equation~\ref{eq:br}) in given stellar layers. Taking the sum of the specific angular momentum lost per unit time $\mathrm{d}j_{\rm B}/\mathrm{d}t$ with respect to mass, we recover the left-hand side term. 

To distribute the total angular momentum lost per unit time, the summation goes over the layers of the entire star in the INT case ($x \approx 3000$ zones), whereas in the SURF case it goes from the photosphere to a lower boundary. This boundary is always in the radiative stellar envelope of our models. However, more massive models have larger convective cores, and thus for very massive stars ($> 60$~M$_\odot$), this condition may need to be revised as we do not expect the fossil field to be able to penetrate into the convective core. In the SURF models, $x \approx 200$ zones undergo magnetic braking. Here, we chose the boundary layer where $q=m/M_\star=0.8$ (the enclosed mass is 80\% of the total mass) since \cite{braithwaite2008} demonstrated that complex, non-axisymmetric fields (around the magnetic axis) can form if the magnetic flux is initially not centrally concentrated, leading to a stable magnetic field configuration in which twisted magnetic field lines spread throughout the stellar surface layers. In the simulations of \cite{braithwaite2008}, a strong toroidal field (enclosed by poloidal field lines) is present in approximately 20\% of the upper mass fraction and this motivates our choice for this parameter. 


\subsubsection{Angular momentum transport}\label{sec:amtr}

The main impact of the angular momentum transport equation in stellar interiors is to change the angular velocity profile $\Omega(r)$, which is also measurable via modern asteroseismology (see, e.g., \citealt{aerts2019} for a comprehensive review). In \textsc{mesa}, angular momentum transport is modelled in a fully diffusive scheme. Note that this approach inadequately models the meridional currents\footnote{Meridional currents are large-scale flows arising from the thermal imbalance between the polar axis and the equatorial regions in a rotating star.}, which are an advective process by nature. \textsc{mesa} solves the angular momentum transport equation following Equation 46 of \cite{heger2000}, which is based on the works by \cite{endal1978} and \cite{pin1989}, that is:
\begin{equation}\label{eq:mam2}
 \frac{ \partial \Omega  }{ \partial t}    =  \frac{\partial}{\partial m} \left[(4 \pi r^2 \rho)^2  \,  D_{\rm AM} \,  \frac{\partial \Omega}{\partial m}  \right] \, ,
\end{equation}
where $D_{\rm AM}$ is the total diffusion coefficient responsible for angular momentum transport, while $r$, $\rho$, and $m$ are the radius, density, and enclosed mass, respectively, and $t$ is the time.

%
%
In the non-magnetic models, we assume that $D_{\rm AM}$ is constructed as a sum of four diffusion coefficients (resulting from dynamical and secular shear, meridional circulation, and GSF instability), which are the same as used for the Mix1 chemical mixing scheme in Equation~\ref{eq:chem1}; however, not scaled by any efficiency parameters for angular momentum transport. For simplicity and a consistent treatment of angular momentum transport, we also use these diffusion coefficients for angular momentum transport when a different chemical mixing scheme is adopted (Mix2, see below).

%
%
In the INT models (see also Figure~\ref{fig:intsurf}), we assume that the magnetic field is capable of establishing radially uniform (solid-body) rotation throughout the entire star. This is representative of an axisymmetric magnetic field that "freezes" rotation along the poloidal field lines following Ferraro's theorem \citep{ferraro1937}. We model this by using the \textsc{mesa} controls  \texttt{set\_uniform\_am\_nu\_non\_rot = .true.}  and
\texttt{uniform\_am\_nu\_non\_rot = 1.d16} such that a high diffusivity ($D_{\rm AM} = 10^{16}$ cm$^2$ s$^{-1}$) leads to efficient angular momentum transport and hence solid-body rotation throughout the entire star. 
Unfortunately, the naming conventions here are somewhat confusing as these controls are applied to the entire star regardless of the convective/radiative nature of given layers. Otherwise "am\_nu\_non\_rot" refers to layers of the star with convective mixing.
The precise value of this quantity is not crucial so long as it achieves solid-body rotation. Above a critical value, the diffusivity can saturate, meaning that an already flat $\Omega$ profile will remain unchanged if an even higher diffusivity is applied. The \textsc{mesa} "default" value for this control is $D_{\rm AM} = 10^{20}$ cm$^2$ s$^{-1}$. Such a high diffusivity would mean a diffusion timescale ($\tau_{\rm D} \approx r^2 / D_{\rm AM} $) of a few hours, which is physically not justified. The saturation, i.e., solid-body rotation for a given diffusivity may happen for diffusivities $>10^{10}$ cm$^2$ s$^{-1}$, depending on model specifics such as mass and evolutionary stage. 

%
%
In the SURF models, we distinguish between three regions of the star i) the stellar core, ii) the envelope from $q=0.8$ to the stellar core, and iii) the envelope above $q = 0.8$ in which magnetic braking is applied (see above). On the main sequence, the cores of massive stars are convective, dominated by strong turbulent mixing. In \textsc{mesa}, this is modelled by a high diffusion coefficient (relying on mixing-length theory) that establishes a constant angular velocity profile, that is, the core is rigidly rotating. 
In the radiative layers between the stellar core and the boundary of $q = 0.8$, the usual hydrodynamical instabilities (dynamical and secular shear, meridional circulation, GSF instability) transport angular momentum. More directly, the assumption here is that there is no magnetic coupling between the stellar core and the envelope. While even for complex surface fields there may be weak dipole components in the deep stellar layers which may contribute to angular momentum transport, we neglect those here to be able to test a limiting, boundary case, in which differential rotation may develop between the core and the surface. For a consistent comparison, in both Mix1/Mix2 chemical mixing schemes (see below), we apply the same treatment of angular momentum transport in this region.

The fossil magnetic field may relax into a non-axisymmetric configuration, strongly impacting the upper stellar layers \citep{braithwaite2008}. In these layers (with 20 per cent of the stellar mass in our models), we apply a high diffusion coefficient of $D_{\rm AM} = 10^{16}$~cm$^2$~s$^{-1}$ via the \texttt{other\_am\_mixing} subroutine to account for the expected effect of the magnetic field. 

In both INT and SURF cases, for layers with increased angular momentum transport attributed to the magnetic field, the angular momentum transport equations are of secondary importance in the sense that we expect an appropriate transport equation to result in a flat angular velocity profile, thereby deviating from non-magnetic models. One would also expect that in those layers where the fossil magnetic field is present, hydrodynamical instabilities could not transport angular momentum.

Further guidance regarding the internal rotation profile and magnetic field properties can also be obtained observationally using (magneto-)asteroseismology (see recently \citealt{lecoanet2022}). For instance, radial differential rotation was observed in several massive stars using the rotational splitting of gravity modes (e.g. \citealt{aerts2003, triana2015}). On the other hand, the nearly identical surface and core rotation of red giant stars requires very efficient transport \citep[e.g.,][]{moyano2022}. Using asteroseismic analysis of Kepler data, it has indeed been attributed to magnetic fields \citep{fuller15}. Due to possible mode suppression by strong magnetic fields, magnetoasteroseismology remains an elusive target, having been performed for only a few massive stars (e.g. HD 43317 and V2052 Oph; \citealt{briquet2012,bram2018}). However, the advent of nearly all-sky high-precision space-based photometry can help further this line of inquiry, with large asteroseismic target lists of OB stars already being assembled (e.g. \citealt{burssens2020}).

\subsection{Rotational mixing of chemical elements}

Following \cite{pin1989}, rotational mixing of chemical elements is commonly applied via the diffusion equation in one-dimensional stellar evolution models:
\begin{equation}\label{eq:diff}
 \frac{ \partial X_i }{ \partial t}    = \frac{\partial}{\partial m} \left[ (4 \pi r^2 \rho)^2  \, D_{\rm chem} \,  \frac{\partial X}{\partial m}   \right] + \left( \frac{\mathrm{d} X_i}{\mathrm{d} t} \right)_{\rm nuc}  \, , 
\end{equation}
\noindent where $X_i$ is the mass fraction of a given element $i$, $t$ is the time, $m$ and $\rho$ are the mass coordinate and mean density at a given radius $r$, $D_{\rm chem}$ is the sum of individual diffusion coefficients contributing to chemical mixing (see also \citealt[][]{salaris2017}), and the last term accounts for nuclear burning. 

In this approach, the central question is how to encapsulate inherently three-dimensional physical processes and apply them via a single parameter $D_{\rm chem}$. In this study, we contrast two commonly used approaches. Spectroscopic studies of massive stars often find discrepancies between the observed and predicted surface abundances from rotating stellar evolution models computed with a given scheme of chemical mixing \citep[e.g.,][]{trundle2004,martins2017,markova2018}. 
Recent works suggest that such discrepancies may be resolved by including additional processes in the calculations, for example, internal gravity waves (and magnetic fields) lead to a more complex physical interplay between various processes and a variety of mixing profiles \citep{aerts2019,bowman2020,michielsen2021,pedersen2021}.

\subsubsection{Basic thermodynamic quantities}

Before introducing the diffusion coefficients, we briefly outline the most important thermodynamic quantities that enter into those equations.
The thermal diffusivity is defined as:
\begin{equation}\label{eq:k} 
K = \frac{4ac}{3\kappa} \frac{T^4 \nabla_{\rm ad}}{\rho P \delta} = \frac{4acT^3}{3 \kappa \rho^2 c_{P}} \, , 
\end{equation}
where $a$ is the radiation constant, $c$ the speed of light, $\kappa$ the mean radiative opacity, $c_P$ the specific heat capacity per unit mass at constant pressure, $T$ the temperature, $\rho$ the density, and $P$ the pressure. The different $\nabla$-s below denote the adiabatic, radiative, and chemical composition ($\mu$) gradients: %
\begin{equation}\label{eq:nab} 
\begin{aligned}
\nabla_{\rm ad}  & = \left( \frac{\partial \ln T}{\partial \ln P} \right)_{S, \mu}  = \frac{P \delta}{T \rho c_{P}}\\ 
\nabla_{\rm rad} & =  \left( \frac{\partial \ln T}{\partial \ln P} \right)_{\rm rad} = \frac{3}{16 a c G} \frac{ \kappa L P}{ m T^{4} } \\
\nabla_{\rm \mu} & =  \frac{\partial \ln \mu}{\partial \ln P} \, , 
\end{aligned}
\end{equation}
where $S$ is the entropy, $\mu$ the mean molecular weight, and $G$ the gravitational constant. The local luminosity $L$ is the rate of energy transported outward through a sphere of radius $r$, and $m$ is the enclosed mass. From Equation 4.22 of \cite{maeder1998}, the derivatives from the equation of state are:
%
\begin{equation}\label{eq:eos} 
\begin{aligned}
\delta &= - (\partial \ln \rho / \partial \ln T )_{P, \mu} \\ 
\phi &= (\partial \ln \rho / \partial \ln \mu )_{P,T} \, .
\end{aligned}
\end{equation}

\subsubsection{Mix1 scheme}

A commonly used scheme of rotational mixing in stellar evolution models was developed by \cite{kippenhahn1974}, \cite{endal1978}, and \cite{pin1989} and applied subsequently by several authors. This scheme (the "default" \textsc{mesa} scheme, "Mix1" hereafter) is typically used in \textsc{mesa} models \citep[e.g.,][]{paxton2013,choi2016}. $D_{\rm chem}$ is constructed as the sum of 6 individual diffusion coefficients, describing dynamical shear instability (DS), Solberg-H\o iland instability (SH), secular shear instability (SS), Goldreich-Schubert-Fricke instability (GSF), Eddington-Sweet circulation (ES), and Tayler-Spruit dynamo (ST) (see \citealt{endal1978,pin1989,eddington1925,sweet1950,solberg1936,hoiland1941,goldreich1967,fricke1968,tayler73,spruit2002}). 

To be able to compare our results to common model grids, in the Mix1 scheme we adopt the diffusion coefficient applied in Equation~\ref{eq:diff} as: 
\begin{equation}\label{eq:chem1}
    D_{\rm chem}^{\rm Mix1} = f_{\rm c} (D_{\rm ES} + D_{\rm SS} + D_{\rm DS} + D_{\rm GSF})
\end{equation}
\noindent where the individual diffusion coefficients are described according to \cite{heger2000}. Transport by dynamo mechanisms and by the Solberg-H\o iland instability are not considered as their contribution to chemical mixing has been debated \citep[e.g.,][]{yoon2006,brott2011}. Of particular interest is the meridional circulation term, which was described via the circulation velocity in the radial direction by \cite{kippenhahn1974} and constructed into a diffusion coefficient by \cite{endal1978}. However, the base formulation of the problem in terms of a steady-state circulation by \cite{vogt1925}, \cite{eddington1925} and \cite{sweet1950} has been disputed by, e.g., \cite{busse1981,busse1982,zahn92} -- see further discussion by \cite{rieutord2006}.

The simple summation of the various processes by \cite{heger2000,heger2005} is often criticised on theoretical grounds as the various processes are not independent of one another \citep[e.g., recently][and references therein]{chang2021}. For example, the dynamical and secular shears act on different timescales, and therefore their mutual use is physically contradictory. \cite{maeder2013} proposed a diffusion coefficient accounting for the interactions between the different physical processes. While several studies have scrutinised these instabilities and resulting diffusion coefficients \citep[e.g.,][and references therein]{caleo2016,goldstein2019,barker2019,barker2020,chang2021,park2021}, a unified description of instabilities in rotating stars is still not fully complete. 

In fully diffusive approaches as described above, two arbitrary scaling factors $f_{\rm c}$ and $f_{\mu}$, introduced by \cite{pin1989}, are commonly adopted.
%
If chemical gradients $\nabla_{\mu}$ (Equation~\ref{eq:nab}) develop, they may inhibit the efficiency of mixing. This is primarily due to $\nabla_\mu$ serving as a stability criterion for the development of rotational instabilities \citep[see e.g.,][]{maeder1997} since it appears directly in several of the individual diffusion coefficients used in Equation \ref{eq:chem1}. To alter the effect of chemical gradients on mixing, the scaling factor $f_{\mu}$ is introduced such that $\nabla_{\mu}$ is replaced by $f_{\rm \mu} \cdot \nabla_{\rm \mu}$ when calculating stability criteria for various instabilities. 

The parameter $f_{\rm c}$, multiplying all individual diffusion coefficients in Equation~\ref{eq:chem1}, was first calibrated to $f_{\rm c} = 0.046$ by \cite{pin1989}. This reduction in the efficiency of chemical mixing (compared to angular momentum transport) was needed to explain the observed lithium depletion in the Sun. However, recent studies (e.g., \citealt{prat2016}) found that, at least, for the shear instability, both chemical mixing and angular momentum transport should have similar efficiencies when using the same diffusion coefficient.

\cite{heger2000} found that $f_{\rm c} = 0.033$ with $f_\mu = 0.05$ (which were the default \textsc{mesa} options until recently) best reproduce the observed nitrogen enrichment in the $10-20$~M$_\odot$ mass range at Solar metallicity\footnote{\citealt{heger2000} also comment that for an initial rotation of 200~km\,s$^{-1}$ and fixed $f_c = 0.033$, $f_\mu \geq 0.25$ is inconsistent with observations in the 30-60~M$_\odot$ mass range.}. 
\cite{yoon2006} concluded that when the angular momentum transport is very efficient (by using the magnetic term $D_{\rm ST}$ accounting for the Tayler-Spruit dynamo), then $f_{\rm c} = 0.033$ should be used with $f_\mu = 0.1$ instead of $f_\mu = 0.05$.
Using similar physical assumptions as \cite{heger2000}
and \cite{yoon2006}, \cite{brott2011} calibrated $f_{\rm c} = 0.0228$ for a 13~M$_\odot$ model based on the surface enrichment of early B stars in the LMC\footnote{These values were also adopted for their Solar and SMC models.} and adopted $f_\mu = 0.1$ from \cite{yoon2006}. Recently, \cite{markova2018} found that this calibration produces insufficient mixing for more massive stars to be compatible with observations. Some subsequent modelling approaches even adopt a mixing efficiency parameter $f_{\rm c}$ that is a factor of 10 higher \citep{ad2020}.

When using similar physics (assuming a purely diffusive equation to model angular momentum transport), \cite{chieffi2013} obtained calibrations for $f_{\rm c} = 0.07$ with $f_\mu = 0.03$ and $f_{\rm c} = 0.2$ with $f_\mu = 1.0$ (correctly noting the degeneracy between these parameters). They also performed calibrations with different physics (using the advecto-diffusive equation of angular momentum transport) which yielded $f_{\rm c} = 1$ with $f_\mu = 0.03$ for chemical mixing.
Recently, also using a physical approach different from the above mentioned ones, \cite{costa2019} used intermediate-mass binary systems and constrained $f_{\rm c} = 0.17$ with $f_\mu = 0.47$.

To be able to compare to previous works which used the same physics, we adopt $f_{\rm c} = 0.033$ and $f_\mu = 0.1$ (\texttt{am\_D\_mix\_factor} = 0.033, \texttt{am\_gradmu\_factor} = 0.1) when using the Mix1 scheme \footnote{
We note that presently there is a growing amount of evidence that such a reduction in the efficiency of chemical mixing caused by hydrodynamical instabilities is likely not needed at all. Instead, there exist other processes that are simply more efficient in transporting angular momentum than chemical elements, the prime candidates being internal gravity waves and internal magnetic fields \citep[e.g.,][and references therein]{aerts2019}.}.

\subsubsection{Mix2 scheme}

Another commonly used mixing scheme ("Mix2" hereafter) was developed by \cite{zahn92}, \cite{chaboyer1992}, \cite{maeder1997,maeder1998}, \cite{maeder2000}. This scheme has been applied in the Geneva stellar evolution code (\textsc{genec}, \citealt[][]{eggenberger2008,ekstroem2012,georgy2013,meynet2013,groh2019,keszthelyi2019,murphy2021}), as well as in modelling approaches using the \textsc{rose} \citep{potter2012a,potter2012b} and \textsc{franec}
codes \citep[][]{chieffi2013}. Here, we adopt it in \textsc{mesa}, which treats angular momentum transport in a fully diffusive scheme, unlike the above mentioned approaches. Therefore, a direct comparison to previous works is not possible. The major difference is that given the efficient diffusive angular momentum transport, strong shear mixing cannot develop. Consequently, in our models with the Mix2 scheme, the main chemical element transport is via meridional currents during most of the main sequence evolution. This is not the case in the models of \cite{ekstroem2012}, where the advective treatment of angular momentum transport allows for shears, which may also become the dominant process of transporting chemical elements.
As we will see (Section~\ref{sec:str}, Section~\ref{sec:appevol}), the Mix2 scheme leads to quasi-chemically homogeneous evolution for the entire main sequence of our non-magnetic models. Since such a behaviour is expected to be rare, we may consider the adaptation of this mixing scheme in our models as a limiting case for very efficient mixing.

The effective diffusion coefficient for chemical mixing combines the effects of meridional currents and horizontal turbulence, 
\begin{equation}\label{eq:deff} 
D_{\rm eff} = \frac{1}{30} \frac{\left\lvert \ r U(r)\right\rvert^2}{D_h} \, , 
\end{equation}

\noindent where the radial component of the meridional circulation is 
\begin{equation}
\begin{aligned}
U(r) & \, = \frac{P}{\rho g C_{P} T}  \quad \frac{1}{\nabla_{\rm ad} - \nabla_{\rm rad}  + \frac{\phi}{\delta} \nabla_\mu}    \\
&  \times \left( \frac{L}{M'}  [E_{\Omega}^{\star} + E_\mu ]  + \frac{c_P }{\delta} \frac{\partial \vartheta}{\partial t}     \right) \, ,
\end{aligned}
\end{equation}
\noindent with $P$ the pressure, $\rho$ the density, $g$ the gravitational acceleration, $T$ the temperature, $L$ the luminosity, $M' = M_\star ( 1 - \Omega^2 / 2 \pi g \rho_{\rm m}) $, $E_{\Omega}^{\star}$ and $E_\mu$ terms which depend on the distribution of angular velocity and mean molecular weight\footnote{The full expression of these terms is given by \cite{maeder1998}. In our approach, we simplify this expression and adopt only the leading term which is the first term of $E_{\Omega}^{\star}$ as described by \citet{maeder1998}. Since it is a smaller term, we set $E_\mu$ to zero.}, and $\vartheta$ the ratio of the variation of the density to the mean density $\rho_{\rm m}$. The horizontal turbulence is adopted as:
\begin{equation}\label{eq:dh} 
D_h = \frac{1}{c_h} \ r \ \lvert 2 V(r) - \alpha U(r) \rvert \, , 
\end{equation}
\noindent where $c_{\rm h}$ is a constant set to unity (see \citealt[][]{chaboyer1992}), and $V(r)$ expresses the radial dependence of the horizontal component of the meridional circulation. The horizontal component is expressed as $V(r) P2(\cos \Theta)$, where $P2$ is the second Legendre polynomial and $\Theta$ is the co-latitude. We set $V(r) = U(r)$ as a reasonable approximation. Then,
\begin{equation}\label{eq:alpha} 
\alpha = \frac{1}{2} \frac{\mathrm{d} \ln \ (r^2 \Omega)}{\mathrm{d} \ln \ r} \, . 
\end{equation}
%
%

\noindent The diffusion coefficient accounting for vertical shear mixing is derived by \cite{maeder1997} as:
\begin{equation}\label{eq:dshear} 
D_{\rm shear} = f_{\rm energ} \frac{H_{P}}{g \delta} \frac{K}{\left[ \frac{\phi}{\delta} \nabla_\mu + (\nabla_{\rm ad} - \nabla_{\rm rad}) \right]} \left( \frac{9 \pi}{32} \ \Omega \ \frac{\mathrm{d} \ln \ \Omega}{\mathrm{d} \ln \ r} \right) ^2 \, , 
\end{equation} 
where $f_{\rm energ}$ is a free parameter set to unity. $H_{P}$ is the local pressure scale height, $g$ the gravitational acceleration, $\Omega$ the angular velocity, and $r$ the radius.
Finally, in the Mix2 scheme, the diffusion coefficient applied in Equation~\ref{eq:diff} is: 
\begin{equation}\label{eq:chem2}
    D_{\rm chem}^{\rm Mix2} = D_{\rm eff} + D_{\rm shear} \, .
\end{equation}

In this case, the free parameters $f_{\rm c}$ and $f_{\mu}$ are not used in \textsc{genec} calculations. Consequently, we do not
apply them in our \textsc{mesa} Mix2 model calculations either (\texttt{am\_D\_mix\_factor} = 1, \texttt{am\_gradmu\_factor} = 1). To our knowledge this mixing scheme is implemented in \textsc{mesa} for the first time. The sum of diffusion coefficients, dominated by meridional currents ($D_{\rm eff}$, Equation~\ref{eq:deff}) in the solid-body rotating case are comparable in shape to the default \textsc{mesa} approach which uses $D_{\rm ES}$ as derived by \cite{kippenhahn1974} and \cite{pin1989}. However, the amplitudes are not equal (as shown in Figure \ref{fig:str1_midMS}).

We note here that there is some confusion in the literature regarding the work of \cite{chaboyer1992}. \cite{heger2000} (and following publications) state that \cite{chaboyer1992} found $f_{\rm c} = 1/30$ based on a theoretical approach. 
The work of \cite{chaboyer1992} does not introduce any scaling factors. $D_{\rm eff}$, describing the transport resulting from the interaction between meridional currents and the strong horizontal turbulence is obtained by integrating the equation for the transport of chemical elements over latitudes. This integration gives rise to the numerical term of 1/30 (in their Equation 16 and our Equation~\ref{eq:deff}), resulting from the decomposition of the meridional velocity in Legendre polynomials. This is not a scaling factor to match observations and it does not apply to any other diffusion coefficient. Similarly, in Equation~\ref{eq:br} the numerical term 2/3 is not an arbitrary scaling factor that one would tailor to observations.

%
%
%
%
\section{Results}\label{sec:three}

In this section, we first consider non-magnetic models, and then a fiducial model with $M_{\rm ini}=$~20~M$_\odot$, $\Omega_{\rm ini}/\Omega_{\rm crit, ini} = 0.5$, $Z_{\rm ini} =~0.014$, and $B_{\rm eq, ini} =~3$~kG within the INT/Mix1 scheme, and follow changes in its stellar structure and evolution. In particular, we will first vary the mixing and braking schemes to investigate the impact on stellar structure models in Section~\ref{sec:str1} and abundances in Section~\ref{sec:str2} and in evolutionary models in Section~\ref{sec:schemeevol}.
Then, a typical HRD evolution of the INT and SURF magnetic models in the full mass range (3 -- 60 M$_\odot$) will be addressed in Section \ref{sec:hrdevol}, followed by predictions for the Kiel diagram in Section \ref{sec:kielevol}. Nitrogen abundances and other schemes are also shown in Appendix~\ref{sec:appevol}.
Finally, the initial magnetic field strength and metallicity will be varied within the evolutionary models in Sections \ref{sec:fieldevol} and \ref{sec:metevol}.

\begin{figure*}
\includegraphics[width=16cm]{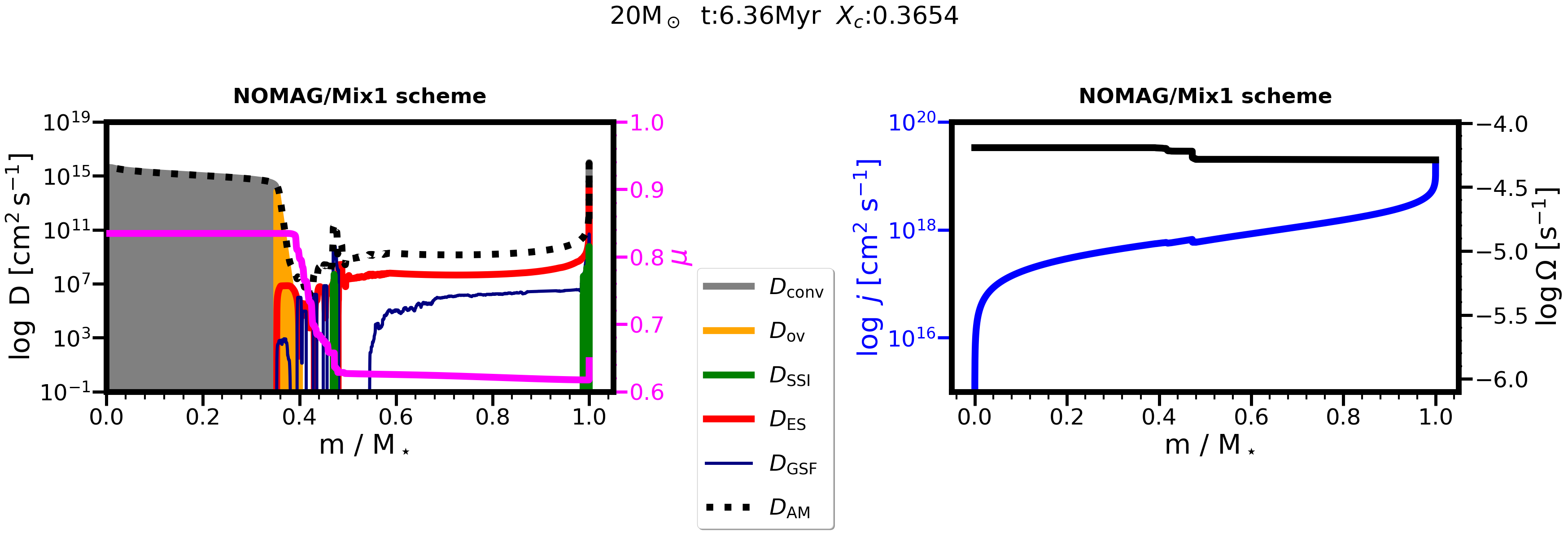}
\includegraphics[width=16cm]{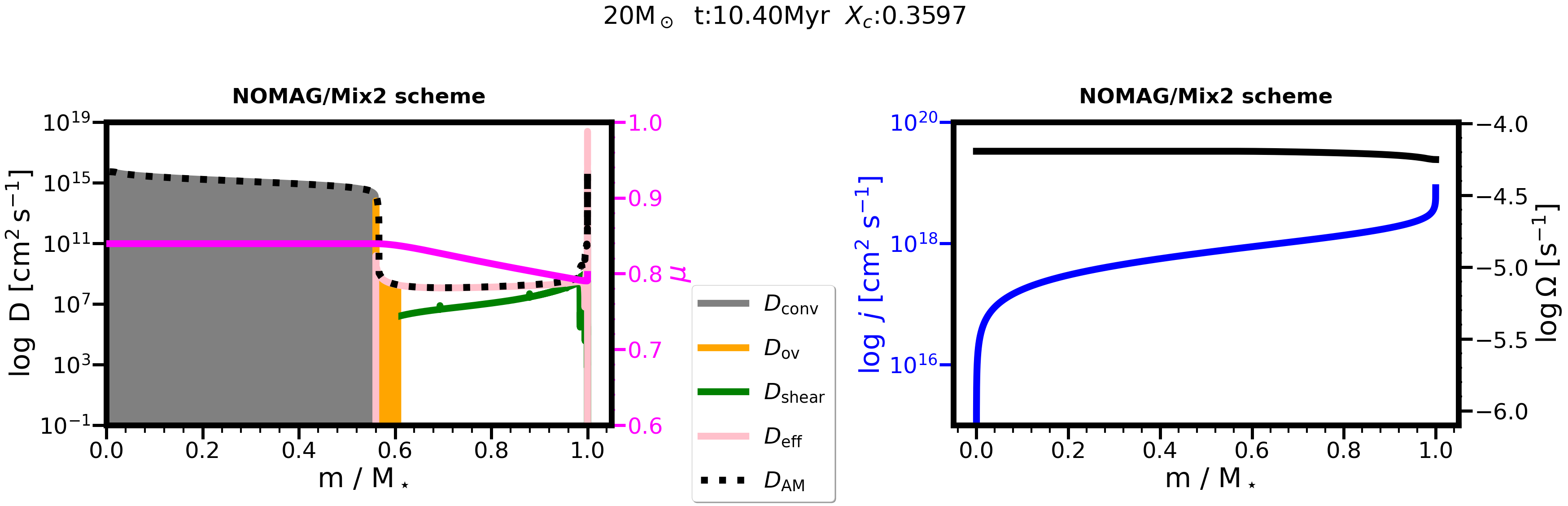}
\caption{We show a non-magnetic model of $M_{\rm ini}$ = 20~M$_\odot$ (at $Z = 0.014$) within the NOMAG/Mix1 scheme (top panel) and an initially identical model within the NOMAG/Mix2 scheme (lower panel) at half-way through its core hydrogen burning phase (with core hydrogen mass fraction of $X_{\rm c}/X_{\rm c, init} \approx 50\%$). \textit{Left}: diffusion coefficients for chemical mixing (solid lines for rotational mixing and shaded grey and orange for convective core mixing and overshooting, respectively; all entering via $D_{\rm chem}$ in Equation~\ref{eq:diff}) and diffusion coefficient for angular momentum transport (dotted line, entering Equation~\ref{eq:mam2}). The right ordinate and magenta line show the mean molecular weight. Due to the stochastic nature of mixing processes and, in some cases, numerical noise, here and hereafter we apply a moderate smoothing of some diffusion coefficients for visualisation purposes. \textit{Right}: Specific angular momentum (left ordinate, blue line) and angular velocity (right ordinate, black line).  }\label{fig:str0_midMS}
\end{figure*}
\begin{figure*}
\includegraphics[width=16cm]{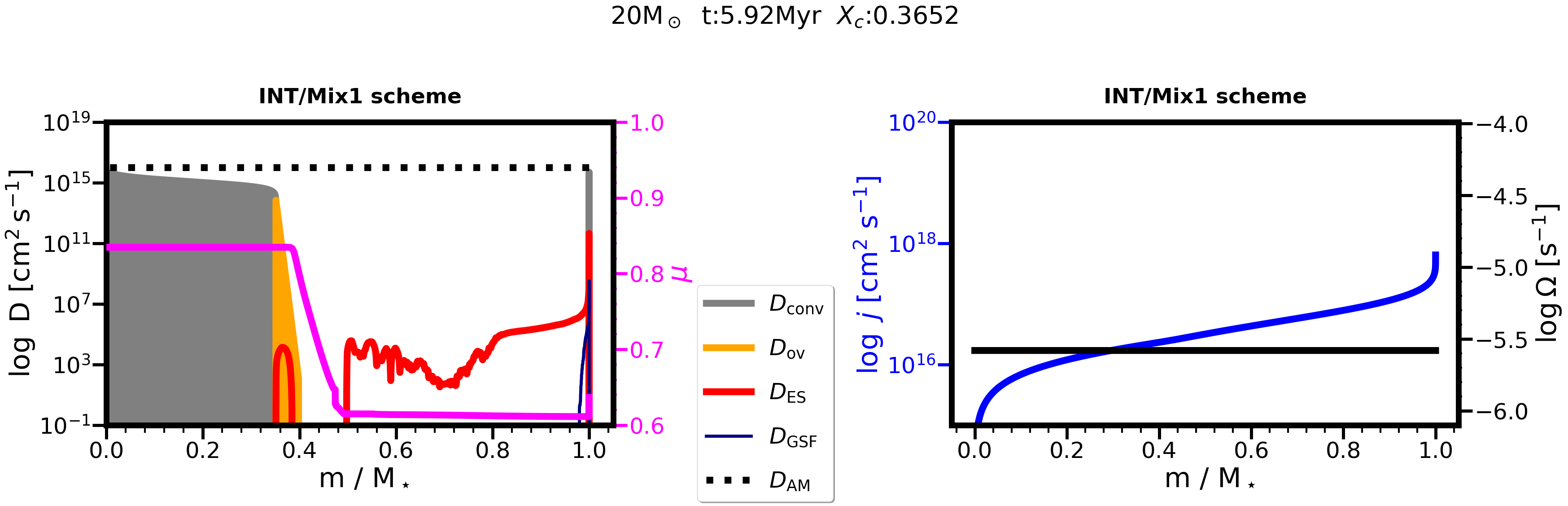}
\includegraphics[width=16cm]{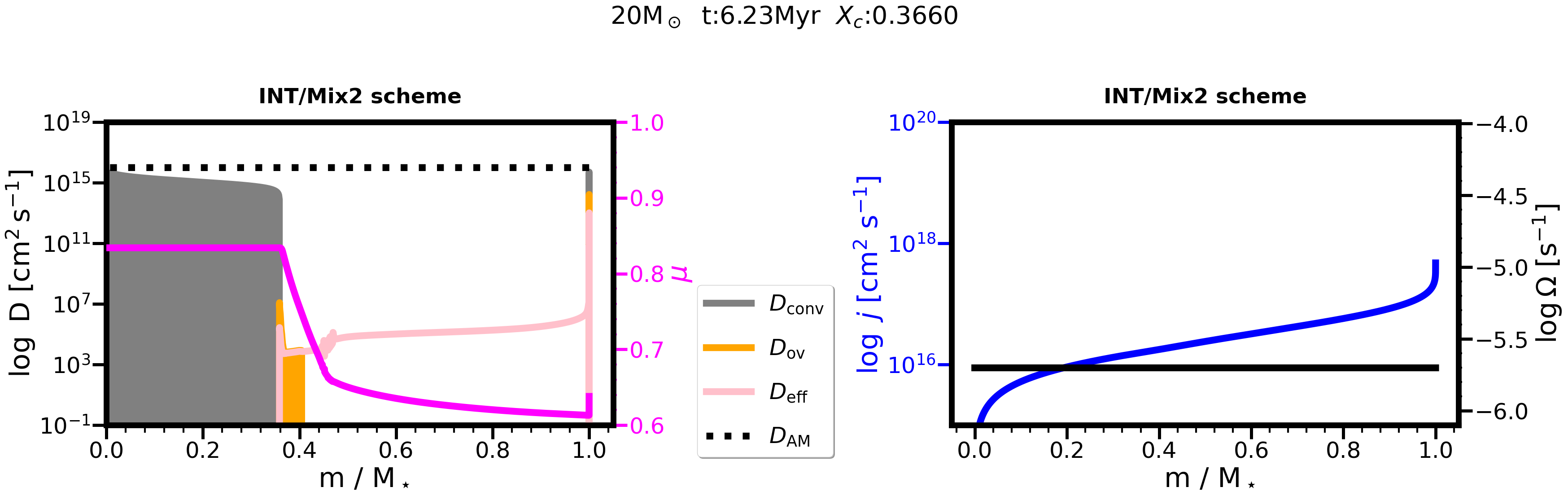}
\caption{Same as Figure~\ref{fig:str0_midMS} but for magnetic models within the INT magnetic braking scheme. }\label{fig:str1_midMS}
\end{figure*}
\begin{figure*}
\includegraphics[width=16cm]{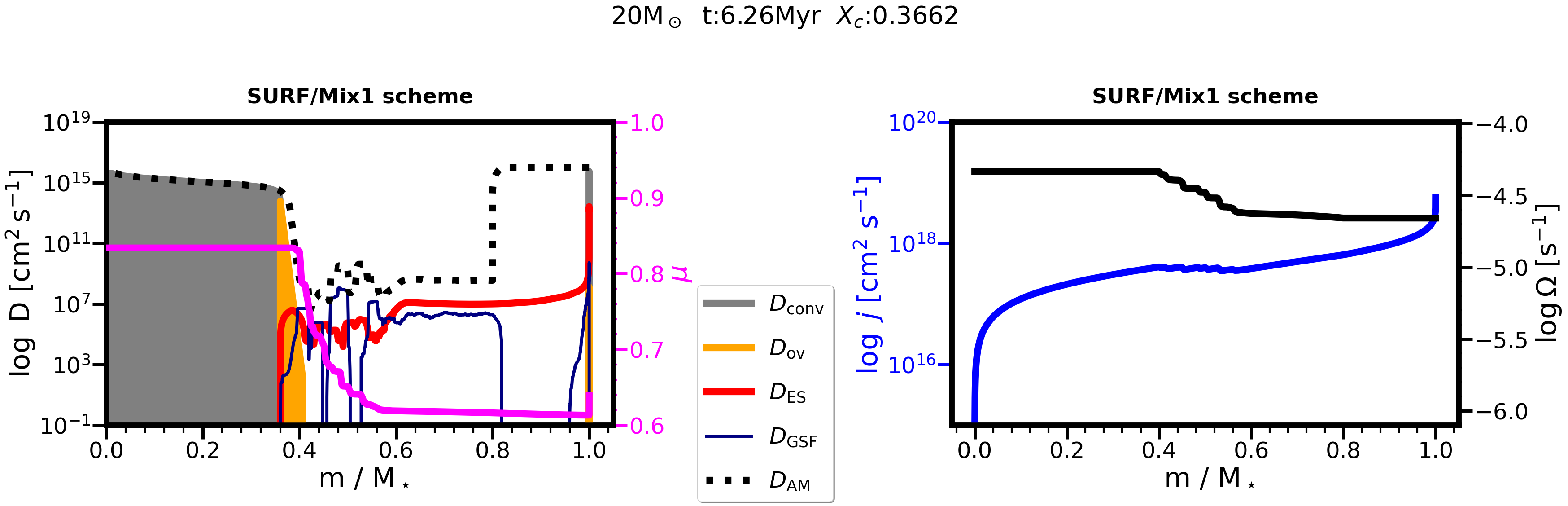}
\includegraphics[width=16cm]{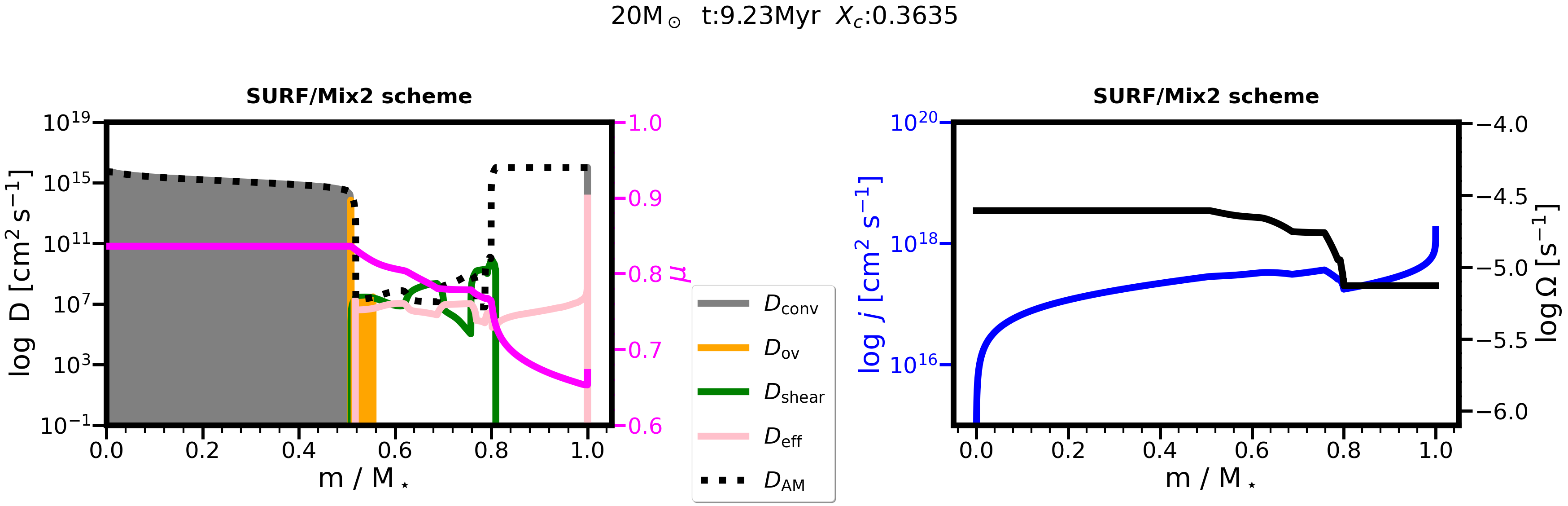}
\caption{Same as Figure~\ref{fig:str0_midMS} but for magnetic models within the SURF magnetic braking scheme.}\label{fig:str2_midMS}
\end{figure*}

%
%
\subsection{Stellar structure models}\label{sec:str}

\subsubsection{Chemical mixing and angular momentum transport}\label{sec:str1}

Figures~\ref{fig:str0_midMS}, \ref{fig:str1_midMS}, and \ref{fig:str2_midMS} show  20~M$_\odot$ models at solar metallicity. These structure models are at half-way through their core hydrogen burning phase\footnote{The ZAMS and TAMS structure models of the four schemes are shown in Figures \ref{fig:app1a} - \ref{fig:app1d} in the Appendix.} (defined as $X_{\rm core}/X_{\rm core, init} = 0.5$). 
In Figure~\ref{fig:str0_midMS}, we show non-magnetic models in the Mix1 (top panel) and Mix2 (lower panel) chemical mixing schemes. (Since magnetic braking is not applied, there is no braking scheme in these cases, hence we refer to these models as "NOMAG".) In Figure \ref{fig:str1_midMS}, the fiducial model (INT/Mix1, top panel) as well as an otherwise initially identical model but within the INT/Mix2 scheme (lower panel) is shown. Models with the SURF/Mix1 and SURF/Mix2 schemes are shown in Figure~\ref{fig:str2_midMS}.
Note that the models with the adopted braking and mixing schemes correspond to different stellar ages when the core hydrogen is half-way depleted in each model (indicated in the title of the panels), given the different evolution resulting from the change in physical assumptions. 
%

When fossil magnetic fields are not considered in the models (NOMAG case; $B_{\rm eq} = 0$), the Mix1 and Mix2 chemical mixing schemes produce drastically different results. In the Mix1 scheme, the mean molecular weight (shown with magenta line and on the right ordinate) drops rapidly at the convective core boundary (the zones with convective overshooting are shown with orange in Figure~\ref{fig:str0_midMS}). Near this region the GSF instability dominates (see, e.g., the recent studies of \citealt{caleo2016,barker2019,barker2020,chang2021}). Once the chemical composition is stabilised, meridional circulation drives chemical mixing (red line) and angular momentum transport (dotted line). Note that the assumption of $f_{\rm c} = 0.033$ reduces the efficiency of all instabilities considered for chemical mixing compared to the efficiency of the same instabilities used for angular momentum transport. The angular velocity profile (right panel) remains completely flat in the stellar envelope, with a small break at the core boundary. Thus the model is very close to solid-body rotation. 
The NOMAG/Mix2 model reveals a very efficient mixing, with an almost flat mean molecular weight profile indicating close-to chemically homogeneous evolution. The dominant transport is via the effective diffusion coefficient (Equation~\ref{eq:deff}). Importantly, the diffusion coefficients at the core-envelope boundary are smooth. This is a critical region that allows for mixing up material from the core to the surface. Note the much larger convective core (in grey) compared to the Mix1 model. The specific angular momentum (blue line, right panel) and angular velocity profiles are smooth throughout the star. $\Omega$ slightly decreases near the surface as a result of mass loss, however, this model is also very close to solid-body rotation.

In the INT braking scheme (Figure~\ref{fig:str1_midMS}) the angular velocity profile is completely flat. The star is rigidly rotating due to the assumed high diffusivity for transporting angular momentum attributed to the magnetic field, albeit the angular rotation is much lower than in the NOMAG model due to magnetic braking (uniformly) lowering the specific angular momentum. The rigid rotation does not allow shears to develop and transport angular momentum or chemical elements. Therefore in these models the chemical enrichment is entirely driven by meridional currents. In the "standard" \textsc{mesa} description (Mix1 scheme) a gap in the transport develops above the overshooting region, corresponding to steep chemical gradients, as seen from the large drop of the mean molecular weight (magenta line, right ordinate) at the core boundary. Despite the mitigating effect of $f_\mu = 0.1$ in these models, the inefficient mixing above the core boundary will prevent a very efficient surface enrichment and overall mixing inside the star. If the gap existed throughout the entire early evolution, it would completely inhibit surface enrichment. However, the gap is not present initially -- see top panels of Figures~\ref{fig:app1a} and \ref{fig:app1b} --, when the mixing and corresponding enrichment are prominent.
With internal magnetic braking, all layers of the star lose angular momentum, therefore the shape of the specific angular momentum profile remains unchanged whereas its overall value decreases over time. In the Mix2 scheme, $D_{\rm eff}$ never becomes zero close to the convective core. This allows for a smoother composition gradient and more overall mixing, therefore differences in surface abundances are expected. On the other hand, the specific angular momenta are not so different between the Mix1 and Mix2 schemes in the INT models. $\Omega$ is smaller in the Mix2 model, but this quantity also depends on the radius of the star. Since the model in the INT/Mix2 scheme takes more time than the INT/Mix1 to deplete hydrogen in its core due to the more efficient chemical mixing, at half-way through its core burning stage it has a larger radius. We also note that the NOMAG/Mix2 model produces a more efficient mixing than the INT/Mix2 model. As a consequence, the INT/Mix2 model results in a smaller convective core than the non-magnetic case.

In the SURF braking scheme (Figure~\ref{fig:str2_midMS}), one major difference is that there is no overall solid-body rotation. 
Let us recall that in the SURF models only the outer layers enclosing the top 20\% of the total mass of the star are assumed to lose angular momentum (see Equation~\ref{eq:br2}) and have an increased diffusivity for angular momentum transport (via Equation \ref{eq:mam2}). 
In a 1D diffusive scheme, angular momentum flows from inner to outer layers.
The angular velocity profile is flat in those layers where the diffusivity is increased. Depending on the mixing scheme, the composition and also the mixing processes are rather different. The magnitude and radial dependence of the diffusion coefficient for angular momentum transport also determines the angular velocity in the rest of the stellar envelope. We see that in the SURF/Mix1 model the angular velocity is also roughly constant between 0.6 and 0.8~$m/M_\star$, and it gradually changes closer to the boundary of the stellar core, as it is the case for $D_{\rm AM}$ (left panel, dotted line). In the SURF/Mix2 model, $D_{\rm AM}$ changes more abruptly at around 0.8~$m/M_\star$, whereas it is roughly constant again closer to the stellar core.
Braking for a given magnetic field strength is less efficient in the SURF scheme than in the INT scheme since the Alfv\'en radius is smaller for a quadrupole field than for a dipole field as defined by Equations~\ref{eq:alf1}-\ref{eq:alf2}. 
Given the shape of the specific angular momentum profile (right panel, blue line), the SUFR/Mix1 model has a break closer to the stellar core, while the SURF/Mix2 model has a break closer to 0.8~$m/M_\star$. This means that the SURF/Mix2 model can more easily exhaust its surface reservoir of angular momentum.

Certainly, further research is required to investigate how angular momentum transport and magnetic braking work for more complex magnetic field configurations and how they could be implemented in 1D stellar evolution models. Overall, the results from the SURF approach may be considered similar to the works of \cite{meynet2011} and \PaperI, where magnetic braking was only applied to the uppermost stellar layer.

In the SURF/Mix1 scheme, the GSF instability can efficiently transport chemical elements near the core boundary. This instability acts on a dynamical timescale and therefore can vary from timestep to timestep. In the upper envelope meridional currents remain efficient. In the Mix2 scheme, the free parameters controlling mixing efficiency are not applied ($f_\mu = 1$, $f_{\rm c} = 1$), and the SURF/Mix2 scheme is thus the most efficient in chemical mixing. This is also evidenced by the larger convective core size compared to the three models in the other magnetic schemes. In fact, the convective core size of the SURF/Mix2 model is similar to that of the NOMAG/Mix2 model.
Strong gradients of chemical elements do not develop near the core boundary. Shear mixing remains efficient in the entire envelope to transport chemical elements.

%
%
%
%
\begin{figure*}
\includegraphics[width=18cm]{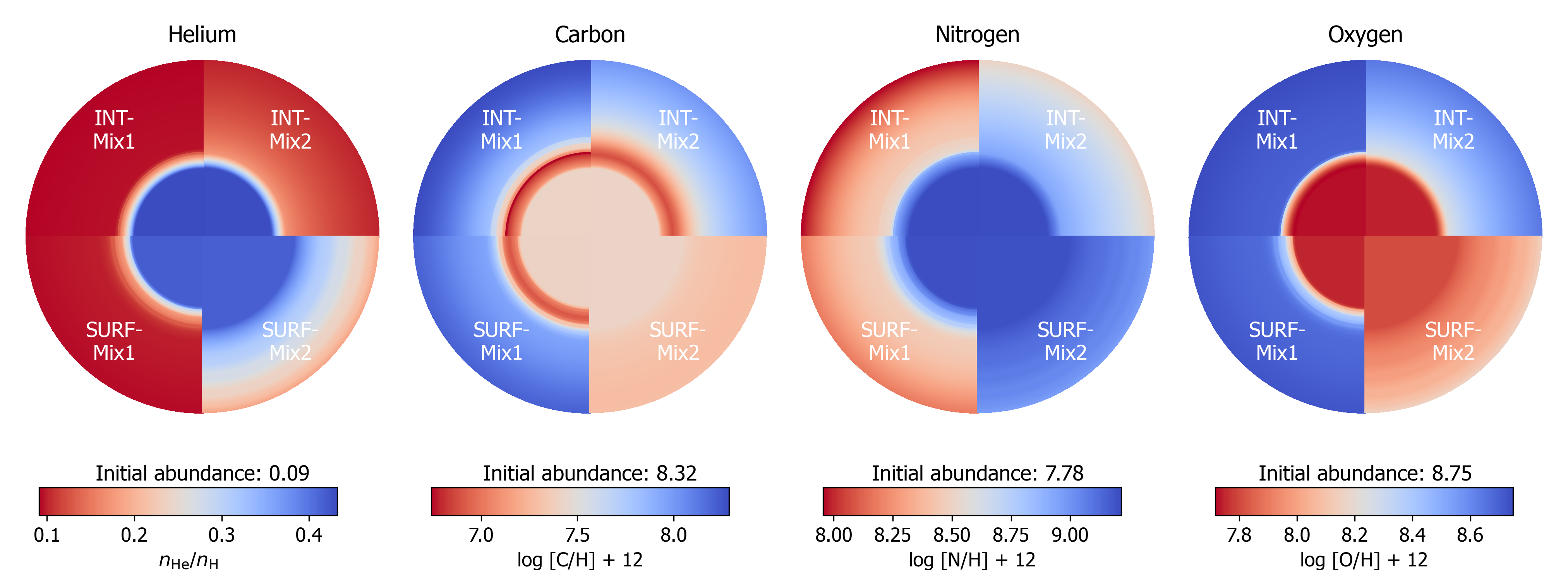}
\caption{He, C, N and O abundances for the same models as in Figures~\ref{fig:str1_midMS} and \ref{fig:str2_midMS} at $X_{\rm c} \approx 0.36$. In each circle, colour maps show the abundance of one element as a function of stellar mass coordinate, where the distance from the centre of the circle corresponds linearly to the mass enclosed within that radius. Each quarter circle contains the abundance profiles of one model. He abundance is in number fraction, the other elements in logarithmic number fractions. These profiles are prepared using \textsc{tulips} \citep{laplace21}.
}\label{fig:str3_midMS}
\end{figure*}
\subsubsection{Abundances of He, C, N, O}\label{sec:str2}
Figure \ref{fig:str3_midMS} shows the abundances of helium, carbon, nitrogen, and oxygen. For three models (INT/Mix1, INT/Mix2, SURF/Mix1) the convective hydrogen core sizes are comparable. The SURF/Mix2 scheme, which is the most efficient in mixing, leads to a much larger core. In this model, the average helium content in the stellar envelope is much higher and the surface will also show this increased abundance already on the main sequence. Carbon is slightly depleted during the CNO-cycle (it becomes most depleted in a thin layer close to the core boundary), however the surface carbon abundance is minimally changed in the first three models. In contrast, the SURF/Mix2 model produces an almost uniform distribution of carbon inside the star, it is completely mixed in the envelope without any gradients. 
Nitrogen excess is produced during the CNO-cycle, and therefore, its surface abundance is a crucial measurement to infer the efficiency of internal mixing. All models produce a surface nitrogen enrichment, except the INT/Mix1 scheme. Here a strong gradient develops between the core and the surface. While a core-surface gradient is also present in the INT/Mix2 and SURF/Mix1 models, their envelopes have a somewhat higher mean nitrogen abundance and their surfaces are slightly enriched in nitrogen. The SURF/Mix2 model has an almost homogeneous nitrogen distribution in its envelope. 
Oxygen is depleted during the CNO-cylce. Similar to carbon, oxygen can still remain abundant in the envelope in the first three models. The SURF/Mix2 model yields a close-to-homogeneous oxygen distribution throughout the star. In contrast to carbon, in all four models oxygen is depleted in the entire stellar core. For example, the core to surface oxygen abundance can differ by an order of magnitude in the INT/Mix1 model.

The reason why strong gradients can develop (and remain in most models) near the core boundary is related to the drop in chemical mixing, identifiable by drops and gaps in the diffusion coefficients (c.f. Figures~\ref{fig:str1_midMS}-\ref{fig:str2_midMS}), which in turn depend on the composition gradients. 
The velocity of the diffusion is zero when there is no composition gradient and it increases when the gradient increases. This is another effect that leads to reducing the diffusion coefficient. For helium, the difference between the core and the envelope grows gently, while for nitrogen it grows faster because nitrogen is enhanced very rapidly in the core. Thus for a given diffusion coefficient, nitrogen will diffuse more rapidly than helium. 
In the Mix2 scheme, the key differences between the INT and SURF models result from their different angular velocity profiles. The INT models lose angular momentum in all layers and thus mixing becomes less efficient overall. The SURF/Mix2 model has a more massive convective core at the same evolutionary stage (c.f. Figure~\ref{fig:str2_midMS}), and the envelope closely reflects on the core composition as this model has an almost homogeneous distribution of chemical elements.
In both INT/Mix1 and INT/Mix2 schemes, envelope mixing has a similar overall efficiency. However, in the INT/Mix2 scheme, the near core mixing is more efficient (see above), greatly impacting the measurable surface abundances of carbon, nitrogen and oxygen.
The surface abundances, especially of nitrogen, are sensitive to the chosen braking and mixing schemes, especially with the Mix2 scheme reflecting more closely the core composition than the Mix1 scheme.

%
%
%
%
\subsection{Evolutionary tracks}\label{sec:evolutionary_tracks}

%
%
%
%
\begin{figure*}
\includegraphics[width=6cm]{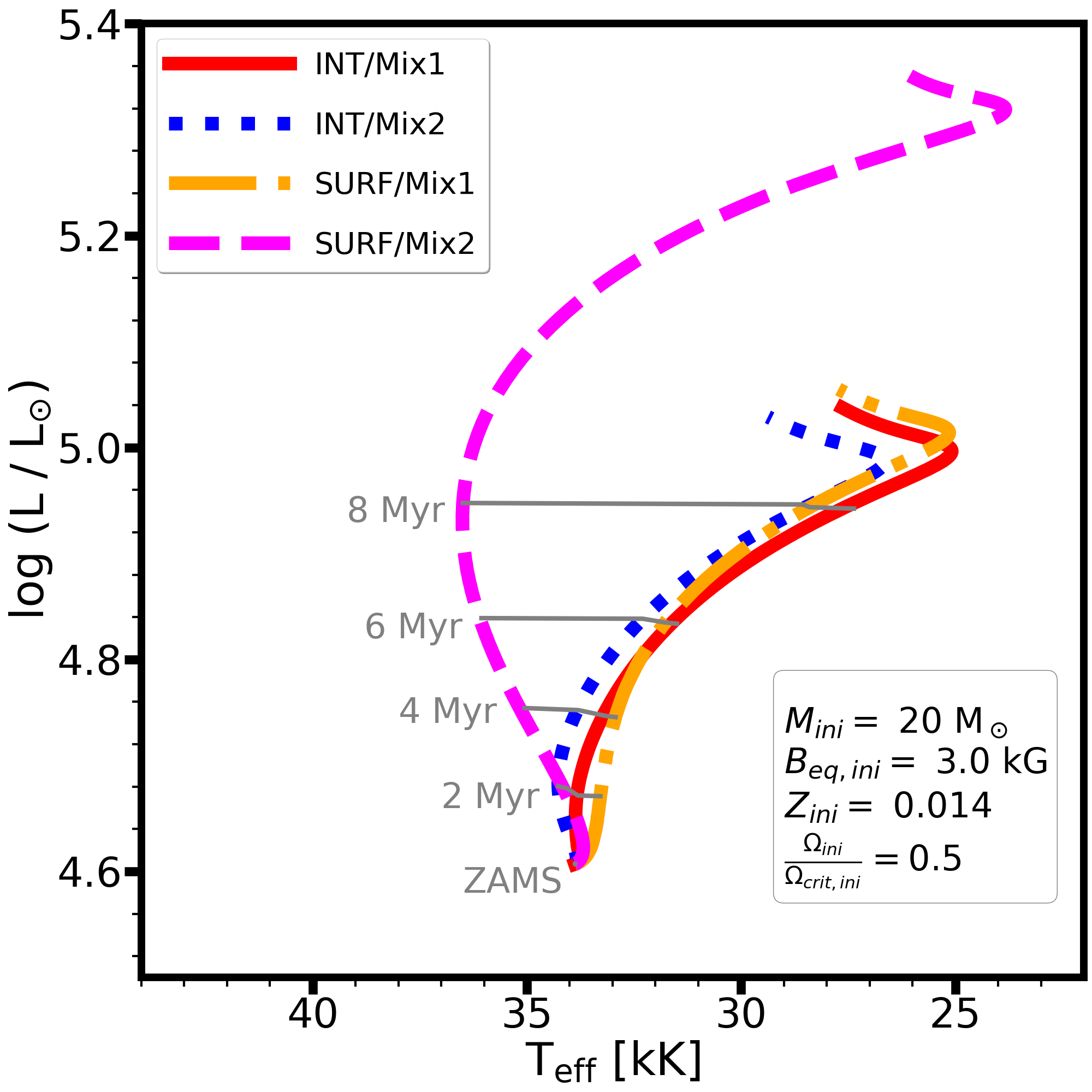}\includegraphics[width=6cm]{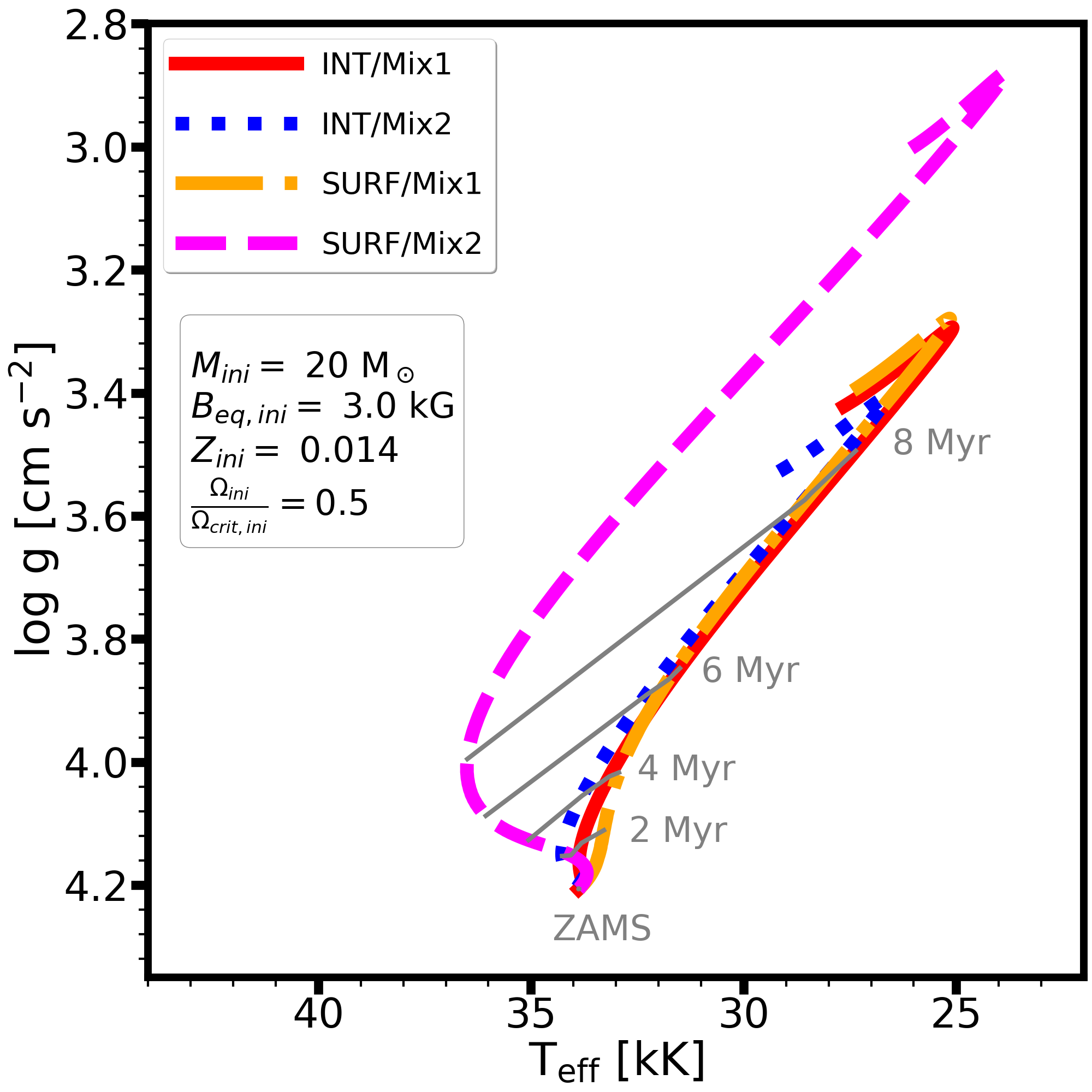}\includegraphics[width=6cm]{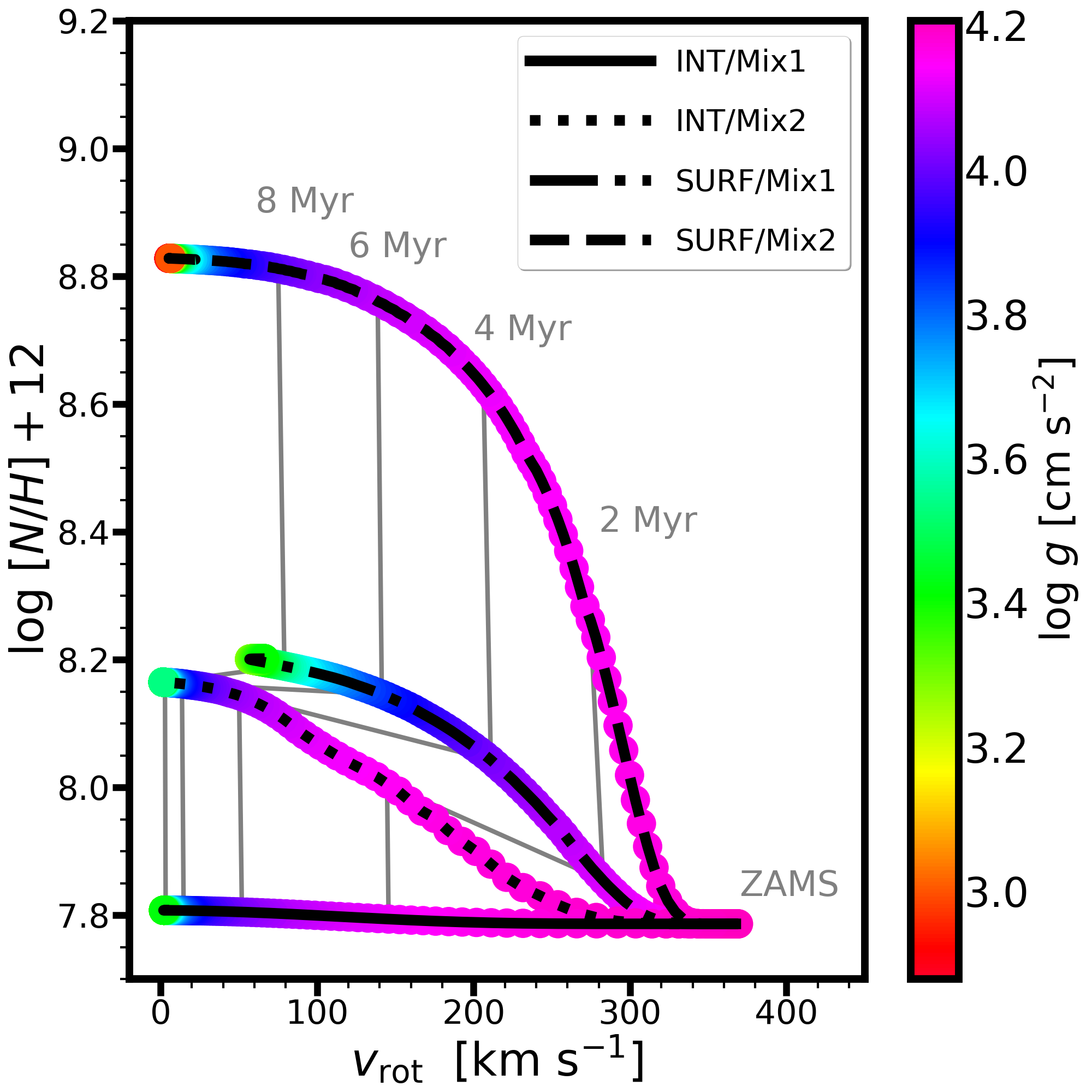} 
\caption{Fiducial evolutionary model with $M_{\rm ini}=$~ 20~M$_\odot$ at $Z = 0.014$ with $B_{\rm eq, ini=}$~3 kG within the INT/Mix1 scheme and the 3 other schemes, indicated with different colours and line-style. Grey lines connect equal ages. Panels from left to right show the HR, Kiel, and Hunter diagrams, respectively. The colour-coding shows the logarithmic surface gravity on the Hunter diagram.}\label{fig:schemeevol1}
\end{figure*}
%
\subsubsection{Impact of magnetic braking and chemical mixing schemes for 20 M$_\odot$ models}\label{sec:schemeevol}

Figure \ref{fig:schemeevol1} shows the fiducial 20 M$_\odot$ model (INT/Mix1) as well as initially identical models but in the other 3 schemes in the HRD, Kiel diagram, and Hunter diagram\footnote{We use the spectroscopic definition of nitrogen abundance for the Hunter diagram, which is $\log (N/H) + 12$, where $N$ and $H$ are the surface number fraction of nitrogen and hydrogen. Note that mass fractions (which are the typical output quantities from evolutionary grids) need to be translated to number fraction by appropriate scaling.} \citep{hunter2008,hunter2009}. Here we address the impact of using the different braking and mixing schemes for otherwise identical models with the same initial mass and initial magnetic field strength. 

The models within the Mix1 chemical mixing scheme result in closely overlapping tracks on the HRD and Kiel diagrams. However, as evidenced from the Hunter diagram (right panel of Figure~\ref{fig:schemeevol1}), the spin-down and chemical enrichment are different when considering the different magnetic braking schemes. The 20 M$_\odot$ INT/Mix1 model (solid line) produces essentially no observable surface nitrogen enrichment in this configuration. The SURF/Mix1 model (dashed-dotted line), on the other hand, maintains a higher angular velocity in the inner regions of the star by braking only the upper layers (see Section \ref{sec:str}). This is why mixing remains more efficient and a larger amount of nitrogen is mixed to the stellar surface (0.45 dex). 
Model predictions within the Mix2 chemical mixing scheme produce a much more efficient chemical mixing than the Mix1 scheme. However, the differences between the INT/Mix2 and INT/Mix1 schemes are relatively modest on the HRD and Kiel diagrams, while the Hunter diagram shows large deviations. The INT/Mix2 model reaches a surface nitrogen abundance that is about 0.4 dex higher than the baseline value. 

Contrary to the first three cases, the SURF/Mix2 model has an extended blueward evolution, shown on the HRD and Kiel diagrams. Such a feature is commonly associated with blue stragglers, merger products, and quasi-chemically homogeneous evolution \citep[e.g.,][]{maeder1987,yoon2006,knigge2009}. Given the very efficient mixing in this model, it is nearly chemically homogeneous (c.f. Figure \ref{fig:str3_midMS}). Indeed, the SURF/Mix2 model produces the highest nitrogen enrichment, more than 1 dex compared to the baseline in this configuration. 

Even without considering magnetic braking, the two mixing schemes are considerably different, which leads to different  evolutionary tracks. The assumptions regarding the mutual effect of the magnetic field and chemical mixing allow for a range of behaviours (see also \PaperIII). Models within the Mix1 scheme have modest differences as a result of using the different braking schemes (INT/SURF). Models within the Mix2 scheme produce very efficient mixing, which is more easily quenched in the INT models than in the SURF models. The INT braking scheme, in contrast to the SURF, decreases the overall rotation rate, and the INT/Mix2 models do not evolve significantly blueward on the HRD. However, quasi-chemically homogeneous evolution is achieved in the case of non-magnetic Mix2 models, as well as in some magnetic SURF/Mix2 models depending on the initial field strength and stellar mass (see Figs.~\ref{fig:field2} and \ref{fig:field4}, in the Appendix). For example, the magnetic model in the SURF/Mix2 scheme leads to quasi-chemically homogeneous evolution for most of the main sequence ($\approx 9$~Myr out of 12~Myr) of a 20 M$_\odot$ model at solar metallicity with a 3 kG magnetic field. However, for the same initial field strength, the initially 60~M$_\odot$ model only experiences a brief phase of quasi-chemically homogeneous evolution and then evolves redwards (Figure~\ref{fig:field4}).

In summary, the braking and mixing schemes can drastically change the main observable characteristics. For example, the 8~Myr isochrone spans over a 10~kK effective temperature range (left panel of Figure \ref{fig:schemeevol1}) despite the models being initially completely identical. (When the "extreme" case of SURF/Mix2, which is assumed to represent stars with complex magnetic fields and very efficient mixing, is not considered the difference is less, around 2~kK.) Thus the uncertainties associated with braking and mixing schemes in evolutionary model predictions are significant. From the Hunter diagram we can conclude that various braking and mixing schemes can cover a wide range of rotation rates and surface nitrogen abundances. Three models already reach slow rotation ($< 50$~km~s$^{-1}$) with high surface gravities. Both INT models take less than 6~Myr to achieve this, while it is slightly over 8~Myr for the SURF/Mix2 model. The SURF/Mix1 model reaches the TAMS with a somewhat higher rotation rate than the other models.

%
%
%
\begin{figure*}
\includegraphics[width=8cm]{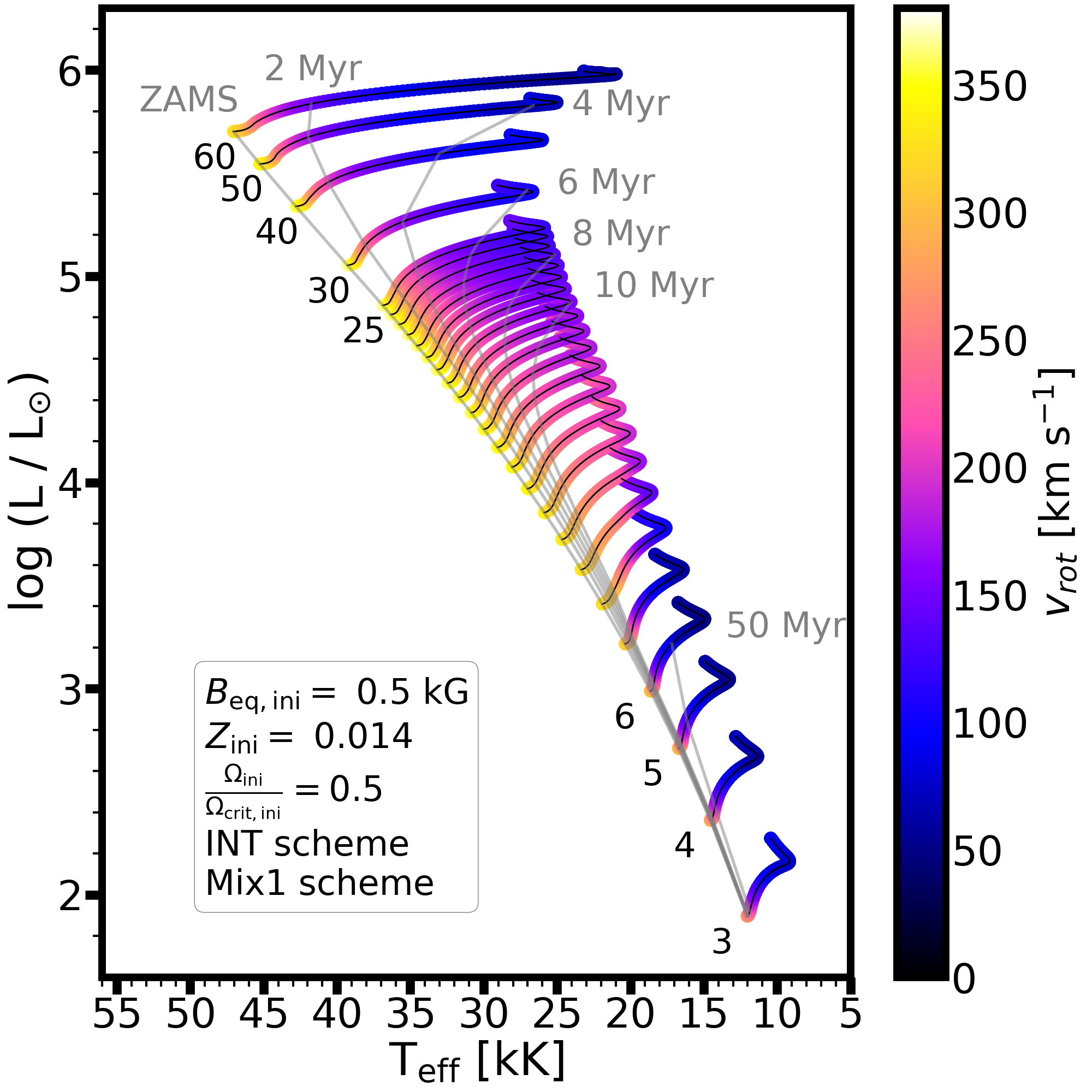}\includegraphics[width=8cm]{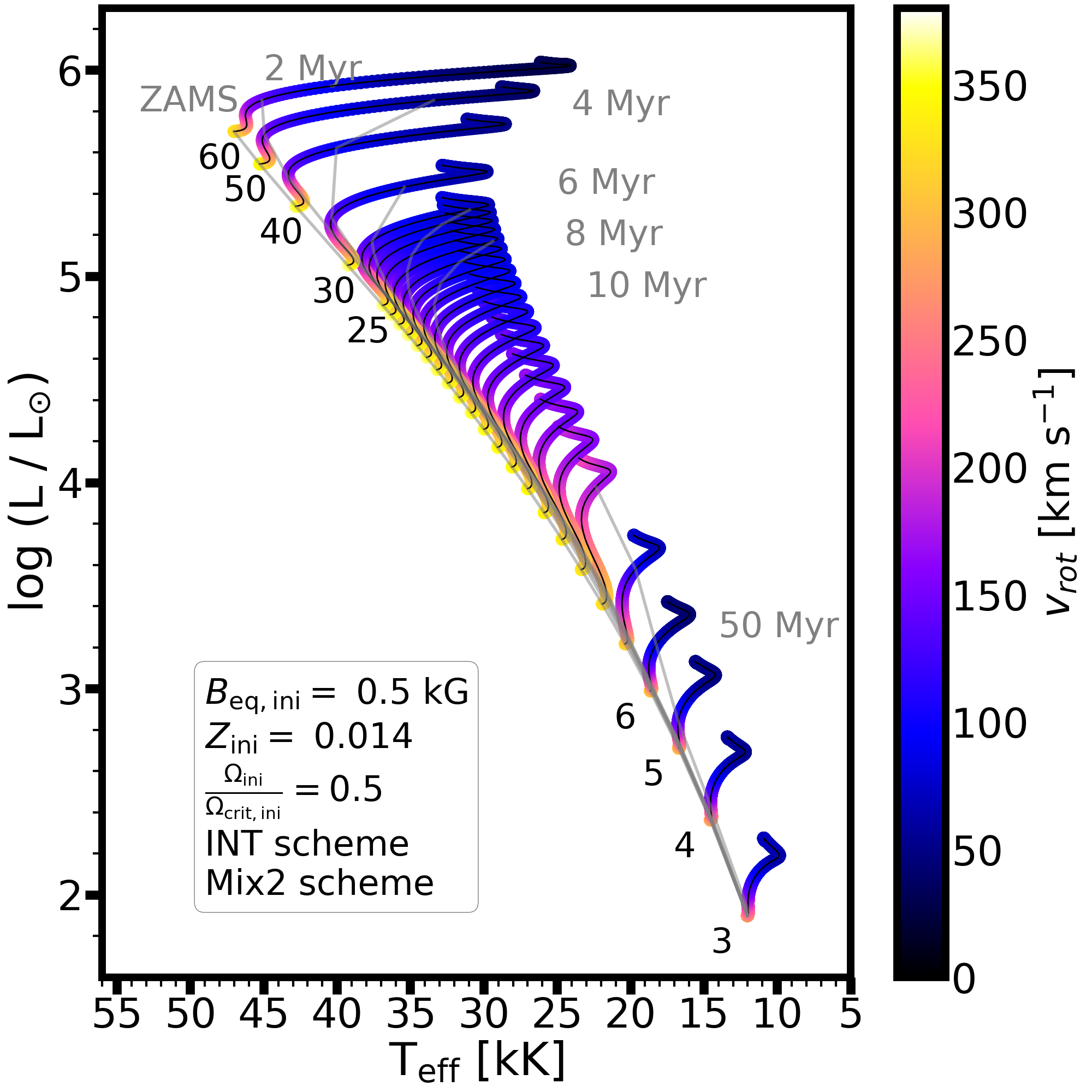}
\caption{Hertzsprung-Russell diagram of magnetic single-star evolutionary models with initial magnetic field strength $B_{\rm eq, ini}= 0.5$~kG, initial rotation rates of $\Omega_{\rm ini}/\Omega_{\rm ini, crit}= 0.5$ at solar $Z_{\rm ini}= 0.014$ metallicity within the INT/Mix1 scheme (left) and INT/Mix2 scheme (right). The SURF schemes are shown in Figs.~\ref{fig:field3} and \ref{fig:field4} in the Appendix. Isochrones with grey lines indicate the time evolution from ZAMS to 50 Myr. The initial masses in solar units are given next to the ZAMS of the models (between 6 and 25M$_\odot$, the increment is 1~M$_\odot$). The colour-bar shows how the equatorial rotational velocity evolves.}\label{fig:hrd1}
\end{figure*}
\subsubsection{HRD evolution of a grid of magnetic models}\label{sec:hrdevol}

Figure \ref{fig:hrd1} shows the model predictions on the HRD colour-coded by the surface equatorial rotational velocity. Here the INT/Mix1 and INT/Mix2 schemes are displayed (see Figs.~\ref{fig:field3} and \ref{fig:field4} in the Appendix for the SURF schemes) and we demonstrate the impact of a magnetic field with an initial equatorial field strength of 500~G.

In the INT models, strong magnetic braking leads to a rapid decrease of surface rotational velocity in the entire mass range from 3 to 60~M$_\odot$. This implies that rapidly-rotating (single) magnetic massive stars are expected to be young and close to the ZAMS on the HRD. 
Some quantitative differences arise from the assumptions of the chemical mixing schemes. Nonetheless, within the INT magnetic braking scheme these differences are small on the HRD, albeit it could affect the parameter determination (current age and mass) of known magnetic stars.

%
%
%
%
%
%
%
%
\begin{figure*}
\includegraphics[width=8cm]{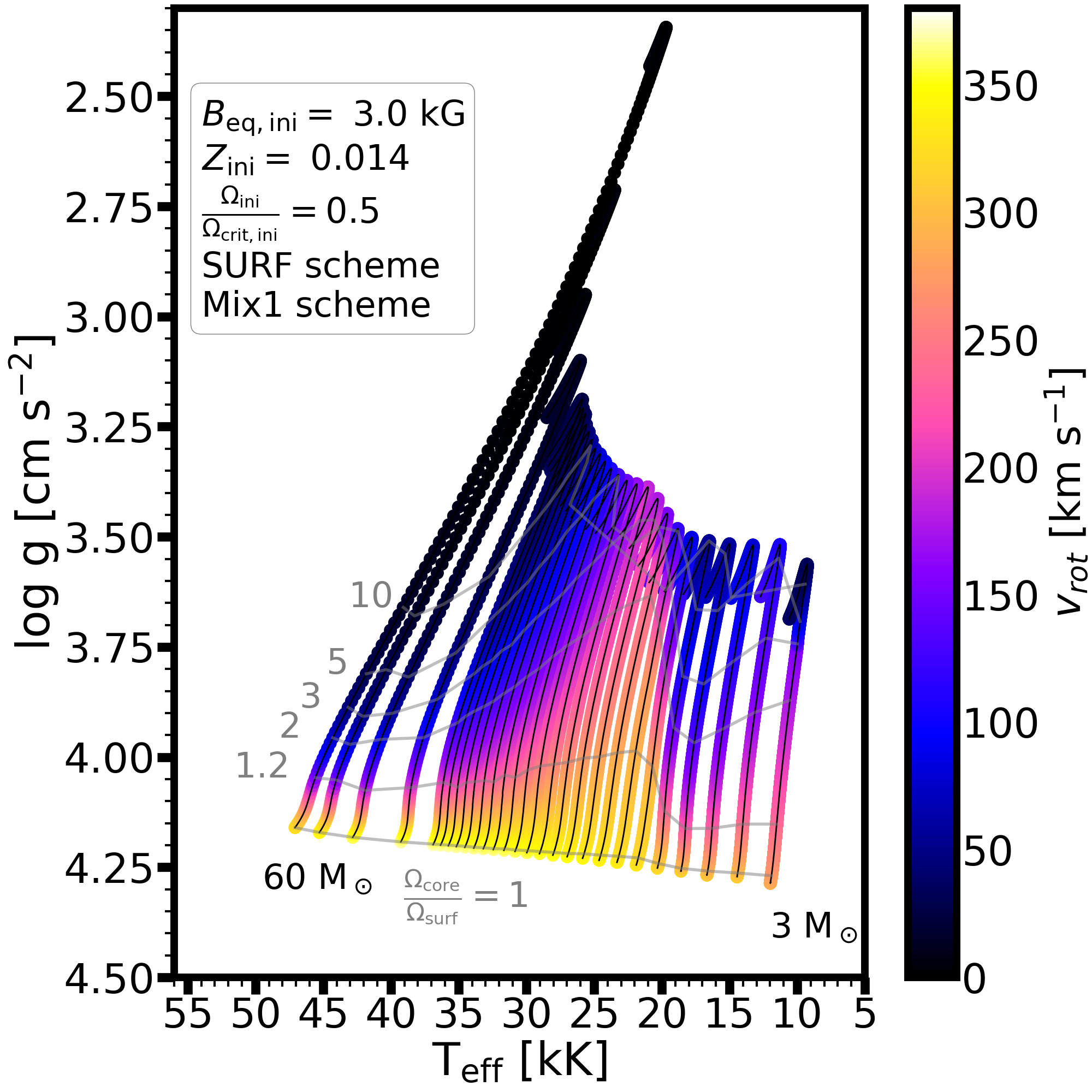}\includegraphics[width=8cm]{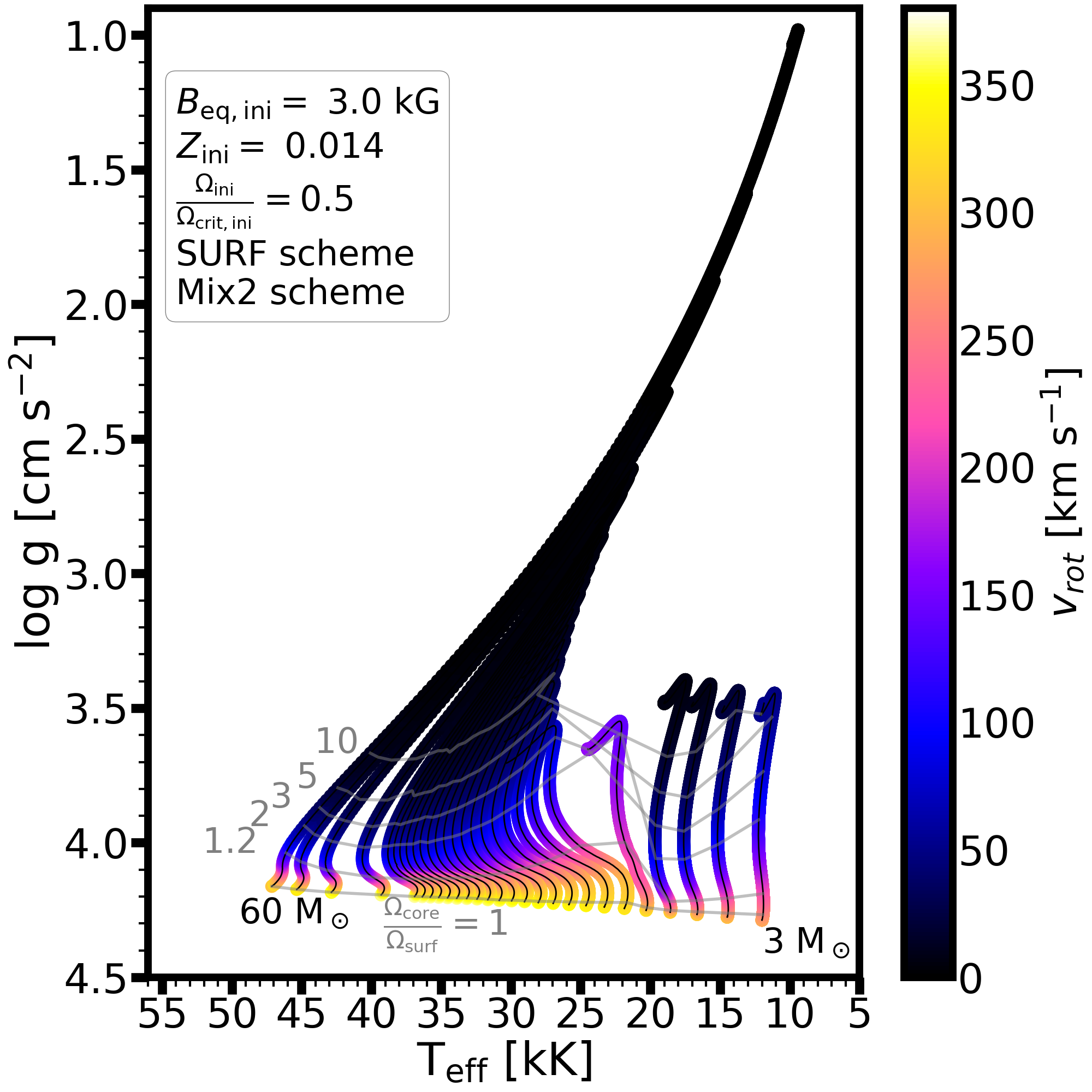}
\caption{Kiel diagram of magnetic evolutionary models with $B_{\rm eq, ini}= 3$~kG, $\Omega_{\rm ini}/\Omega_{\rm ini, crit}= 0.5$ at solar metallicity $Z_{\rm ini}= 0.014$. within the SURF/Mix1 (left) and SURF/Mix2 (right) schemes. Grey contour lines indicate the unitless degree of differential rotation, quantified as the ratio of core to surface angular velocity. Note that Figure~\ref{fig:hrd1} shows INT models where solid-body rotation ($\Omega_{\rm core}/\Omega_{\rm surf}=1$) is achieved throughout the main sequence (see also Figure~\ref{fig:app1a}). The initial masses of the models decrease from left to right. The colour-coding shows the surface equatorial rotational velocity.}\label{fig:kiel1}
\end{figure*}

\subsubsection{Differential rotation on the Kiel diagram}\label{sec:kielevol}

Figure \ref{fig:kiel1} shows the Kiel diagram. In the SURF models, radial differential rotation develops between the stellar core\footnote{Here, we use the \textsc{mesa} output quantity of core angular velocity "\texttt{center\_omega}" to determine the rotation of the stellar core. The exact radius to obtain $\Omega_{\rm c}$ is not essential since the entire stellar core has the same angular velocity.} and surface, as indicated with the grey contour lines. This is because we assume that a magnetic field with a complex geometry would only exert a strong torque on the near-surface layers by considering that organised, strong magnetic flux is not present in deeper stellar layers (see also \citealt{braithwaite2006,braithwaite2008}). Thus angular momentum transport in the deepest layers of the stellar envelope is dominated by -- less efficient -- hydrodynamical instabilities. On the other hand, in the INT models complete solid-body rotation is maintained throughout the main sequence. 
Similar to Figure~\ref{fig:schemeevol1}, where the 20 M$_\odot$ models are discussed, we see that the Mix1 and Mix2 chemical mixing schemes (within the SURF braking scheme here) result in notable differences in observable stellar parameters, which are the most prominent in the 5-10 M$_\odot$ range.

If the assumed magnetic field geometry remained unchanged throughout the main-sequence evolution, the prediction for magnetic massive stars is that differential rotation could be best identified in hotter, more massive, and more evolved (lower $\log g$) stars. An important caveat nevertheless is the time evolution of complex magnetic fields. According to \cite{braithwaite2008}, complex fields may simplify to a dipolar form since higher-order harmonics diffuse more rapidly. Indeed, \cite{shultz2019b} finds that evolved magnetic stars tend to have simpler geometries (however, see also \citealt{kochukhov2019}). Insofar it remains unclear what the exact diffusion time of complex magnetic flux tubes would be (perhaps still longer than the main-sequence lifetime) and whether the corresponding effects which are modelled in evolutionary codes (angular momentum transport, magnetic braking, mass-loss quenching) would change significantly. 
We evaluated the models with different initial field strengths and found that a stronger magnetic field is able to achieve a higher degree of differential rotation in comparison to models with lower field strengths. It would therefore be of great importance to obtain seismic data of a sample of magnetic massive stars. 

%
%
%
%
\subsubsection{Mass-dependent rotational evolution}\label{sec:feature}

Both the HRD and Kiel diagrams presented above\footnote{See Figs.~\ref{fig:field3}-\ref{fig:field4} in the Appendix for the other schemes.} show the same distinctive feature. Namely, irrespective of how fast the spin-down per given scheme is, the mid-mass range models ($\approx$ 5-10 M$_\odot$, the transition typically taking place at around 5-7 M$_\odot$ depending on model assumptions) always maintain a higher rotational velocity than models with other masses. This is the most striking on the left panel of Figure~\ref{fig:kiel1} for the SURF/Mix1 scheme. (The exact model behaviour was also recognised for a 5 and 10 M$_\odot$ model in \PaperII.)
The distinctive feature is a consequence of the spin-down of the models, which does not scale linearly with mass. For higher-mass models, wind mass loss contributes to the spin-down. 
The spin-down time to reach a given surface rotation takes a lower fractional age when the mass increases for initial masses higher than 10~M$_\odot$.
Since stars below about 5~M$_\odot$ have much longer nuclear timescales than higher-mass stars, even if magnetic braking is less efficient due to weaker winds, the available timescale allows for braking the rotation at a lower fractional main-sequence age. This results in stars below 5~M$_\odot$ rotating more slowly at the TAMS than in the 5-10~M$_\odot$ range. 
We emphasise that the extent of the tracks on the HRD is not linear with stellar age: a significant time evolution can take place in a narrow location on the HRD close to the ZAMS as demonstrated by the isochrones. 
For example, the 50~M$_\odot$ model spends the first 2 Myr of its evolution while decreasing its $T_{\rm eff}$ by only about 5 kK, whereas in the second 2 Myr of its evolution, its $T_{\rm eff}$ decreases by about 15 kK (Figure~\ref{fig:hrd1}). Moreover, from 0 to 4 Myr, the 25~M$_\odot$ track evolves roughly 0.2 dex in luminosity and a few kK in $T_{\rm eff}$, which is a typical range of observational uncertainties depending on data quality and knowledge of distance and extinction. The precise age determination of young stars especially at initial masses below 25~M$_\odot$ becomes rather challenging with uncertainties well exceeding 1~Myr. 

In summary, stars evolve slowly in the HRD at the beginning of the main sequence phase, and thus suffer strong magnetic braking while not evolving away from the ZAMS (in effective temperature and luminosity). However the interplay between the dependence on the initial mass of the meridional current velocity, the evolutionary timescale near the ZAMS and the evolution of the radius produces a small bump of the surface rotational velocity in the initial mass range of 5-10 M$_\odot$ (see also \PaperII). At the TAMS, this gives rise to a slower rotational velocity in models below 5~M$_\odot$ than models in the mid-mass range.

%
%
\begin{figure*}
\includegraphics[width=6cm]{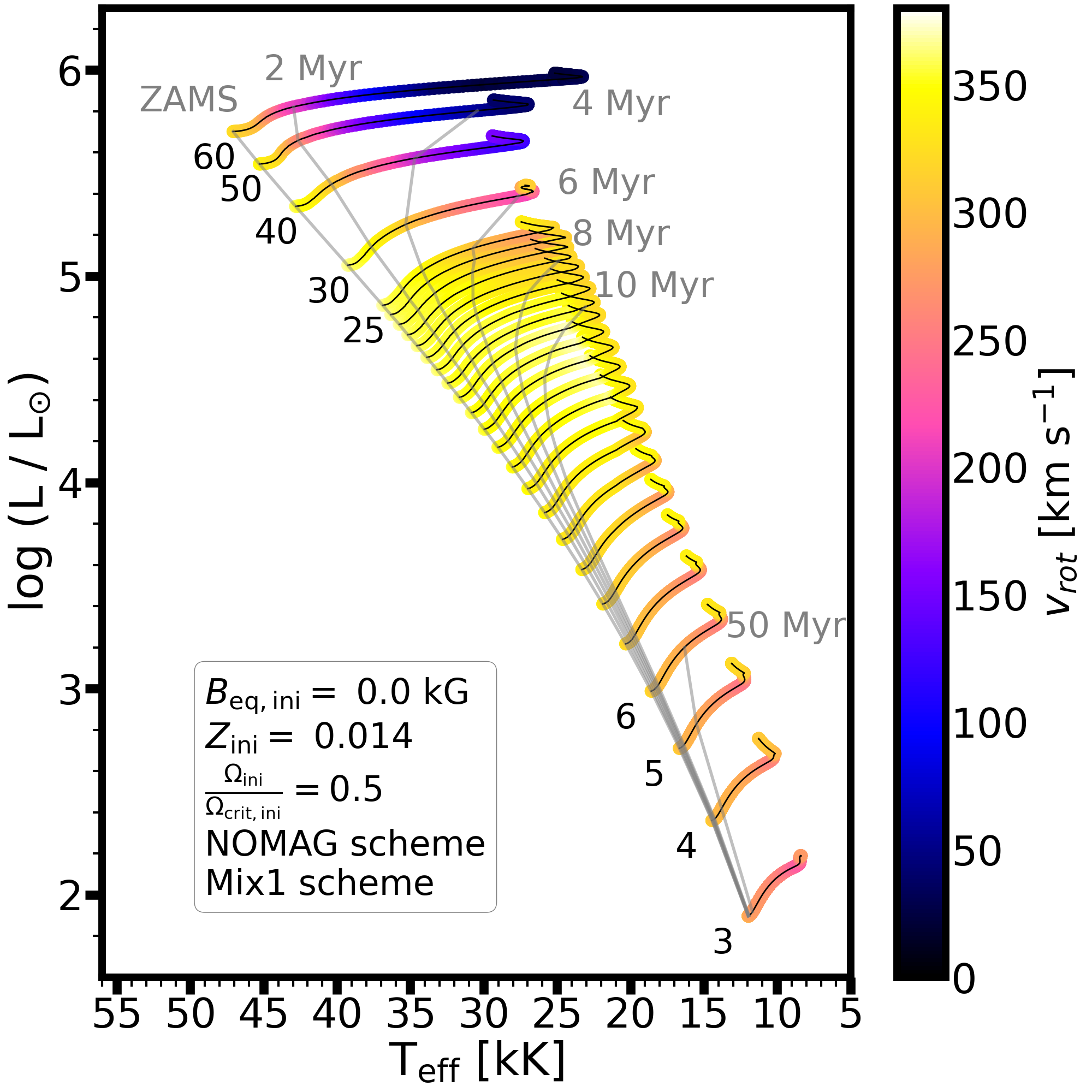}\includegraphics[width=6cm]{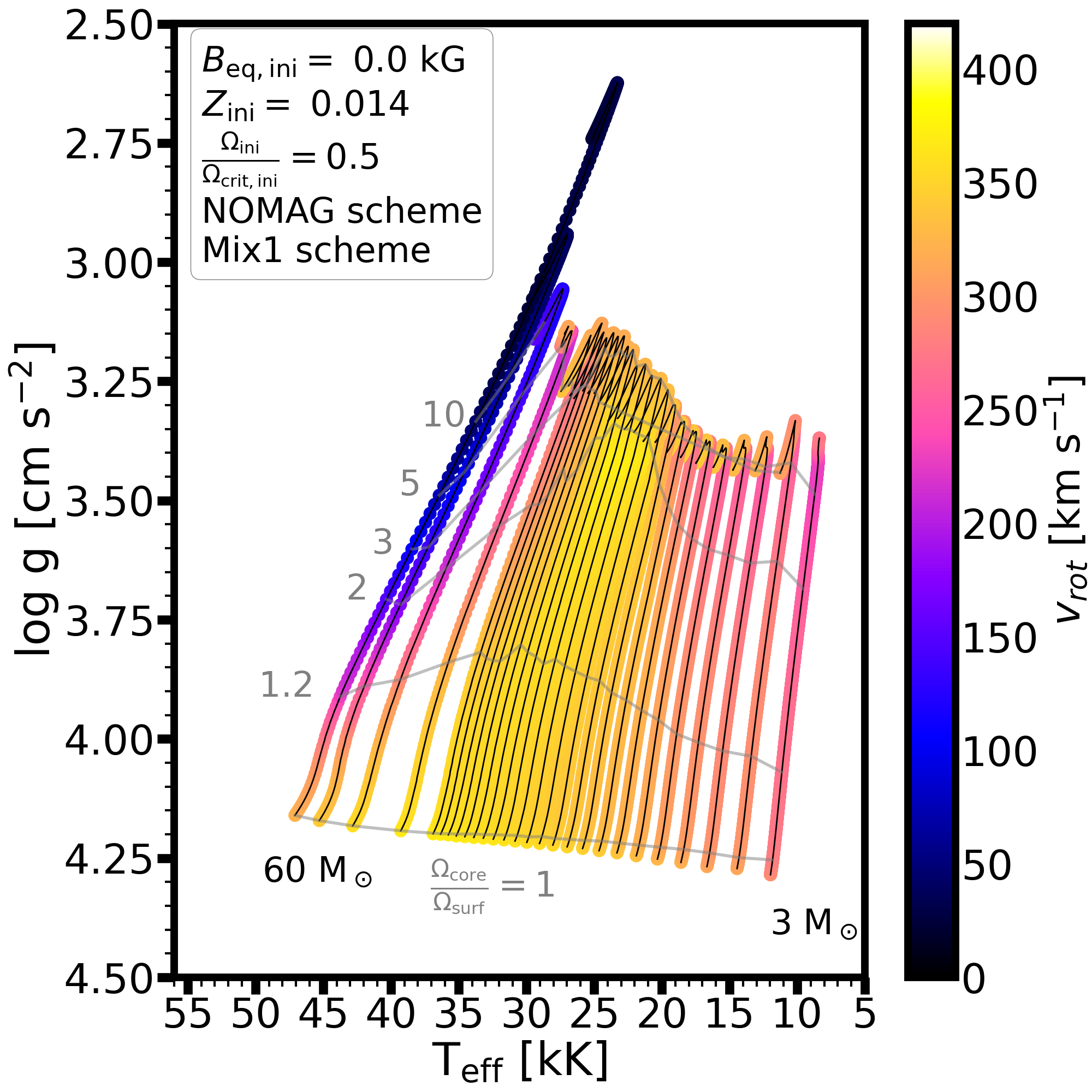}\includegraphics[width=6cm]{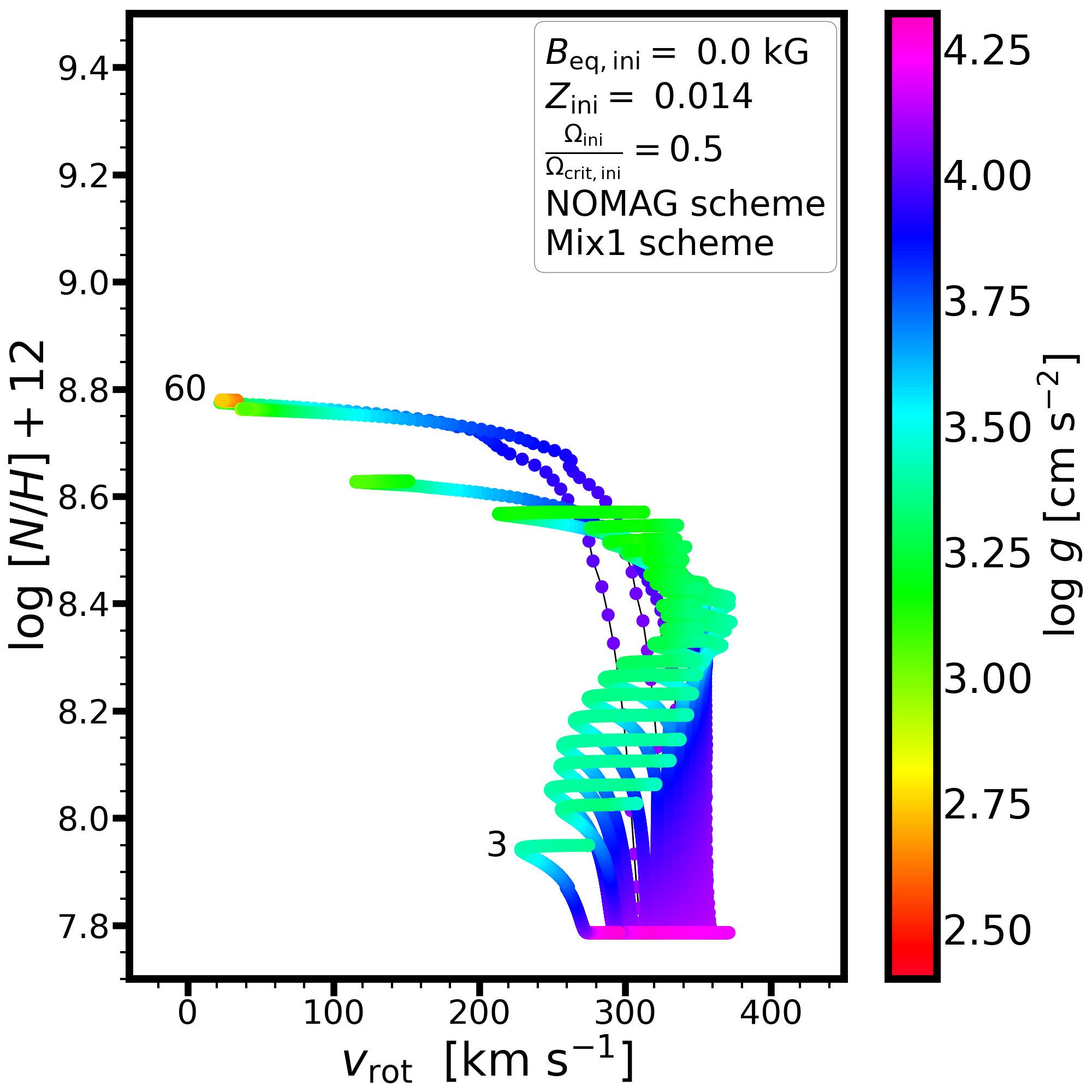} 
\includegraphics[width=6cm]{fig/Z14INT/Mix1/hrd500.png}\includegraphics[width=6cm]{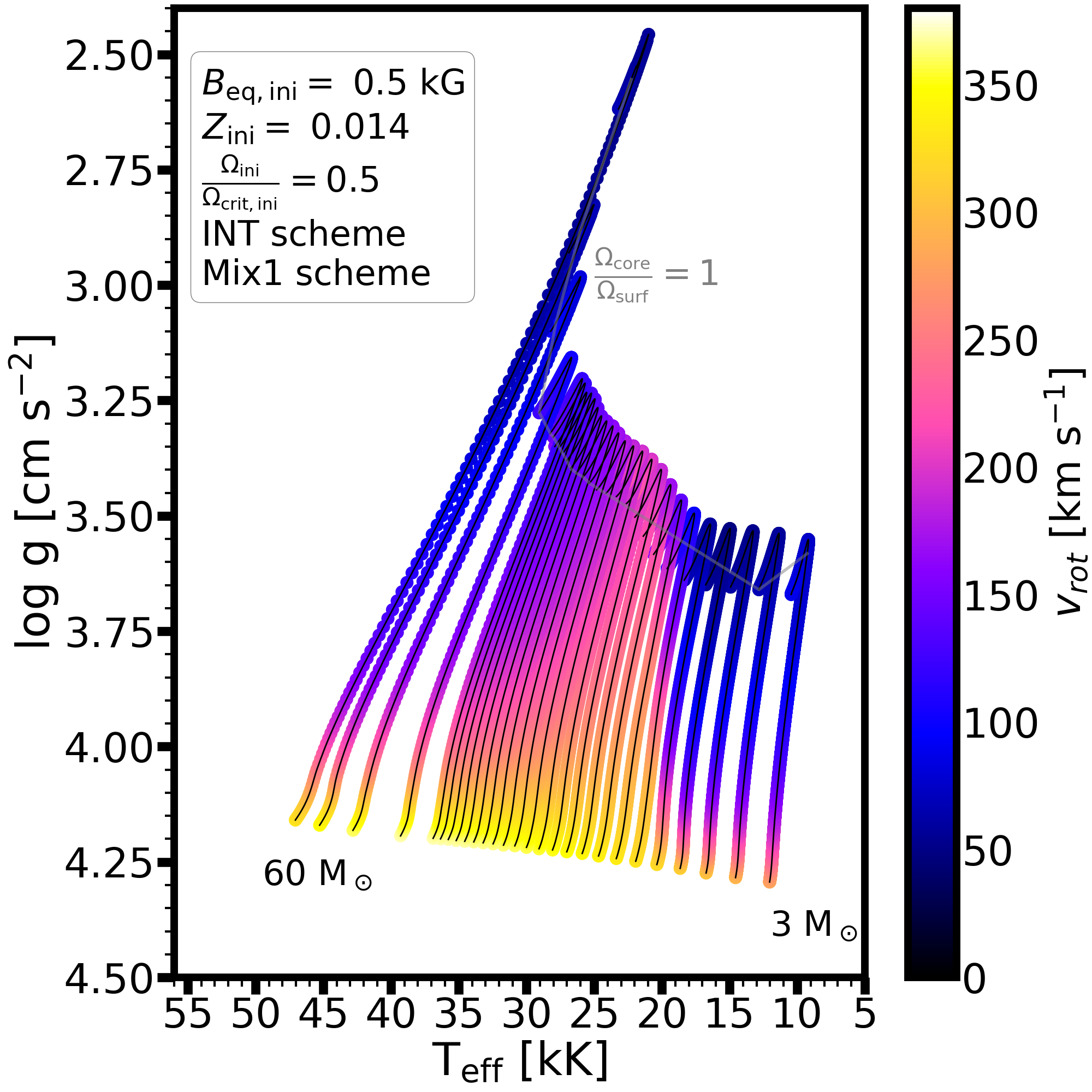}\includegraphics[width=6cm]{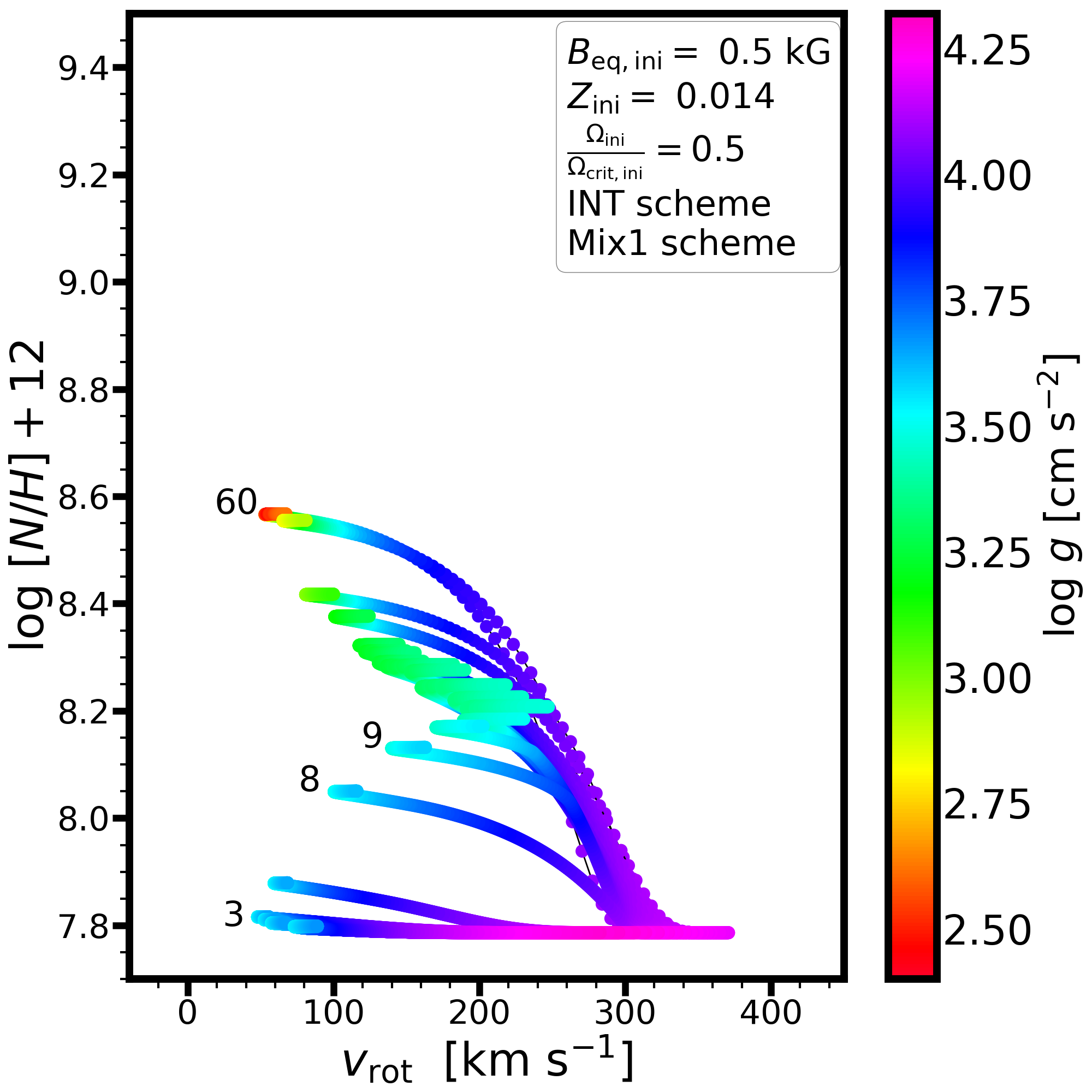}
\includegraphics[width=6cm]{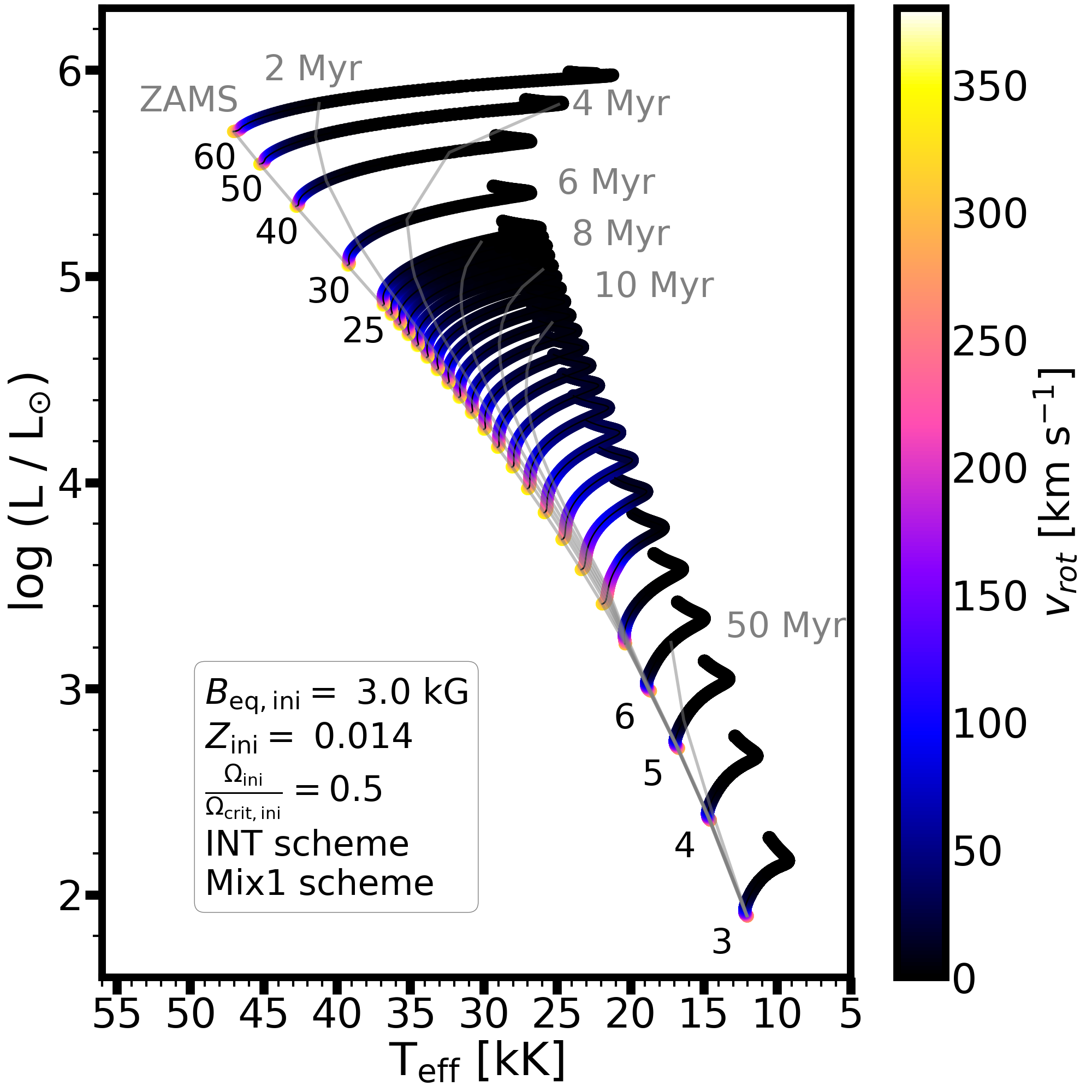}\includegraphics[width=6cm]{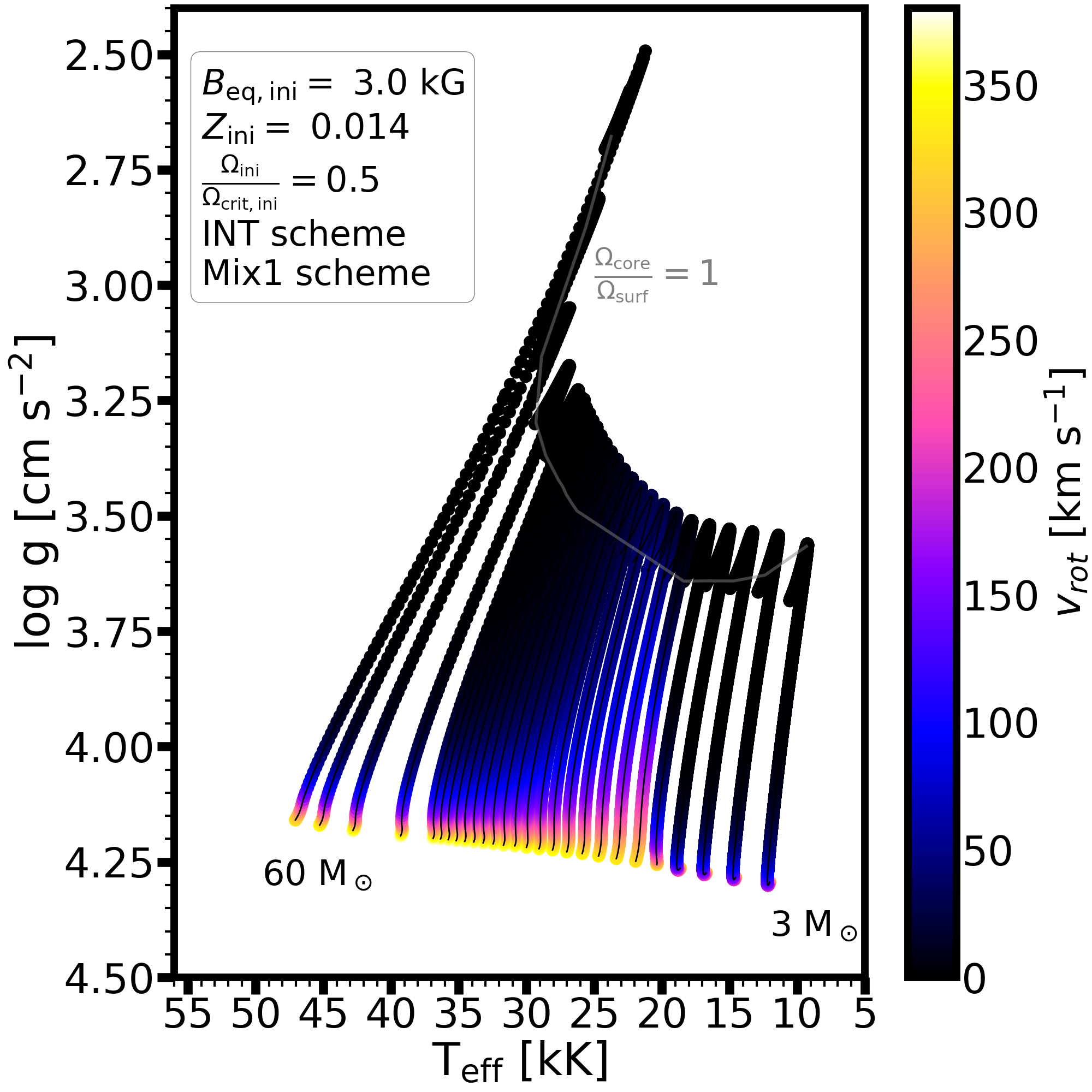}\includegraphics[width=6cm]{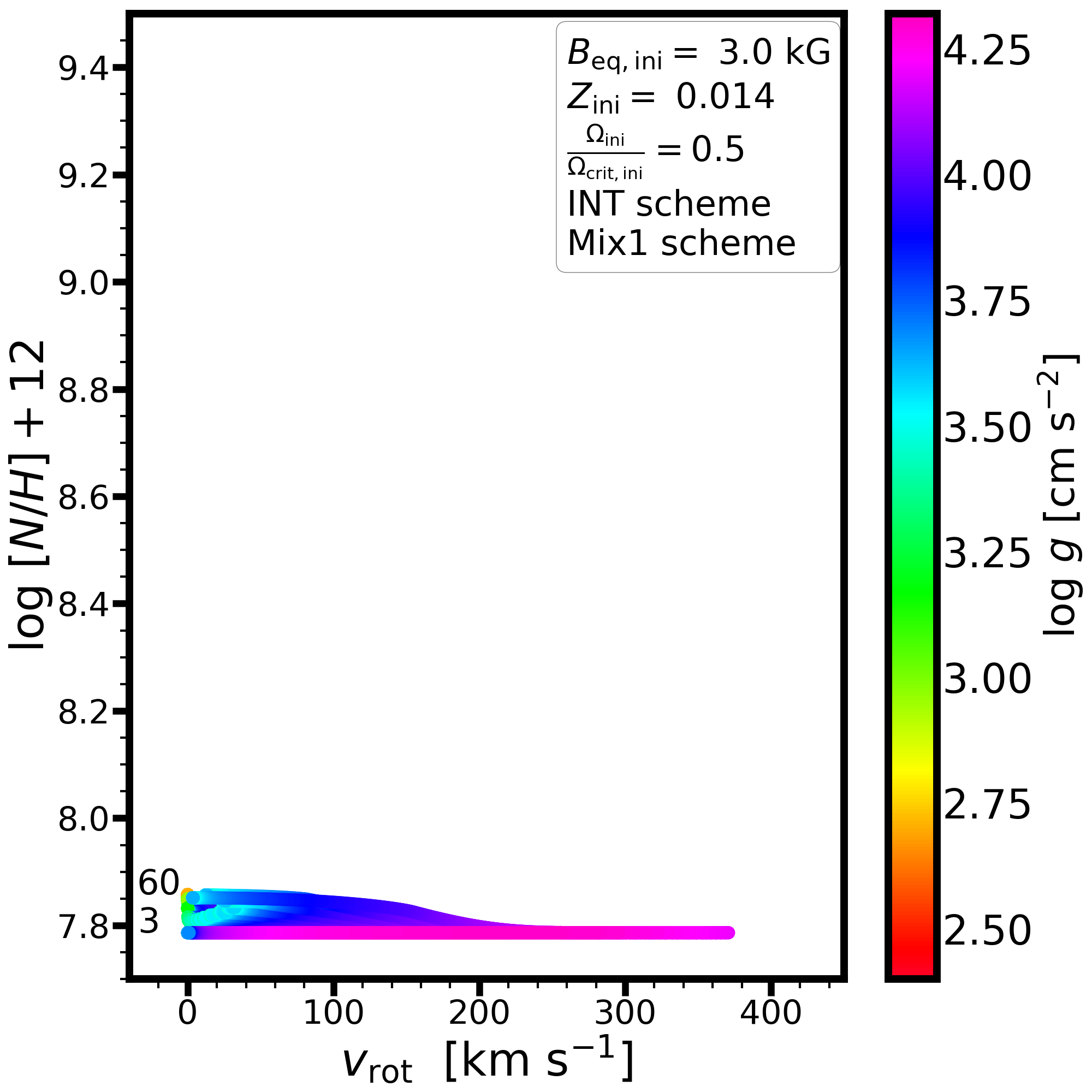}
\caption{Evolutionary models from 3 to 60 M$_\odot$ at $Z = 0.014$ with varying the initial equatorial magnetic field strength (0, 0.5, 3 kG from top to bottom), within the INT/Mix1 magnetic braking and chemical mixing schemes. Panels from left to right show the HRD, Kiel, and Hunter diagrams. The colour-coding denotes surface rotational velocity on the first two panels, while it denotes the logarithmic surface gravity on the right panel.}\label{fig:bfield1}
\end{figure*}
%
\subsubsection{Impact of varying the initial equatorial magnetic field strength}\label{sec:fieldevol}

Figure \ref{fig:bfield1} shows the impact of the initial equatorial magnetic field strength within the INT/Mix1 scheme on the HRD, Kiel, and Hunter diagrams for models from 3 to 60~M$_\odot$. 
%
The colour-coding of the surface rotational velocity on the HRD and Kiel diagrams show that for initial masses above 30~M$_\odot$, stellar winds play a significant role in depleting the angular momentum reservoir, and thus even without magnetic fields those stars can significantly spin down on the main sequence. However, in the mass range from 3 to 30~M$_\odot$ single-star models would not undergo a dramatic angular momentum loss unless they were strongly magnetised. As demonstrated in the figure, the stronger the magnetic field, the more rapidly the surface rotation brakes. In particular, already a 3 kG equatorial field would produce a (sub)population of stars whose surface rotation is less than 50~km~s$^{-1}$ throughout essentially the entire main sequence.

The Kiel diagram shows yet another consequence of magnetic fields. Apart from the highest-mass models ($>$~30M$_\odot$), the models with stronger magnetic fields tend to reach the end of the main sequence with higher surface gravities. For example, for a 10~M$_\odot$ model, the TAMS value of $\log g $ increases from 3.3 to 3.4 and to 3.5 from the 0 kG to the 0.5 kG and to the 3 kG models, respectively. Non-magnetic models maintain a higher rotation and thus the mixing remains more efficient. This mixing (if not too strong to keep the star in a bluer position in the HRD) tends to enlarge the convective core and consequently, extend the width of the main sequence towards lower effective temperatures and higher surface gravities.  

The Hunter diagram reveals that the magnetic models may strongly deviate from non-magnetic model predictions. The stronger the magnetic field, the more rapidly rotation brakes, and the less nitrogen can be mixed to the stellar surface. We identify and demonstrate that there exists a cutoff magnetic field strength, above which no surface enrichment is expected given that it leads to a shorter magnetic braking timescale compared to the rotational mixing timescale. However, this cutoff field strength is strongly model and parameter dependent, and thus the exact value varies for given stellar mass, initial rotation, metallicity, and mixing scheme, amongst others. We evaluate this in Section~\ref{sec:cutoff}.

%
%
\begin{figure*}
\includegraphics[width=6cm]{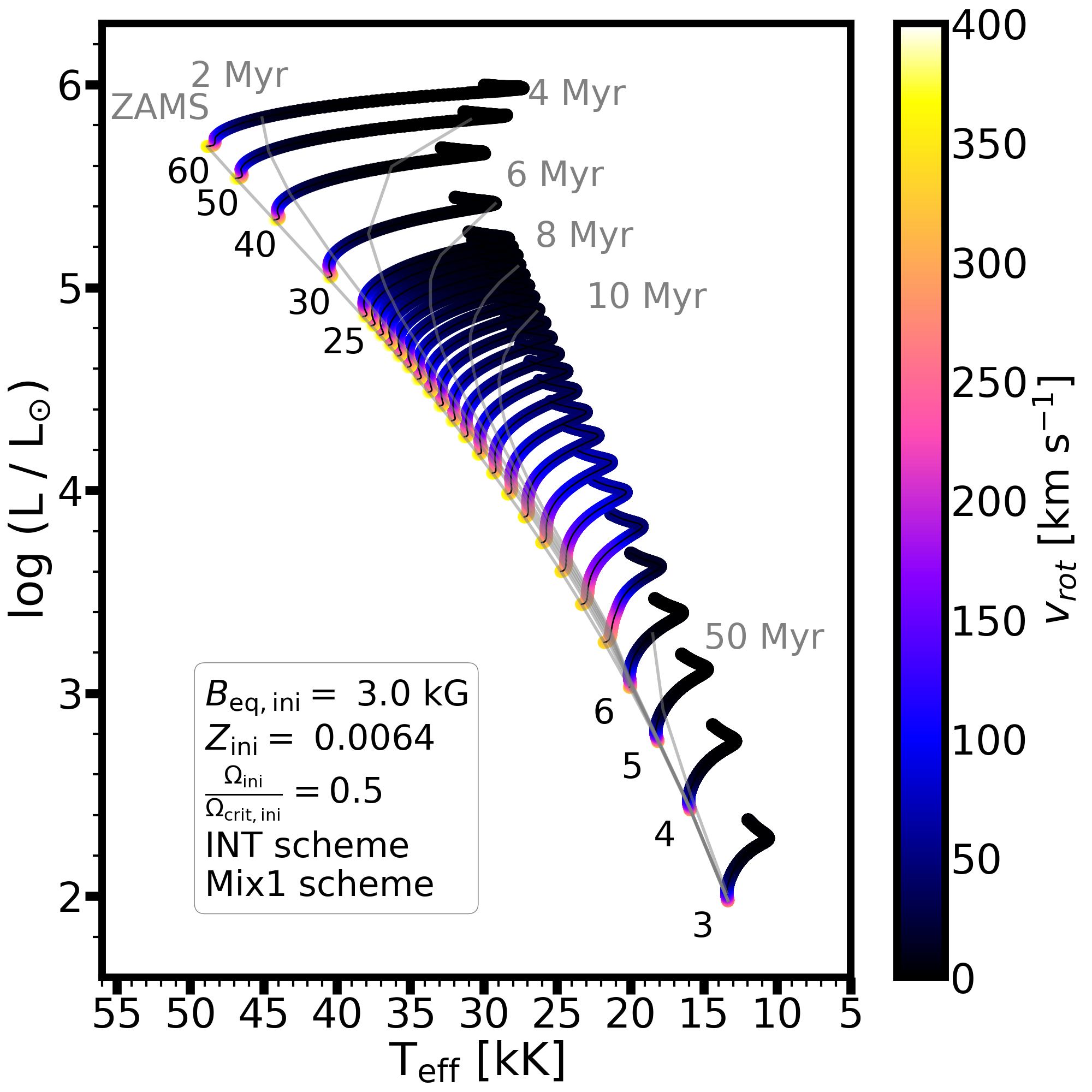}\includegraphics[width=6cm]{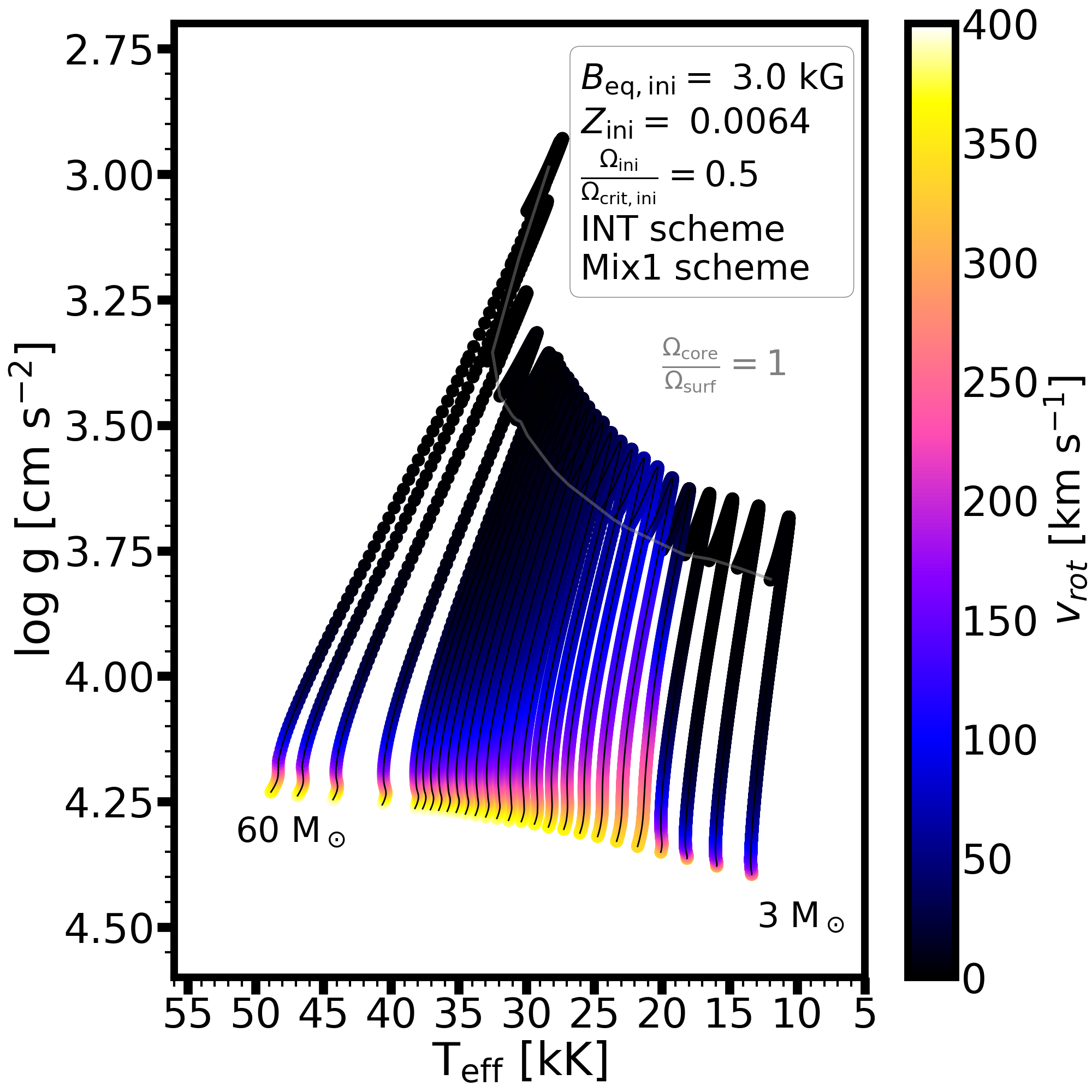}\includegraphics[width=6cm]{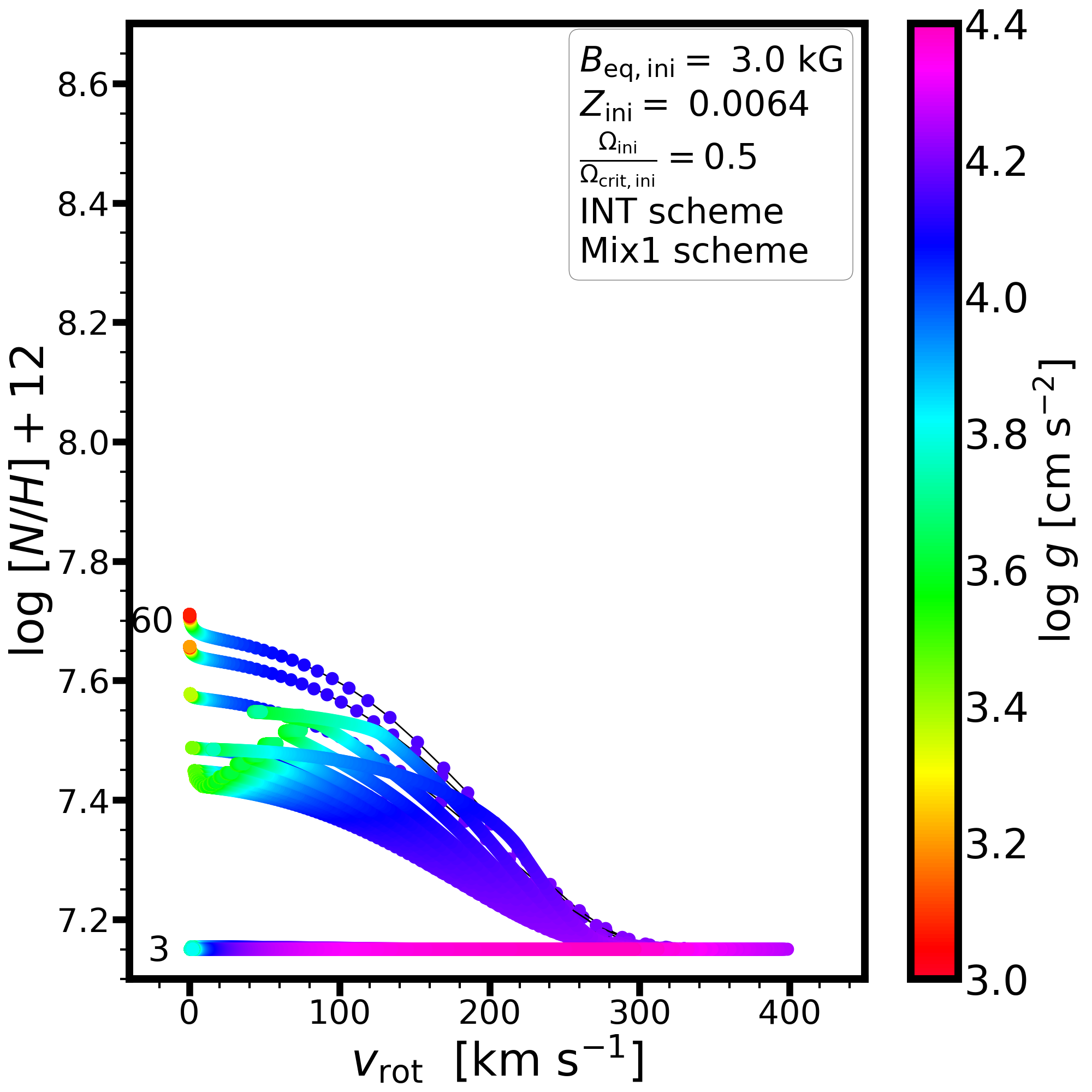}
\includegraphics[width=6cm]{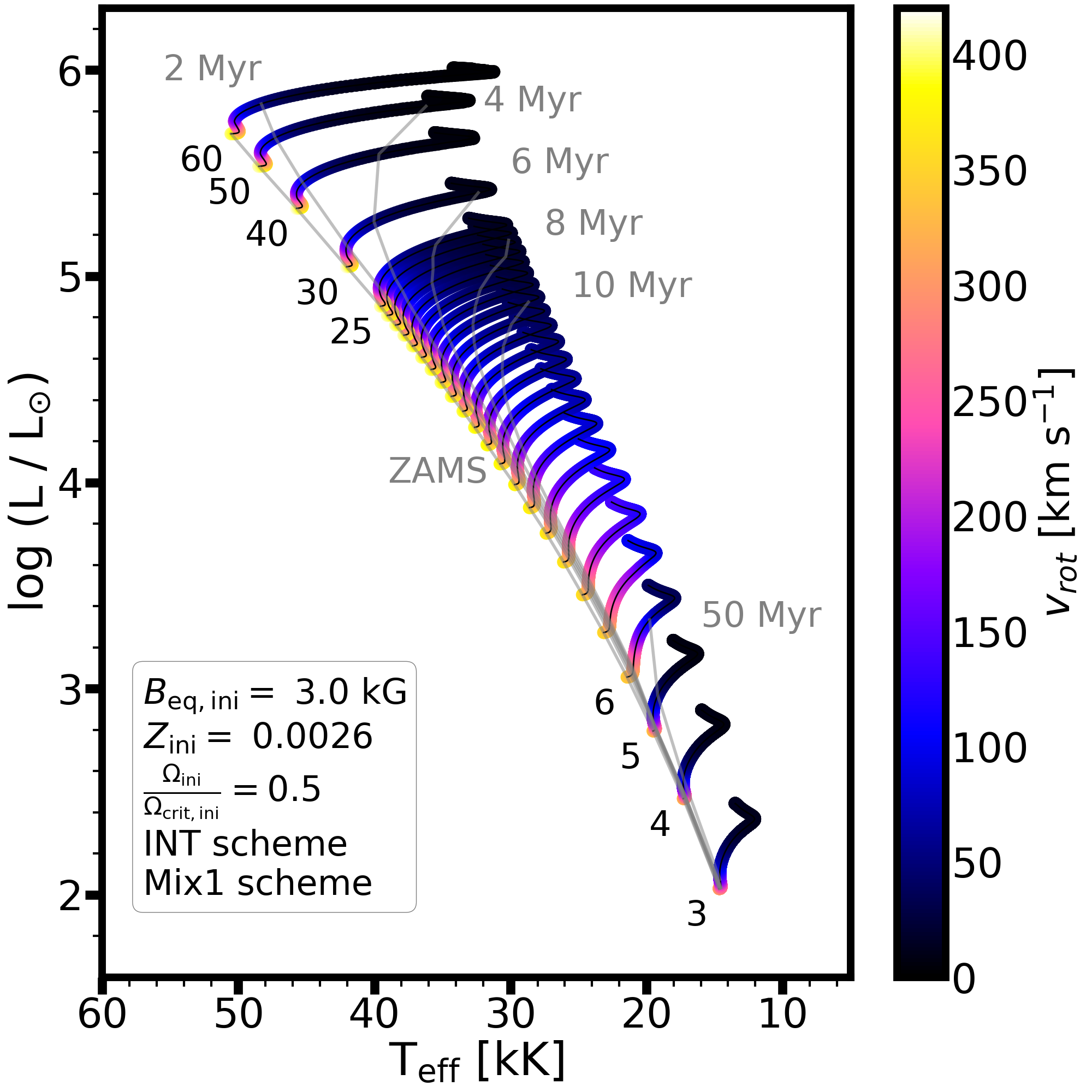}\includegraphics[width=6cm]{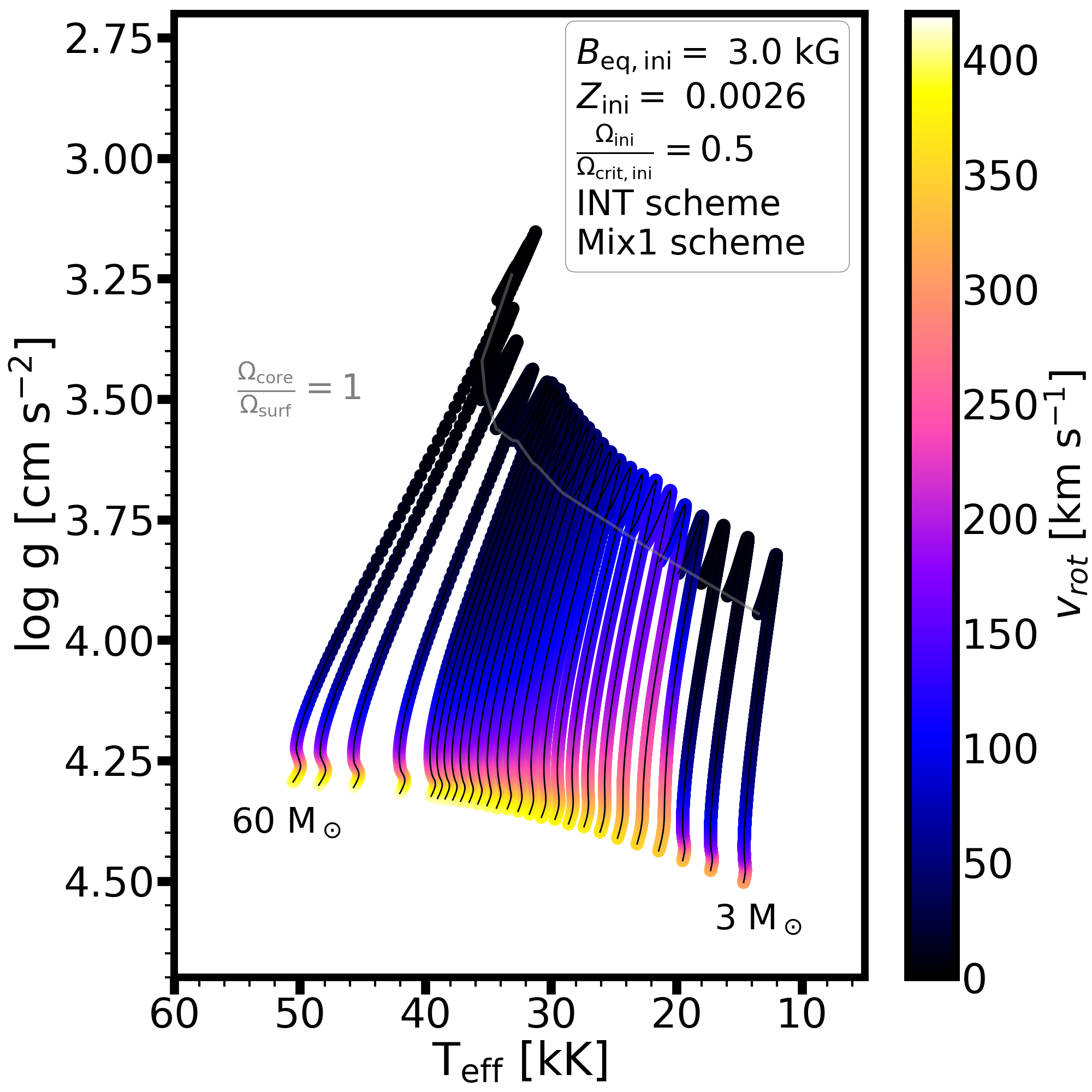}\includegraphics[width=6cm]{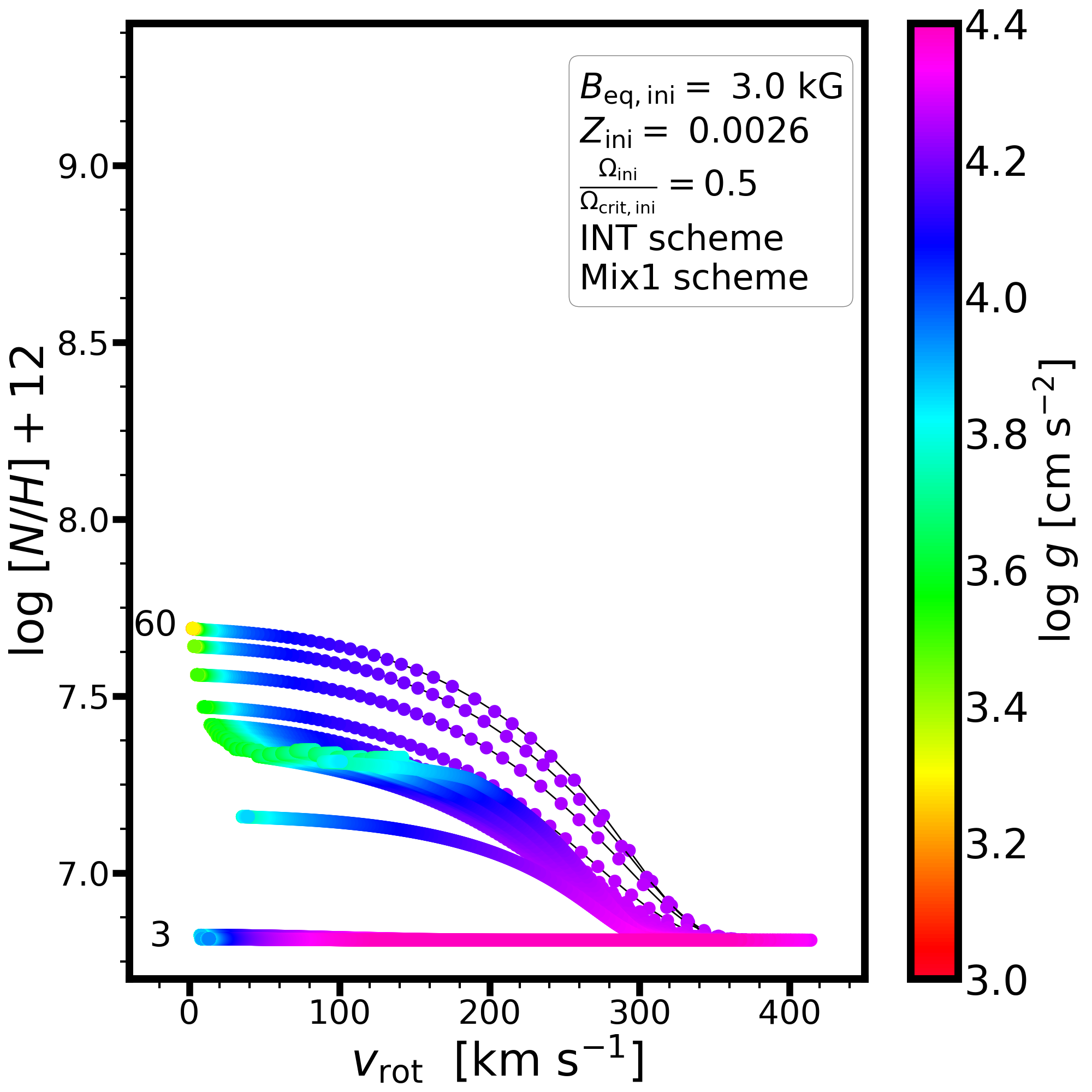}
\caption{Same as Figure~\ref{fig:bfield1} but varying the metallicity (upper - LMC, lower - SMC), within the INT/Mix1 scheme for an initial equatorial magnetic field strength~of~3~kG. }\label{fig:metale1}
\end{figure*}
\subsubsection{Impact of metallicity}\label{sec:metevol}

Figure \ref{fig:metale1} shows the impact of the initial metallicity (here, for LMC and SMC values; the Solar metallicity models with the same input are shown in the lower panels of Figure \ref{fig:bfield1}) for 3 to 60~M$_\odot$ models with initial equatorial magnetic field strengths of 3~kG within the INT/Mix1 scheme on the HRD, Kiel, and Hunter diagrams.
%
At lower metallicity, the ZAMS is shifted to higher effective temperatures given that the stellar models are more compact due to the lower opacity and lower mean molecular weight. Specifically, the lower CNO abundances impose further contraction of a star to initiate core burning.

Magnetic braking, in our formalism, is metallicity independent. However, rotational mixing is not. The various mixing prescriptions depend on chemical composition and their gradients, which in turn affects the evolution of surface rotational velocity as most prominently revealed on the Kiel diagrams. For example, for a given value of a diffusion coefficient, the mixing timescale is $\tau \approx R^2/D_{\rm chem}$. Since stars are more compact in lower $Z$, the timescale becomes shorter. Consequently, the changes in rotation and surface abundances can be more impacted in lower metallicity stars. 

Similarly, the highest relative nitrogen enrichment is seen when metallicity is the lowest (see also e.g. \citealt[][]{brott2011,georgy2013}). Let us recall that Figure~\ref{fig:metale1} shows the INT/Mix1 models, which -- in our approach -- are the lowest estimates for the surface enrichment (c.f. Figure~\ref{fig:schemeevol1}). The other schemes predict higher surface nitrogen enrichment when combining the effects with low metallicity.
The trends produced in magnetic massive star models are unique since they lead to simultaneous surface nitrogen enrichment and a rapid spin-down of the stellar surface (c.f. \PaperI, \PaperII). The rapid spin-down could otherwise only be expected for very massive stars with extremely strong stellar winds.

%
%
%
%
\section{Discussion}\label{sec:four}

%
%
%
%
%
%
%
%
\begin{figure*}
\includegraphics[width=18cm]{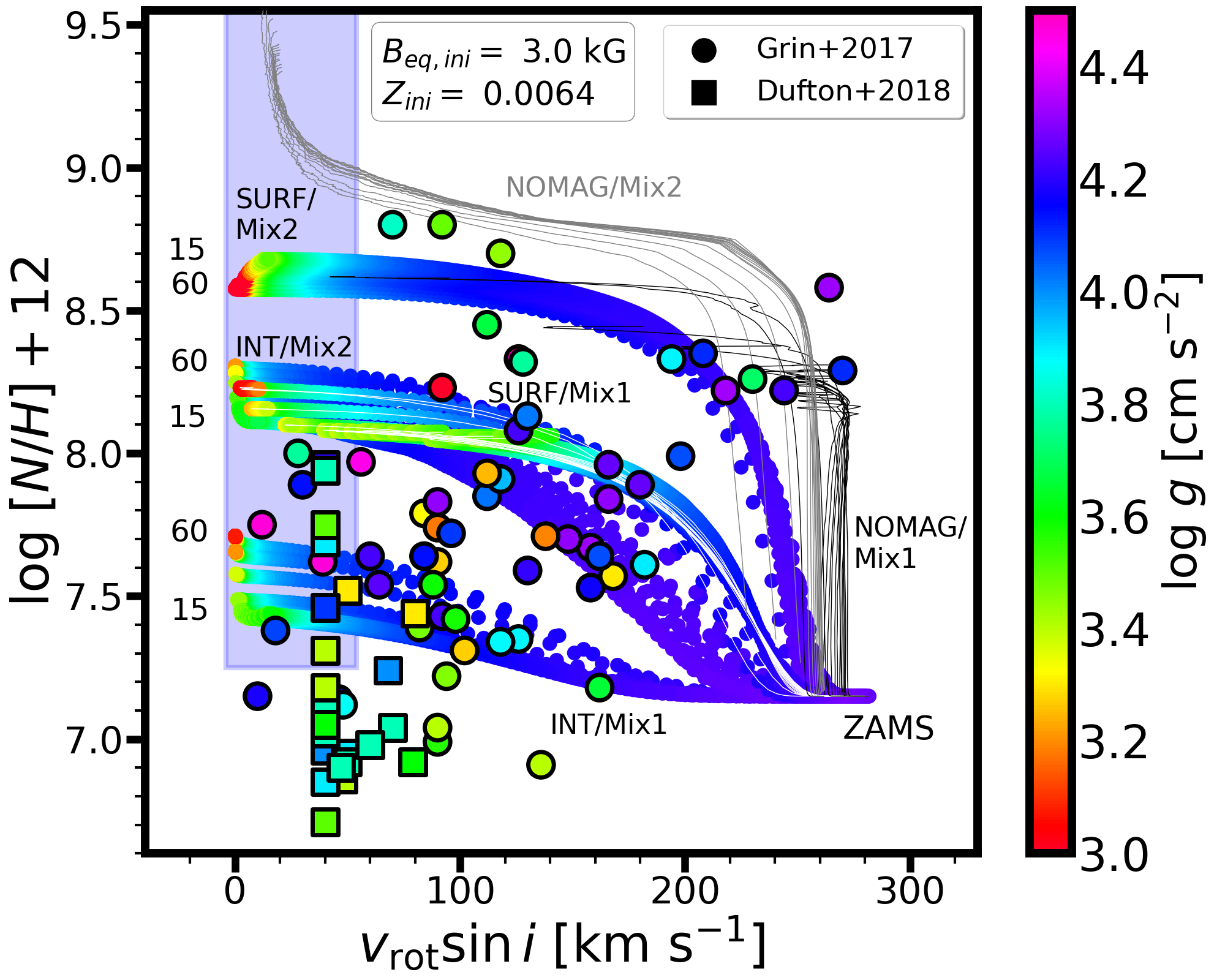}
\caption{Hunter diagram of magnetic single-star evolutionary models with $B_{\rm eq, ini}= 3$~kG, $\Omega_{\rm ini}/\Omega_{\rm ini, crit}= 0.5$ at LMC ($Z_{\rm ini}= 0.0064$) metallicity within the four schemes. Models within the SURF/Mix1 scheme are also shown with white lines since they overlap with the INT/Mix2 models. Additionally, two sets of non-magnetic (NOMAG) models are shown within the Mix1 (grey) and Mix2 (black) schemes. For visualisation purposes, we reduced the numerical noise in the latter case. Models with initial masses from 15 to 60~M$_\odot$ are shown. The actual surface equatorial rotational velocity of the models is scaled by $\sin (\pi/4)$ to account for an average inclination angle. The coloured area corresponds to our definition of Group 2 stars. The colour-coding of the models shows the logarithmic surface gravity. Observations are shown with circles and squares, respectively. A typical reported uncertainty in the observed nitrogen abundances is about 0.1 dex.}\label{fig:hunterobs1}
\end{figure*}

\subsection{Slowly-rotating nitrogen enriched stars in the LMC}\label{sec:huntobs}

The projected rotational velocities of massive stars in the Magellanic Clouds appear to follow a bi-modal distribution \citep[e.g.,][]{dufton2013,dufton2018,dufton2019,dufton2020,riverogonzalez2012,ramireza2013,ramireza2015}.
The bi-modality is also observed for intermediate-mass stars up to about 5~M$_\odot$ \citep[e.g.,][]{bastian2020,sun2021}. The observed slowly-rotating red main sequence stars and rapidly-rotating blue main sequence stars are thought to be evidence for main sequence splitting \citep[e.g.,][]{bastian2020} and an extended main sequence turn-off \citep[e.g.,][]{dantona2015}.
It has been suggested that the low-velocity peak might be caused by magnetic braking \citep[e.g.,][]{wolff1982,sun2021}.
\cite{shultz2018} demonstrated that the dichotomy in $v \sin i$ between Galactic B-type stars with and without magnetic fields is at least qualitatively consistent with the lower $v \sin i$ values observed in the magnetic population.
For observed massive stars in the Magellanic Clouds, a notable fraction of slow-rotators were  found to show measurable nitrogen enrichment, which challenges typical, non-magnetic single-star evolutionary models \cite[][]{lennon1996,lennon2003,dufton2006,dufton2013,dufton2018,dufton2019,dufton2020,riverogonzalez2012,ramireza2013,ramireza2015,mcevoy2015,grin2017}. 

The nitrogen-enriched slow-rotators (also known as "Group 2" stars, \citealt[][]{hunter2008}) correspond to roughly 20\% of the population in the LMC \citep[e.g.,][]{hunter2008,brott2011b,grin2017,dufton2018}. In the Galaxy, the observed incidence rate of fossil magnetism is found to be $\approx 10\%$ \citep[e.g.,][]{fossati2016,grunhut2017,sikora2019} and it has previously been suggested that at least some of the Group~2 stars could be explained by magnetism (\citealt[][]{meynet2011,potter2012b}, \PaperI). This would require an incidence rate of fossil magnetism in the LMC that is likely higher than the 10\% observed in the Galaxy\footnote{For example, lower metallicity environments might favour a higher incidence rate of stars with fossil fields if the convective expulsion scenario, due to the sub-surface iron opacity bump, regulates the incidence rate \citep[][]{jermyn2020}.}. In addition, the (initial) magnetic field strength distribution is not yet known in our galaxy or in other metallicity environments; however, see \cite{petit2019} and \cite{pinar2020} for theoretical models. 
It could also be that the Group~2 stars require an additional channel to explain all observations. In fact, binarity has been suggested by, e.g., \cite{song2018b}. 

Figure~\ref{fig:hunterobs1} shows the Hunter diagram with models representative of LMC metallicity ($Z=0.0064$). Let us recall that we specifically adopted an initial nitrogen abundance in our models of $\log (N/H) + 12 = 7.15$ from \cite{dopita2019} to produce evolutionary models guided by available empirical baseline abundance determinations.
Here we demonstrate some of the complex parameter-space dependences that magnetic single-star models produce, albeit strongly depending on the model assumptions, especially the mixing and braking schemes (see also \citealt{meynet2011,potter2012b}, \PaperI, \PaperIII). We display the non-magnetic models as well as magnetic models in the two braking and two mixing schemes.
The initial equatorial magnetic field strength is 3 kG and the initial rotation is set by $\Omega_{\rm ini}/\Omega_{\rm ini, crit}= 0.5$. These assumptions produce evolutionary models, which over time reasonably approximate mean values measured from observations (magnetic field strengths from, e.g., \citealt{shultz2018} for Galactic magnetic B-type stars, and rotational velocities from, e.g., \citealt{dufton2013} for massive stars in the Magellanic Clouds). For example, the 3 kG initial ZAMS magnetic field strength weakens by roughly an order of magnitude (since $B_{\rm eq} \propto R_\star^{-2}$ over time) at the TAMS to 300~G. Only models with initial masses from 15 to 60~M$_\odot$ are shown given that the available instrumentation allows magnitude limited observation of bright LMC stars that are more massive than $\approx$~15~M$_\odot$ \citep[e.g.,][]{schneider2018}. Thus the models are the most representative of O (and early B-type) stars.

Non-magnetic models (black and grey lines for the Mix1 and Mix2 schemes, respectively) mostly show high rotational velocities. Close to the TAMS, the NOMAG/Mix2 models spin down efficiently and yield a high N/H ratio. However, this prediction is associated with producing Helium stars since these models experience Wolf-Rayet type mass loss due to their quasi-chemically homogeneous main sequence evolution (see Section \ref{fig:app1b}). This helps to decrease the surface hydrogen abundance significantly, and hence the N/H ratio can further increase. Given the high effective temperatures that these models show close to the TMAS, we do not expect that the non-magnetic models should match the observations of typical Group~2 stars.

Chemical mixing remains challenging to constrain. The schemes that we assume in this work cover a large area on the Hunter diagram, which represents modelling uncertainties. In our models, for a given initial rotation and initial magnetic field strength, the INT/Mix1 and SURF/Mix2 models lead to the smallest and highest amount of nitrogen enrichment, respectively (see also Figure~\ref{fig:schemeevol1}). The INT/Mix2 and SURF/Mix1 models produce similar results, although the latter models cover a narrower domain. In this sense, the Mix1 and Mix2 schemes are limiting cases in terms of the produced nitrogen enrichment from our models. In nature, the situation might be much more complex since, for example, the study of slowly-pulsating B-type stars reveals a diverse range of mixing profiles \citep{pedersen2021}. The variation of these profiles over time is not yet quantified.

The magnetic properties of stars in the LMC remain unknown. For this reason, the INT and SURF braking schemes represent assumptions and uncertainties that could only be resolved if, at least, upper limits on the magnetic field strengths were constrained. From the models we see that for a given initial magnetic field strength, the INT models produce less enrichment than the SURF models in a given mixing scheme.

In Figure \ref{fig:hunterobs1} we show abundance measurements\footnote{Although the observations of supergiants by \cite{mcevoy2015} are available, our models only cover the main sequence evolution and thus we refrain from a direct comparison to more evolved stars.} of observed massive stars at the LMC made by \cite{grin2017} and \cite{dufton2018}. Since the observations only allow derivation of the projected rotational velocity $v~\sin~i$, we scale the actual rotational velocity in our models with $\sin(\pi/4)$ to account for an average inclination angle. We only consider here observations with $v \sin i < 300$~km\,s$^{-1}$. 

The observations reveal a large scatter, which likely represents a range of initial conditions (mass, rotation rates, magnetic field strength, binarity, etc.) and current age. The data from \cite{grin2017} and \cite{dufton2018} indicate stars in their main sequence evolutionary stages with surface gravities systematically decreasing towards lower rotation rates. Our models show that magnetic braking typically yields slowly-rotating stars early on in the evolution, still with high surface gravities. Once the rotation is slow (and $\log g$ is still $\approx$ 4.0), chemical mixing becomes inefficient and no further surface enrichment may be expected on the main sequence. This is the primary reason why magnetic models produce less surface enrichment than non-magnetic models with the same mixing assumptions (see also Figure~\ref{fig:fidu}). However, at this point, the magnetic models still evolve further in time, albeit their location does not change in the Hunter diagram. Thus the magnetic models decrease their surface gravities (to $\log g \approx$~3.0) in a narrow region in the Hunter diagram, when their (assumed projected) surface rotational velocities are below 50~km~s$^{-1}$ in all cases, except for the SURF/Mix1 models where the spin-down is the least efficient (c.f. Section~\ref{sec:schemeevol}).

There are several other caveats, which hamper a quantitative comparison between the models and observations. For example, the mass determinations are uncertain and often rely on (non-magnetic) evolutionary models. As we demonstrate in Figure~\ref{fig:hunterobs1}, even for an idealised situation where the braking and mixing schemes were known for a given magnetic field strength, the produced nitrogen enrichment is still a function of initial mass (see also \citealt{aerts2014,maeder2014a}). In the INT/Mix1 scheme, the final nitrogen abundance becomes a factor of 2.5 higher when increasing the stellar mass from 15 to 60 M$_\odot$. This difference is less in the INT/Mix2 and SURF/Mix1 schemes (factor of $\approx$1.5), while this trend reverses for the SURF/Mix2 scheme.

Despite all these uncertainties, the models incorporating the effects of surface fossil magnetic fields can cover the region on the Hunter diagram where the "anomalous" Group 2 stars (slow rotation along with surface nitrogen enrichment, see \citealt{hunter2008}) are located, which is not possible with standard main sequence evolutionary models of single stars (see \citealt{martins2017} for non-magnetic $\approx$~30~M$_\odot$ models in the Galaxy). 
In particular, for the slowly-rotating non-magnetic Mix2 models the produced nitrogen enrichment seems to be larger than indicated by observations, whereas the non-magnetic Mix1 models do not spin down sufficiently.

Since the parameter space is degenerate, not only the mixing and braking scheme could produce results that cover Group~2 stars but also the variation of initial magnetic field strength in a given scheme. A stronger initial field would yield less enrichment (Figure~\ref{fig:bfield1}), possibly explaining the less nitrogen enriched stars, whereas a weaker initial field could be compatible with the most highly enriched stars (see Figure~\ref{fig:fidu}). To quantify this, we introduce and discuss the cutoff magnetic field strength in the next section.

%
%
%
%
%
%
\begin{figure*}
\includegraphics[width=9cm]{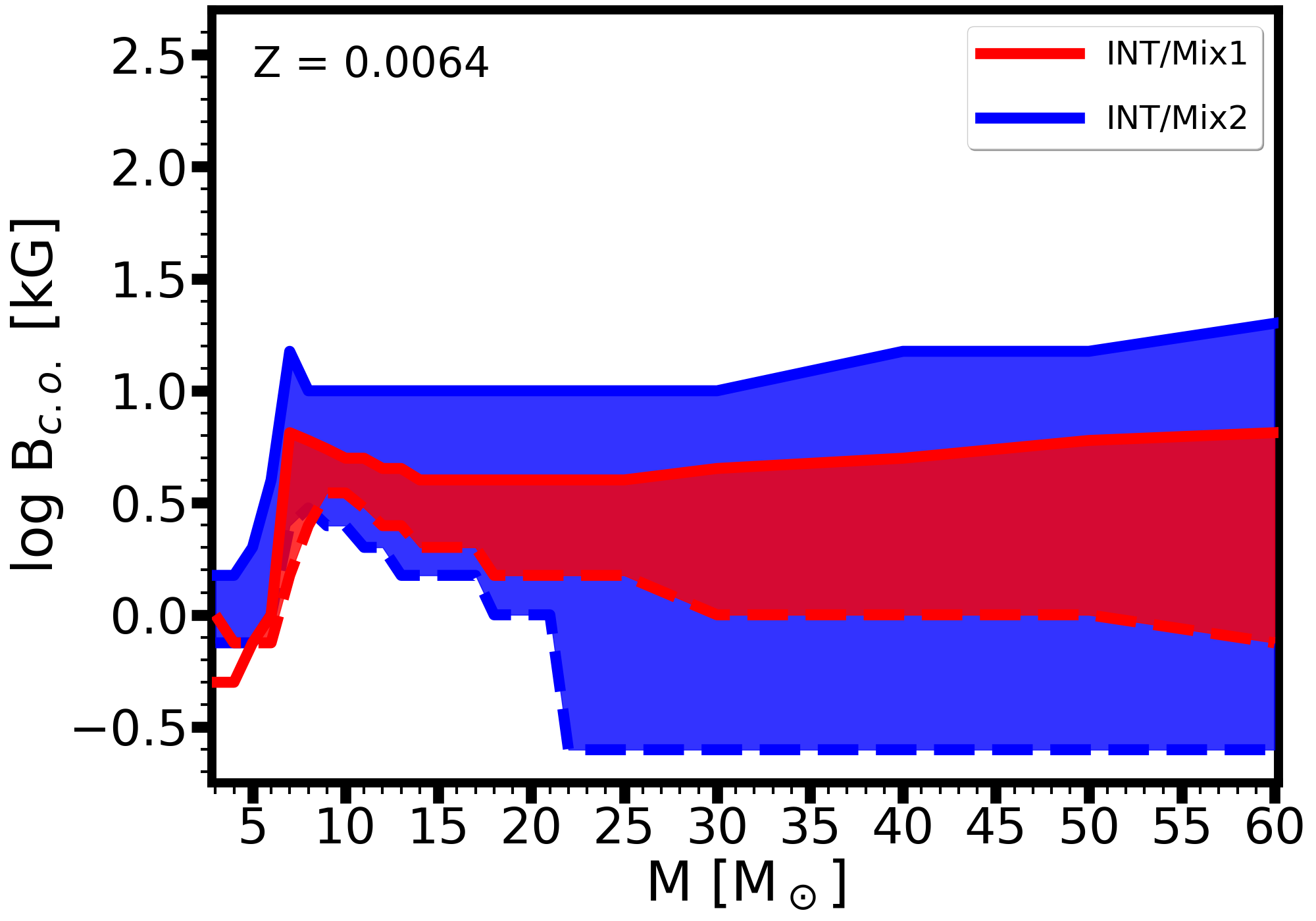}\includegraphics[width=9cm]{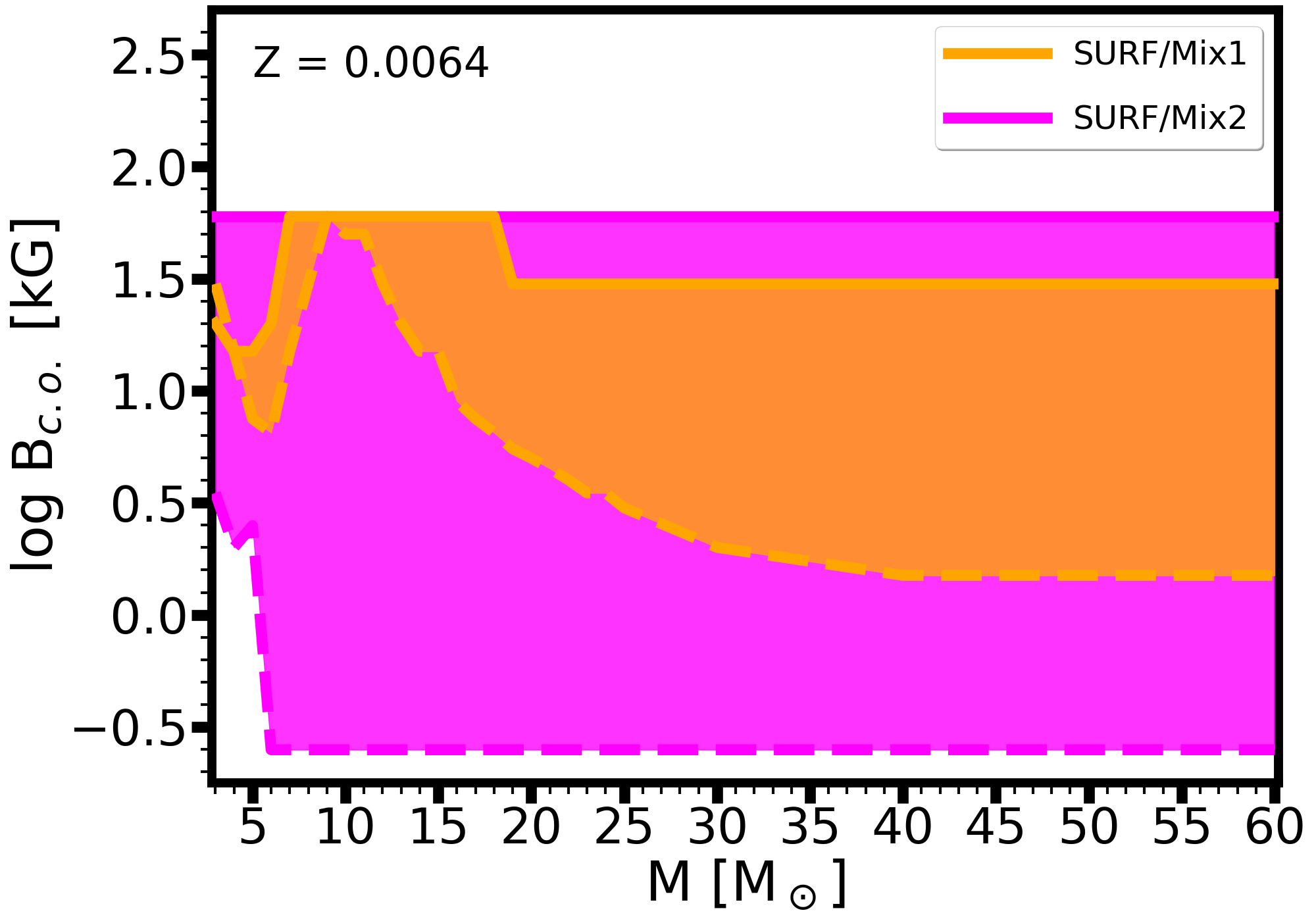}
\caption{Cutoff magnetic field strengths as a function of mass to produce surface nitrogen enrichment ($B_{\rm max, N}$, solid line) and slow rotation ($B_{\rm min, v}$, dashed line). Values above $B_{\rm max, N}$ will inhibit surface nitrogen enrichment, whereas values below $B_{\rm min, v}$ will not spin down the star sufficiently. Left/right panels show the INT/SURF models. The models are considered for initially $\Omega/\Omega_{\rm crit} = 0.5$ at LMC ($Z = 0.0064$) metallicity. }\label{fig:cutoff}
\end{figure*}
\subsection{Cutoff magnetic field strengths in the LMC}\label{sec:cutoff}

It is of interest to evaluate a critical value of the magnetic field that strongly impacts observable properties. In particular, what range of initial magnetic field strengths allow for producing Group 2 stars? To this extent, we define a cutoff (maximum) field strength $B_{\rm max, N}$ as the initial equatorial magnetic field strength in a given model that allows for producing more than 0.1 dex of surface nitrogen enrichment (in spectroscopic units) during its main sequence evolution. If the initial magnetic field strength is higher than $B_{\rm max, N}$, then the mixing is inefficient due to the magnetic spin-down, and no nitrogen enrichment can be observed. Similar to the above definition, we may also define a cutoff (minimum) magnetic field strength that is the initial equatorial magnetic field strength needed to produce sufficiently slow rotators ($v_{\rm rot}<71$~km s$^{-1}$) by the end of the main sequence evolution. Any value higher than $B_{\rm min, v}$ will yield slow-rotating models; however, values below $B_{\rm min, v}$ will still result in considerable rotational velocities at the TAMS.
Figure~\ref{fig:cutoff} shows the range of possible initial equatorial magnetic field strengths that are able to produce Group 2 stars given the constraints given above. $B_{\rm max, N}$ is shown with solid line and $B_{\rm min, v}$ with dashed line. The range is shown as a function of initial mass for the LMC (the Solar and SMC metallicity models are discussed in Appendix~\ref{sec:cutoff23}). The cutoff field strengths depend on the  initial rotation rates, metallicity, mass, and chemical mixing and magnetic braking schemes. We discuss now the latter three.  


Models within the INT/Mix1 scheme result in the lowest value of $B_{\rm max, N}$ since this scheme is the least efficient in chemical mixing. However, stronger magnetic fields would brake the rotation faster than the timescale of rotational mixing. In contrast, models within the SURF/Mix2 scheme have such strong mixing that even a 50~kG equatorial field strength is insufficient to inhibit mixing\footnote{For visualisation purposes, we assigned 60~kG to those models where the maximum value in our grid of models (50~kG) was still insufficient to prevent nitrogen enrichment.}. 
A particular feature, a jump in $B_{\rm min, v}$ around 5 to 10 M$_\odot$, can be explained by the mass-dependent rotational behaviour of the models discussed in Section~\ref{sec:feature}. In most models, the decreasing value of $B_{\rm min, v}$ for stars more massive than 10 M$_\odot$ implies that stronger stellar winds aid the spin-down thus a weaker magnetic field is sufficient to achieve slow-rotating stars. 

We find that equatorial magnetic fields of initially a few kG are able to produce Group~2 stars in the INT/Mix1 scheme. Nevertheless, the range of allowed initial field strengths is the most limited in this case. In fact, models below 6~M$_\odot$ are in a "forbidden" range where the minimum field strengths needed to brake rotation are higher than the maximum field strengths allowed to produce nitrogen enrichment. 
In the INT/Mix2 scheme, a much larger range of initial field strengths are allowed to produce Group~2 stars, particularly from 22~M$_\odot$, where the lower limit drops to 250~G. For initial masses higher than 7~M$_\odot$, the upper limit is of the order of 10~kG to produce Group~2 stars. 

Models within the SURF/Mix1 and SURF/Mix2 schemes (right panel of Figure \ref{fig:cutoff}) cover a wide range of possible initial field strengths. 
The SURF/Mix1 scheme produces a similar mass-dependent pattern as the models in the INT scheme. Namely, models from 10~M$_\odot$ have systematically decreasing values of $B_{\rm min, v}$. However, for initial masses lower than 17~M$_\odot$ a 10~kG initial equatorial field strength is still needed to achieve slow rotation. Interestingly, for initial masses $\geq$~19~M$_\odot$ there is a dip in $B_{\rm max, N}$, staying constant at 30~kG in contrast to the mass range of 7-18~M$_\odot$, where the 50~kG initial equatorial field strength still allows for producing nitrogen enrichment.
The SURF/Mix2 models have a constant upper limit given by $B_{\rm max, N}$, meaning that even the strongest initial magnetic field strength we considered in this study is not sufficient to prevent nitrogen enrichment on the main sequence. The lower limit given by $B_{\rm min, v}$ is constant for initial masses higher than 6~M$_\odot$.
In general, we can conclude that the INT scheme favours lower values for the cutoff magnetic field strengths of $B_{\rm max, N}$ to produce Group 2 stars and the SURF scheme allows for higher values of  $B_{\rm max, N}$. It is quite remarkable that the upper limit remains roughly constant in the INT cases for stars more massive than about 10~M$_\odot$. In the SURF/Mix2 case, likely an unrealistically strong magnetic field would be needed to prevent nitrogen enrichment (as the 50~kG field is still insufficient).  
The Mix1 scheme tends to allow for a narrow range of values and Mix2 scheme covers a wide range of possible solutions. 

In the LMC, both quantities used in our criteria, surface nitrogen abundance and (projected) rotational velocity, can be measured. Large-scale surveys dedicated to magnetic field measurements are not yet available in lack of high-resolution spectropolarimetry (however, see \citealt{bagnulo2017,bagnulo2020}). This means that our predictions can be used as constraints on the strengths of magnetic fields that might exist in slowly-rotating nitrogen-enriched ("Group 2") stars in the Magellanic Clouds. An initial equatorial magnetic field strength above $B_{\rm min, v}$ and below $B_{\rm max, N}$ will produce stars that can be identified as Group 2 stars.

%
%
%
%
\subsection{Future work}

The models presented in this work cover the main sequence phase of single stars. Logical extensions include calculating pre-main sequence models and continuing the computations to the post-main sequence phase to be able to scrutinise connections with end-products of stellar evolution, such as strongly magnetised white dwarfs and neutron stars (magnetars). The recently discovered link between magnetars and fast-radio bursts \citep[][]{chime2020,boc2020} further supports investigations of the magnetic field origin of magnetars \citep[e.g.,][]{spruit2009,makarenko2021}.  
Our grid of models could be further extended to cover initial masses below 3~M$_\odot$ and thus to compare with, for example, Ap stars. This requires some further considerations about the winds of these objects and the inclusion of atomic diffusion in the models. 
We assumed single-star models in this work. Some magnetic massive stars, such as $\tau$ Sco  \citep[][]{schneider2016,schneider2019,schneider2020,keszthelyi2021}, may challenge this scenario. Nonetheless, the vast majority of OBA stars with fossil fields have characteristics that do not require invoking a merger event. In particular, fossil magnetic fields are detected in young massive stars, for example, in $\sigma$~Ori~E \citep[][]{landstreet1978,town2010,oksala2012,song2021}. Certain binary systems also present challenges to the merger scenario, such as the doubly-magnetic binary $\epsilon$ Lupi \citep{shultz2015} and the eclipsing late B-type binary HD 62658, which comprises two young nearly-identical stars in a circularised orbit, only one of which is magnetic \citep{mobster3}. Thus single-star models presented in this work are a reasonable first approach; however, future work remains to address binarity and mergers in combination with fossil field effects. In particular, multiplicity is common among massive stars \citep[][]{sana2012,sana2014,demink2013,demink2014} and while close magnetic binary systems are rare \citep[e.g.,][]{alecian2013,shultz2018}, the mutual impact of magnetism and tidal interactions need to be further studied \citep[e.g.,][]{song2018,vidal2018}.

The models computed in this work can be confronted with observations of known magnetic massive stars. They will complement previous approaches which relied on grids of stellar evolution models that did not include surface magnetic field effects \citep[e.g.,][]{brott2011,ekstroem2012,chieffi2013,choi2016} to infer stellar parameters and ages of magnetic stars. The differences are expected to be most pronounced for higher-mass stars \citep{petit2017}, whereas -- within the framework considered here -- the lower-mass, A-type stars should be less impacted \citep{deal2021}. Nonetheless, the available TESS data and continuous spectropolarimetric monitoring can be used to constrain accurate rotation periods of such stars and directly compare with evolutionary models incorporating magnetic braking.

The INT models represent stars with strong, predominantly dipolar fields that are commonly identified in the sample of known magnetic massive stars. The SURF models are a limiting case motivated by stars with complex magnetic fields. In the future, the implementation of different magnetic field configurations and their time evolution could be considered to improve the present models.

The internal mixing efficiency remains uncertain in evolutionary modelling. In our models, quasi-chemically homogeneous evolution during the main sequence develops in the non-magnetic Mix2 case. In the INT/Mix2 models this behaviour is prevented by the efficient overall spin-down already when initially weak magnetic fields are considered, whereas in the SURF/Mix2 models a very efficient mixing still remains. The occurrence and duration of quasi-chemical evolution depends on the initial field strength and stellar mass; some SURF/Mix2 models evolve significantly bluewards, whereas some models turn to a redward evolution after mixing becomes less efficient (see Figure \ref{fig:app1d}).
Quasi-chemically homogeneous evolution is expected to be rare in nature; however, it may be a crucial channel, for example, for some supernova events, gamma-ray bursts, or gravitational wave sources \citep[e.g.,][]{georgy2012,martins2013,demink2016,szecsi2017}. 

Observational studies should help constrain the mixing efficiency and ultimately the physical mixing processes. As such, it would be beneficial to further study our models and confront them with measurements of surface nitrogen abundances in magnetic massive stars \citep[][]{morel2008,morel2015,aerts2014,martins2012,martins2015}, as well as studies which identified anomalous trends on the Hunter diagram in the LMC and SMC \citep[e.g.,][and see Section~\ref{sec:huntobs}]{dufton2020}. 

Finally, our current understanding of magnetic field evolution is still incomplete and, in particular, how different field geometries evolve over time is largely unconstrained. It will therefore be valuable to explore various field evolution scenarios, for example, magnetic flux decay (\citealt{shultz2019b}, \PaperIII). This might also lead to a time-dependent magnetic braking scheme depending on the relative dissipation timescales of various complex field components.

%
%

\section{Conclusions}\label{sec:concl}

In this work we present the most extensive grid of stellar structure and evolution models taking into account the effects of surface fossil magnetic fields. The grid is publicly available on Zenodo and we recommend that, while acknowledging the uncertainties, it could be used to infer stellar parameters of known magnetic massive stars. No particular braking (INT/SURF) or mixing (Mix1/Mix2) scheme can be preferred at this time, although we do assume that if the field geometry is known, the INT case is applicable for dipolar fields and the SURF case for more complex geometries. 
Thus we consider the four schemes as limiting cases and, as we demonstrate in this work, the differences between these subgrids can substantially impact the determination of stellar parameters. It is therefore essential to confront the Mix1/Mix2 mixing schemes with spectroscopic studies even for stars where surface magnetic fields are not detected. In all cases (2 mixing schemes and 3 metallicities), we provide a subgrid of non-magnetic evolutionary models.
Furthermore, the grid of models is suitable for population synthesis studies, which thus far have neglected magnetic field effects and magnetic massive stars within stellar populations \citep[however, see][]{potter2012b}. The impact of magnetic fields nonetheless may have important consequences on stellar populations and stellar-end products, for example, considering progenitors of magnetars \citep[e.g.,][]{schneider2019}. 

We demonstrate that magnetic braking by a fossil field leads to efficient spin-down. For example, an initial equatorial field of 3~kG strength at solar metallicity is sufficient in most models to decrease an initial surface equatorial rotational velocity of 300~km\,s$^{-1}$ below 50~km\,s$^{-1}$ within the early stages of the main sequence evolution (e.g., Figure~\ref{fig:kiel1}). For a given magnetic field strength, the spin-down of high-mass stars (> 10~M$_\odot$) is further aided by mass-loss, whereas the spin-down of lower-mass stars in our grid (< 5~M$_\odot$) is identifiable due to the long nuclear timescale. The intermediate-mass range (5-10 M$_\odot$) has the least efficient spin-down over the main sequence evolution. 

The "magnetic population" is thus far only identified within the Galaxy by spectropolarimetry and it is unknown what fraction of massive stars possesses strong, surface magnetic fields in extragalactic environments. Generally, the spin-down of the stellar surface for a given magnetic field strength is the most rapid at high metallicity (due to stronger winds), whereas the measurable surface nitrogen abundances are more impacted at lower metallicity (as chemical mixing effects are more pronounced). In the Large Magellanic Cloud, about 20\% of stars follow an anomalous pattern on the Hunter diagram, which can be covered with magnetic stellar evolution models.
We identify the existence of a range of initial magnetic field strengths (the exact values depending on metallicity, mixing schemes, etc.) that allow for producing slowly-rotating nitrogen enriched Group~2 stars. The lower limit is constrained by a field strength that is needed to brake rotation and produce slow rotators. The upper limit is constrained by a field strength that is needed to allow for rotational mixing and still produce nitrogen enrichment. The range of possible field strengths for the INT models is much narrower than for the SURF models, however, it is compatible with typically measured values. In the LMC and SMC almost all (except some of the lowest mass) models lead to a solution. Contrary, we find that in the Galaxy the formation of Group~2 stars may essentially be prevented for initial masses from 6 to 23~M$_\odot$ in the INT/Mix1 scheme (Figure~\ref{fig:cutoff3}).

Overall, we find significant differences between the braking and mixing schemes. With internal magnetic braking caused by a strong dipolar field, differential rotation cannot develop. With surface magnetic braking caused by a complex magnetic field the physical scenario remains much less clear; the results strongly depend on the chosen assumptions regarding chemical mixing.

\section*{Acknowledgements}

We thank the anonymous referee for providing us with constructive comments.
We thank the \textsc{mesa} developers for making their code publicly available. 
We also acknowledge the open-source code and detailed description of \cite{fuller2019}, which helped us with developing our implementation for the SURF models.
Furthermore, we greatly appreciate discussions with Henk Spruit, \nobreak{Fabrice} \nobreak{Martins}, and Philip Dufton.
Support for Y.G. was provided by NASA through the NASA Hubble Fellowship Program grant \#HST-HF2-51457.001-A awarded by the Space Telescope Science Institute, which is operated by the Association of Universities for Research in Astronomy, Inc., for NASA, under contract NAS5-26555.
G.M. acknowledges support from the Swiss National Science Foundation (project number 200020-172505). G.M. and C.G. have received funding from the European Research Council (ERC) under the European Union’s Horizon 2020 research and innovation program (Grant Agreement No. 833925).
A.D.-U. is supported by NASA under award number 80GSFC21M0002.
S.G. acknowledges support from a NOVA grant for the theory of massive star formation.
R.H. acknowledges support from the World Premier International Research Centre Initiative (WPI Initiative), MEXT, Japan; the IReNA AccelNet Network of
Networks, supported by the National Science Foundation under Grant No. OISE-1927130; the ChETEC COST Action
(CA16117), supported by COST (European Cooperation in Science and Technology) and  ChETEC-INFRA (grant No 101008324) supported by the European Union’s Horizon 2020 research and innovation programme.
V.P. is supported by the National Science Foundation under Grant No. AST-2108455.
M.E.S. acknowledges support from the Annie Jump Cannon fellowship, endowed by the Mount Cuba Astronomical Observatory and supported by the University of Delaware.
A.u.-D. acknowledges support by the National Aeronautics and Space
Administration through Chandra Award Number TM1-22001B issued by the
Chandra X-ray Center which is operated by the Smithsonian Astrophysical
Observatory for and on behalf of NASA under contract NAS8-03060.
This work was carried out on the Dutch national e-infrastructure with the support of SURF Cooperative.

\section*{Data Availability}

A full reproduction package is available on Zenodo at \url{https://doi.org/10.5281/zenodo.7069766}, in accordance with the Research Data Management Plan of the Anton Pannekoek Institute for Astronomy at the University of Amsterdam.

The data used in this paper amounts to the order of 1/3 TB. A typical evolutionary model ("history" file in \textsc{mesa} nomenclature) is a few MBs, whereas a typical structure model ("profile" file in \textsc{mesa} nomenclature) is 10 MB. Each evolutionary model has three structure models saved at the ZAMS, mid-MS, and TAMS, respectively. In addition to the evolutionary and structure files, we save and provide the ".mod" files of each run when available. This allows for continuing the computations on the post-main sequence. We also generated isochrones for each sub-grid, these are included in the Zenodo record.

Given the large range of covered parameter space, the output data in this paper is particularly useful for stellar evolution and population synthesis studies, as well as to compare with observational results of even individual stars. However, we emphasise that when interpreting observational results the modelling assumptions and uncertainties should be considered.


\bibliographystyle{mnras}
\bibliography{ref}


\appendix

\section{\textsc{mesa} microphysics}\label{sec:micro}

%
%
The MESA EOS is a blend of the OPAL \citep{Rogers2002}, SCVH
\citep{Saumon1995}, FreeEOS \citep{Irwin2004}, HELM \citep{Timmes2000},
PC \citep{Potekhin2010}, and Skye \citep{Jermyn2021} EOSes.

Radiative opacities are primarily from OPAL \citep{Iglesias1993,
Iglesias1996}, with low-temperature data from \citet{Ferguson2005}
and the high-temperature, Compton-scattering dominated regime by
\citet{Poutanen2017}.  Electron conduction opacities are from
\citet{Cassisi2007}.

Nuclear reaction rates are from JINA REACLIB \citep{Cyburt2010}, NACRE \citep{Angulo1999} and
additional tabulated weak reaction rates; \citet{Fuller1985, Oda1994,
Langanke2000}.  Screening is included via the prescription of \citet{Chugunov2007}.
Thermal neutrino loss rates are from \citet{Itoh1996}.

\section{Mass-loss quenching for quadrupole field geometries}\label{sec:quad}
\begin{figure}
\includegraphics[width=\linewidth]{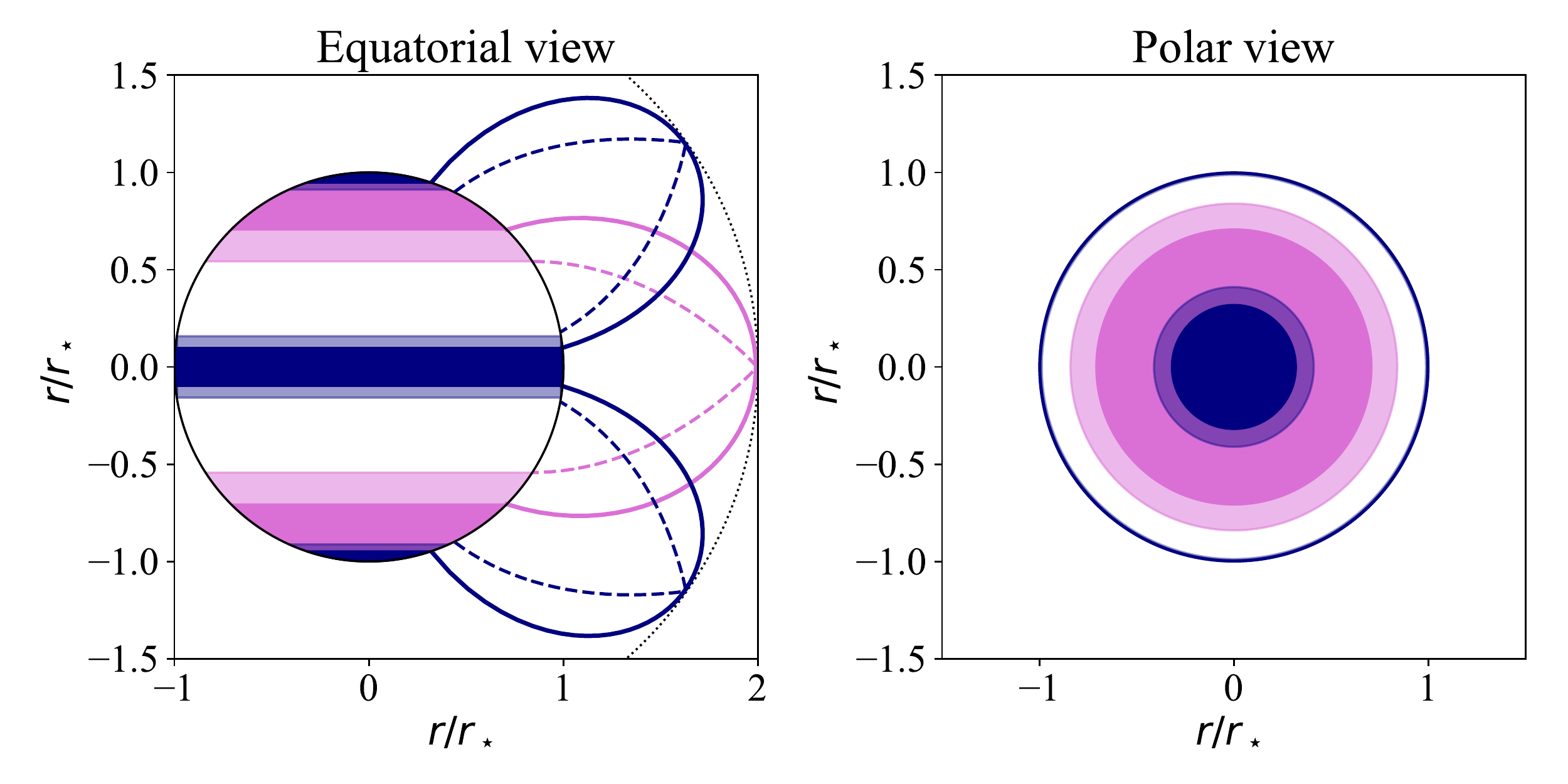}
\includegraphics[width=\linewidth]{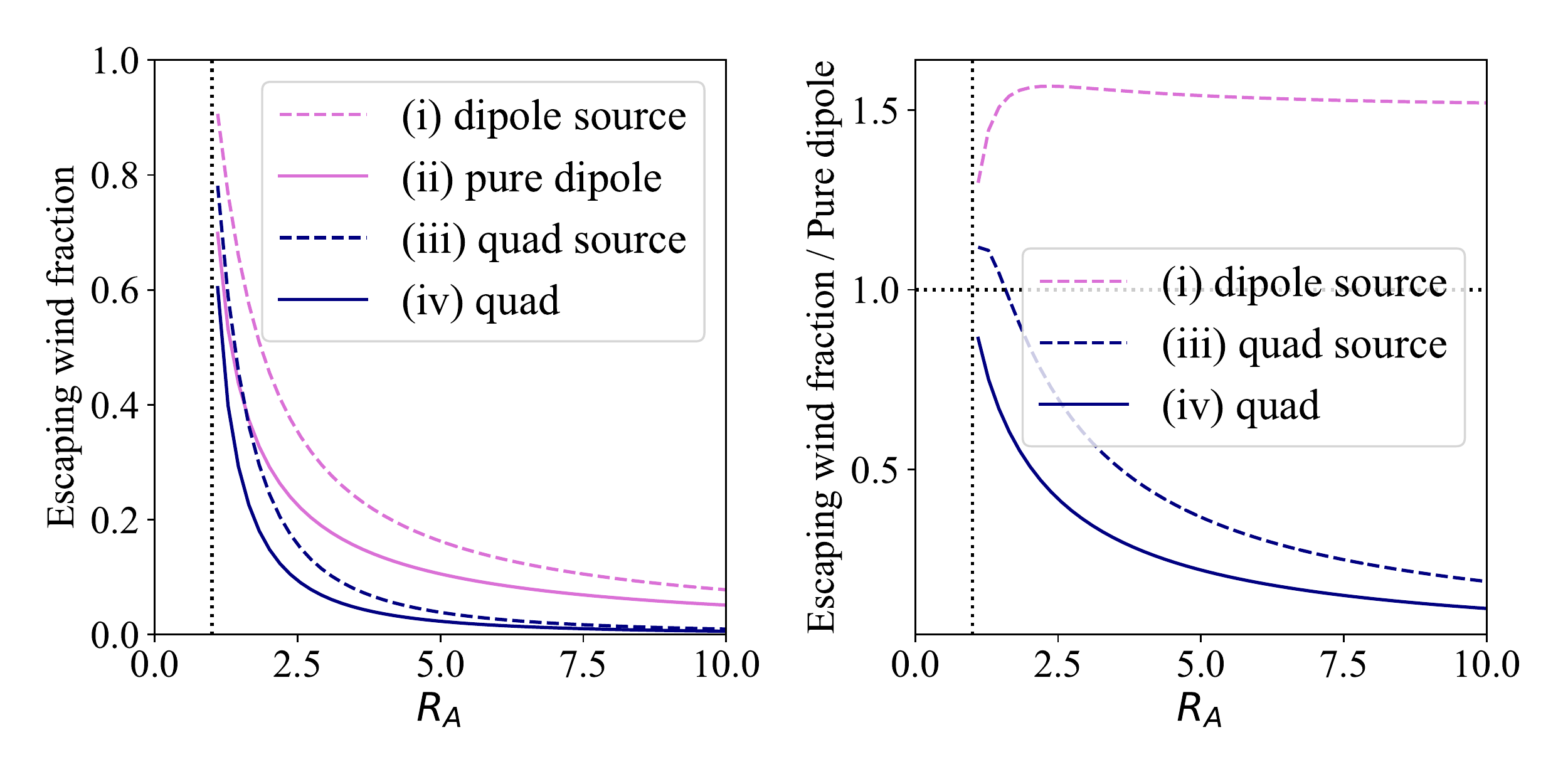}
\caption{Top: illustration of the surface area covered by open field loops for (i) a dipole extrapolated with a source radius ($2R_\star$), (ii) a pure dipole, (iii) a quadrupole extrapolated with a source radius ($2R_\star$), and (iv) a pure quadrupole. The radial extent of the largest closed loop is $2R_\star$ (grey dotted circle) in all cases. Bottom Left: Escaping wind fraction as a function of closure radius (which is of the order of the Alfv\'en radius) for the 4 cases above. Bottom right: Same as Left panel, but normalised to the escaping wind fraction of a purely dipolar configuration.}\label{fig:mquad}
\end{figure}

As a simplifying assumption, we consider that even for more complex magnetic fields (represented by the SURF models), a scaling based on the Alfv\'en radius of a quadrupole field may be adopted to calculate the mass-loss quenching effect via $f_{\rm B}$ (Equations~\ref{eq:fb1}-\ref{eq:fb2}). Here, we outline the actual geometrical effects of a quadrupolar magnetic field configuration. 

Following \cite{ud2008}, the escaping wind fraction is the fraction of the stellar surface covered with open field loops. For a purely dipolar geometry, the region of open field loops corresponds to a polar cap extending from $\mu=1$ to $\mu=\mu_m$, where $\mu$ is the cosine of the co-latitude $\theta$, and $\mu_m=\cos\theta_m$ is the location at which the largest loop connects to the stellar surface (see pink shaded regions in Fig \ref{fig:mquad}). 

This assumes that the magnetic field lines retain their pure dipolar geometry, even though in principle the field gets distorted by the outflowing stellar wind. Another approach is to solve for the magnetic field in the magnetosphere using a force-free extrapolation \citep[e.g.][]{1999MNRAS.305L..35J}. Assuming that the magnetic field can be expressed as the gradient of a potential $\Phi$ such that $\nabla^2\Phi(r, \theta,\varphi)=0$, the magnetic field can be found by imposing a dipolar (or quadrupolar, see below) boundary condition at the surface and imposing that the magnetic field must become radial at a certain distance from the stellar surface (at the so-called source radius, $r_s$). Following the description of \citet{2011AmJPh..79..461G} (their Equation~32 with $l=1$), in this case the largest field loops (illustrated in the top left panel of \ref{fig:mquad} with dashed curves) are connected to the surface at:
\begin{equation}
    \sin^2\theta_m = \frac{1}{r_m} \frac{3r_m^3}{1 + 2r_m^3} \, .
\end{equation}
\noindent For a quadrupolar configuration, open field loops are located in two polar caps ($\mu_\mathrm{cap}$) and an equatorial belt ($\mu_\mathrm{belt}$) (see \ref{fig:mquad} top panels in blue shading). Thus the total fractional area of open field lines is:
\begin{equation}
    f_B = [1-\mu_\mathrm{cap}] + [\mu_\mathrm{belt}] \, .
\end{equation}
\noindent For a purely quadrupolar configuration, these footpoints of the largest closed loop can be found by solving for the $\mu_m$ roots of:
\begin{equation}
     \mu_m^3 - \mu_m + \left[ \frac{2}{3^{3/2}}\frac{1}{r_m^2} \right]  = 0 \, , 
\end{equation}
and for a quadrupole extrapolated with a source radius:
\begin{equation}
    \mu_m^3 - \mu_m + \left( \frac{1}{r_m^2} \frac{5r_m^4}{2+3r_m^4} \frac{2}{3\sqrt{3}} \right) = 0 \, .
\end{equation}
In general, for the same closure radius the escaping wind fraction (surface covered by open field loops) decreases from extrapolated dipole $\rightarrow$ pure dipole $\rightarrow$ extrapolated quadrupole $\rightarrow$ pure quadrupole. We also recall that for a given field strength the closure radius is smaller for a quadrupole field than for a dipole. The lower panels of Fig~\ref{fig:mquad} display the corresponding change in magnetic mass-loss quenching between the dipolar and quadrupolar geometries as a function of closure radius. In a future work we will implement and study magnetic mass-loss quenching for field geometries that deviate from a pure dipole. As shown here, we expect that a quadrupole field will result in a lower escaping wind fraction, i.e., stronger mass-loss quenching for a given magnetic field strength at a given time. However, the time evolution may sensitively impact the results in the evolutionary context.

%
%
%
%
\section{Convective expulsion and magnetic suppression of convective layers}\label{sec:convex}

"Magneto-convection", the interaction between the magnetic field and turbulent motions in convectively unstable stellar layers, has been extensively studied and recent progress warrants a brief discussion of this topic. 
Modifying Schwarzschild's criterion for convection, the stability criterion for magneto-convection was first given by \cite{gough1966} and recently revised by \citet{macdonald2019}, specifically considering radiation pressure to account for sub-surface layers of massive stars, in the form of:
\begin{equation}\label{eq:magconv}
    Q (\nabla - \nabla_{\rm ad}) < \frac{v_{A}^2}{v_{A}^2 + c_{s}^2} \left(1+ \frac{\mathrm{d}\ln\Gamma_1}{\mathrm{d}\ln P}\right) \, , 
\end{equation}
\noindent where $Q$ is the thermal expansion coefficient, $v_{A}$ and $c_{s}$ are the Alfv\'en and sound speeds, $\Gamma_1$ is the first adiabatic exponent, and $P$ is the local pressure. This criterion leads to a critical magnetic field strength, which, if reached, suppresses convection \citep[e.g.,][]{lydon1995,feiden2012}, whereas, if it is not reached, the field lines are prevented from reconnecting by strong convective turbulence, which over time leads to diffusing the magnetic flux out of those regions \citep[e.g.,][]{parker1963}.
In the one-dimensional stellar evolution models presented in this study, we refrained from using Equation~\ref{eq:magconv}. Both \citet{macdonald2019} and \citet{jermyn2020} showed that the key (surface) characteristics during main sequence evolution are not expected to change significantly -- when using Equation~\ref{eq:magconv} -- for the stellar mass domain considered in this work, except for the 50-60~M$_\odot$ models. We anticipate that, in general, the post-main sequence evolution needs particular attention due to the structural changes which could lead to the development of extended convection zones.

The cores of main sequence massive stars are strongly convective, such that a critical magnetic field strength, which is needed to suppress convection (according to the stability criterion given in Equation \ref{eq:magconv}) may not be reached. In this case, if the magnetic field strength is below its critical strength throughout the core, it is expected that convection will destroy any large scale, organised magnetic flux. This is commonly known as convective expulsion \citep{parker1963,weiss1966,spruit1979,tao1998,schussler2001}. While the convective core should expel any fossil magnetic fields, the development of small-scale magnetic fields via convective dynamos \citep[e.g.,][]{stello2016,augustson2016} may lead to a complex physical picture near the core boundary \citep[][]{feat2009}.
It is possible that the fossil magnetic field has a strong impact, for example, by suppressing the radiatively unstable overshooting regions (or even the stellar core, to some extent, see, e.g., \citealt[][]{petermann2015}). Thus far, magneto-asteroseismology of two magnetic intermediate-mass stars are consistent with a low value of overshooting \citep[][]{briquet2012,bram2018}. This gives some support of using a modest overshooting, at least, in our INT models, where we assume dipoles embedded deep in the stellar radiative zone. However, this question needs to be studied in more detail in a future work. 

A domain of interest is the sub-surface layer of massive stars where inefficient convection (accounting for a few percent of the total luminosity transport) is predicted to arise due to the increase in the opacity of iron, known as iron opacity bumps \citep{cantiello2009,cantiello2011,graefener2012}. In these radiation-dominated layers, some excess luminosity is transported by convective flux.
Using equipartition arguments, \citet{auriere2007} and \cite{sundqvist2013} argue that sub-surface convection in radiation-dominated stellar layers can be suppressed by a critical magnetic field when the magnetic pressure reaches the thermal pressure. 
The exact value of the critical field strength needed to suppress or even inhibit convection remains somewhat uncertain -- a problem that is mostly explored for the dynamo fields of low-mass stars \citep[][]{mullan2001,feiden2012,feiden2013,feiden2014,feiden2016,mcdonald2017}. Nevertheless, observations clearly show that the Ap phenomenon disappears for stars less massive than about 1.5~M$_\odot$, coinciding with the development of efficient convection (due to the hydrogen opacity bump) in the stellar envelope \citep{landstreet1991,braithwaite2017}. We might therefore assume that extended convective zones cannot develop in regions of massive stars where the fossil field (which reaches the critical strength) permeates through. A fossil magnetic field, observable at the stellar surface, should certainly reach sufficient critical strength in these near-surface layers to inhibit convection. All OB stars show evidence for measurable macroturbulent broadening \citep[][]{simon2017}, including magnetic OB stars (\citealt{grunhut2017,shultz2018}, with the only exception of NGC~1624-2, see \citealt{sundqvist2013}), making them a valuable laboratory to further our understanding regarding the origin of macroturbulence. The observed macroturbulence and the putative lack of subsurface convection in magnetic OB stars would support an origin of macroturbulence (at least, in magnetic stars) that is unrelated to sub-surface convection, and could be due to, for example, wave-propagation from the stellar core \citep[][]{aerts2009,bowman2020}.

\section{ZAMS and TAMS stellar structure models}
%
\begin{figure*}
\includegraphics[width=16cm]{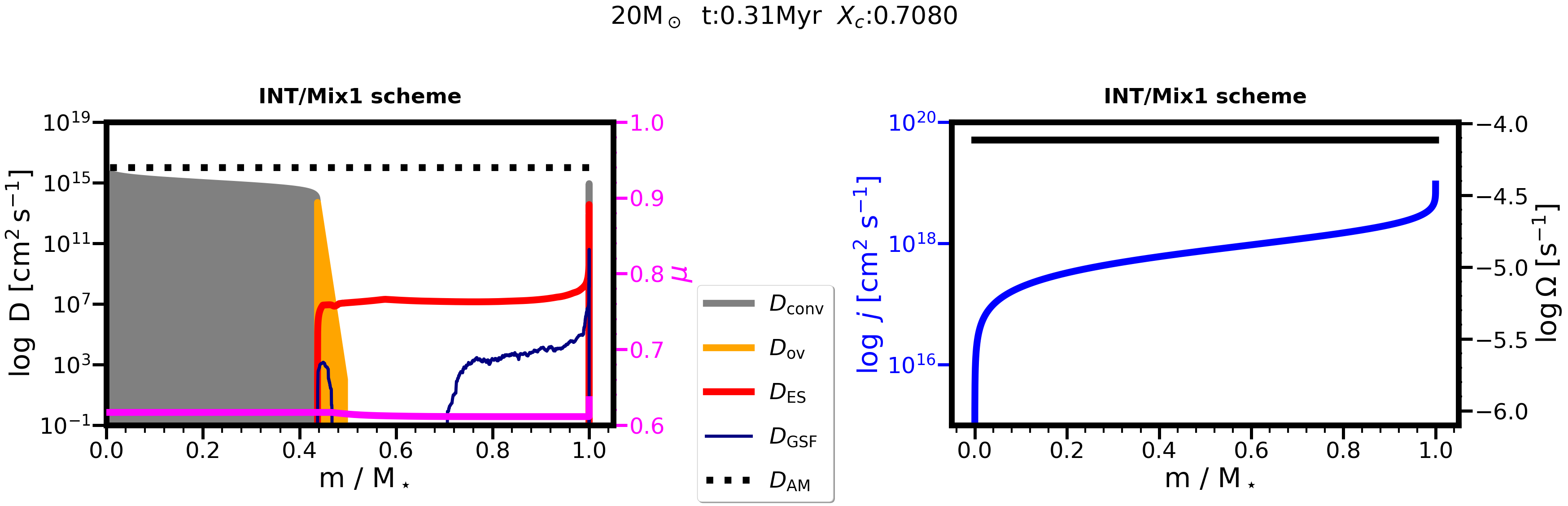}
\includegraphics[width=16cm]{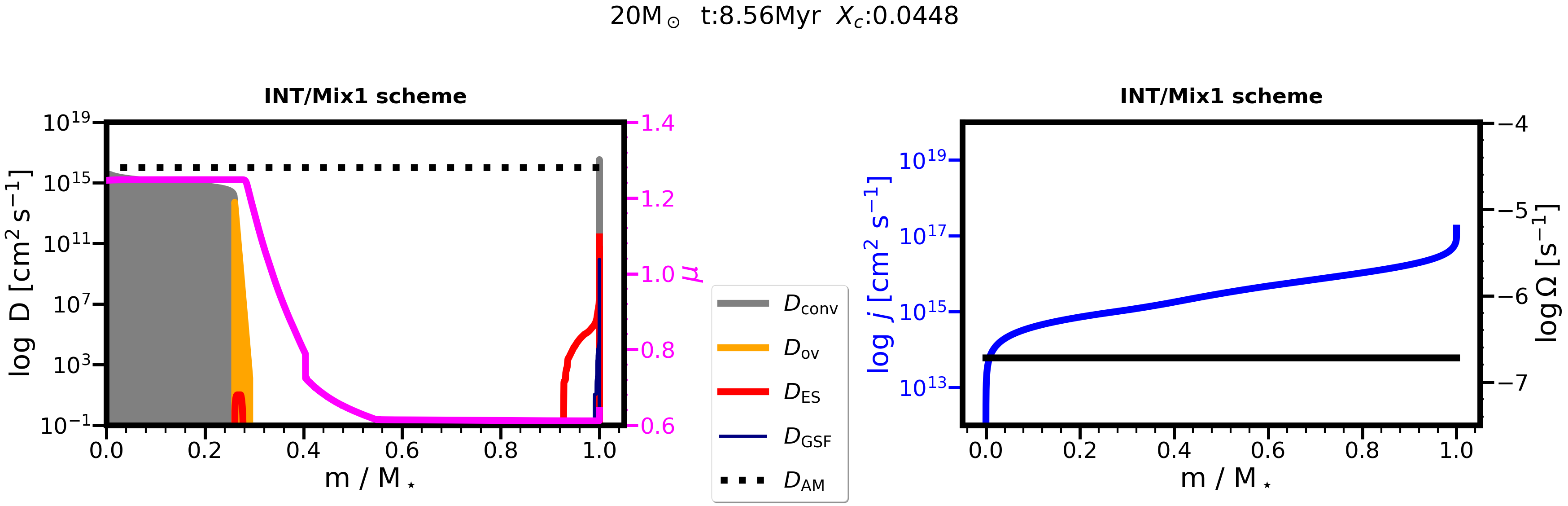}
\caption{Fiducial model with initial mass of $M_{\rm ini}$ = 20~M$_\odot$ (and initial equatorial magnetic field strength of 3 kG at $Z = 0.014$) in the INT/Mix1 scheme. \textit{Left} and \textit{Right} panels are as described in Figure~\ref{fig:str0_midMS}. \textit{Top}: near ZAMS. \textit{Bottom}: near TAMS.}\label{fig:app1a}
\end{figure*}
\begin{figure*}
\includegraphics[width=16cm]{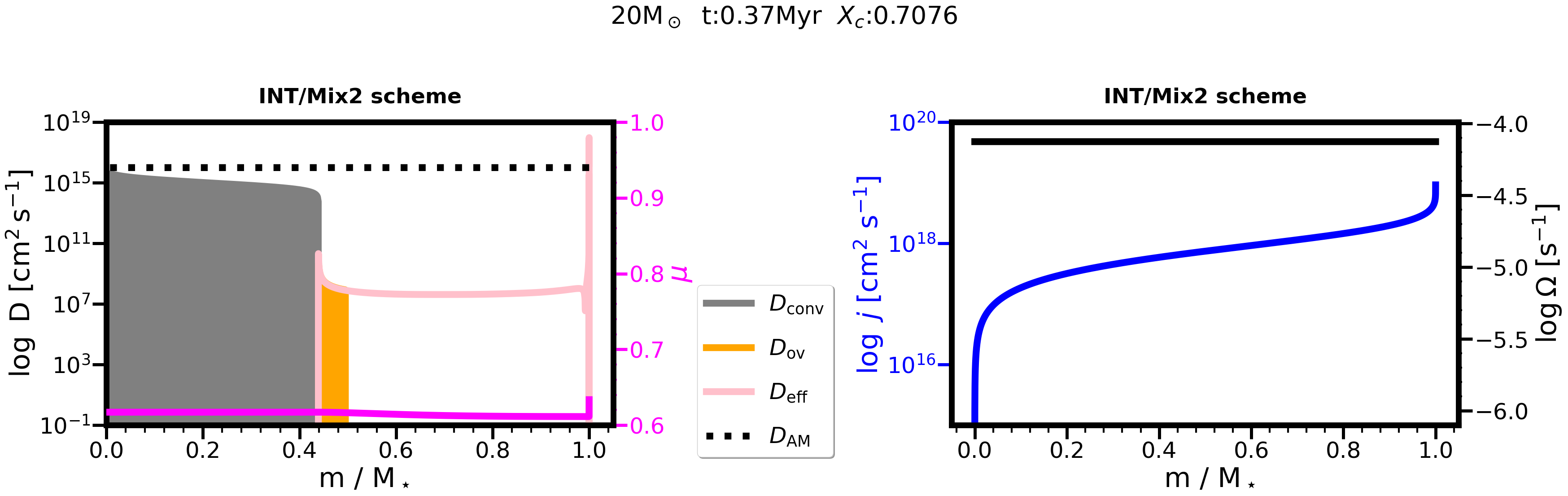}
\includegraphics[width=16cm]{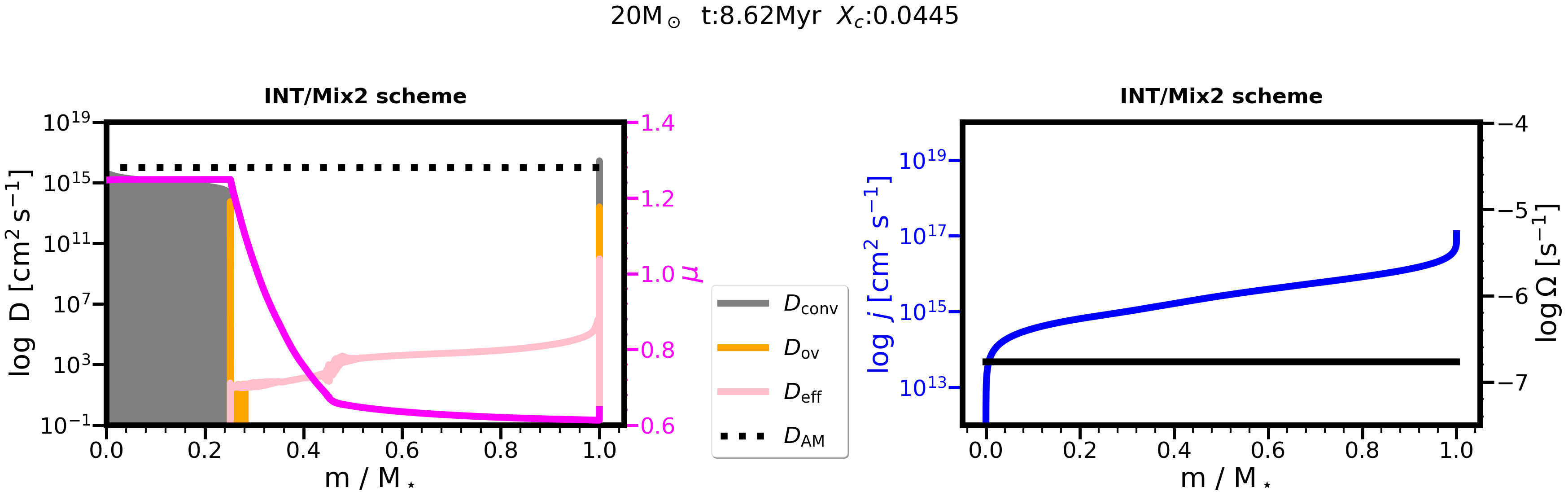}
\caption{Same as Figure \ref{fig:app1a} but for the INT/Mix2 scheme.}\label{fig:app1b}
\end{figure*}
\begin{figure*}
\includegraphics[width=16cm]{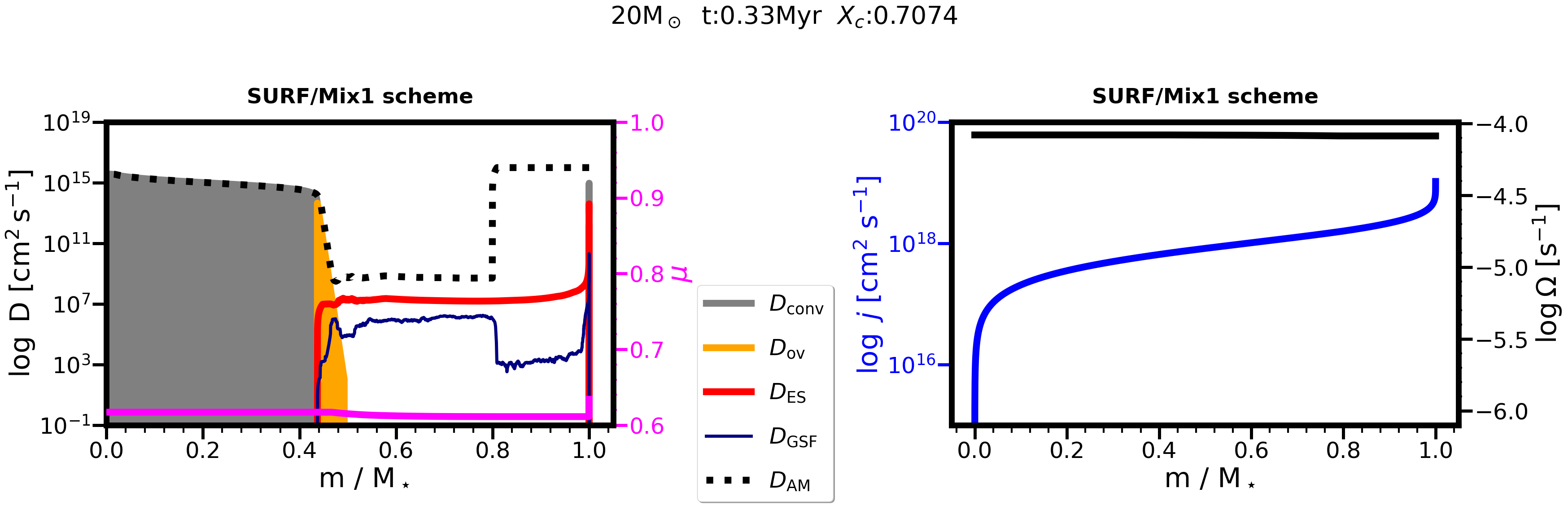}
\includegraphics[width=16cm]{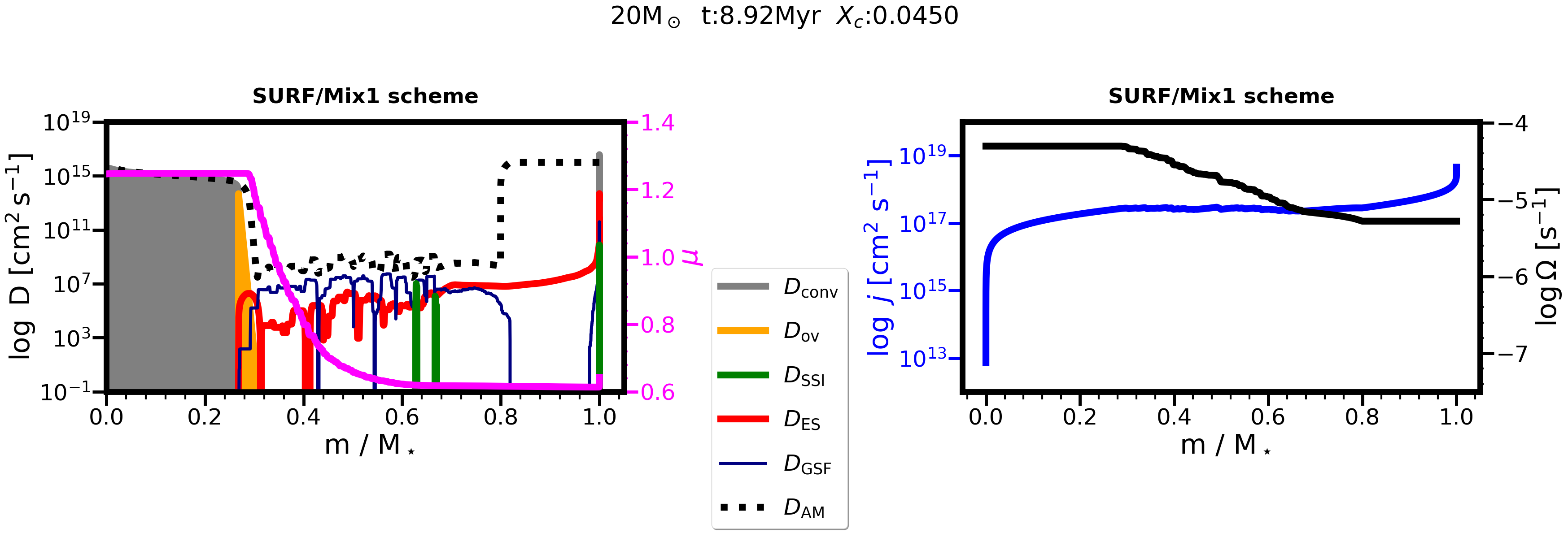}
\caption{Same as Figure \ref{fig:app1a} but for the SURF/Mix1 scheme.}\label{fig:app1c}
\end{figure*}
\begin{figure*}
\includegraphics[width=16cm]{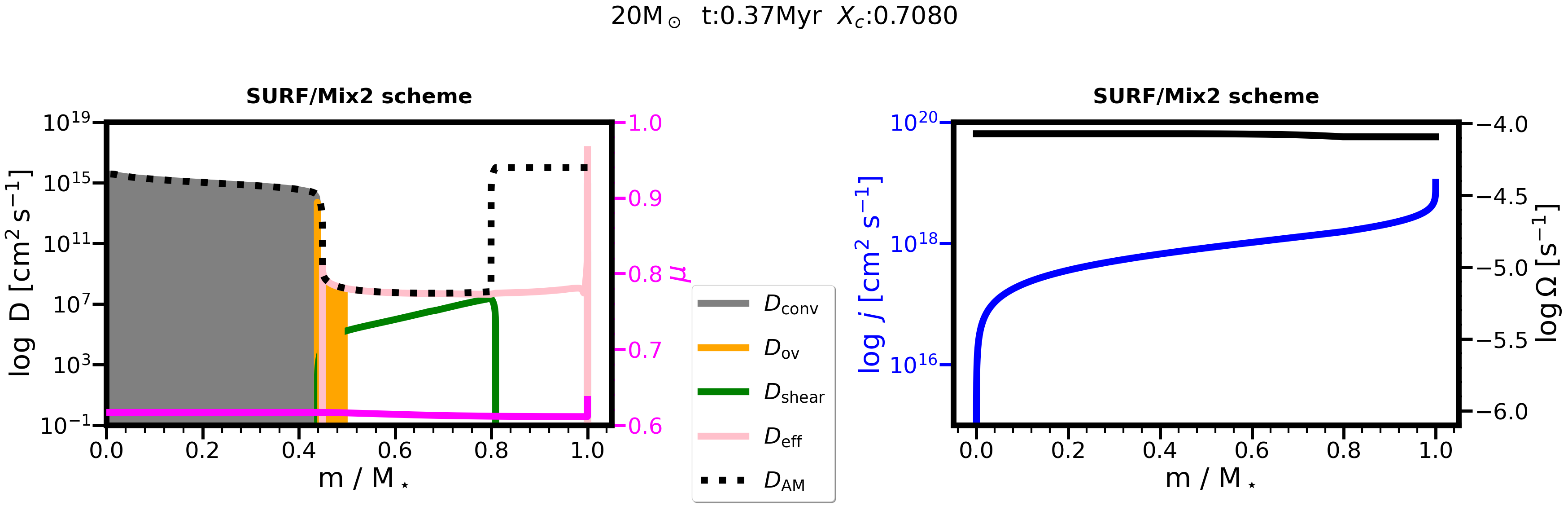}
\includegraphics[width=16cm]{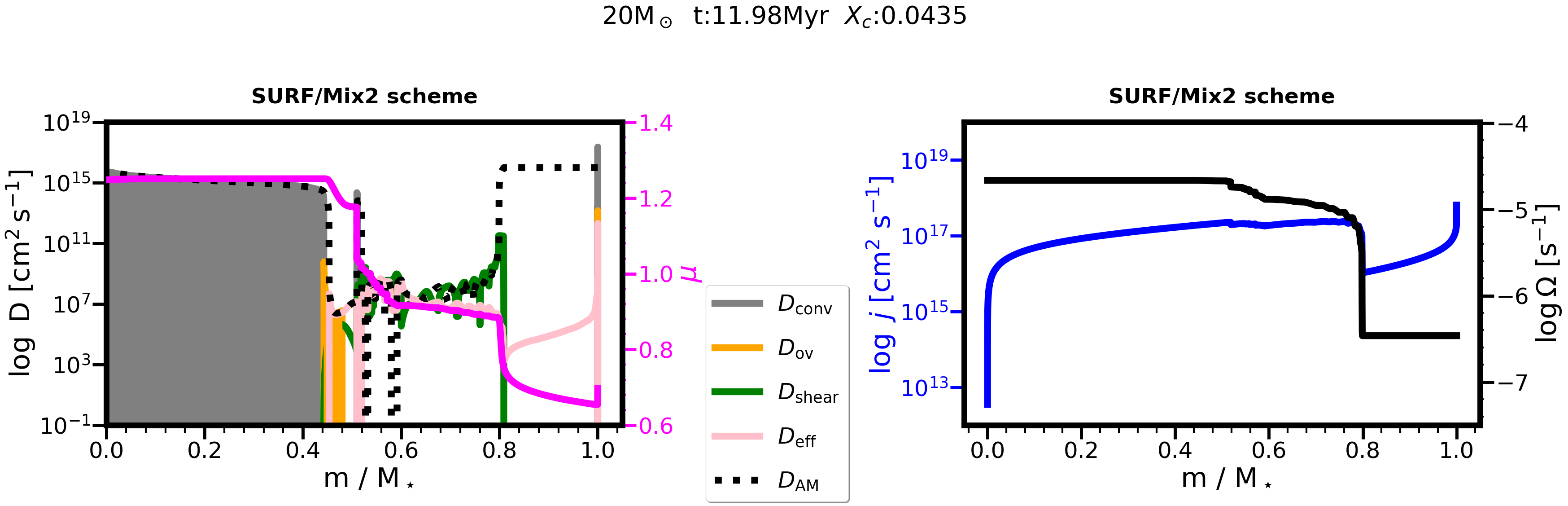}
\caption{Same as Figure \ref{fig:app1a} but for the SURF/Mix2 scheme.  }\label{fig:app1d}
\end{figure*}

Here, we describe the stellar structure models of the fiducial model and the corresponding models with initial masses of 20\,M$_\odot$ in the other braking and mixing schemes at times close to the ZAMS and TAMS. 
In Figure~\ref{fig:app1a} the INT/Mix1 models are shown. Initially, meridional currents dominate chemical mixing. Given the very efficient magnetic braking in all layers and the solid-body rotation, the angular velocity decreases more than an order of magnitude in all stellar layers from ZAMS to TAMS. Correspondingly, chemical mixing becomes drastically inefficient, leaving a gap between the stellar core and the surface. In Figure~\ref{fig:app1b} the INT/Mix2 models are shown. The results are similar to the INT/Mix1 scheme; however, at the TAMS the effective diffusion coefficient (combining meridional currents and horizontal turbulence) remains to transport chemical elements, although with relatively low values ($D\approx10^2$\,cm$^2$s$^{-1}$) at the core boundary. 

In Figure~\ref{fig:app1c} the SURF/Mix1 models are shown. Similar to the INT/Mix1 scheme, at the ZAMS meridional circulation is the most efficient process driving chemical mixing, while the GSF instability is also present throughout the radiative envelope of the star. At the TAMS, this instability takes over and becomes the dominant process for chemical mixing. Unlike in the INT models, the SURF models show a break in the angular velocity profile, and correspondingly a change in the distribution of specific angular momentum. The stellar core has much more angular momentum (close to its initial value) than in the INT models at the TAMS. In Figure~\ref{fig:app1d} the SURF/Mix2 models are shown. Similarly to the SURF/Mix1 scheme, chemical mixing is initially dominated by meridional circulation before becoming mostly driven by another instability, here by shear mixing. In this scheme, magnetic braking changes the angular momentum profile, and the surface layers have much less angular momentum than the other models. Consequently, the surface rotation of these models at the TAMS is very slow, while again the core maintains a significant fraction of angular momentum. 
Finally, let us note how the stellar age systematically increases in the TAMS models from the INT/Mix1 (8.56 Myr) to SURF/Mix2 (11.98 Myr) schemes as a consequence of the different braking and mixing assumptions.


\section{Evolutionary models}\label{sec:appevol}

\subsection{Fiducial model on the Hunter diagram}\label{sec:fidhunt}
%
%
\begin{figure*}
\includegraphics[width=9cm]{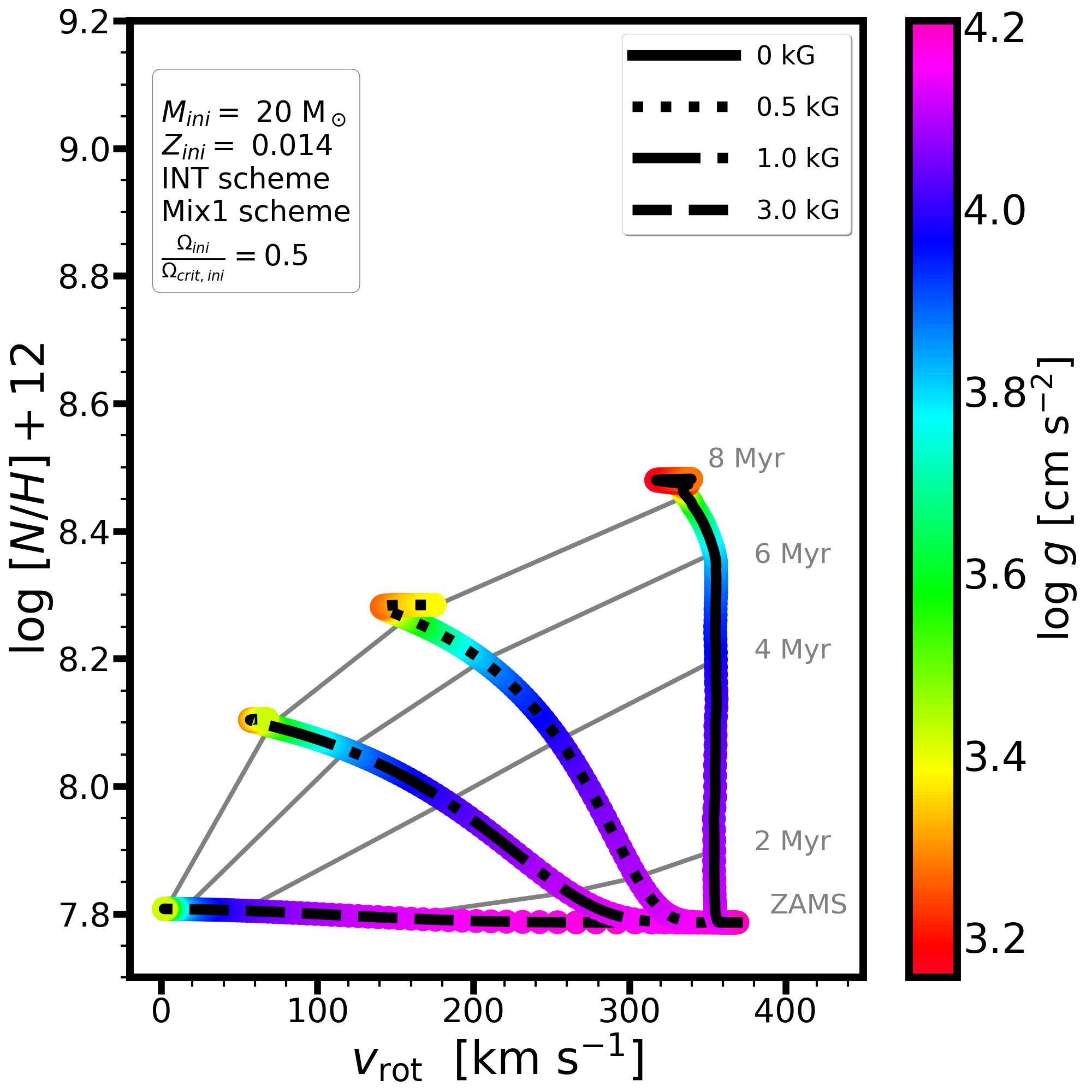}\includegraphics[width=9cm]{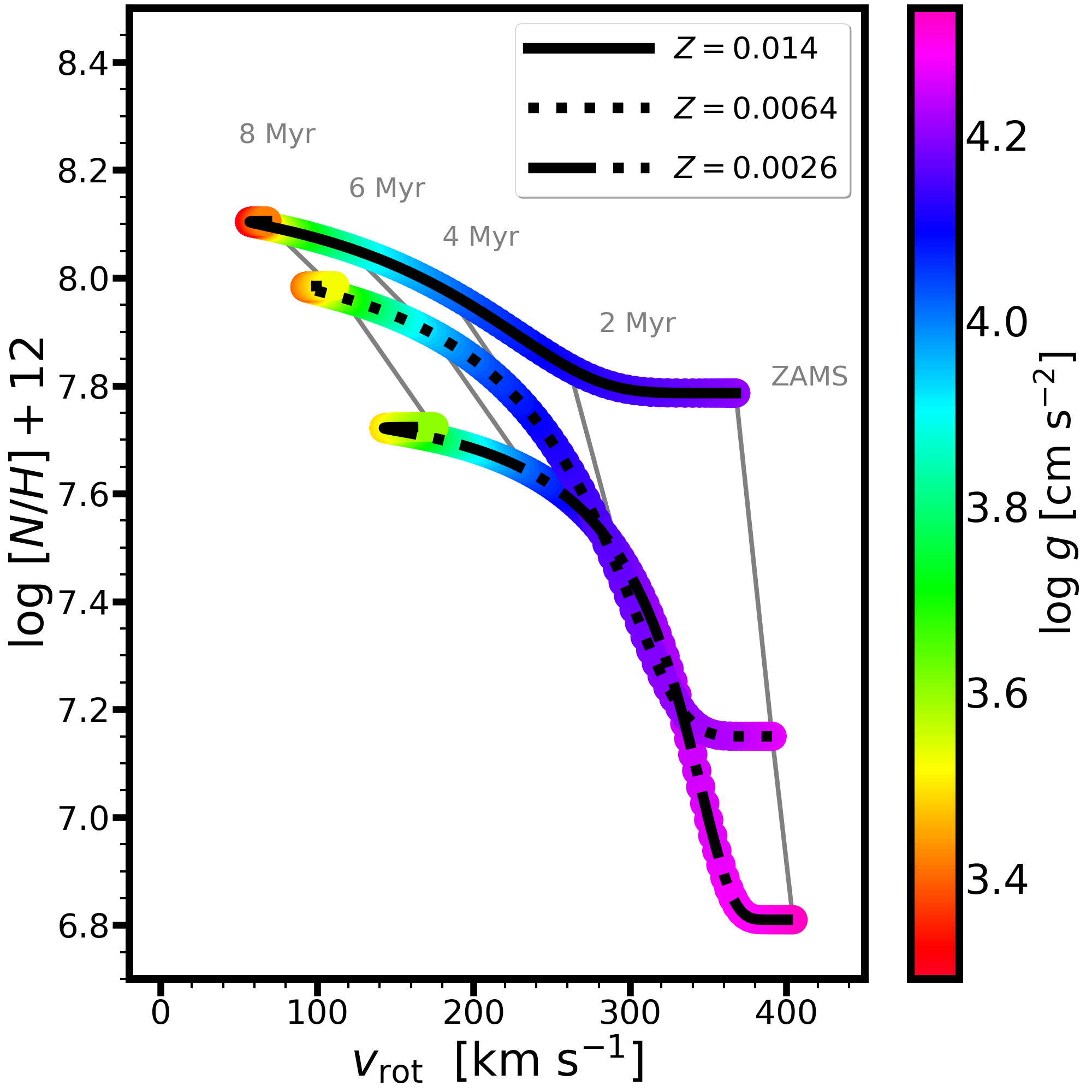}
\caption{Hunter diagrams of the fiducial 20~M$_\odot$ models. \textit{Left}: varying the initial magnetic field strength for given metallicity, braking, and mixing schemes. \textit{Right}: varying the initial metallicity representative of solar, LMC, and SMC environments for 1 kG initial field strength.}\label{fig:fidu}
\end{figure*}
%
%
%
%
In Figure \ref{fig:fidu}, we follow the evolution of our 20~M$_\odot$ fiducial models on the Hunter diagram \citep{hunter2008,hunter2009}, showing surface nitrogen abundance against the rotational velocity. The ZAMS is at the lower right corner of the diagrams with high surface gravities (colour-coded).  

%
First, we demonstrate how a competition between magnetic braking and rotational mixing alters the evolutionary tracks (see also \PaperIII). The left panel of Figure~\ref{fig:fidu} shows how the magnetic field strength results in drastic changes of the model predictions. The stronger the magnetic field, the more rapidly rotation brakes, and the less nitrogen can be mixed to the stellar surface. Above a given magnetic field strength, which strongly depends on other stellar and mixing parameters, the braking is efficient enough that nitrogen is not mixed to the surface and the observable nitrogen abundance remains the initial value throughout the main sequence. For the case of the 20~M$_{\odot}$ solar-metallicity model with the INT/Mix1 scheme shown in Figure~\ref{fig:fidu}, this threshold occurs roughly at 3~kG.

%
The right panel of Figure~\ref{fig:fidu} reveals the impact of the initial metallicity on the evolution of the 20~M$_{\odot}$ INT/Mix1 models in the Hunter diagram. The initial equatorial magnetic field strength is chosen to be 1~kG. Consistently with previous findings (see Section \ref{sec:metevol}), we obtain that chemical mixing is easier to trace in lower metallicity environments where the fractional change between the initial and final abundances are much larger (factor of $\sim~8$ at SMC metallicity) than at higher metallicity (factor of $\sim~2$ at solar metallicity) for the same fiducial model.

%
%
%
%

%
%
\begin{figure*}
\includegraphics[width=6cm]{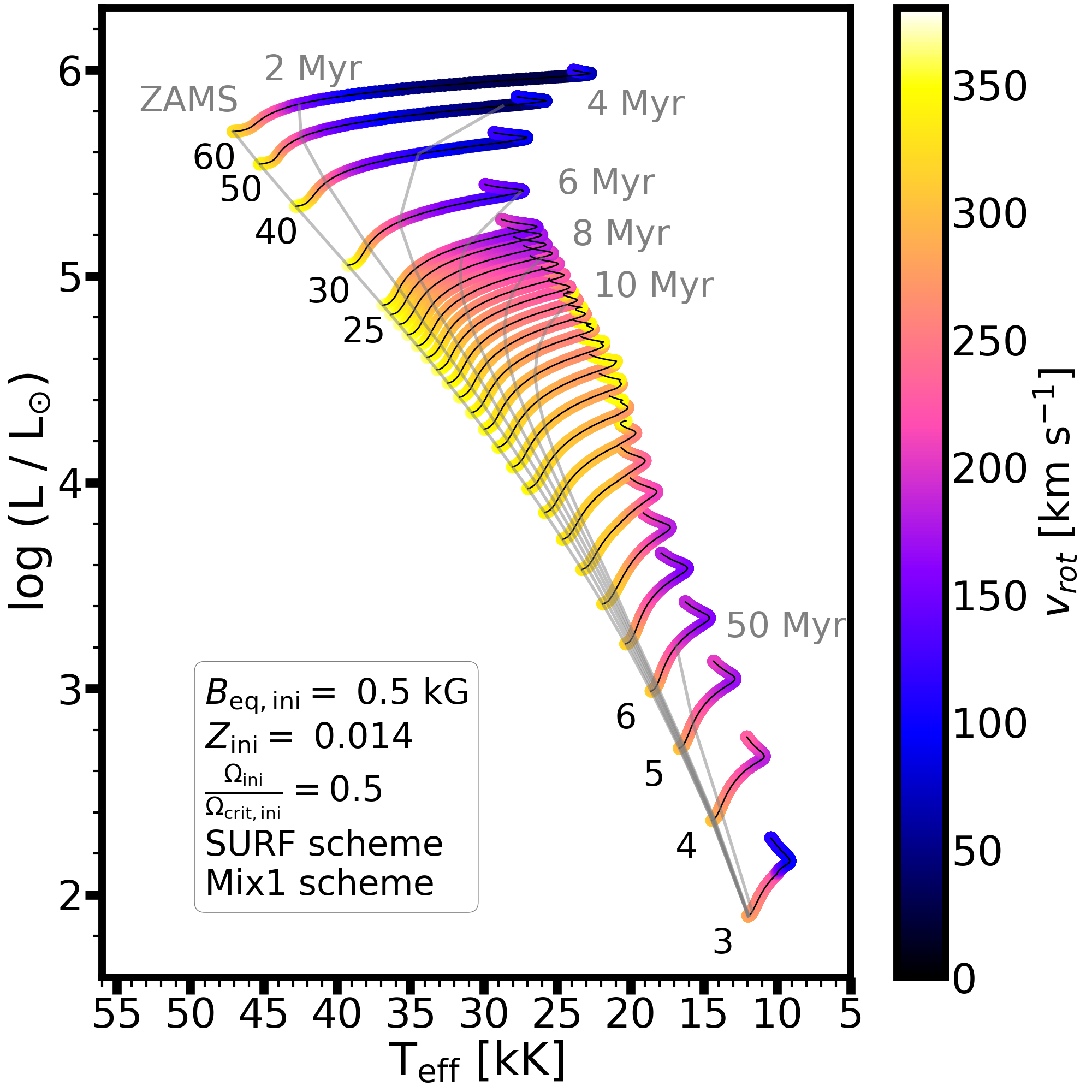}\includegraphics[width=6cm]{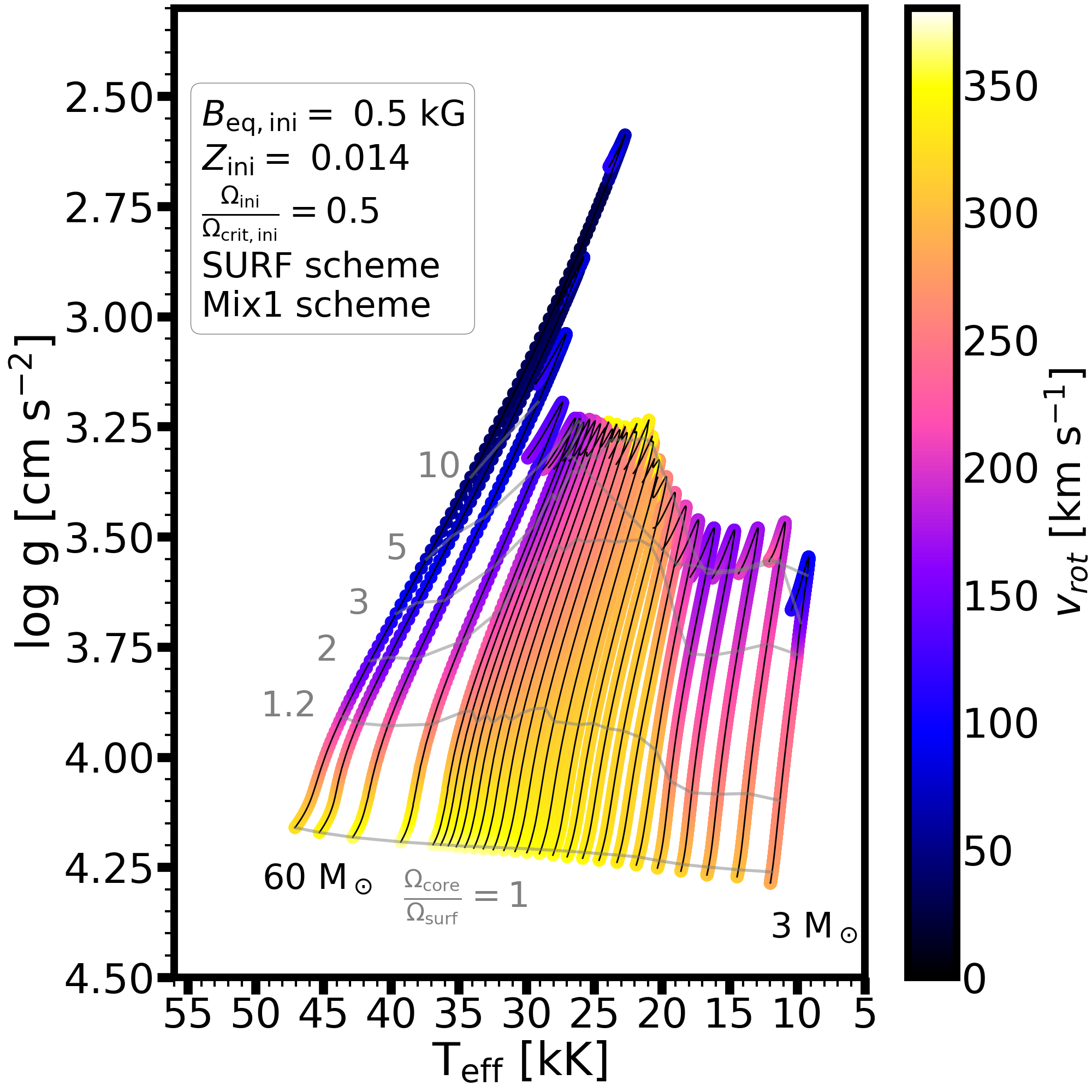}\includegraphics[width=6cm]{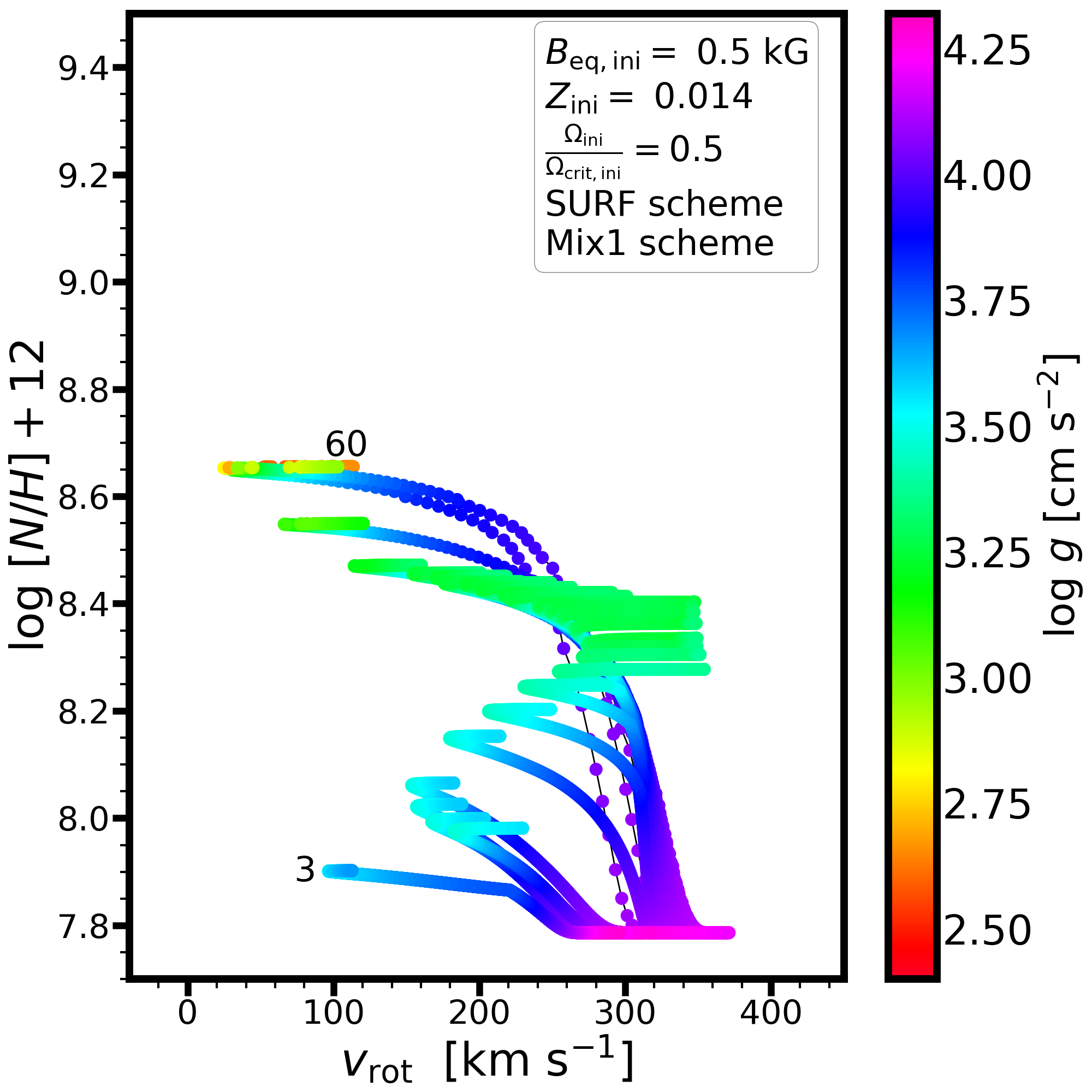}
\includegraphics[width=6cm]{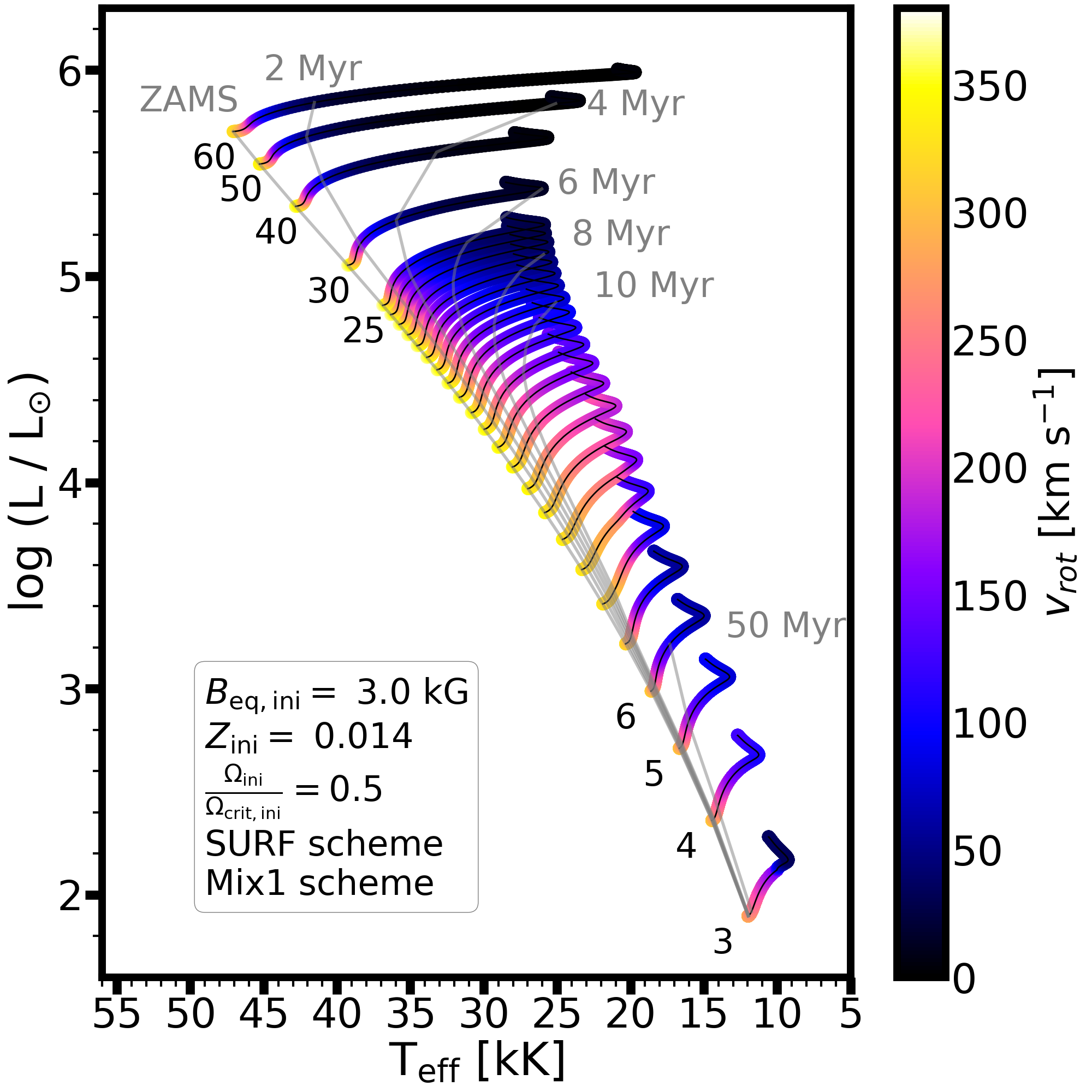}\includegraphics[width=6cm]{fig/Z14SURF/Mix1/kiel3000.png}\includegraphics[width=6cm]{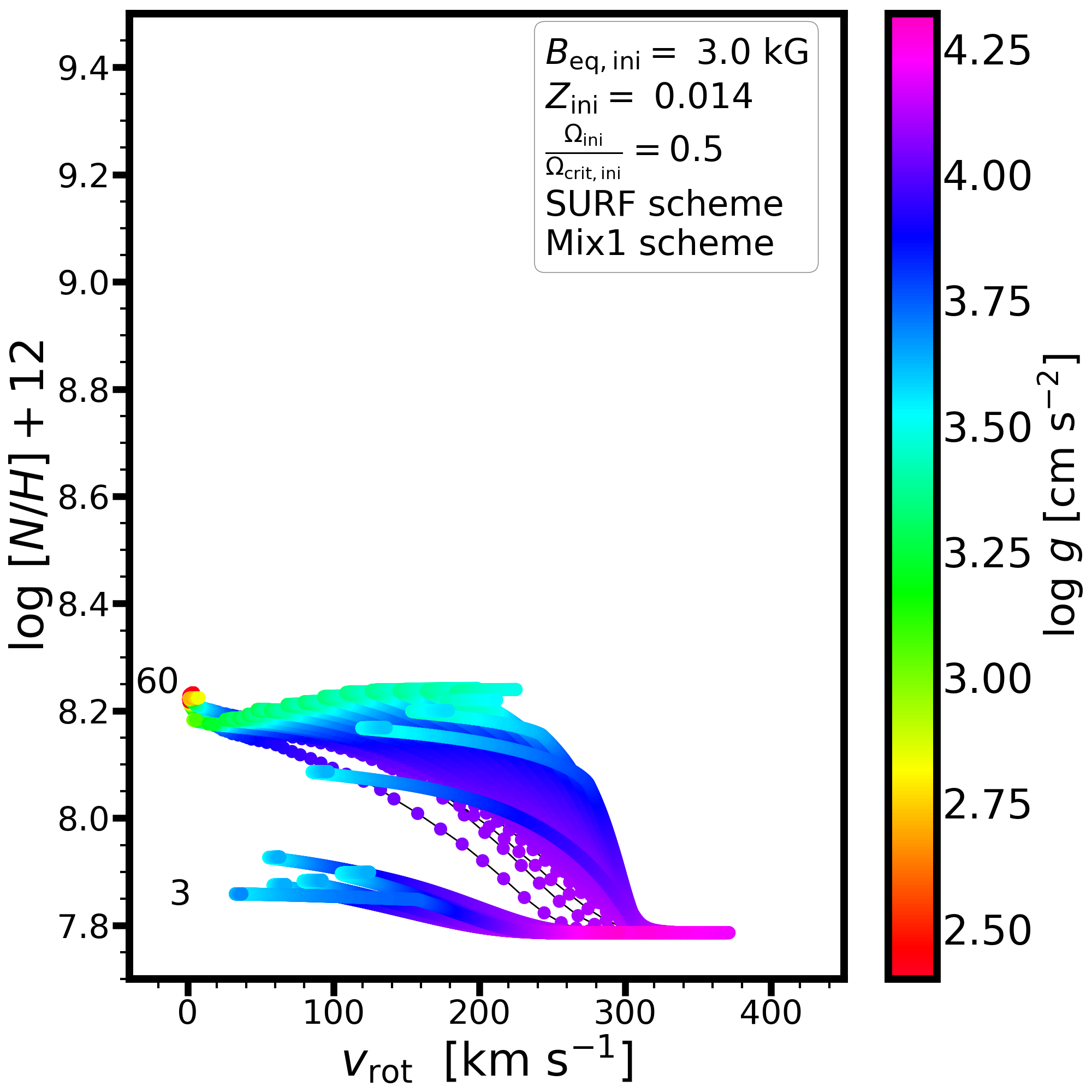}
\caption{Same as Figure \ref{fig:bfield1} but for the SURF/Mix1 scheme. Top panels show models with an initial equatorial magnetic field strength of 0.5 kG, whereas the lower panels show models with 3 kG. The NOMAG/Mix1 model is presented in Figure \ref{fig:bfield1}. }\label{fig:field3}
\end{figure*}
%
%
\begin{figure*}
\includegraphics[width=6cm]{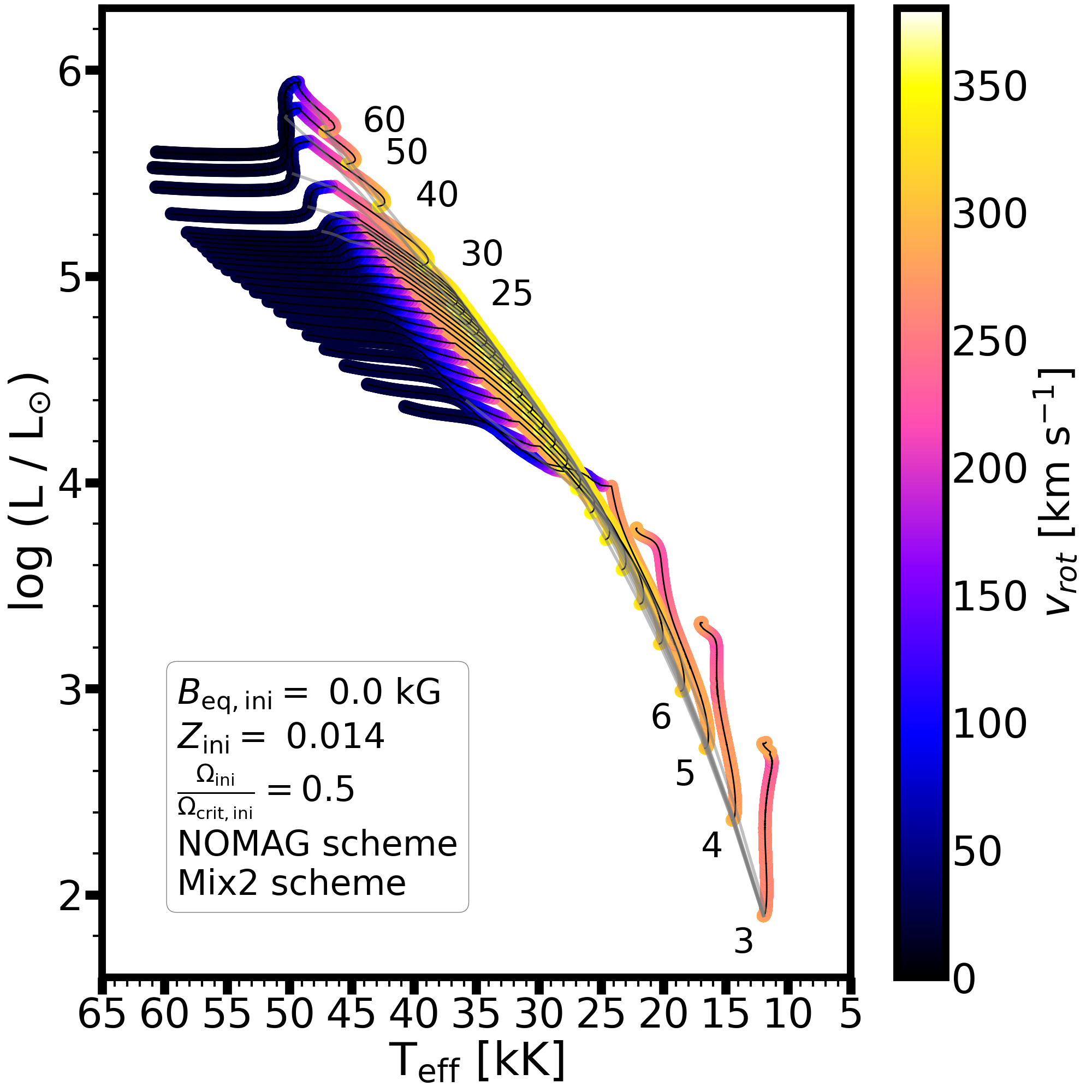}\includegraphics[width=6cm]{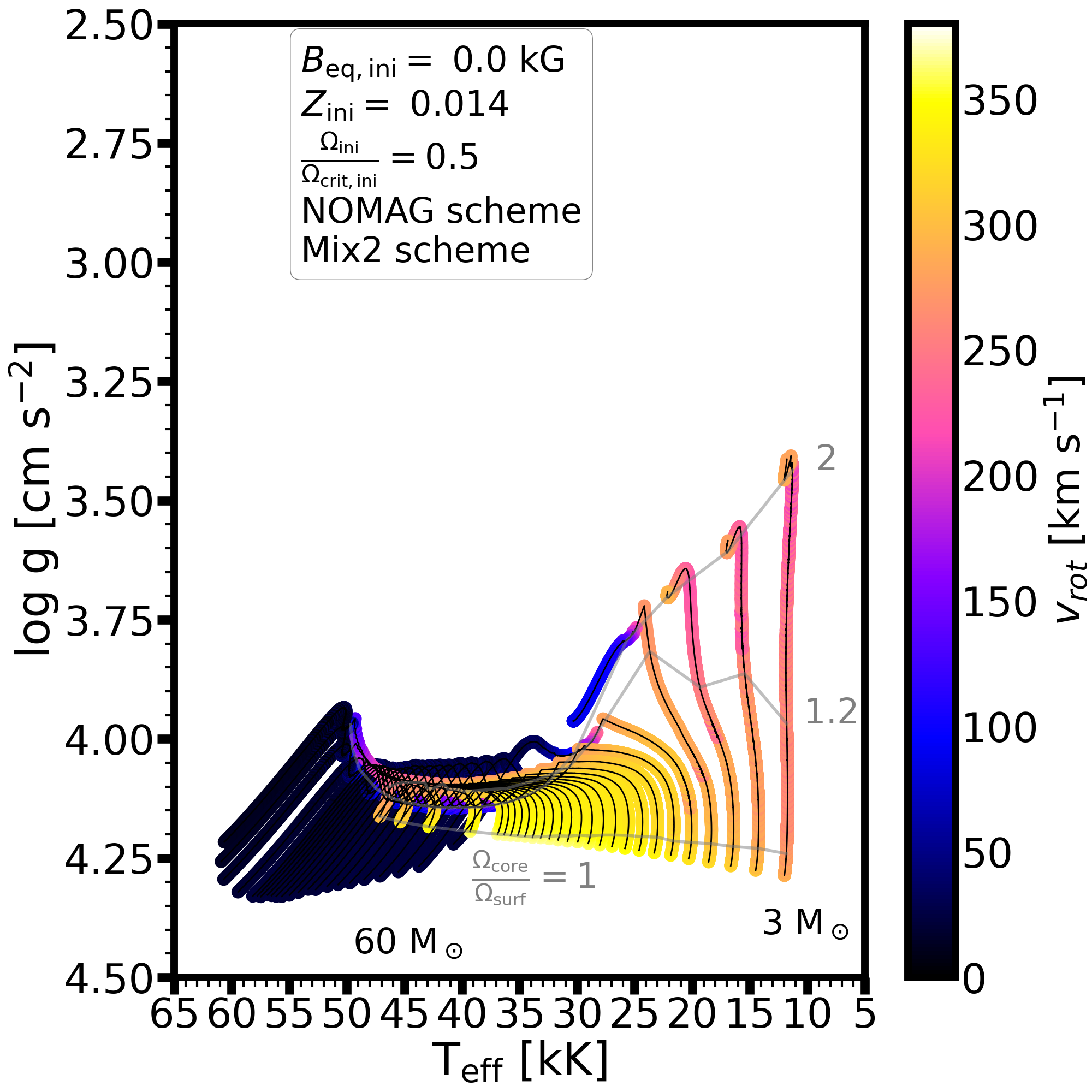}\includegraphics[width=6cm]{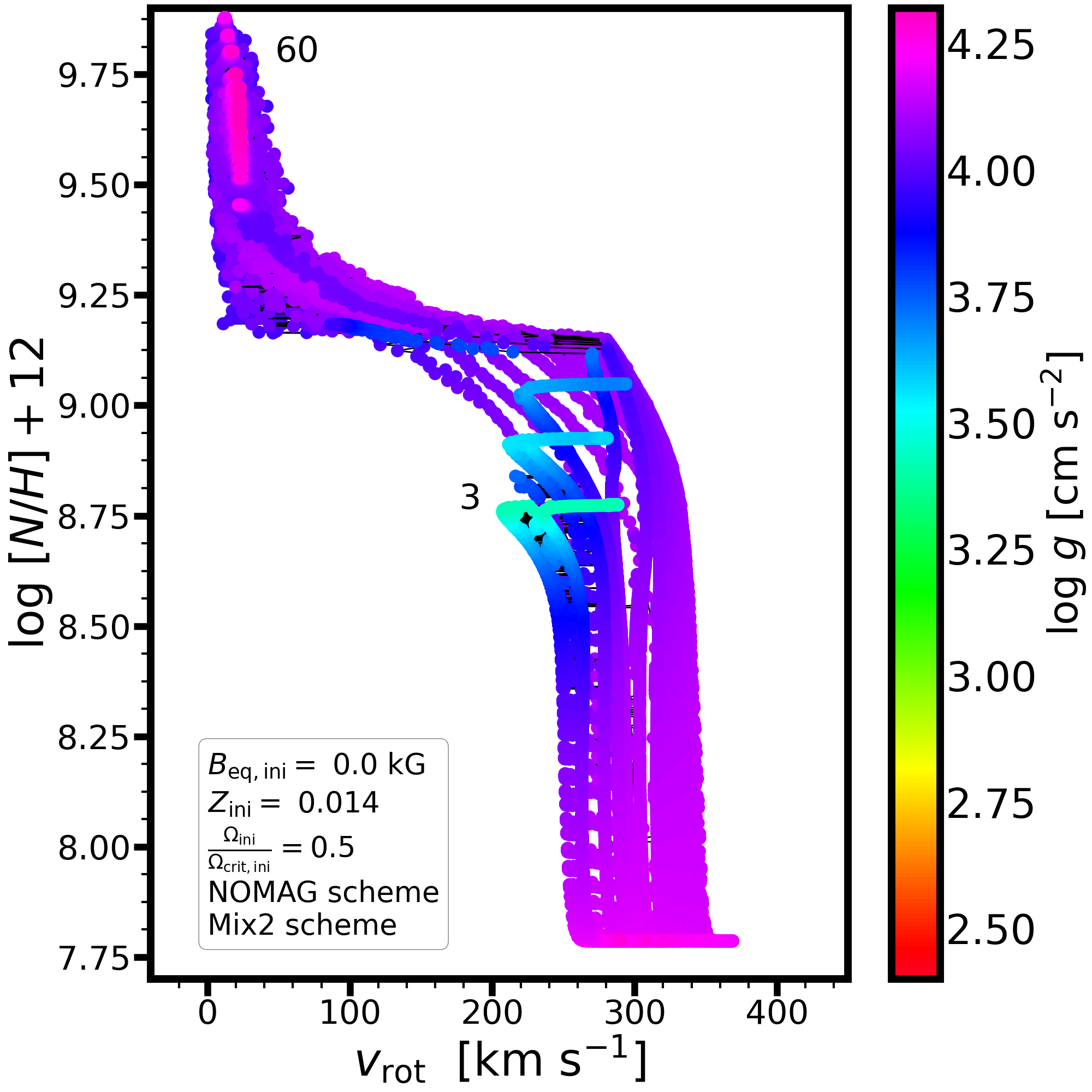} 
\includegraphics[width=6cm]{fig/Z14INT/Mix2/hrd500.png}\includegraphics[width=6cm]{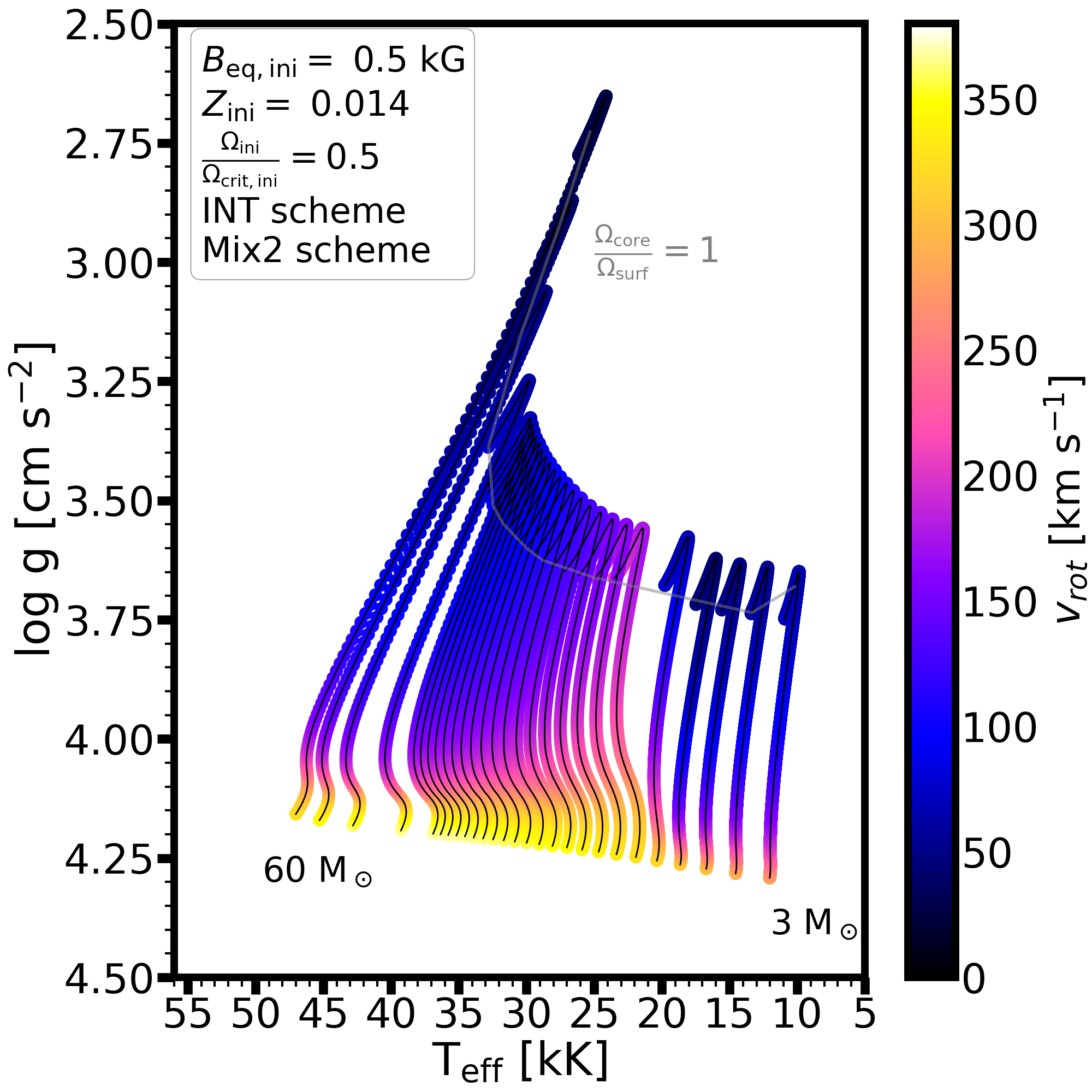}\includegraphics[width=6cm]{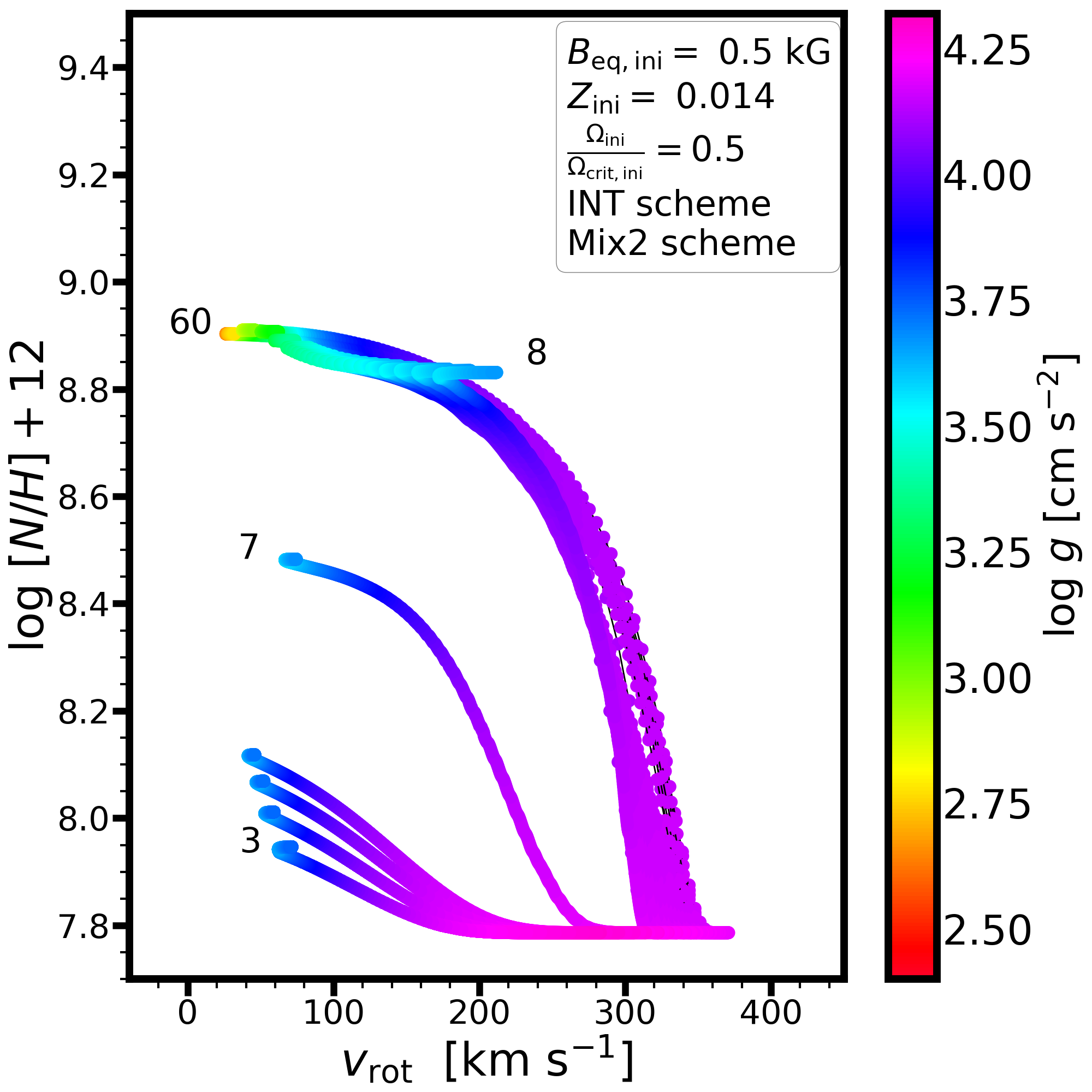}
\includegraphics[width=6cm]{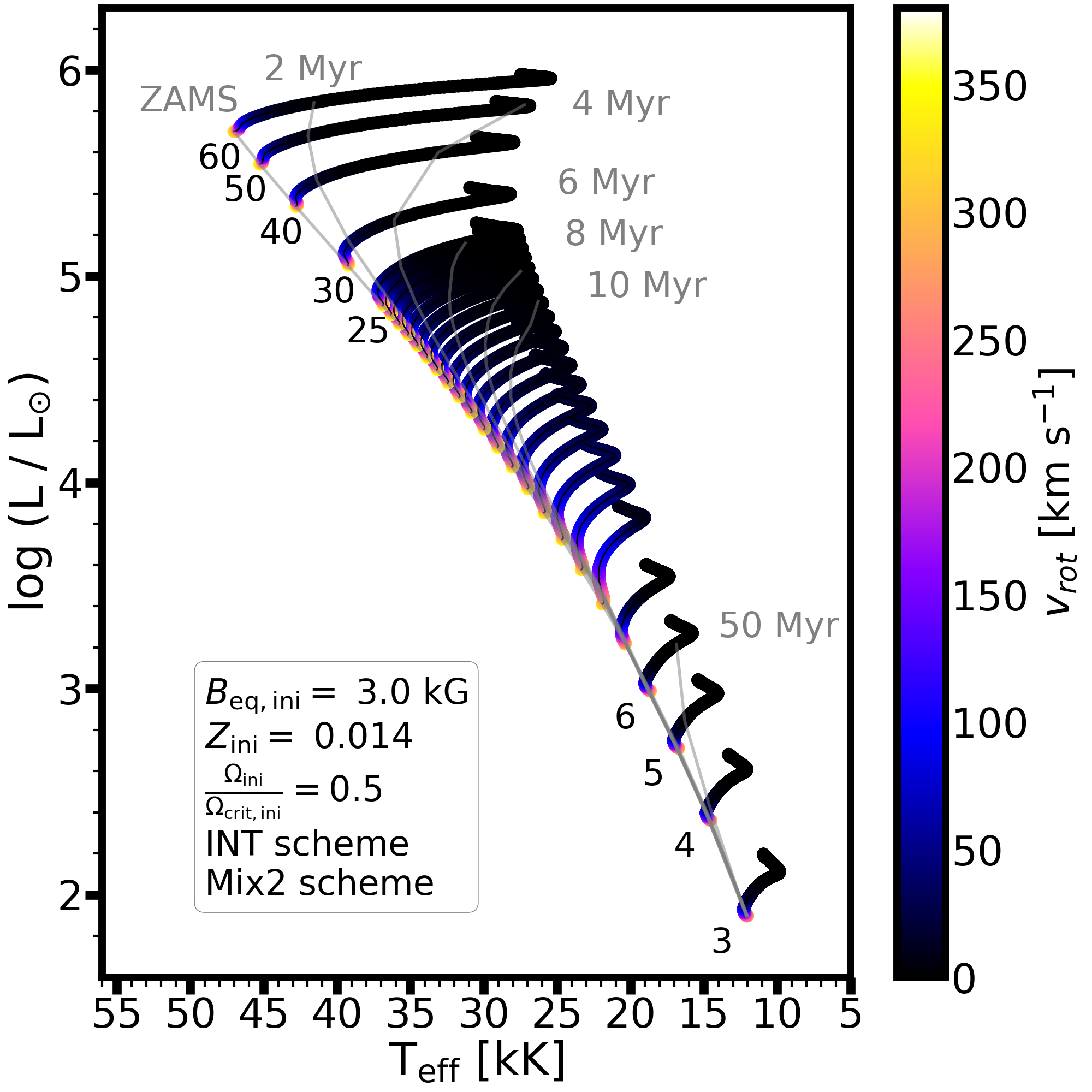}\includegraphics[width=6cm]{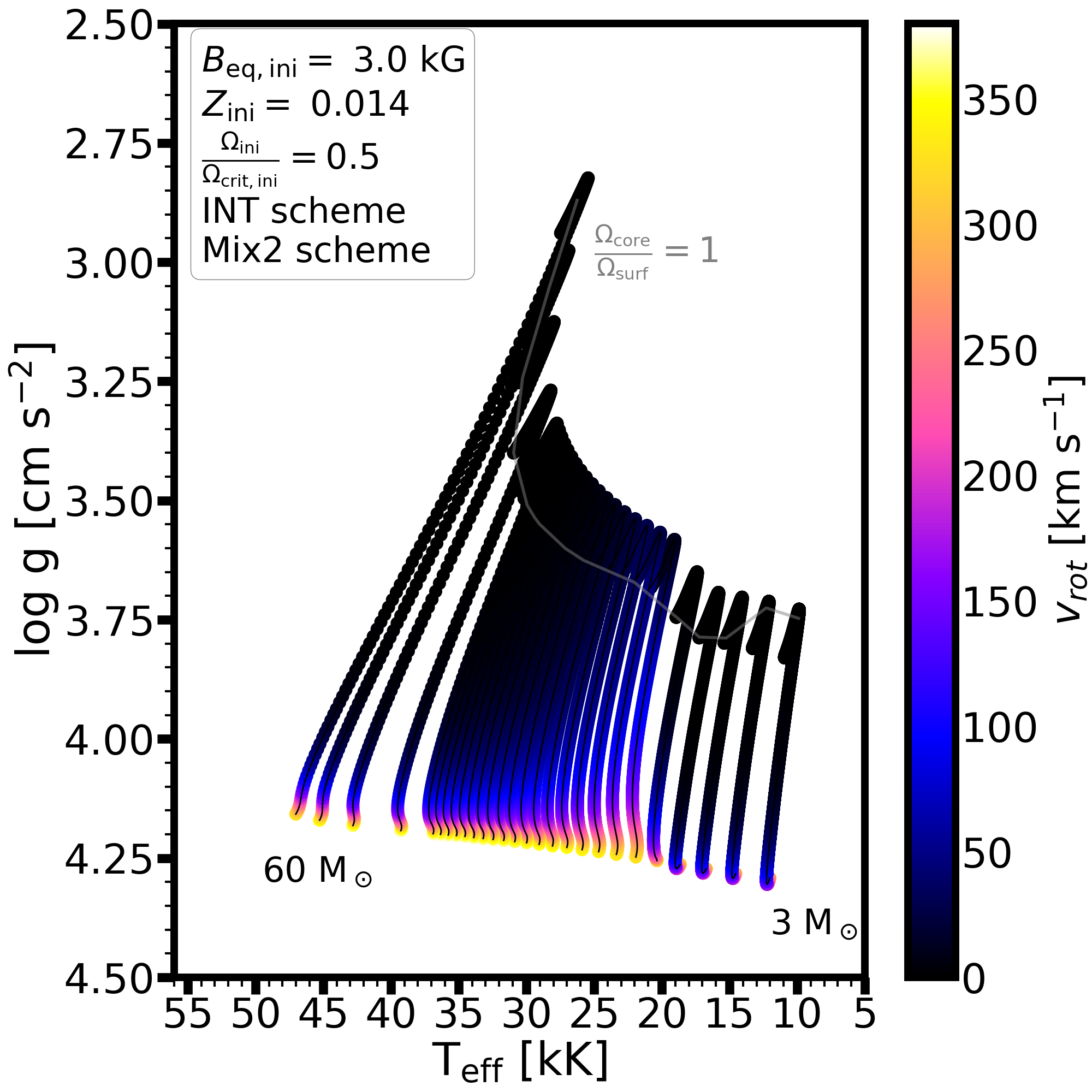}\includegraphics[width=6cm]{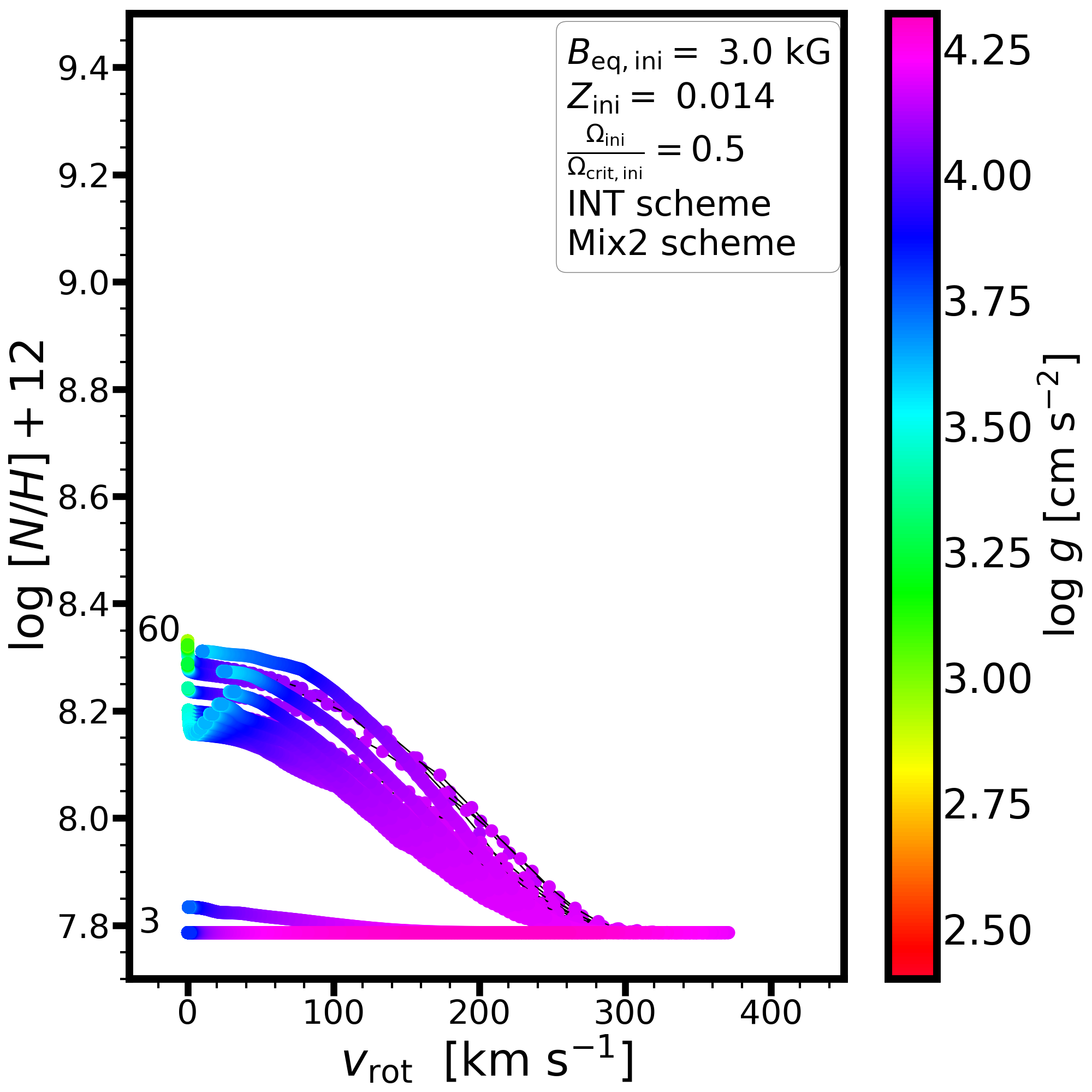}
\caption{Same as Figure \ref{fig:bfield1} but for the NOMAG/Mix2 (top panels) and INT/Mix2  (middle and lower panels) schemes.}\label{fig:field2}
\end{figure*}
%
%
\begin{figure*}
\includegraphics[width=6cm]{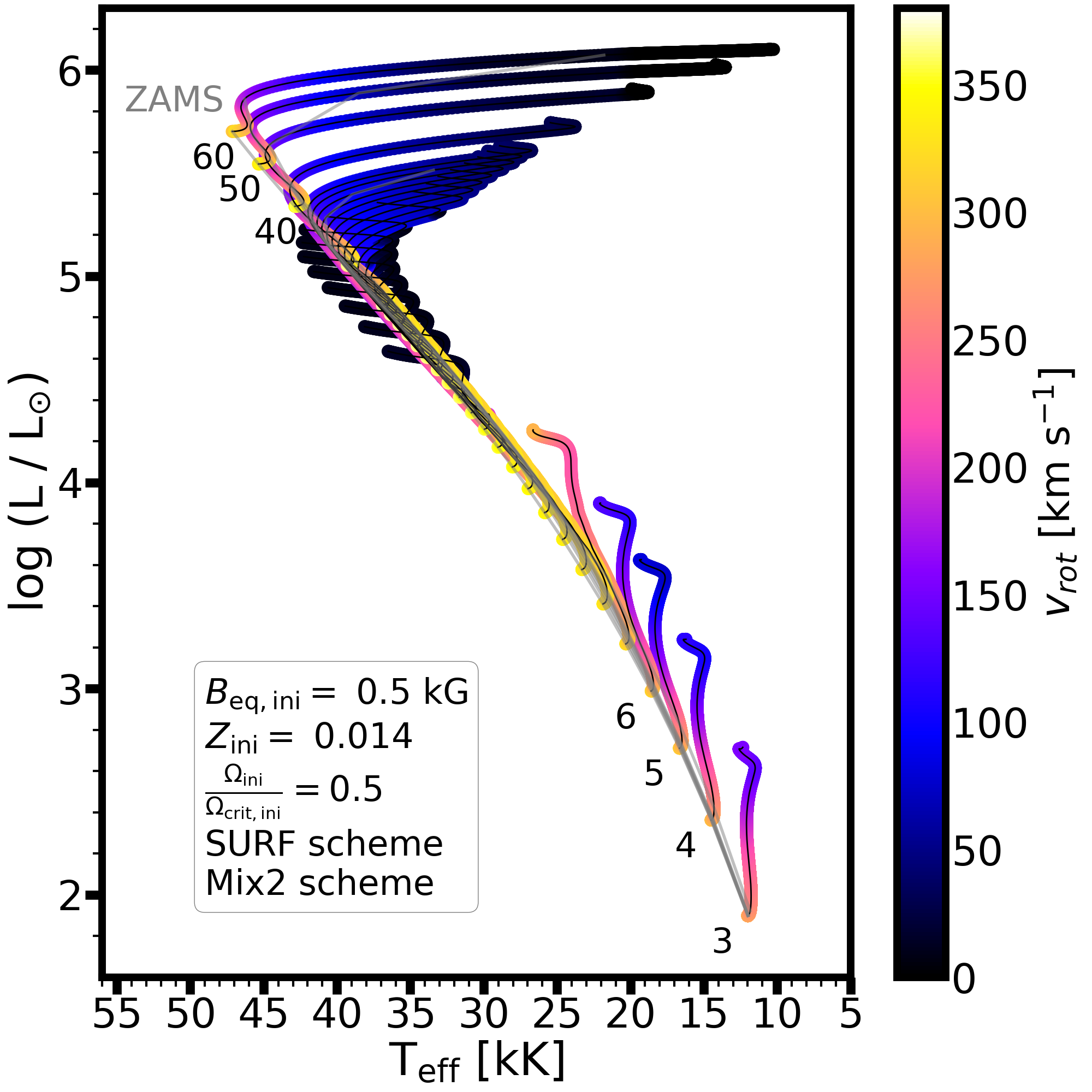}\includegraphics[width=6cm]{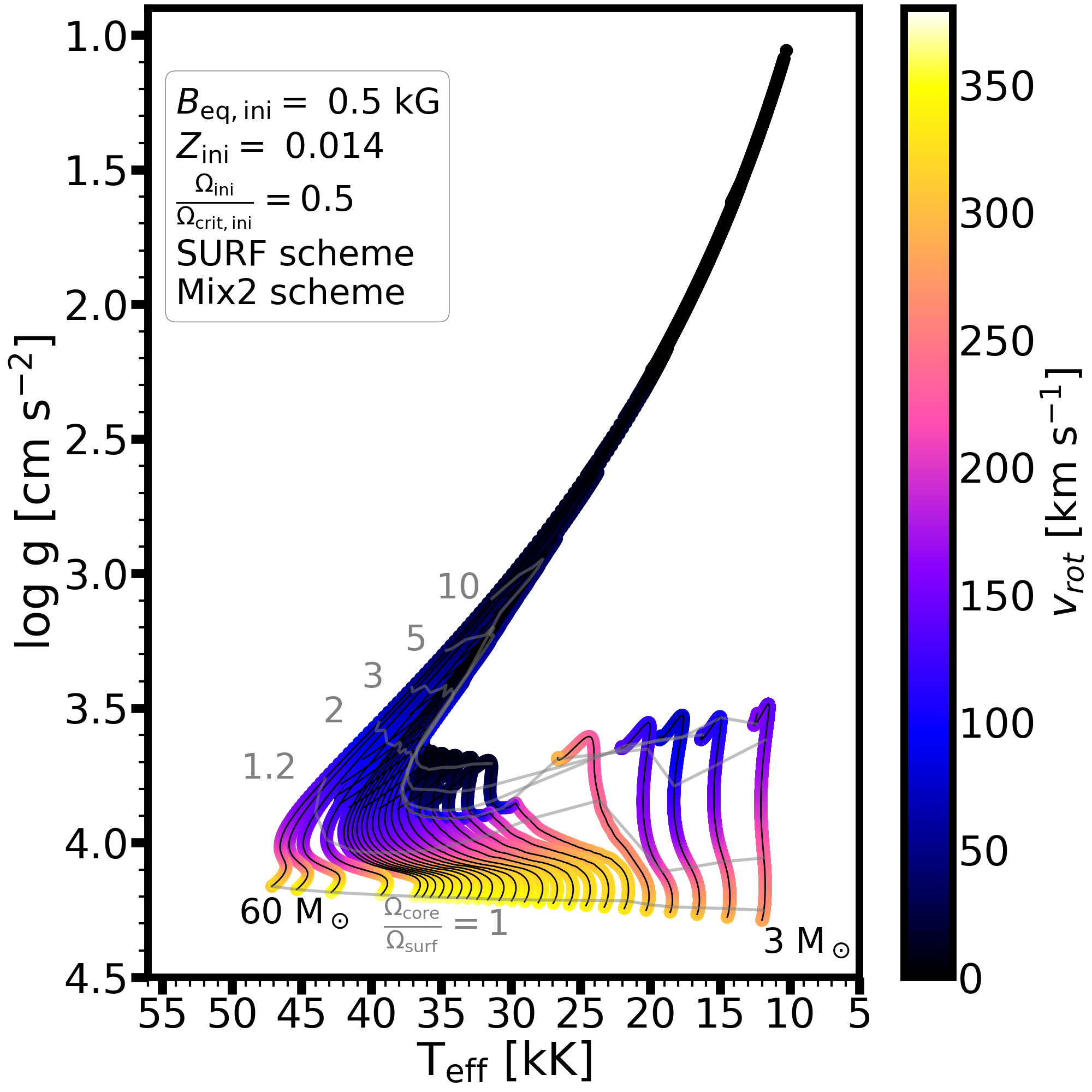}\includegraphics[width=6cm]{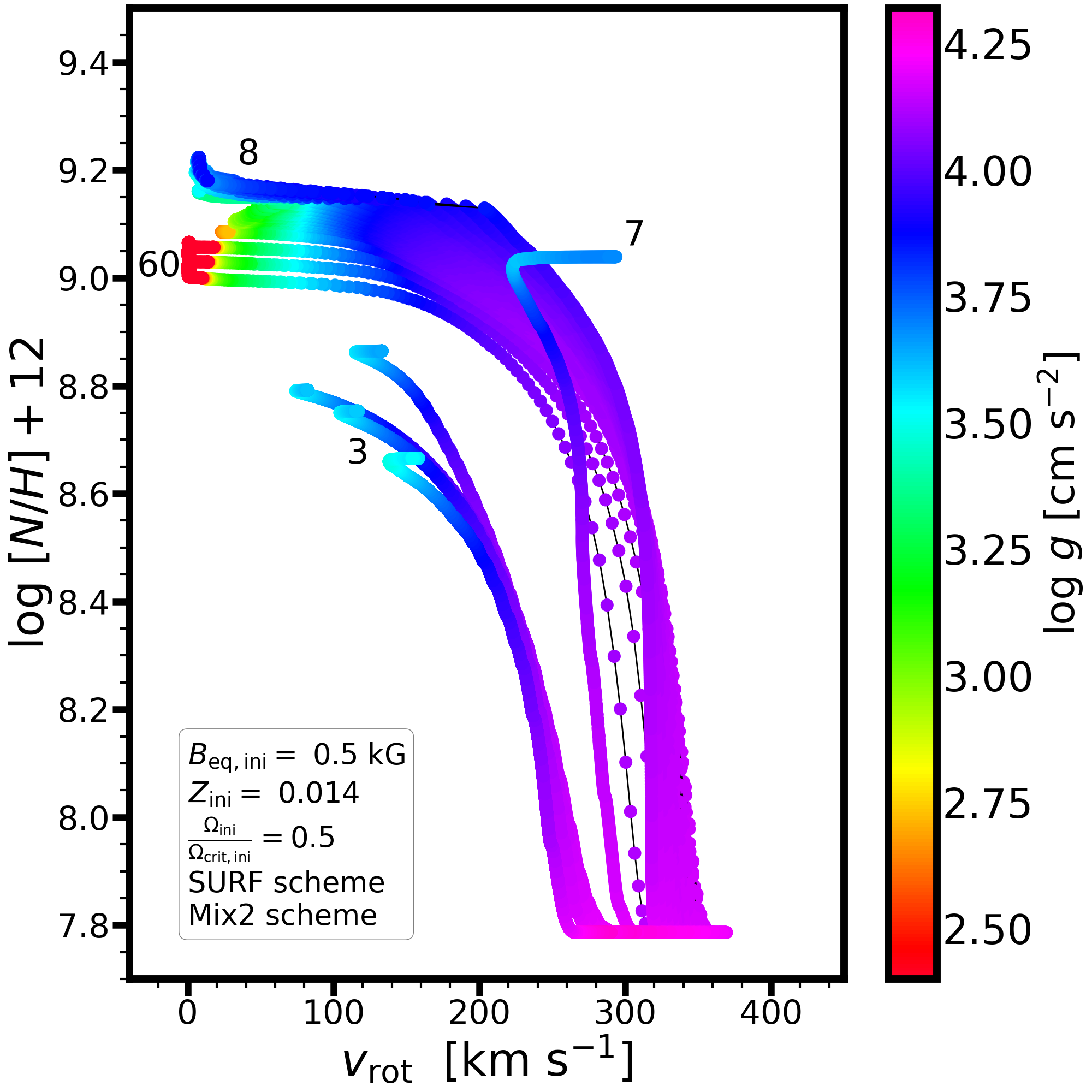}
\includegraphics[width=6cm]{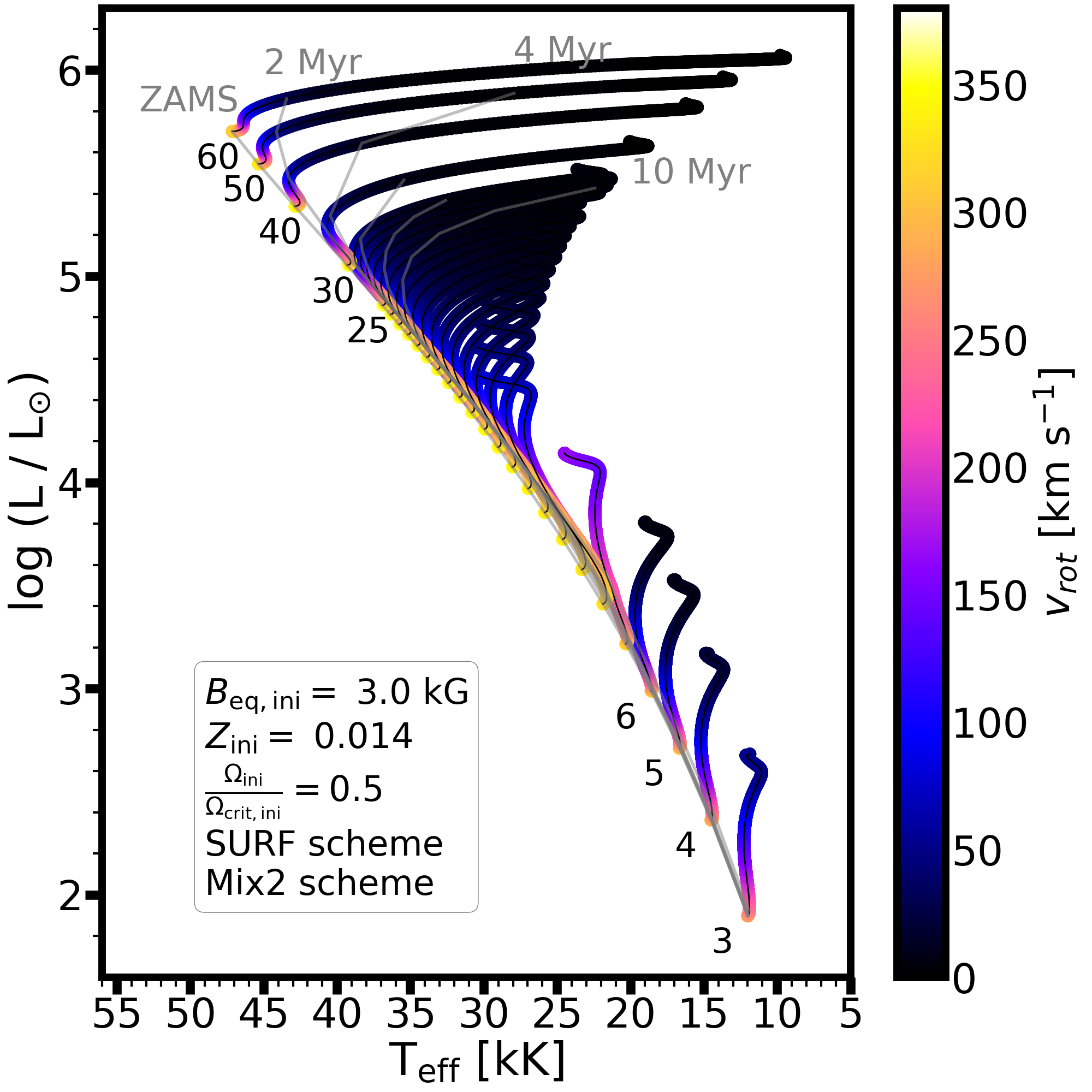}\includegraphics[width=6cm]{fig/Z14SURF/Mix2/kiel3000.png}\includegraphics[width=6cm]{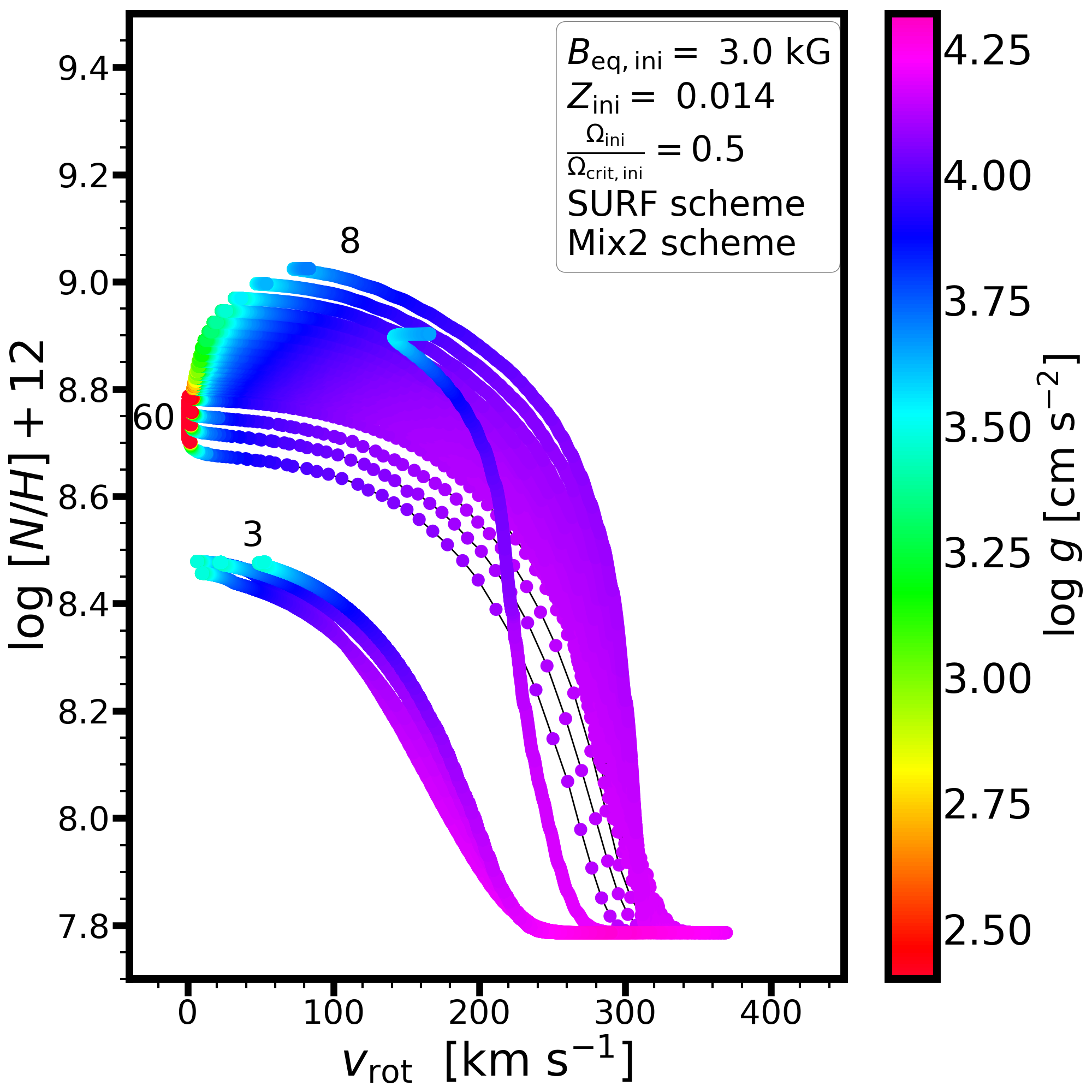}
\caption{Same as Figure \ref{fig:bfield1} but for the SURF/Mix2 scheme. See top panels of Figure \ref{fig:field2} for the NOMAG/Mix2 models.}\label{fig:field4}
\end{figure*}

\subsection{Impact of varying magnetic field strength within the other braking and mixing schemes at solar metallicity}

%
Here, we describe the effect of varying the magnetic field strength for the stellar evolution when accounting for the INT magnetic braking scheme in combination with the Mix2 chemical mixing scheme and the SURF magnetic braking scheme in combination with both the Mix1 and Mix2 chemical mixing schemes at solar metallicity. We display the evolutionary tracks for these models with initial magnetic field strengths of 0.5 and 3 kG in Figures \ref{fig:field3}, \ref{fig:field2}, and \ref{fig:field4}. The NOMAG/Mix1 models are shown in Figure \ref{fig:bfield1} and the NOMAG/Mix2 models are shown in Figure \ref{fig:field3}. 
This is complementary to the description in Section \ref{sec:fieldevol} and Figure \ref{fig:bfield1} for varying the magnetic field strength when using the INT magnetic braking scheme and the Mix1 chemical mixing scheme. 
Below, we first describe how the stellar evolution is affected by accounting for magnetic fields for each of the SURF/Mix1, INT/Mix2, and SURF/Mix2 cases.

\subsubsection{SURF/Mix1}

As described in Section~\ref{sec:evolutionary_tracks} and seen from Figure~\ref{fig:bfield1}, when the Mix1 rotational mixing scheme is used without including magnetic fields, the stars with initial masses $<30$~M$_{\odot}$ remain rapidly rotating ($>200$~km~s$^{-1}$) until the end of the main sequence. The nitrogen abundance increases smoothly (compared to the initial value, see Table~\ref{tab:met}) with the initial mass, stretching from $\log (N/H) + 12 = 8.0$ up to $8.8$. The stars develop mild differential rotation as the main sequence progresses, resulting in cores spinning roughly twice as fast as their envelopes towards the end of the main sequence. 

When treating magnetic fields using the SURF scheme, angular momentum is very efficiently transported in the outermost 20\% of the mass. In deeper layers of these models efficient shear mixing is invoked by differential rotation (see Section~\ref{sec:str}). 
As a result, mixing is enhanced in magnetic models treated with the SURF scheme compared to those treated with the INT scheme, which in contrast are rigidly rotating. Nitrogen is mixed out more efficiently, reaching surface abundances up to $\log (N/H) + 12 = 8.7$ in the SURF models compared to 8.6 in the INT models of stars with initial equatorial magnetic field strength of 0.5 kG. This can be seen when comparing the middle row of Figure~\ref{fig:bfield1} with the top row of \ref{fig:field3}. In fact, for stars with initial equatorial magnetic field strength of 0.5 kG in the SURF/Mix1 scheme, the surface enrichment of nitrogen is very similar throughout the main-sequence as for the corresponding non-magnetic models. However, the stars spin down somewhat more when magnetic fields are implemented. 
 
In the bottom row of Figure~\ref{fig:field3}, models with initial equatorial magnetic field strength of 3 kG are displayed. From these models it is even more evident that the SURF scheme gives rise to slower spin-down than the INT scheme does (cf.\ the bottom row of Figure~\ref{fig:bfield1}). While some stars still have a considerable rotational velocity at the end of the main-sequence, magnetic braking is sufficient to produce less enriched stars with nitrogen abundances of $\log (N/H) + 12 = 8.3$. When the INT scheme is implemented instead, none of the models with initial magnetic field strength of 3 kG exhibit nitrogen-enriched surfaces.

\subsubsection{INT/Mix2}

The Mix2 chemical mixing scheme is an efficient rotational mixing scheme, resulting in quasi-chemically homogeneous main-sequence evolution when magnetic fields are not taken into account (see Section~\ref{sec:str} and Figure~\ref{fig:str3_midMS}).  
The evolution of the non-magnetic models is displayed in the upper row of panels of Figures~\ref{fig:field2}. We can see that non-magnetic stars are expected to experience modest differential rotation (since a high diffusivity, which we attributed to the magnetic field, is not assumed in these models). All of the non-magnetic models efficiently mix nitrogen out to the surface, reaching values of $\log (N/H) + 12$ of 8.8-9.5. All but the most massive stellar models remain rapidly rotating throughout most of the main sequence, with surface rotation speeds of $\sim 200-350$~km~s$^{-1}$. 

%
%
In the middle row of Figure~\ref{fig:field2}, we display models initialised with 0.5 kG magnetic field strengths according to the INT scheme. This magnetic field strength is sufficient to brake the rotation of the stars, but only after a significant amount of nitrogen has been mixed out to the surface. As a result, all stars rotate slower than $\sim 50$~km~s$^{-1}$ once the TAMS is reached, and stars with initial masses above 7~$M_{\odot}$ have surface nitrogen abundances of $\log (N/H) + 12 \sim 8.9$. As visible from the colour-coding in the Kiel diagram, both the most massive and least massive stars spin down early during the main-sequence evolution, while the models in the intermediate-mass range of $\sim 5-10$~M$_{\odot}$ spin down later during the main-sequence evolution (see Section~\ref{sec:feature}). 

As seen from Figure~\ref{fig:field2}, a modest initial magnetic field strength of 0.5 kG is sufficient to completely alter the evolution in the Hertzsprung-Russell diagram. If magnetic fields are not accounted for, the stars are expected to evolve quasi-chemically homogeneously during the entire main sequence, which is not the case when magnetic fields are taken into account.
In the Kiel diagrams, the ratios of the core to envelope spins are shown. When magnetic field effects are accounted for in the INT scheme, the models rotate rigidly throughout the main sequence (cf.\ Section \ref{sec:kielevol}). 

In the case of models with initial magnetic field strength of 3 kG (bottom row of Figure~\ref{fig:field2}), magnetic braking is even more efficient compared to the case of the 0.5 kG models. As a result, rotational mixing can only act briefly before the stars have spun down, leading to slow rotation rates of $\lesssim 50 $~km~s$^{-1}$ already in the beginning of the main sequence (for example when $\log g ~\sim~4.0$). The maximum surface nitrogen abundances reached are also correspondingly low in comparison to the non-magnetic and weakly magnetic cases, reaching $\log (N/H) + 12 \sim 8.4$.

\subsubsection{SURF/Mix2}

Since stars spin down slower when the SURF scheme is used compared to the INT scheme to implement magnetic fields, the quasi-chemically homogeneous main sequence evolution that the non-magnetic models in the Mix2 rotational mixing scheme exhibit is not as easily quenched by the implementation of magnetic fields in the SURF/Mix2 combination. The effect of the weaker magnetic braking is evident when comparing Figures~\ref{fig:field2} and \ref{fig:field4}. 

As a result of the slower spin-down, even stars with initial magnetic field strength of 3 kG still substantially enrich their surfaces with nitrogen ($\log (N/H) + 12 = 8.7-9.2$). They also reach rotation rates close to zero at the end of the main sequence. As seen in Figure~\ref{fig:str3_midMS}, the combination of SURF and Mix2 results in such efficient mixing that even the surface helium abundance significantly increases during the main sequence.

%
%
%
%

%
\begin{figure*}
\includegraphics[width=6cm]{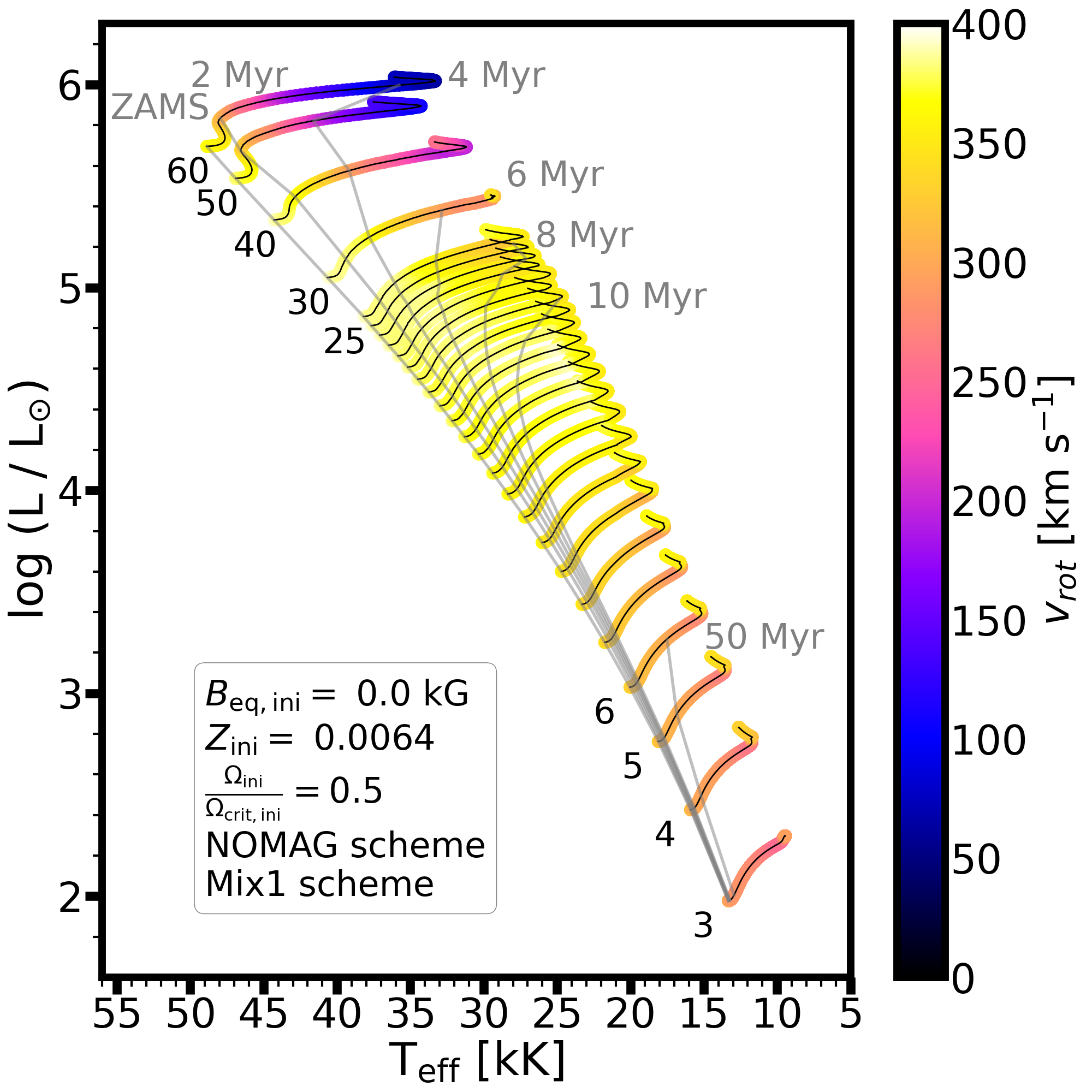}\includegraphics[width=6cm]{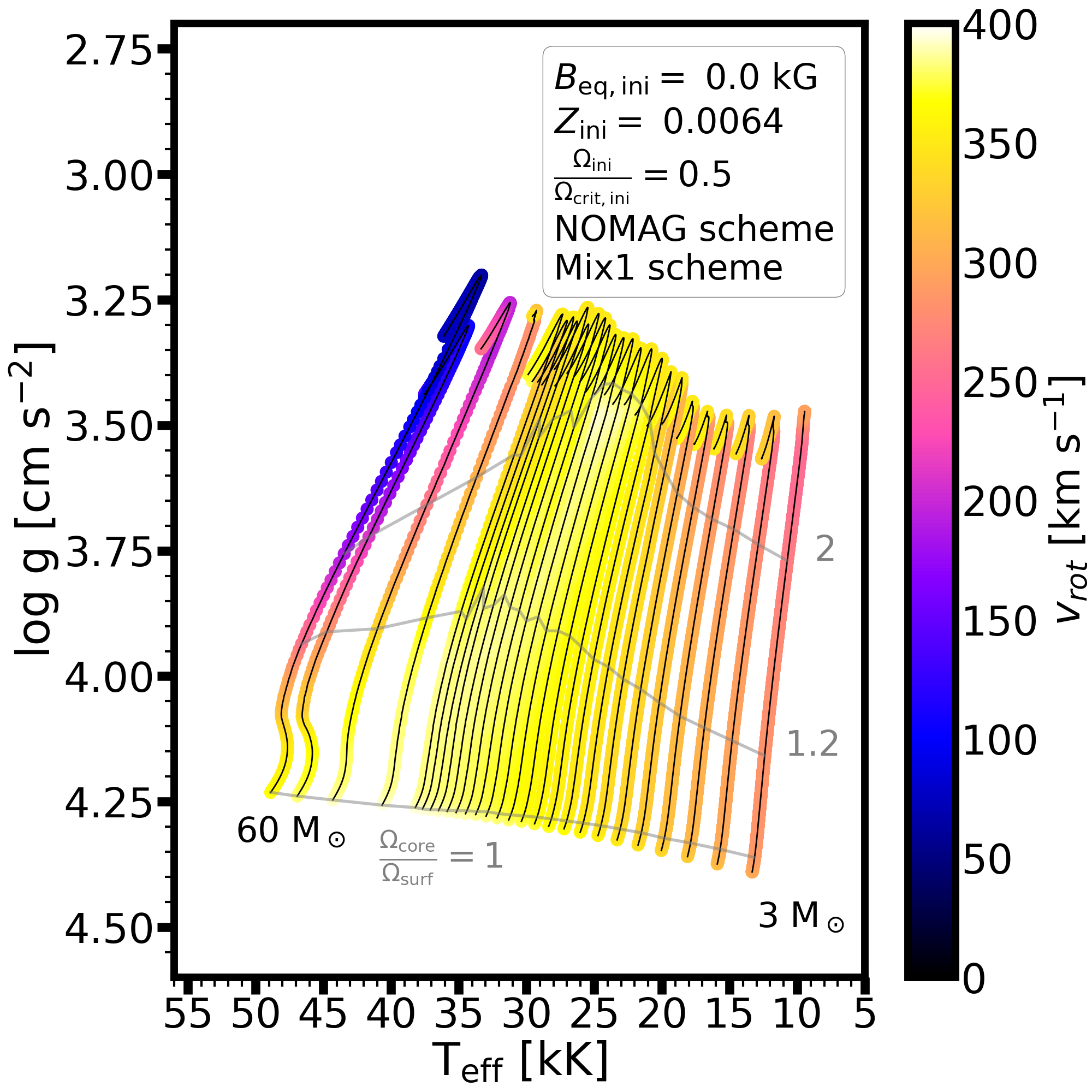}\includegraphics[width=6cm]{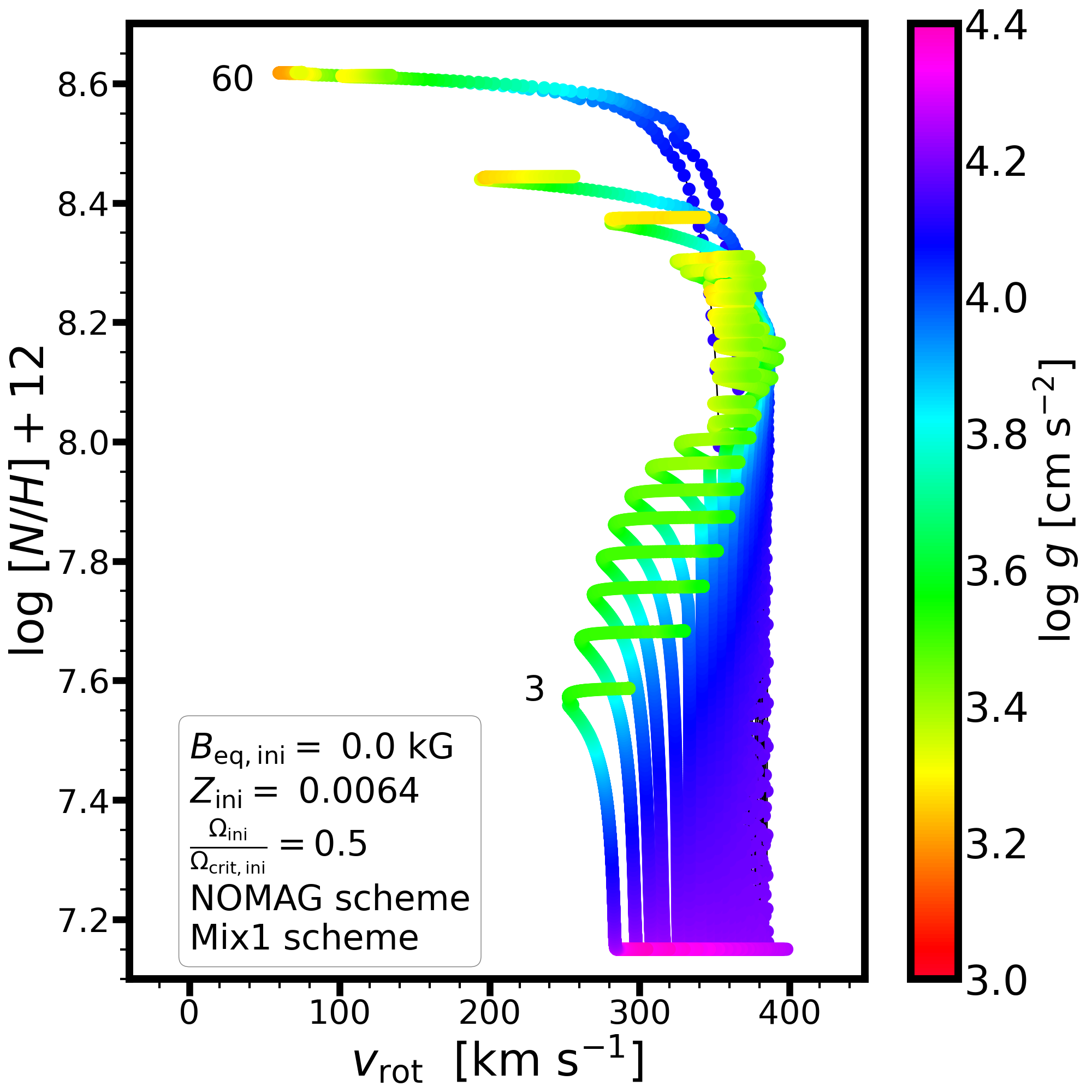} 
\includegraphics[width=6cm]{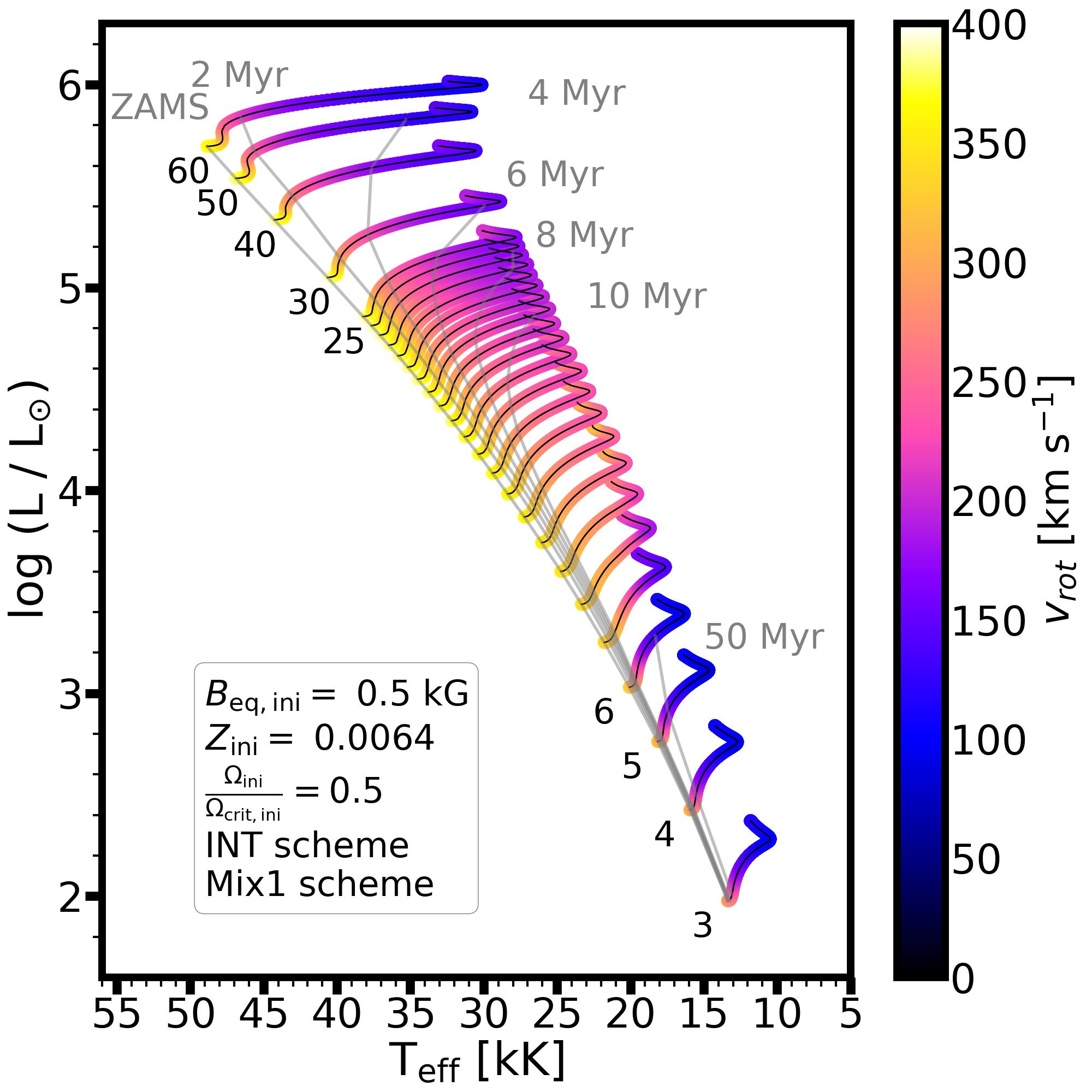}\includegraphics[width=6cm]{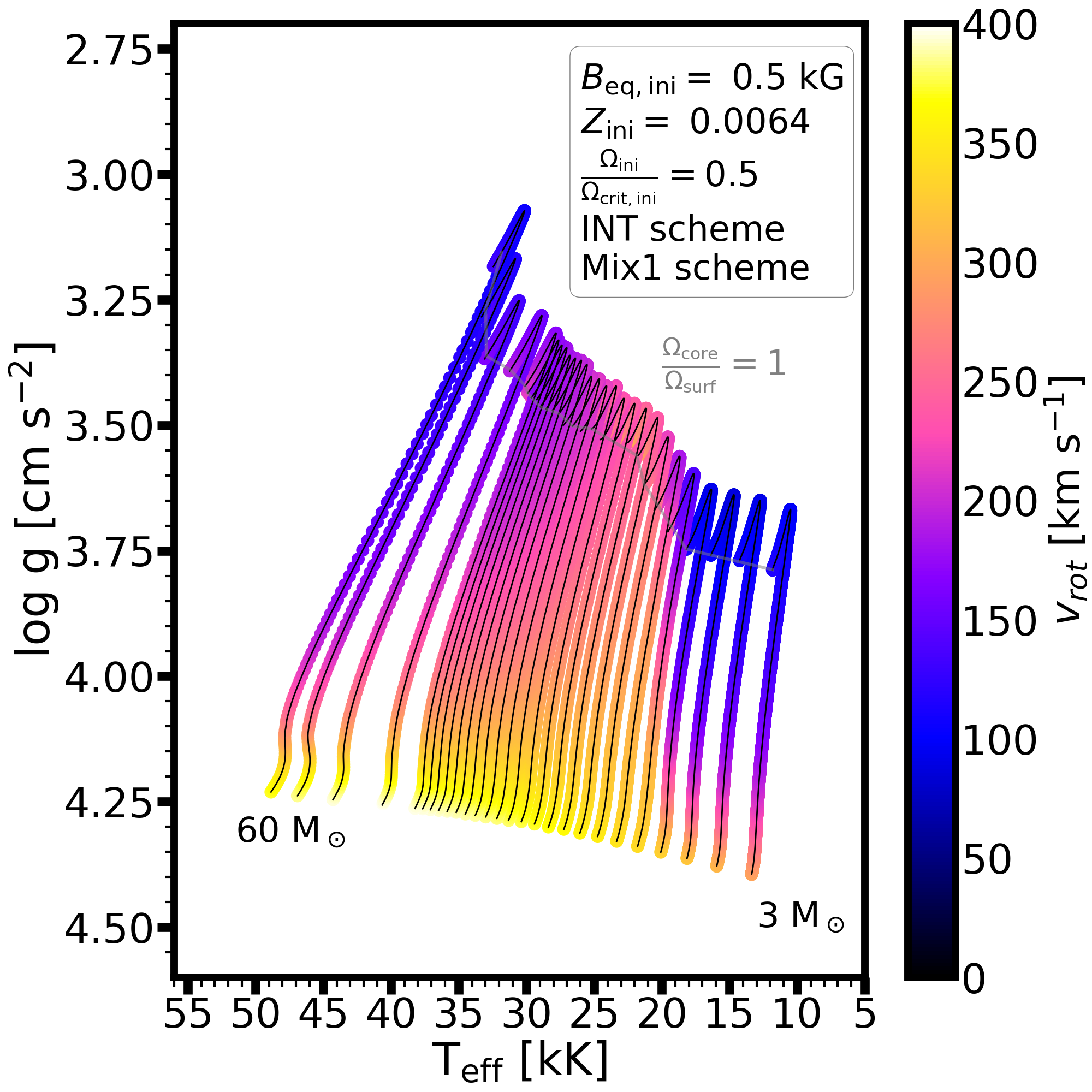}\includegraphics[width=6cm]{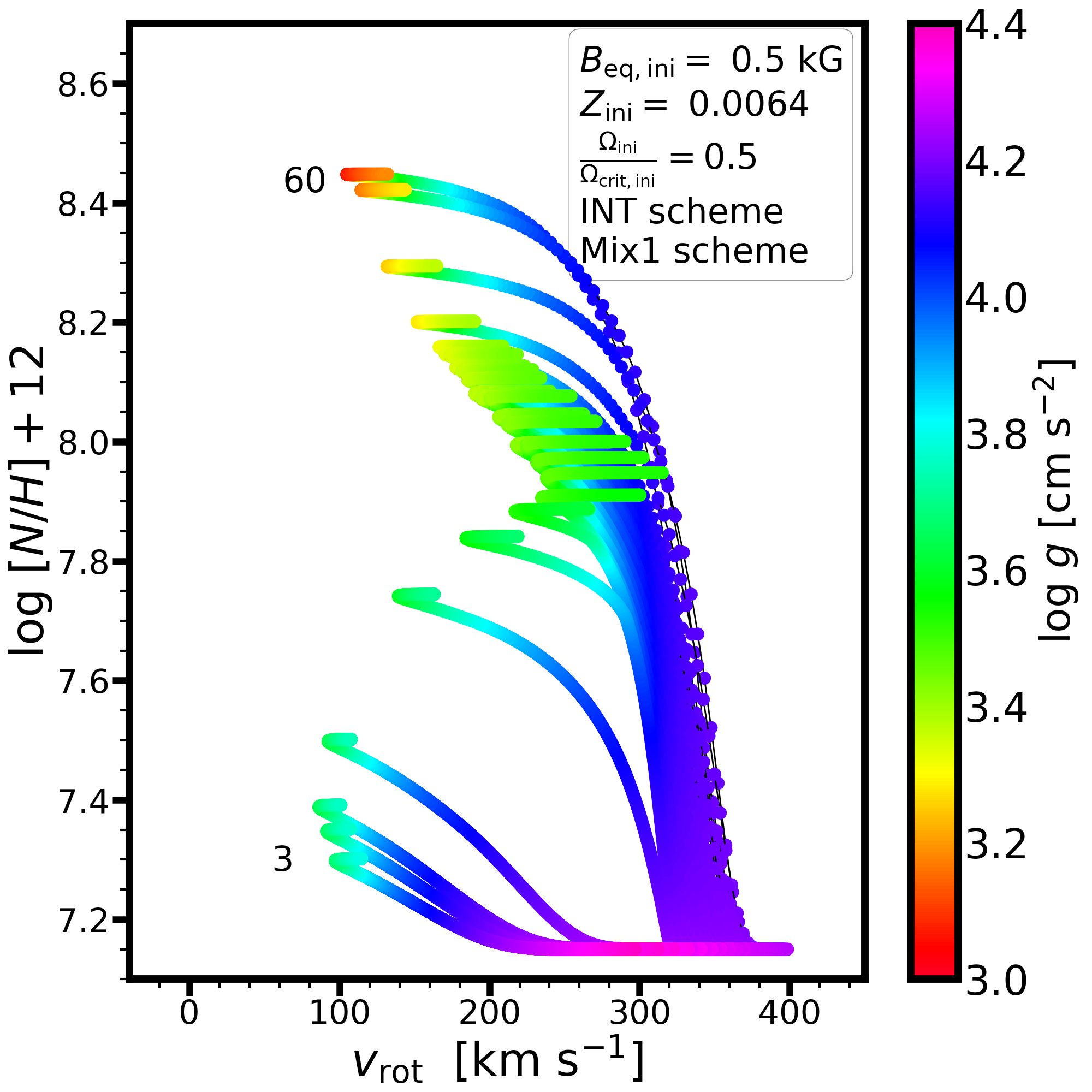}
\caption{Same as Figure \ref{fig:bfield1} but for the NOMAG/Mix1 scheme (top) and INT/Mix1 scheme with an initial equatorial magnetic field strength of 0.5 kG (lower panel) at $Z = 0.0064$. The INT/Mix1 model with initial equatorial magnetic field strength of 3~kG is presented in Figure \ref{fig:metale1}. }\label{fig:lmc_intmix1}
\end{figure*}
%
%
%
\begin{figure*}
\includegraphics[width=6cm]{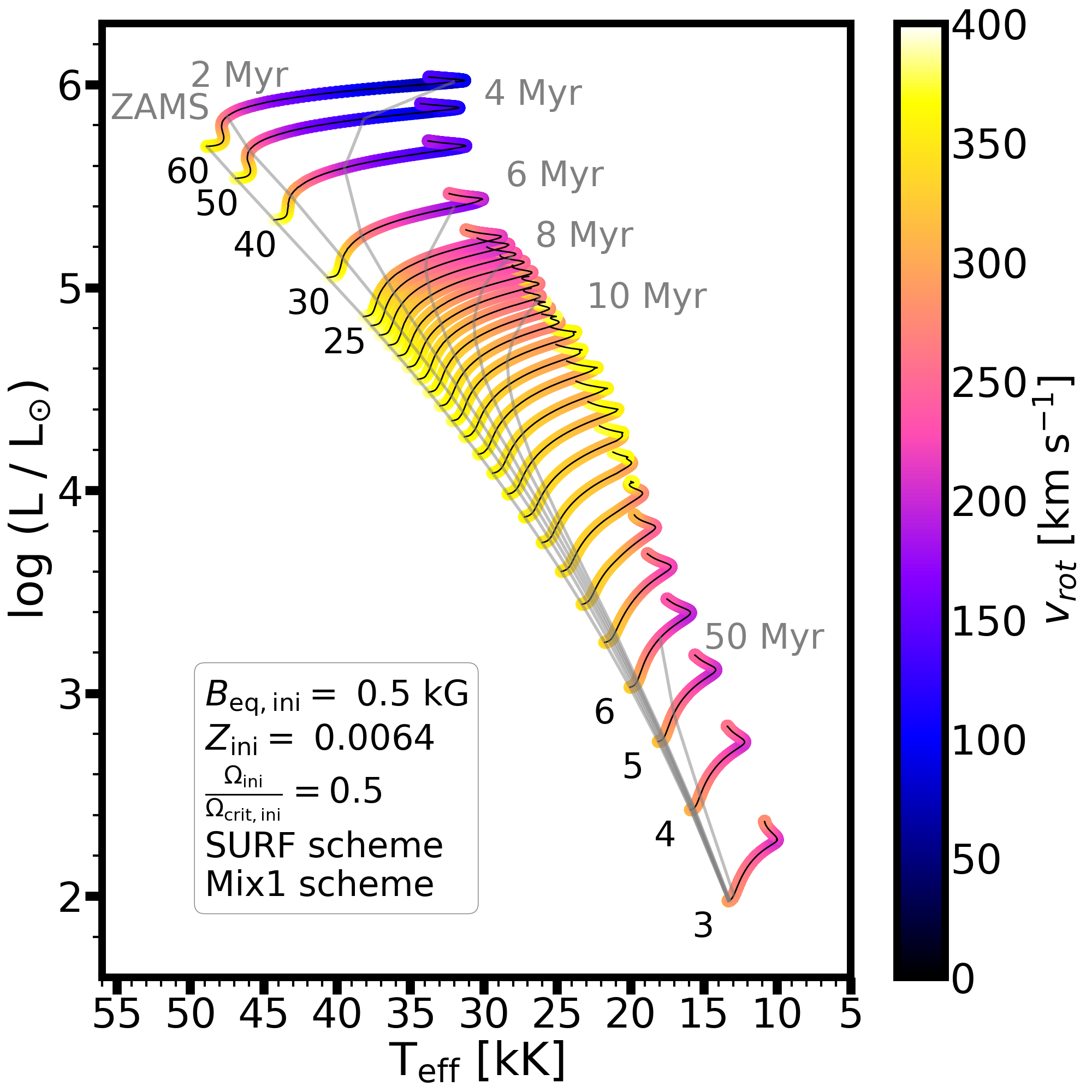}\includegraphics[width=6cm]{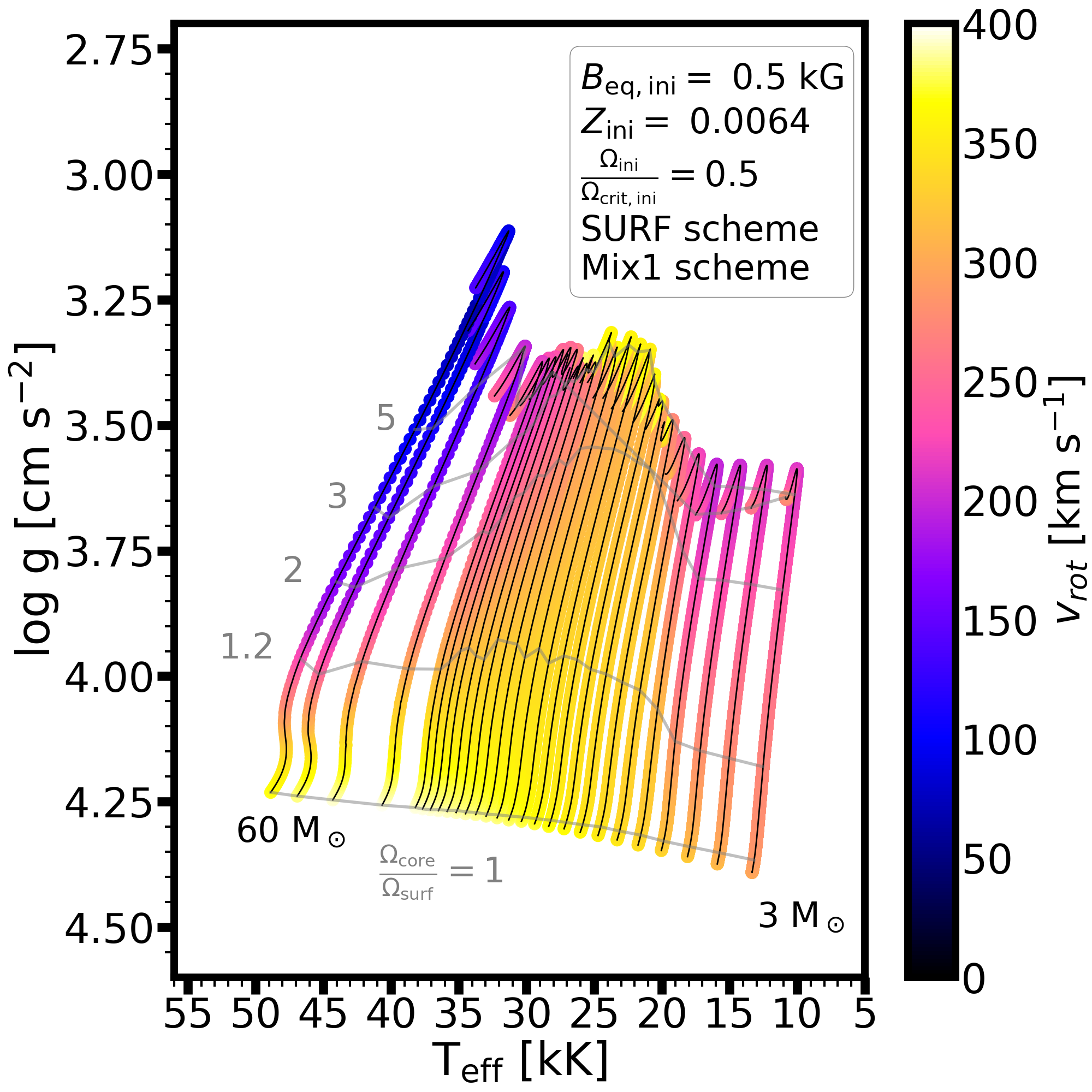}\includegraphics[width=6cm]{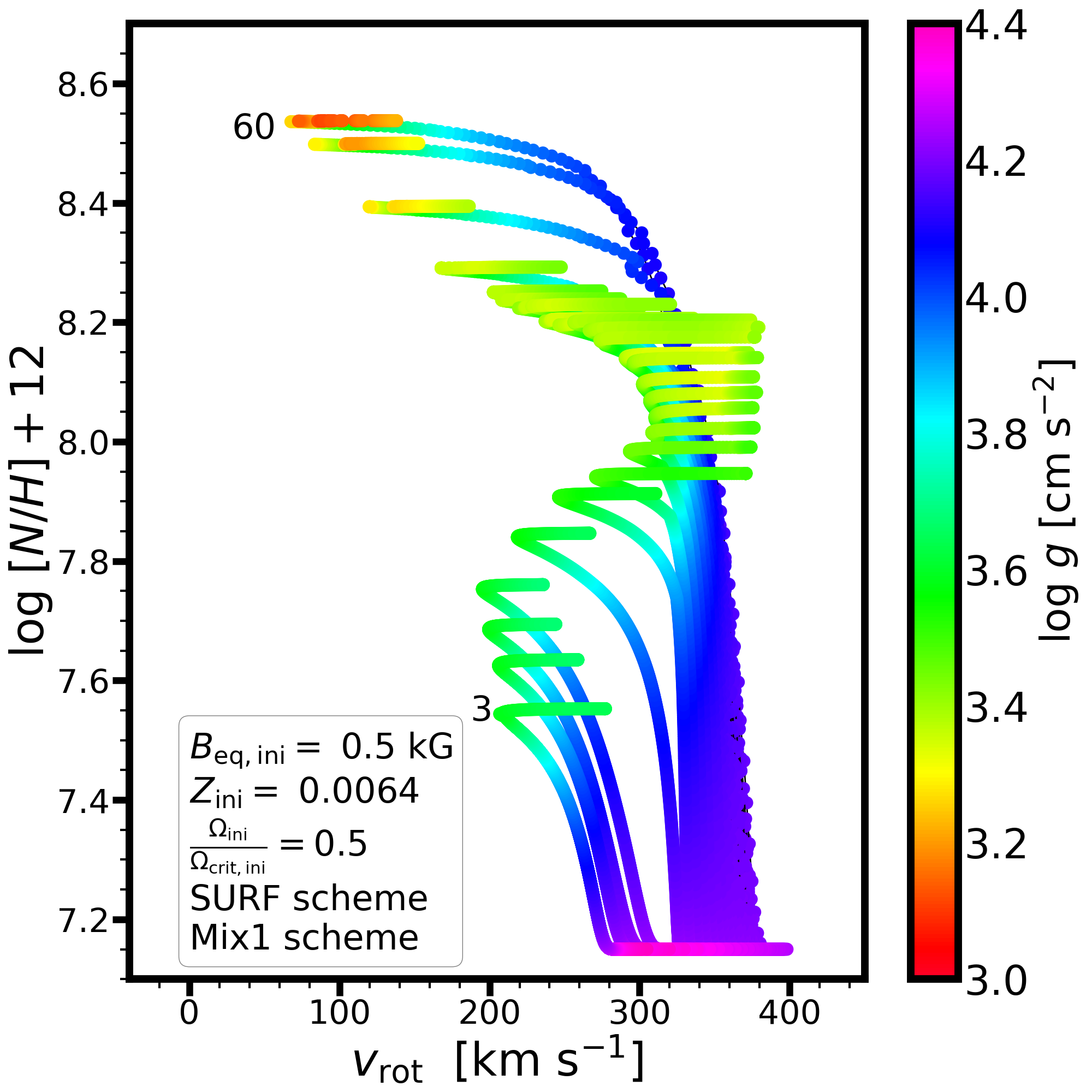}
\includegraphics[width=6cm]{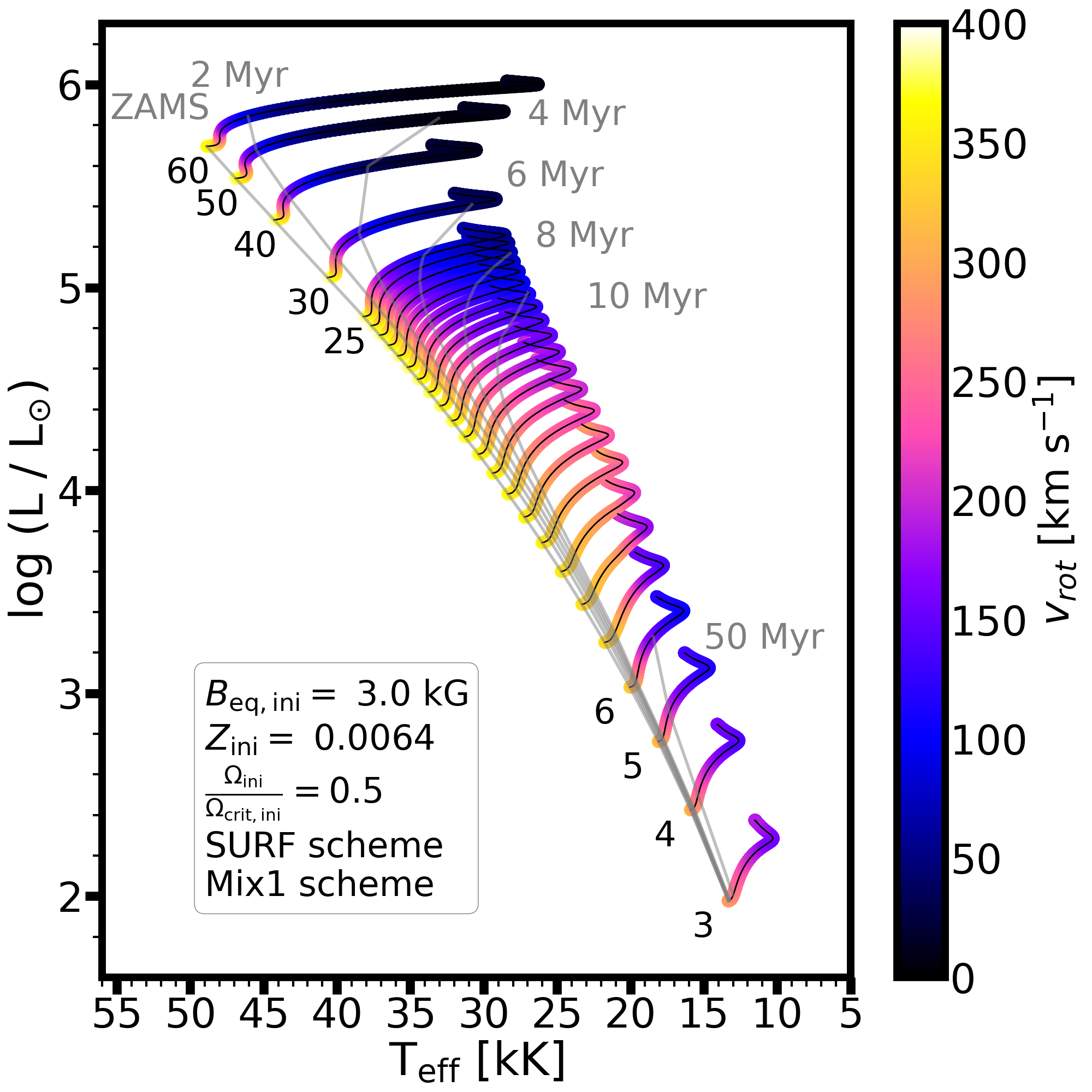}\includegraphics[width=6cm]{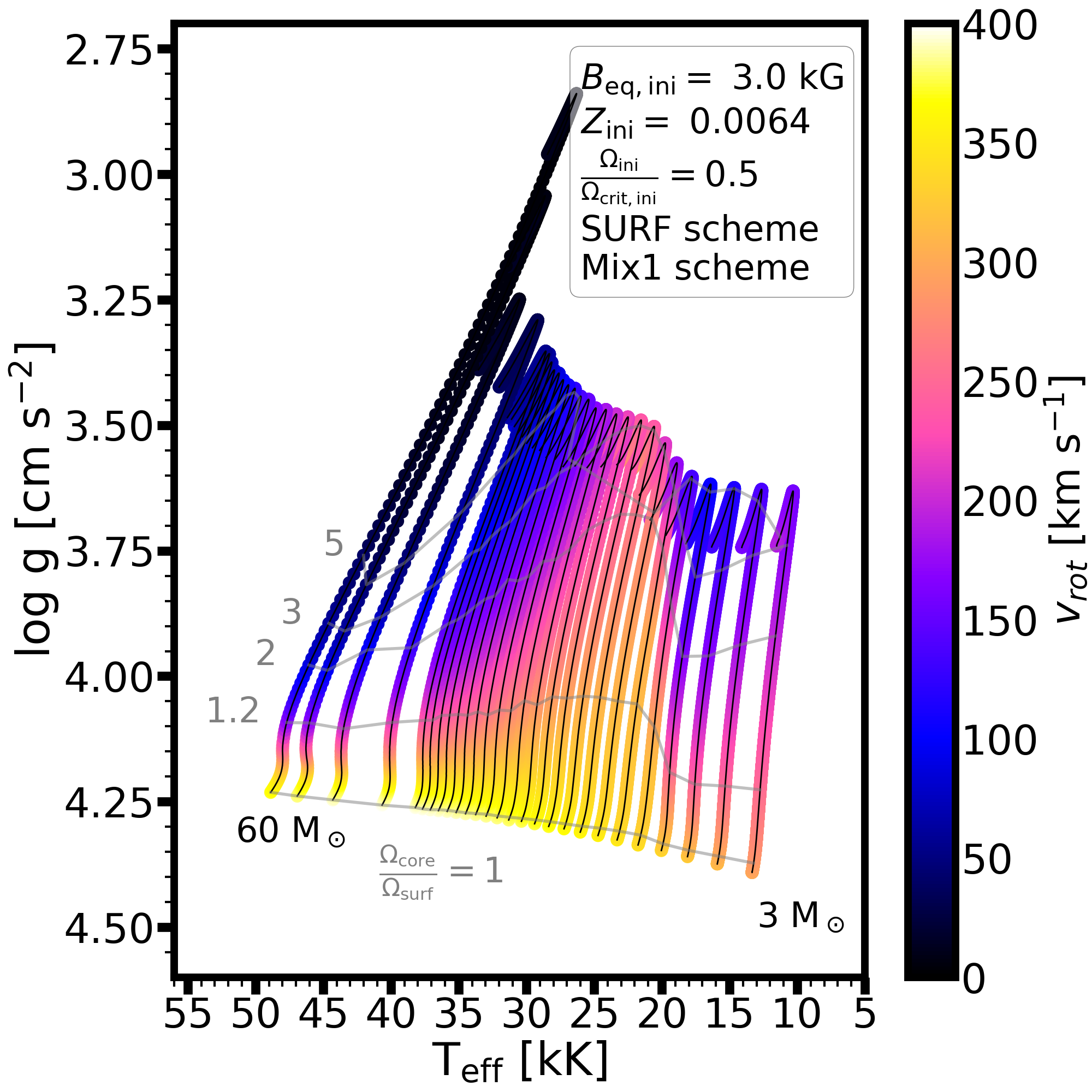}\includegraphics[width=6cm]{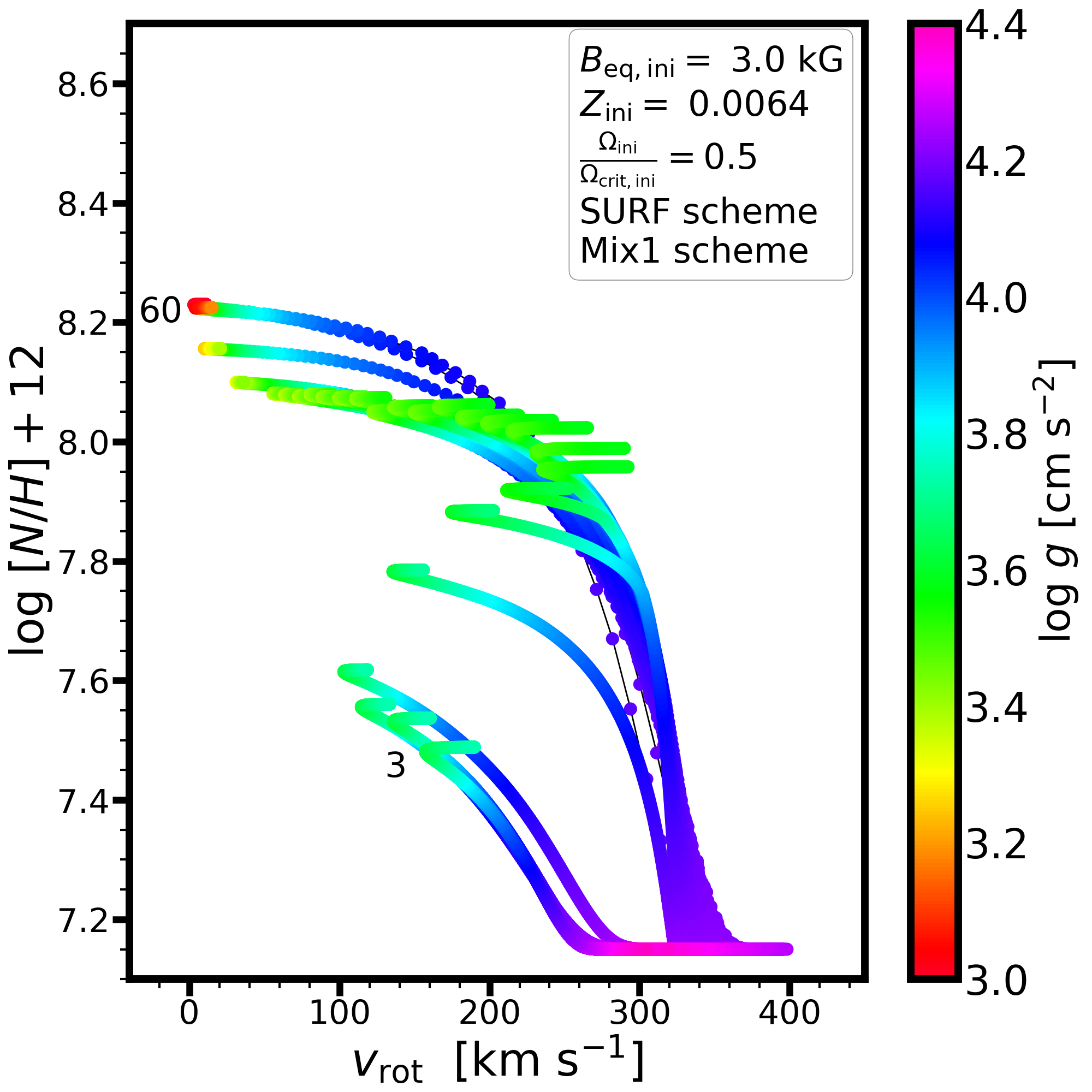}
\caption{Same as Figure \ref{fig:lmc_intmix1} but for the SURF/Mix1 scheme. Top panels show models with an initial equatorial magnetic field strength of 0.5 kG, whereas the lower panels show models with 3 kG. The NOMAG/Mix1 model is presented in Figure \ref{fig:lmc_intmix1}. }\label{fig:lmc_surfmix1}
\end{figure*}
%
%
\begin{figure*}
\includegraphics[width=6cm]{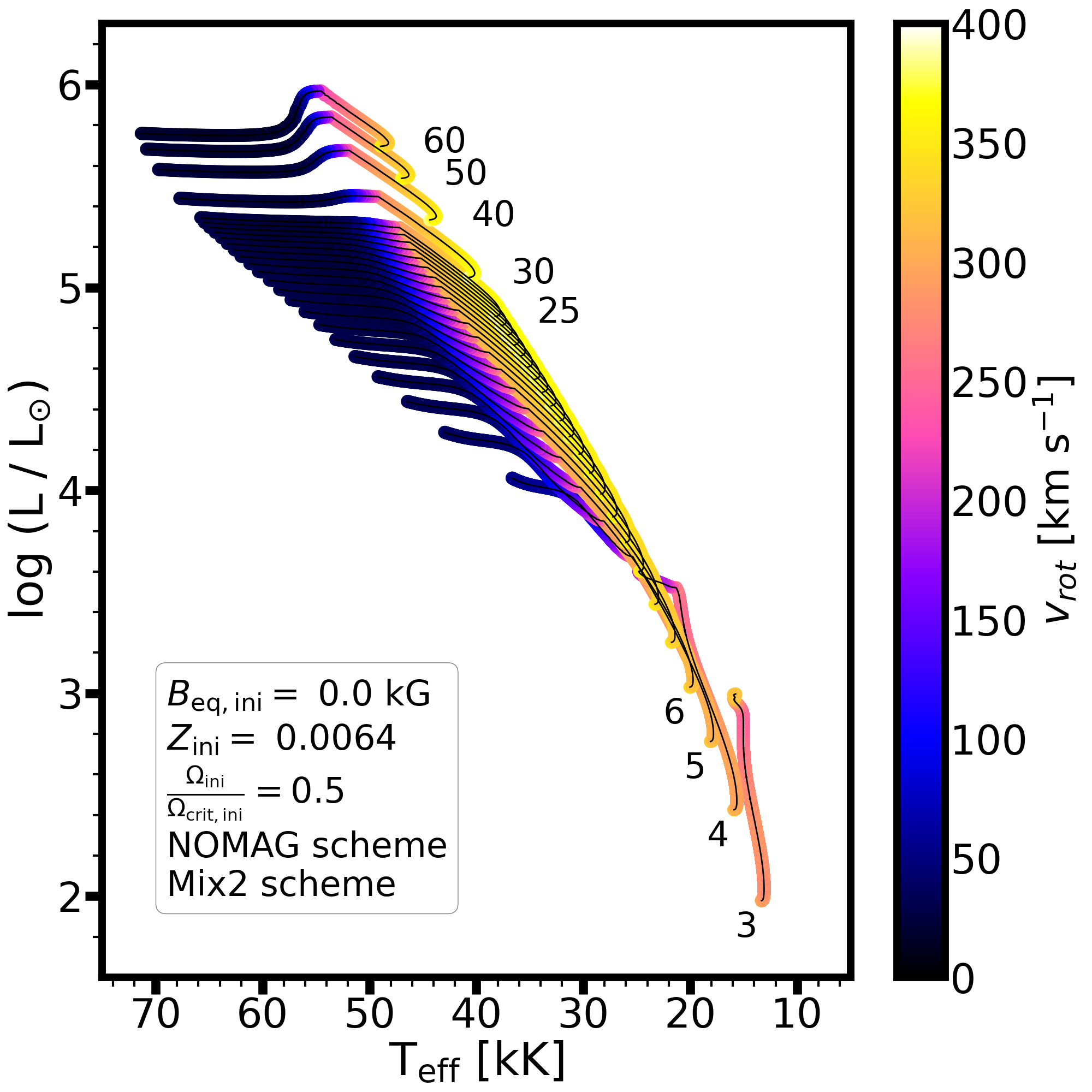}\includegraphics[width=6cm]{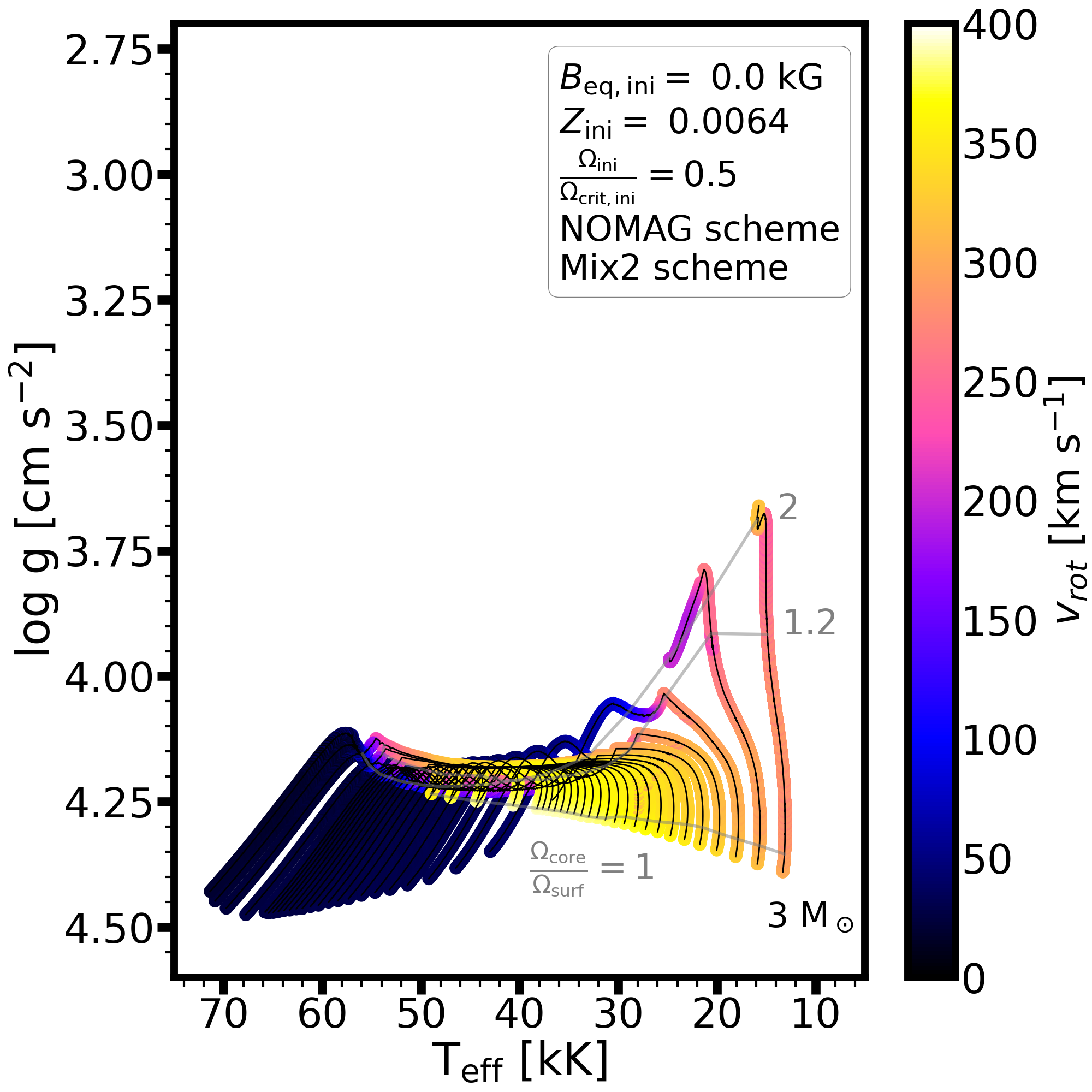}\includegraphics[width=6cm]{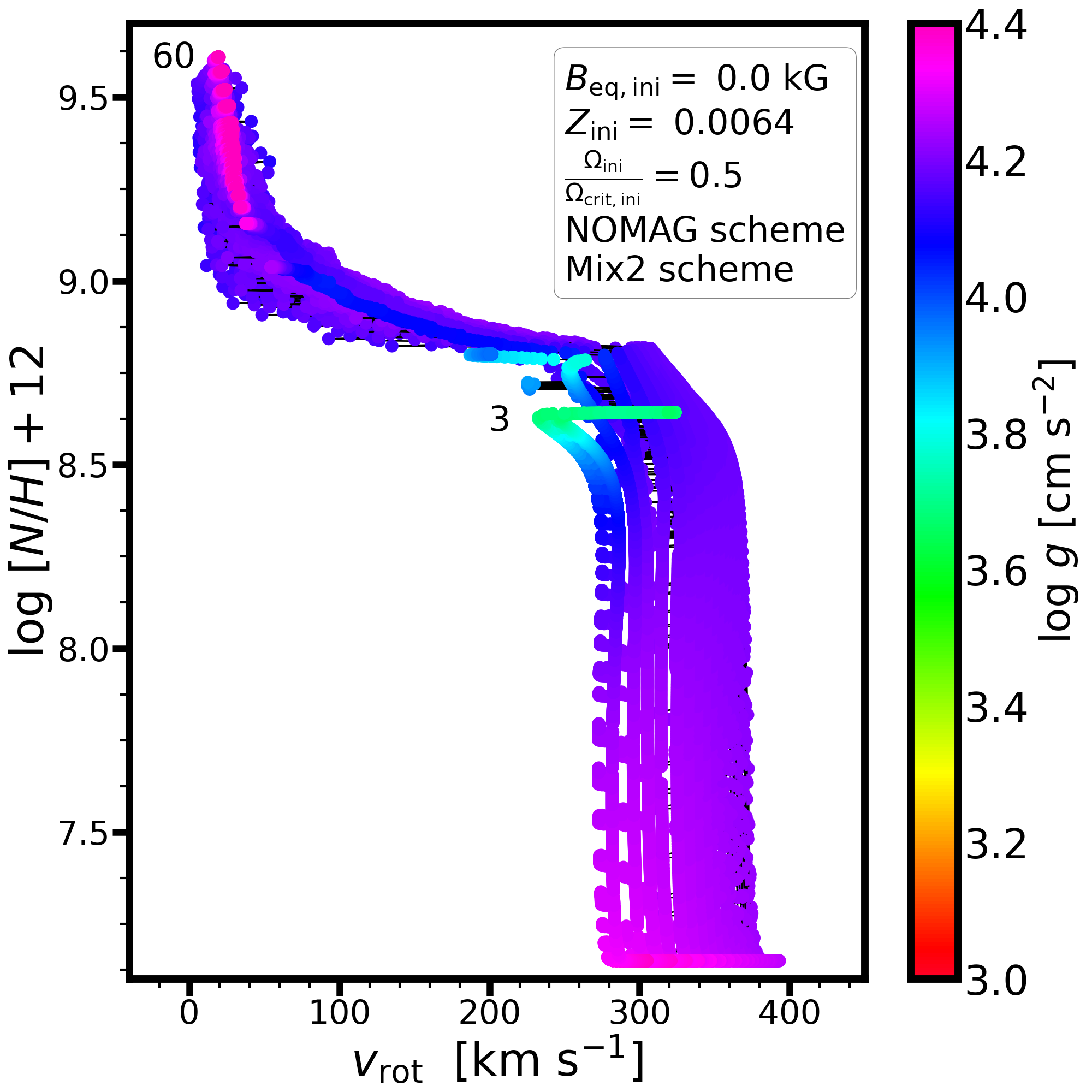} 
\includegraphics[width=6cm]{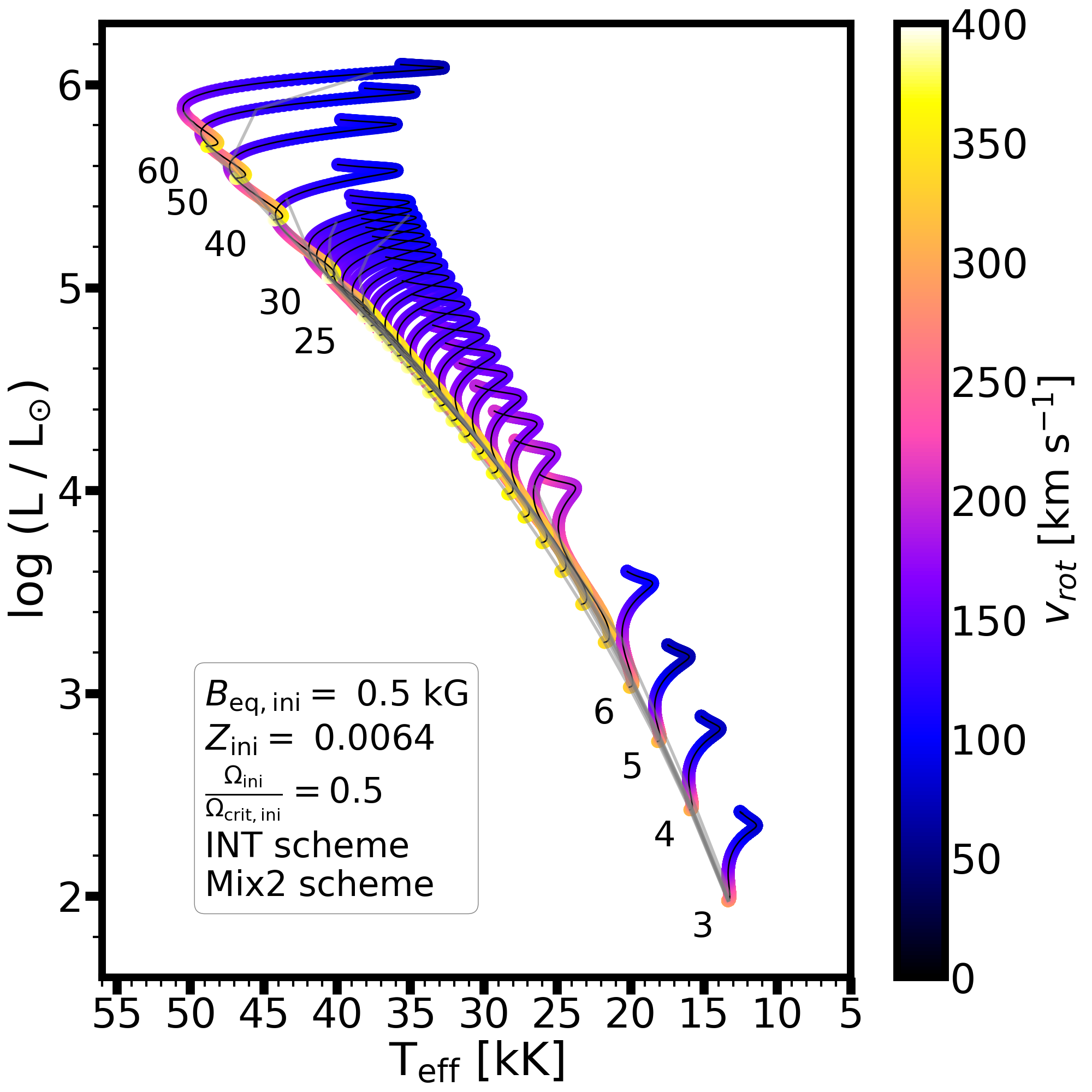}\includegraphics[width=6cm]{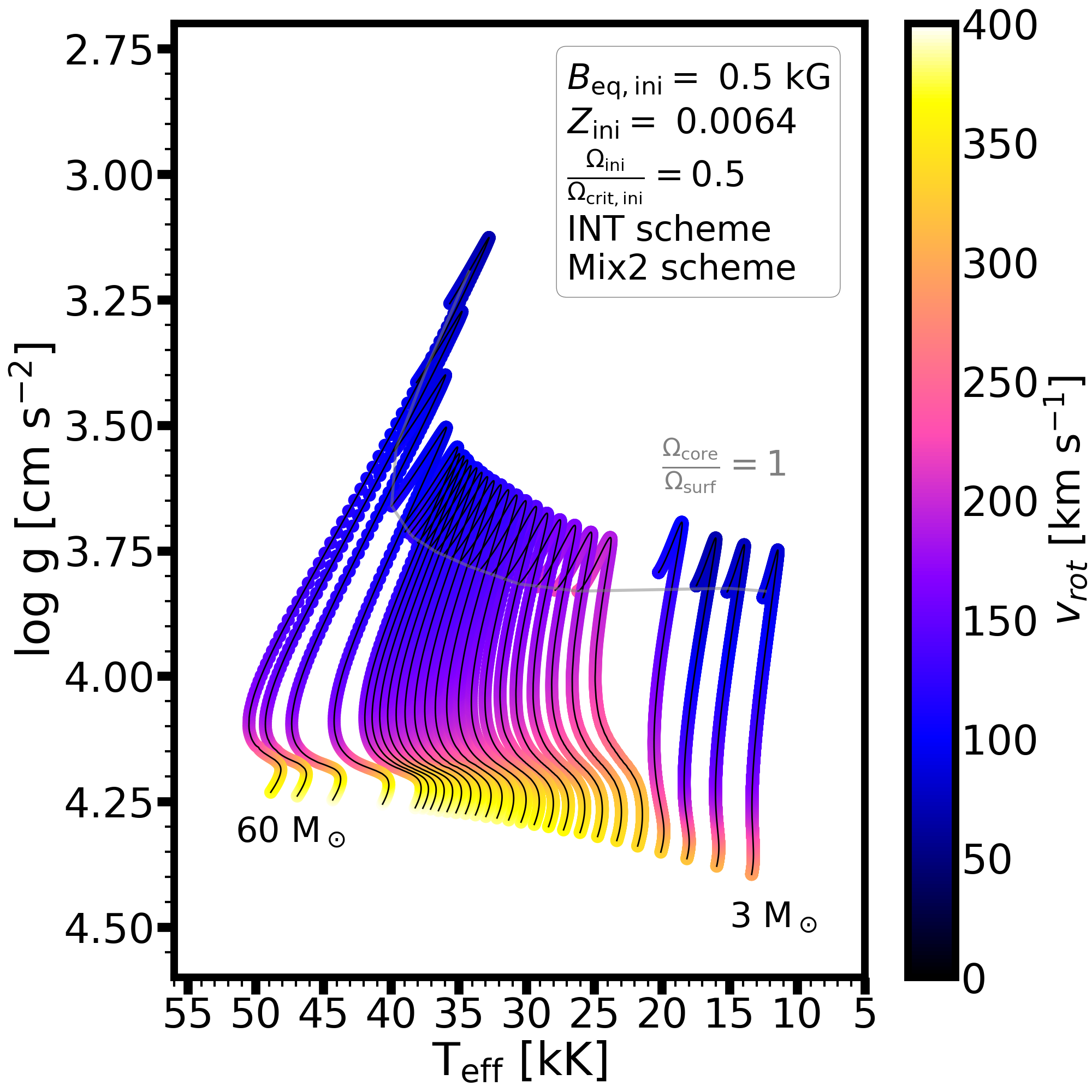}\includegraphics[width=6cm]{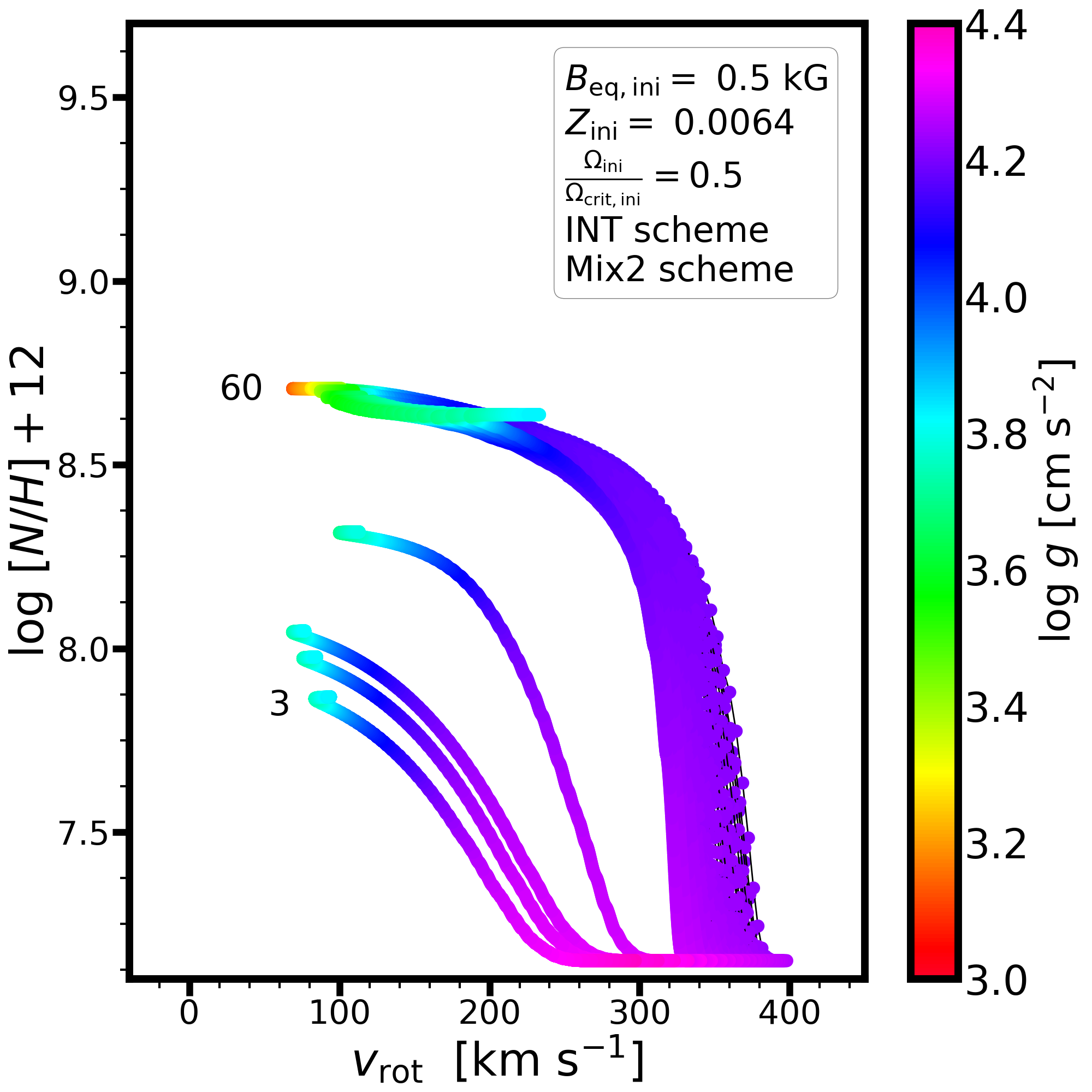}
\includegraphics[width=6cm]{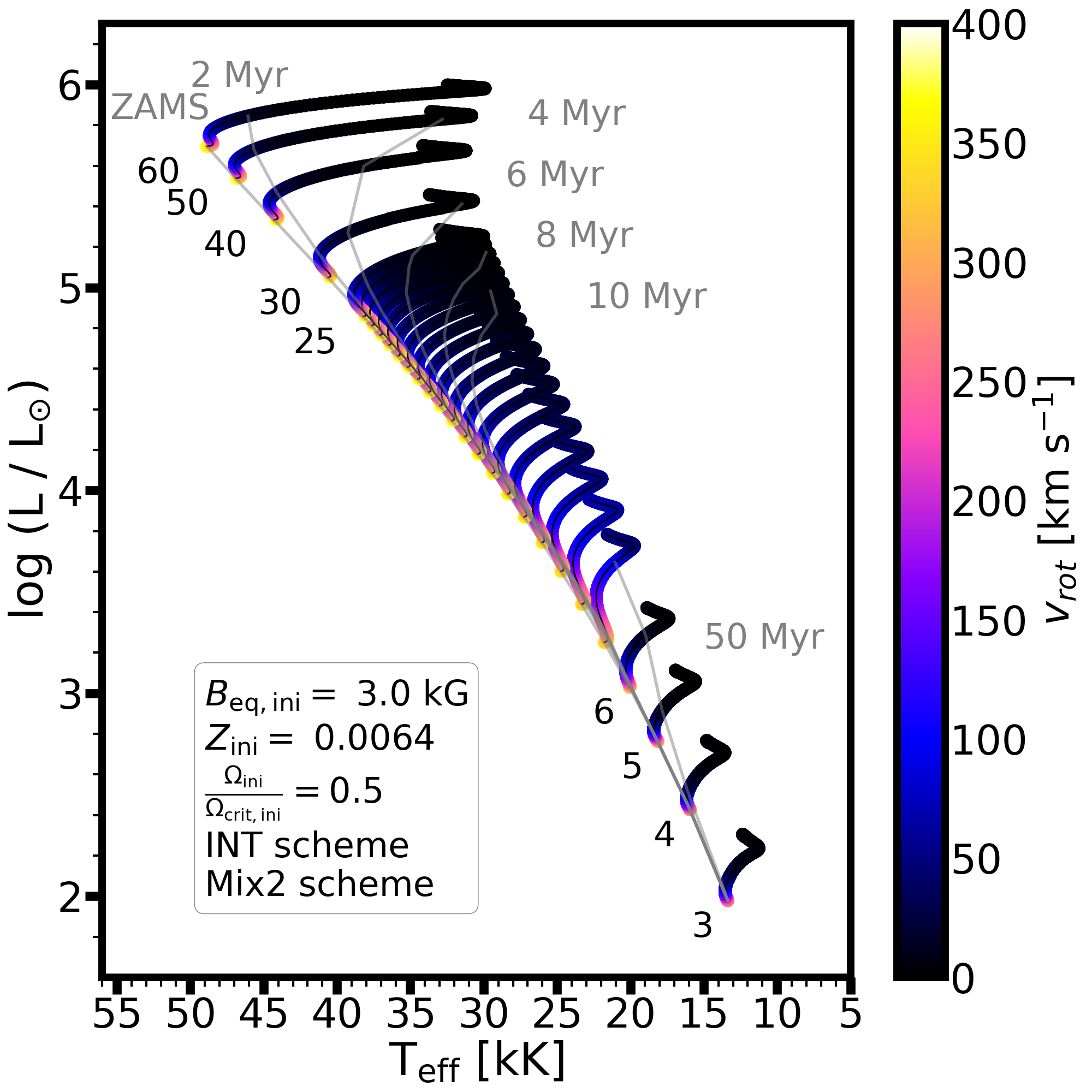}\includegraphics[width=6cm]{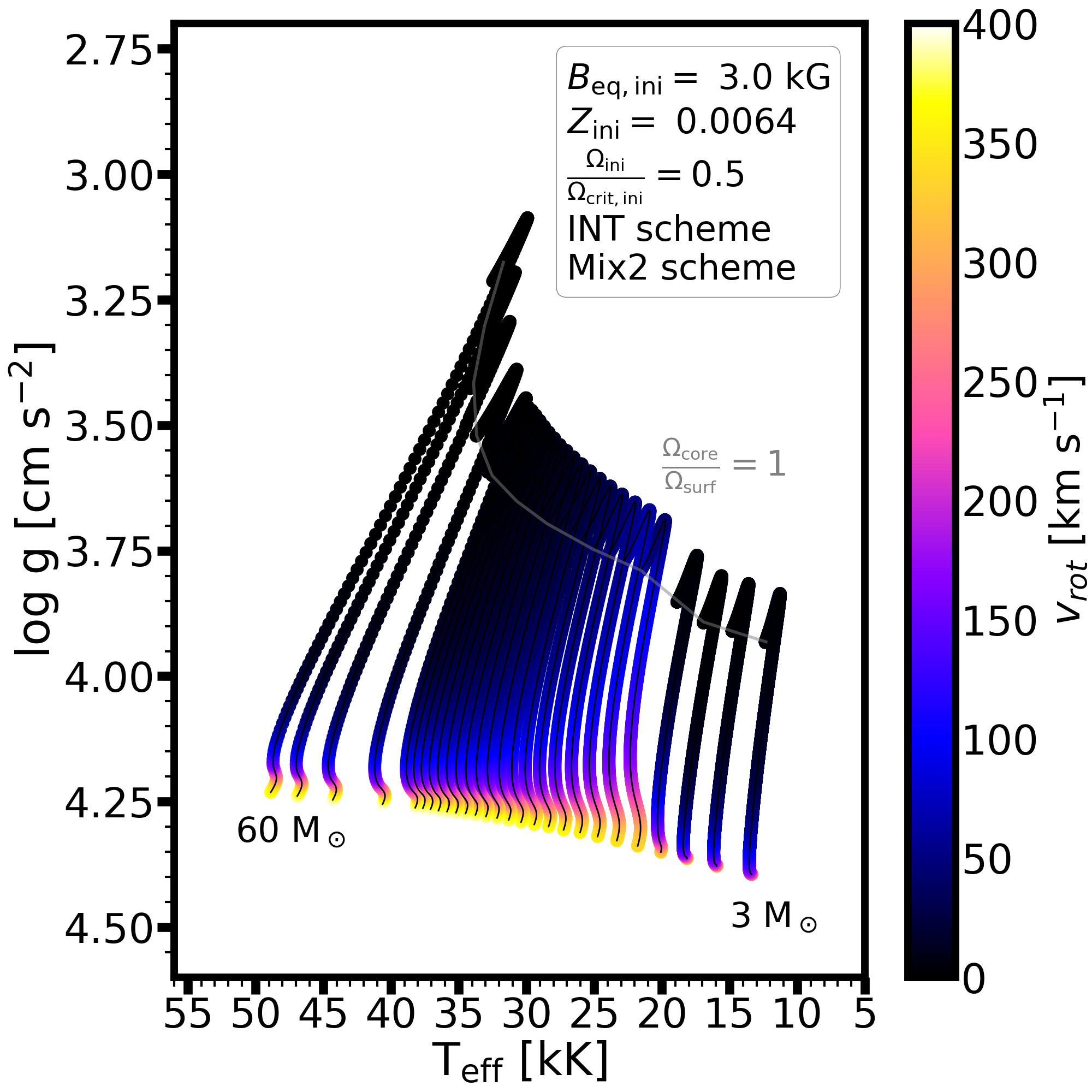}\includegraphics[width=6cm]{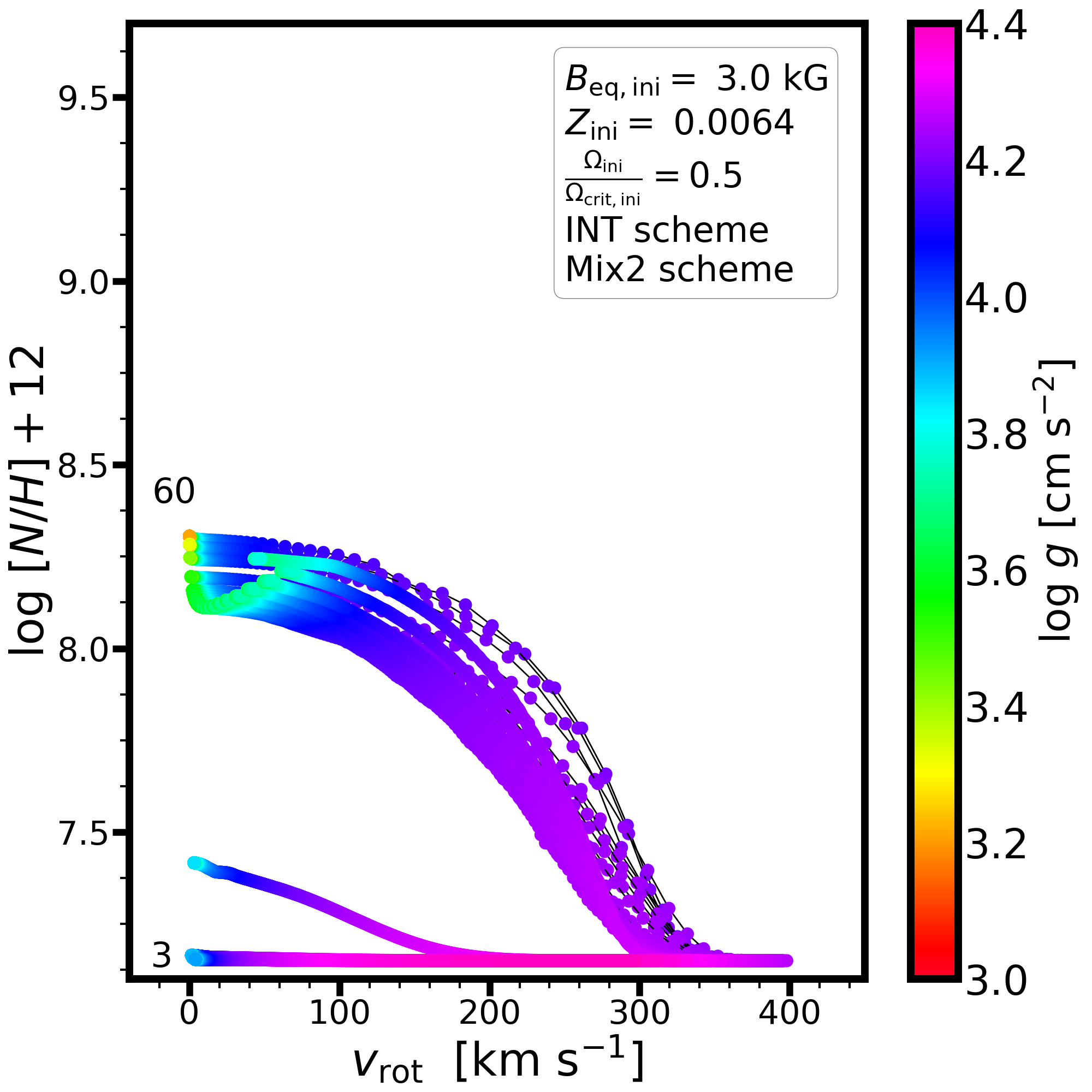}
\caption{Same as Figure \ref{fig:lmc_intmix1} but for the NOMAG/Mix2 (top panels) and INT/Mix2 (middle and lower panels) schemes.}\label{fig:lmc_intmix2}
\end{figure*}
%
%
\begin{figure*}
\includegraphics[width=6cm]{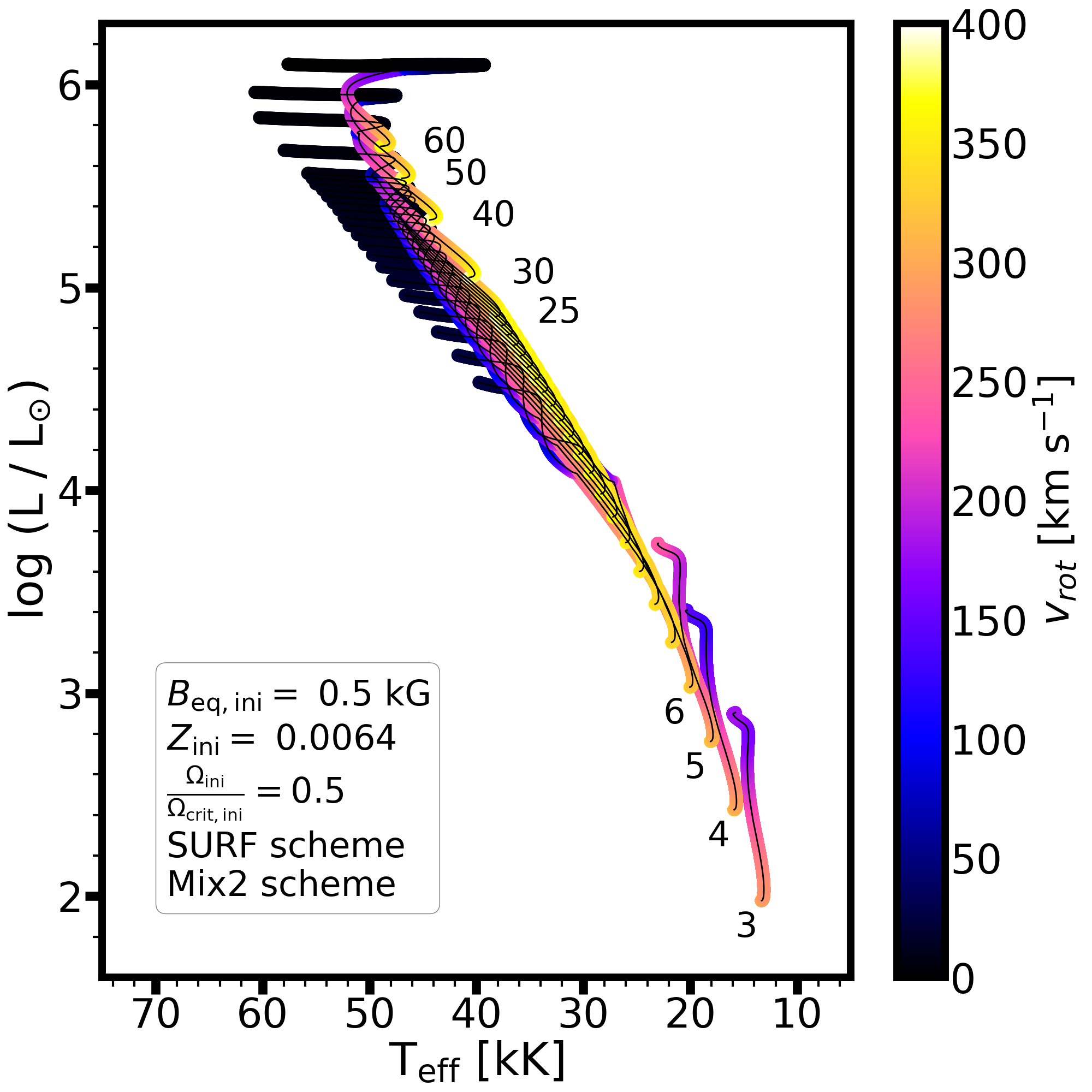}\includegraphics[width=6cm]{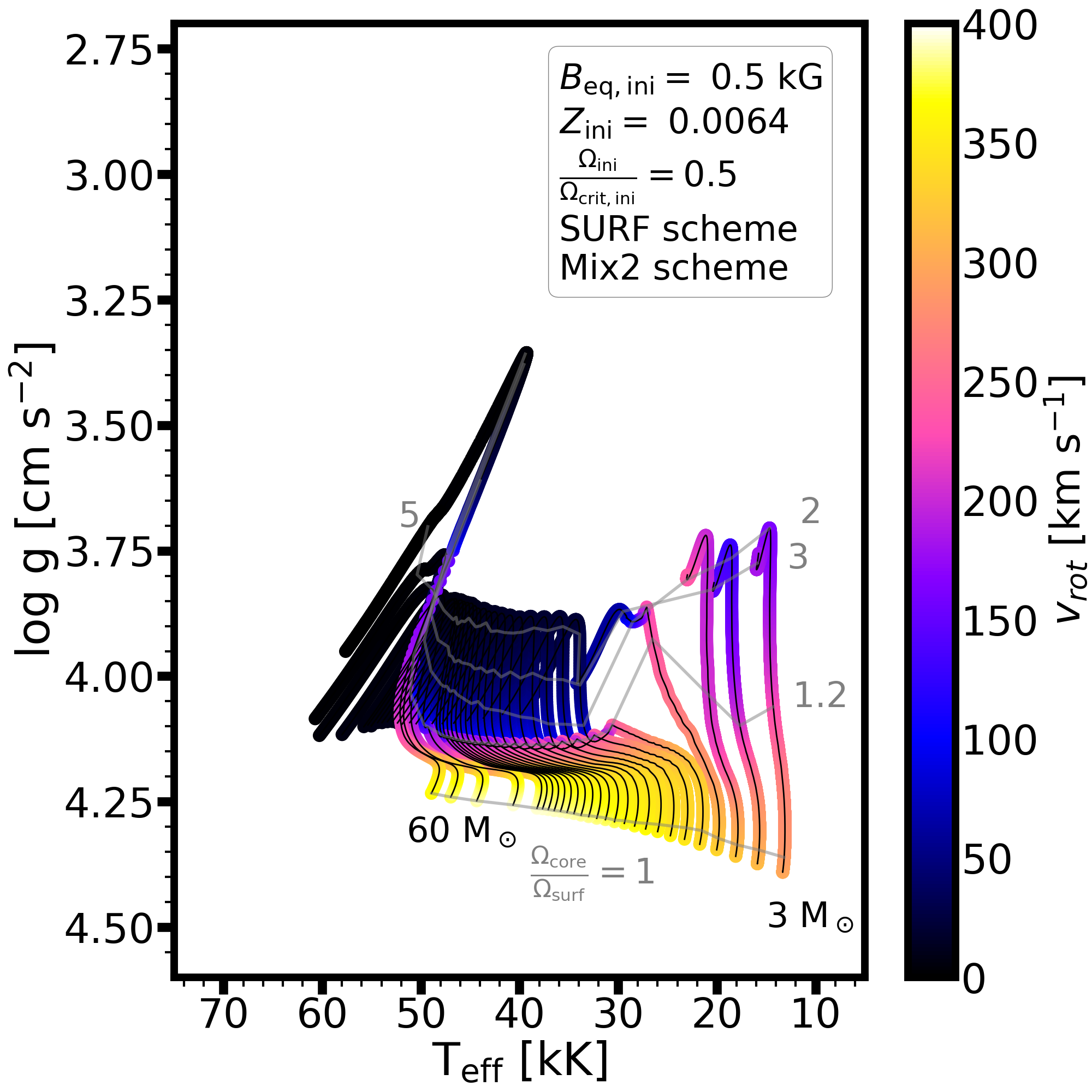}\includegraphics[width=6cm]{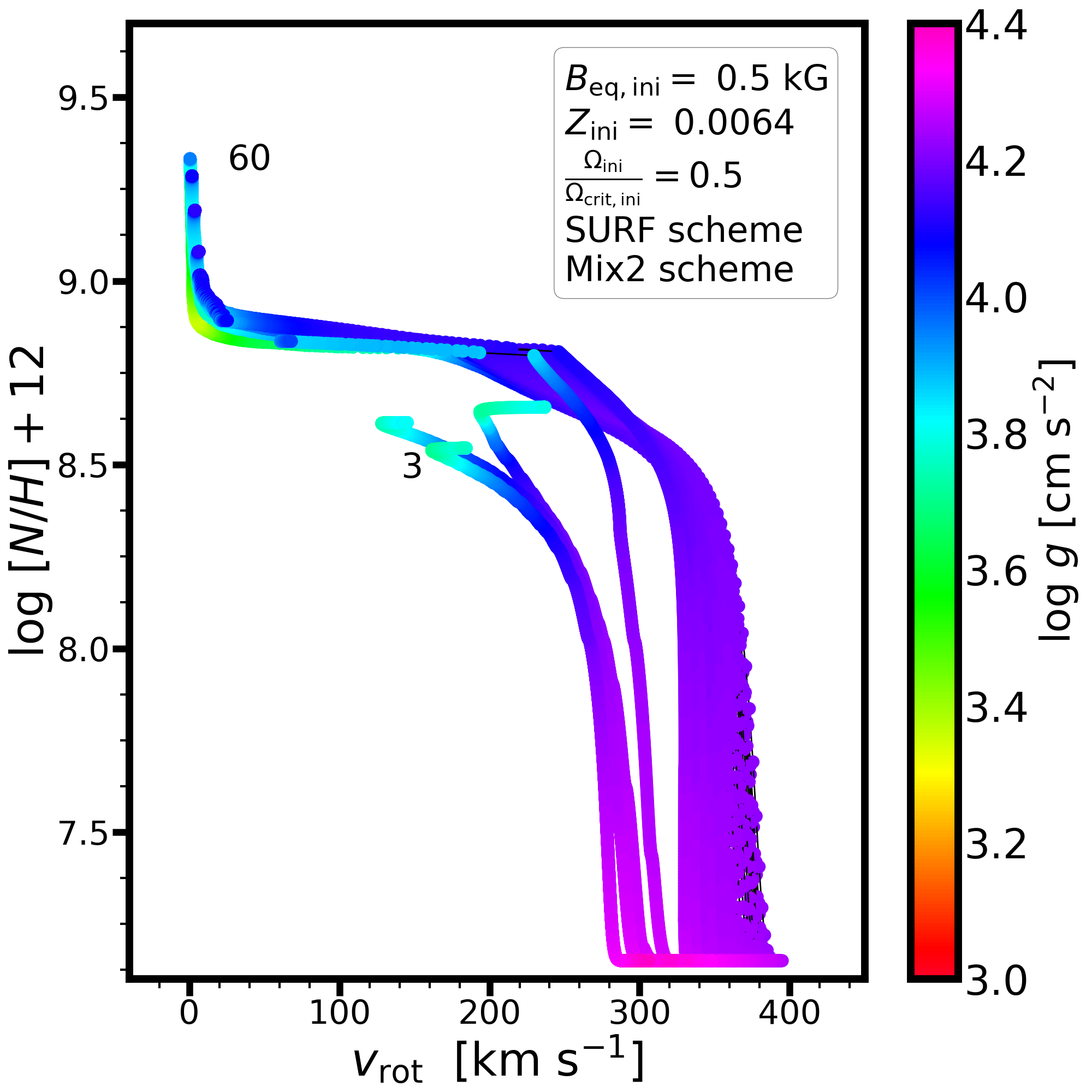}
\includegraphics[width=6cm]{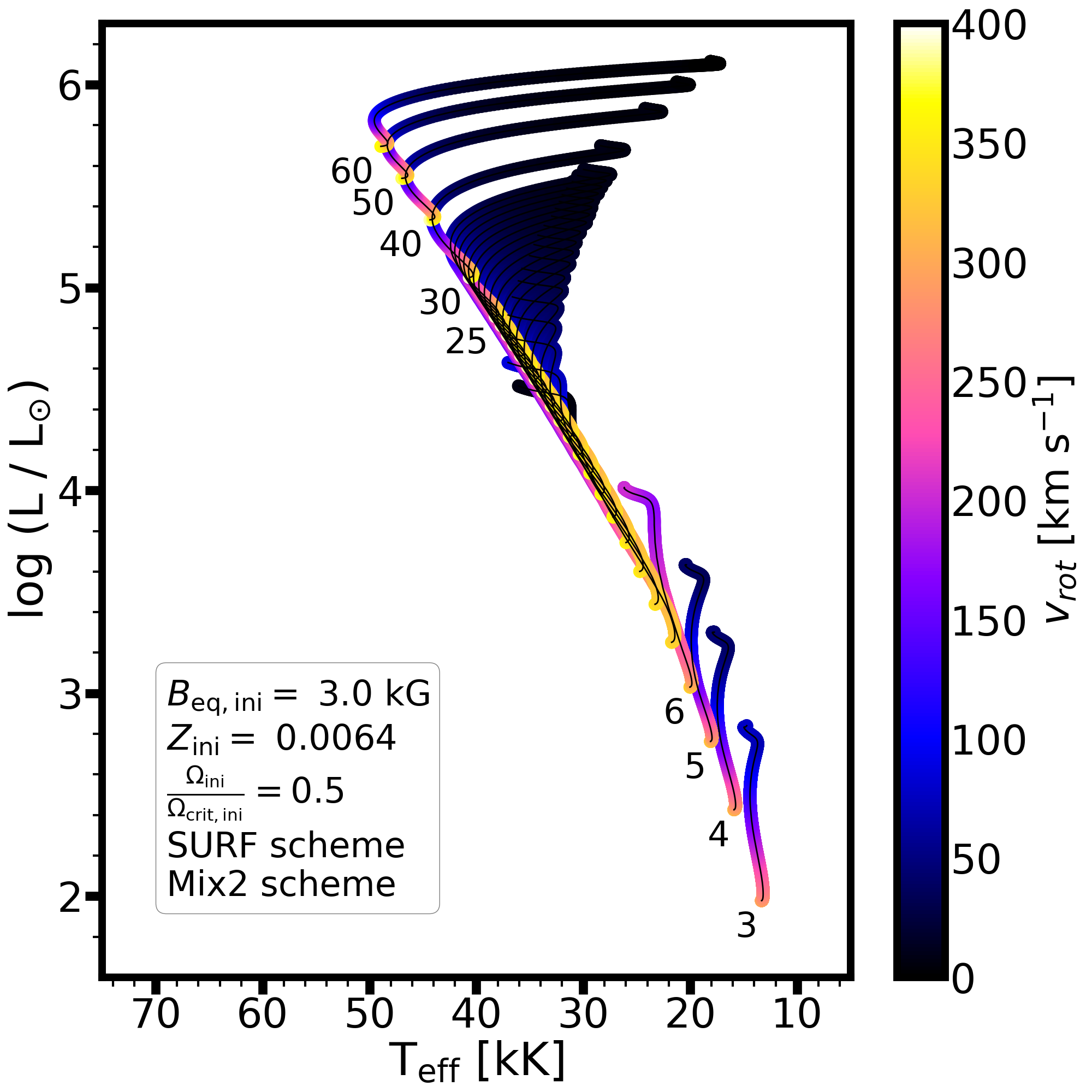}\includegraphics[width=6cm]{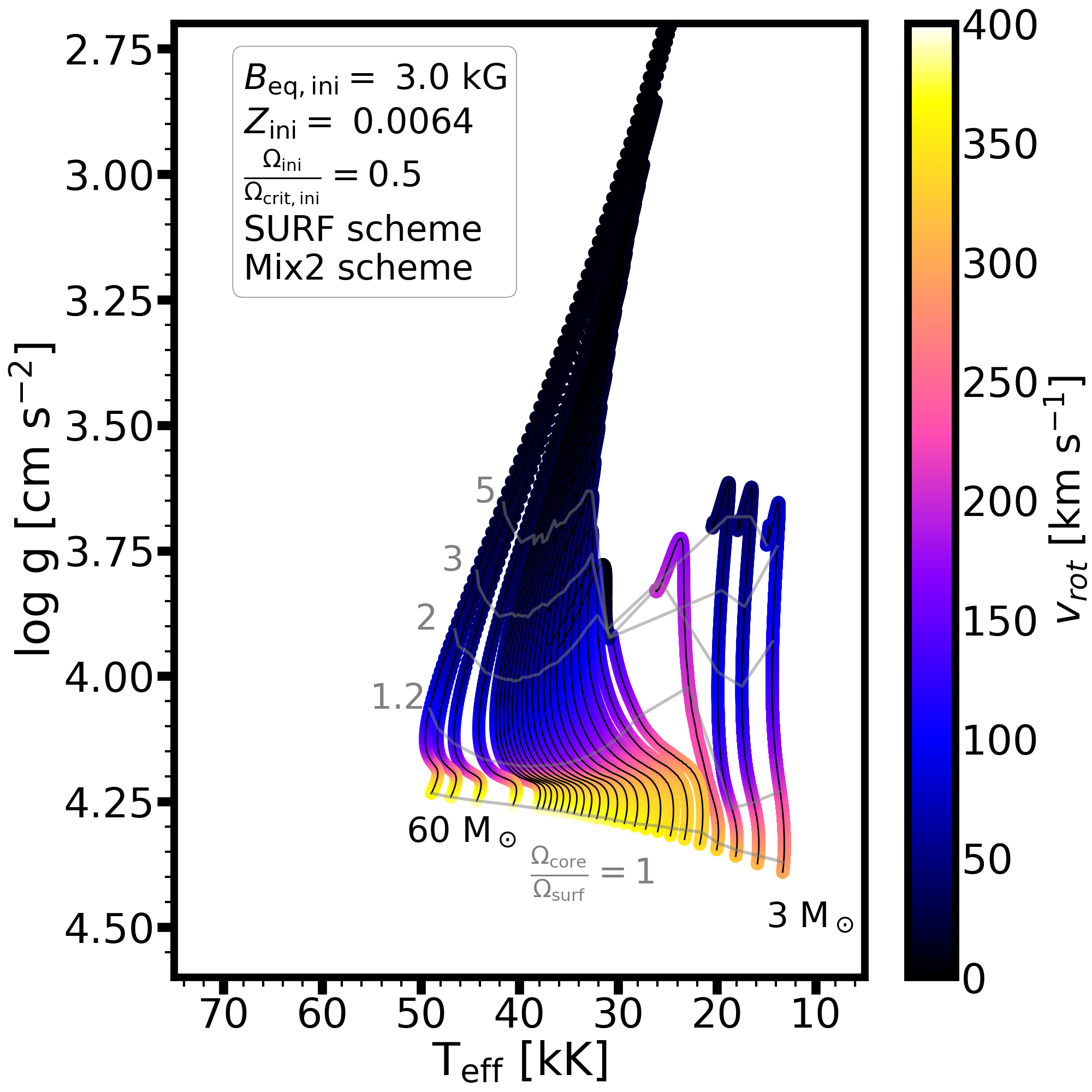}\includegraphics[width=6cm]{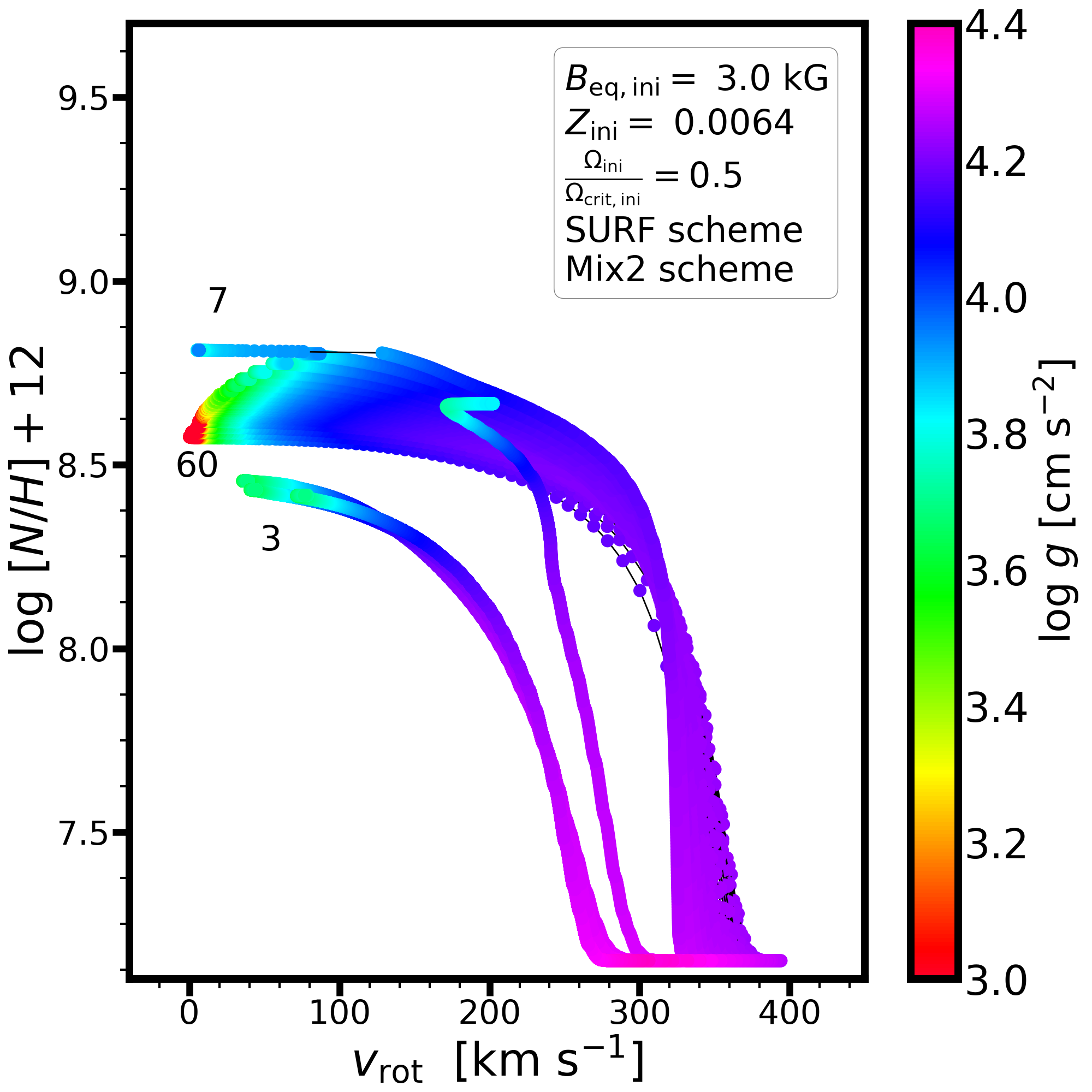}
\caption{Same as Figure \ref{fig:lmc_intmix2} but for the SURF/Mix2 scheme. See top panels of Figure \ref{fig:lmc_intmix2} for the NOMAG/Mix2 models.}\label{fig:lmc_surfmix2}
\end{figure*}

\subsection{Impact of varying magnetic field strength at $Z = 0.0064$}

\subsubsection{INT/Mix1}

At LMC metallicity, the non-magnetic models have several main differences compared to solar metallicity models. First, at lower metallicity the position of the ZAMS is shifted to higher effective temperatures. Second, the relative amount of surface nitrogen enrichment is much higher. For comparison, the 60~M$_\odot$ model in the Mix1 chemical mixing scheme at solar metallicity produces an order of magnitude enrichment in its surface nitrogen to hydrogen ratio over its main sequence evolution, whereas the LMC model produces almost a factor of 30 (see Figure~\ref{fig:lmc_intmix1}). 

With a 0.5 kG initial equatorial magnetic field strength within the INT/Mix1 scheme, the surface equatorial rotational velocity is reduced in the models due to magnetic braking. This produces slight differences in the evolutionary tracks on the HR and Kiel diagrams. As a consequence of the reduced rotation in the star, chemical mixing is also somewhat less efficient compared to the non-magnetic case.

\subsubsection{SURF/Mix1}

With a 0.5 kG initial equatorial magnetic field strength within the SURF/Mix1 scheme (Figure \ref{fig:lmc_surfmix1}), the overall reduction in surface equatorial rotational velocities is somewhat less than in the INT/Mix1 case. On the other hand, the most massive models in this configuration tend to achieve lower rotational velocities at the TAMS. 

With a stronger, 3 kG initial equatorial magnetic field strength, the numerical simulations show that the position of the models on HR and Kiel diagram are similar to the non-magnetic and 0.5 kG cases, except for the most massive stars, which can reach much lower surface gravities towards the end of their main sequence evolution. As expected, the evolution of surface equatorial rotational velocities is also impacted, showing that a spin-down is achieved in all models. (However, see Section \ref{sec:feature} for the discussion on models in the mass range of 8-15 M$_\odot$.) Consequently, the surface nitrogen enrichment is more modest compared to the non-magnetic case but still notable in these models. 

\subsubsection{INT/Mix2}

When chemical mixing is assumed in the Mix2 scheme, the non-magnetic models at LMC metallicity evolve bluewards, quasi-chemically homogeneously for their entire main sequence evolution. Within this configuration the $[N/H]$ ratio reaches extremely high values (Figure \ref{fig:lmc_intmix2}). This is in part due to the efficient mixing of nitrogen, and in part due to losing a significant amount of hydrogen from the stellar surface via winds. In this case, the low surface rotational velocities at the TAMS are achieved due to Wolf-Rayet type mass loss taking away angular momentum from the star.

When a 0.5 kG initial equatorial magnetic field strength is considered within the INT/Mix2 scheme the models return to a redward evolution on the main sequence following a rapid blueward evolution. This is because the spin-down due to magnetic braking reduces the overall efficiency of chemical mixing and prevents a quasi-chemically homogeneous evolution for the entire main sequence. Note that in this case the TAMS surface rotational velocities are actually higher than in the non-magnetic case as magnetic braking weakens over time and stellar winds are less efficient compared to the non-magnetic case.

With a 3 kG initial equatorial magnetic field strength, the main differences result from a stronger magnetic braking. This almost immediately prevents a blueward evolution, producing slowly-rotating tracks in which the surface nitrogen enrichment is significantly reduced compared to the non-magnetic case.

\subsubsection{SURF/Mix2}

In the SURF/Mix2 scheme both set of models with 0.5 and 3 kG initial equatorial magnetic field strength undergo a rapid initial blueward evolution as in the INT/Mix2 case (Figure \ref{fig:lmc_surfmix2}). With weaker initial magnetic fields, the models above \~10~M$_\odot$ turn to a short redward evolution before spinning down and reaching their TAMS again at hotter effective temperatures. With stronger initial magnetic fields, the spin-down leads to preventing the models from turning back to a blueward evolution by the end of their main sequence. Similarly to the SURF/Mix1 case, the most massive models with 3~kG initial equatorial magnetic field strength may reach low surface gravities. 
In this configuration, it is also possible that the most massive models produce a final $[N/H]$ ratio that is lower than that of lower mass models (see also Figure \ref{fig:hunterobs1}). However, given the very efficient Mix2 scheme, the predicted surface nitrogen enrichment is still expected to be observable as it is over an order of magnitude compared to the baseline.

%
%
%
%

%
\begin{figure*}
\includegraphics[width=6cm]{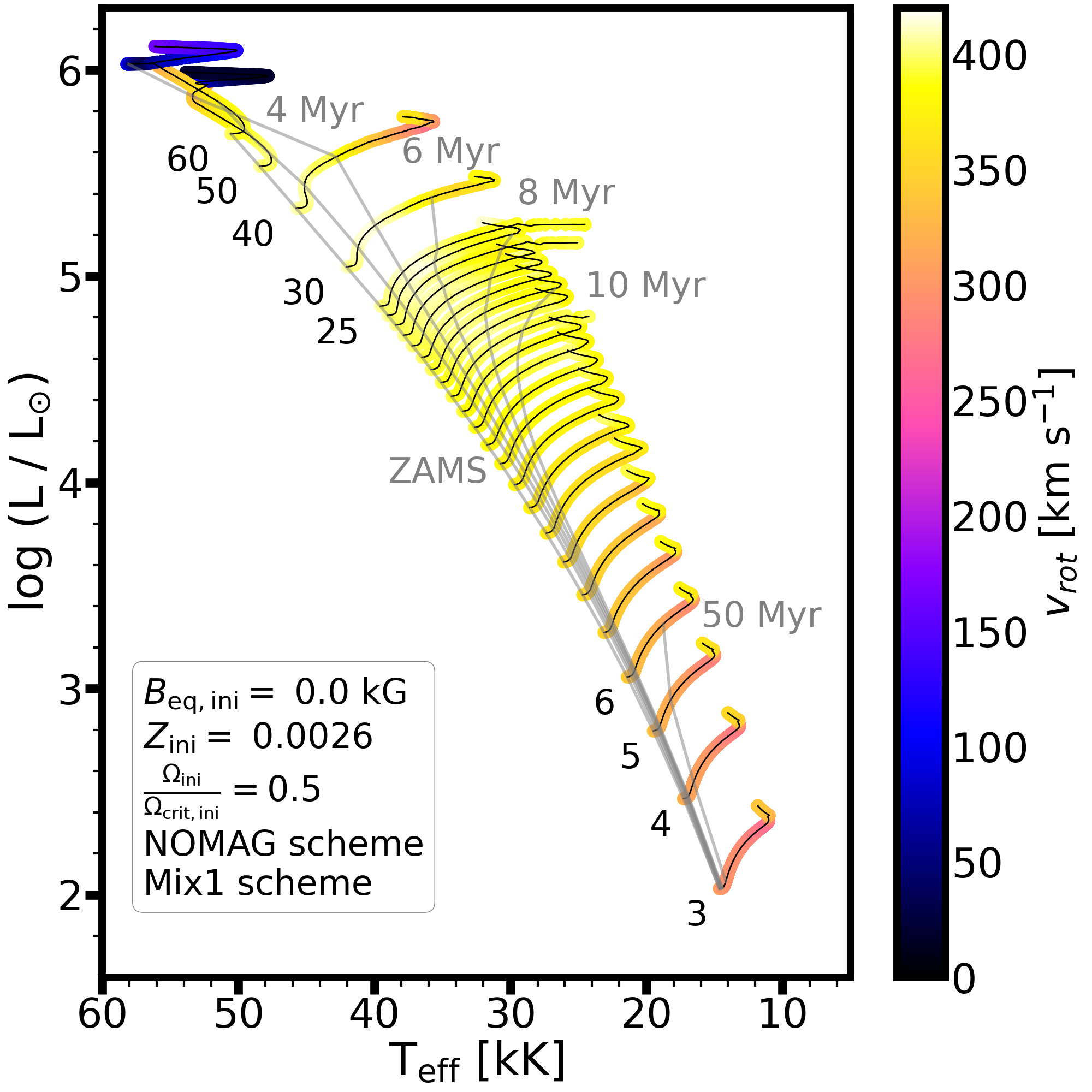}\includegraphics[width=6cm]{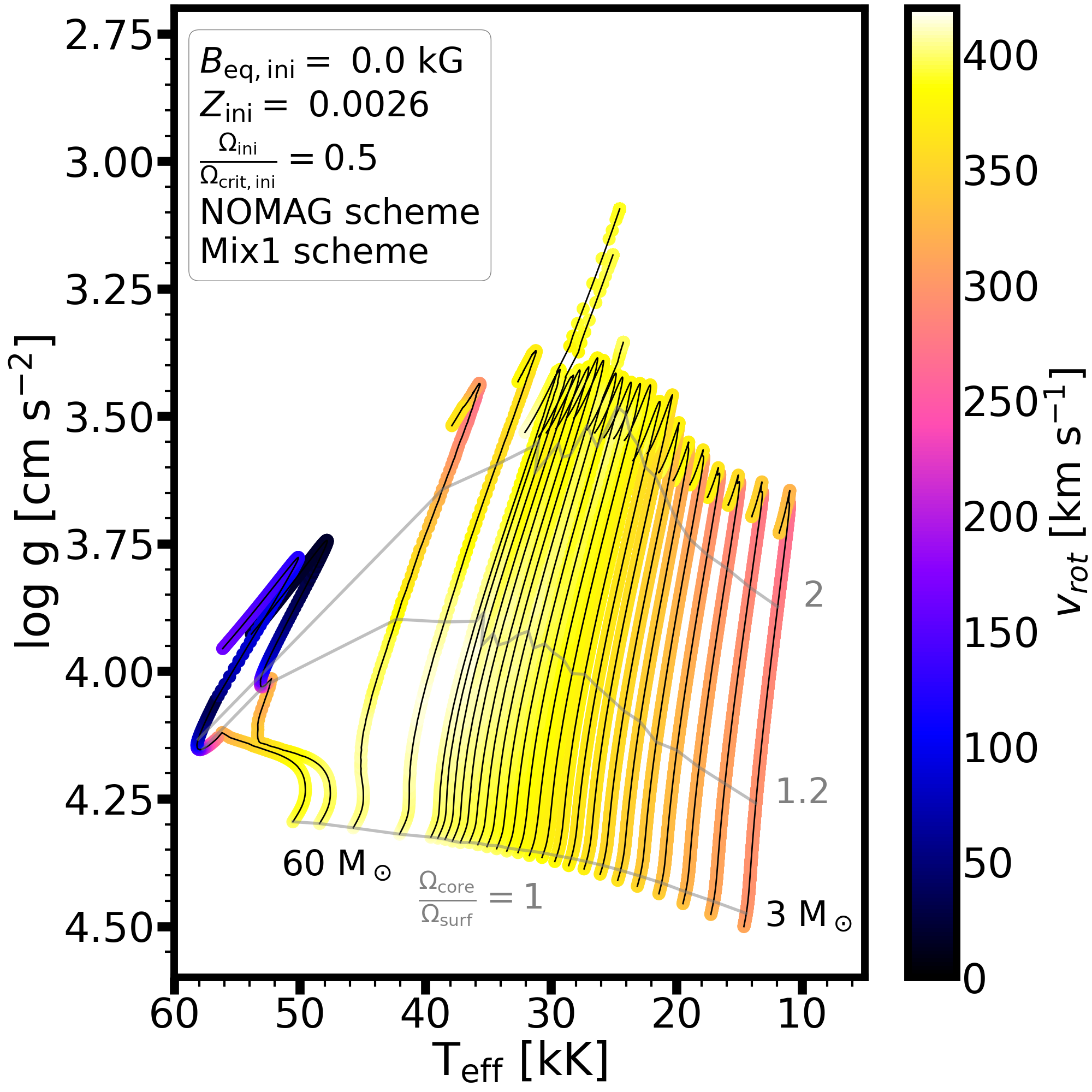}\includegraphics[width=6cm]{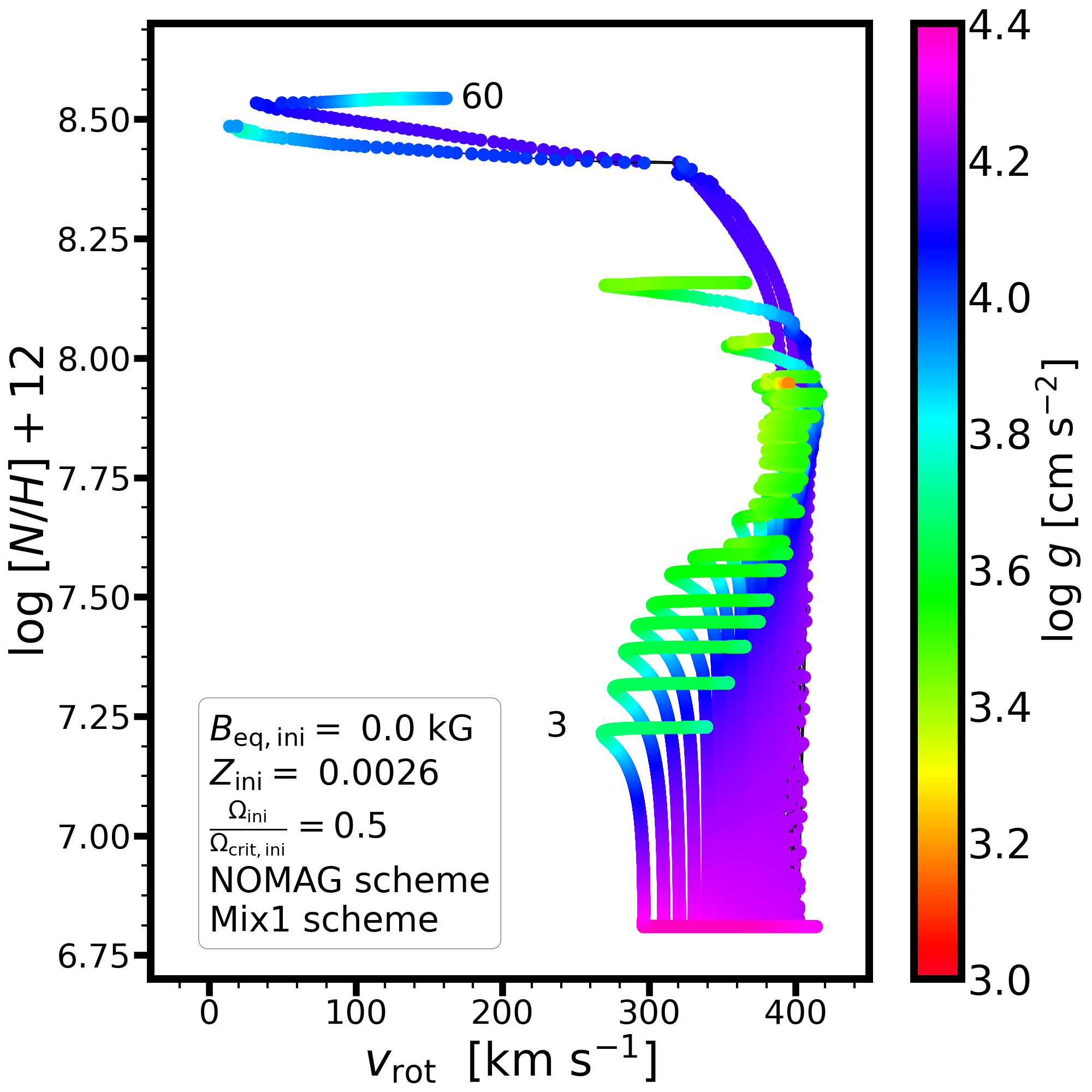} 
\includegraphics[width=6cm]{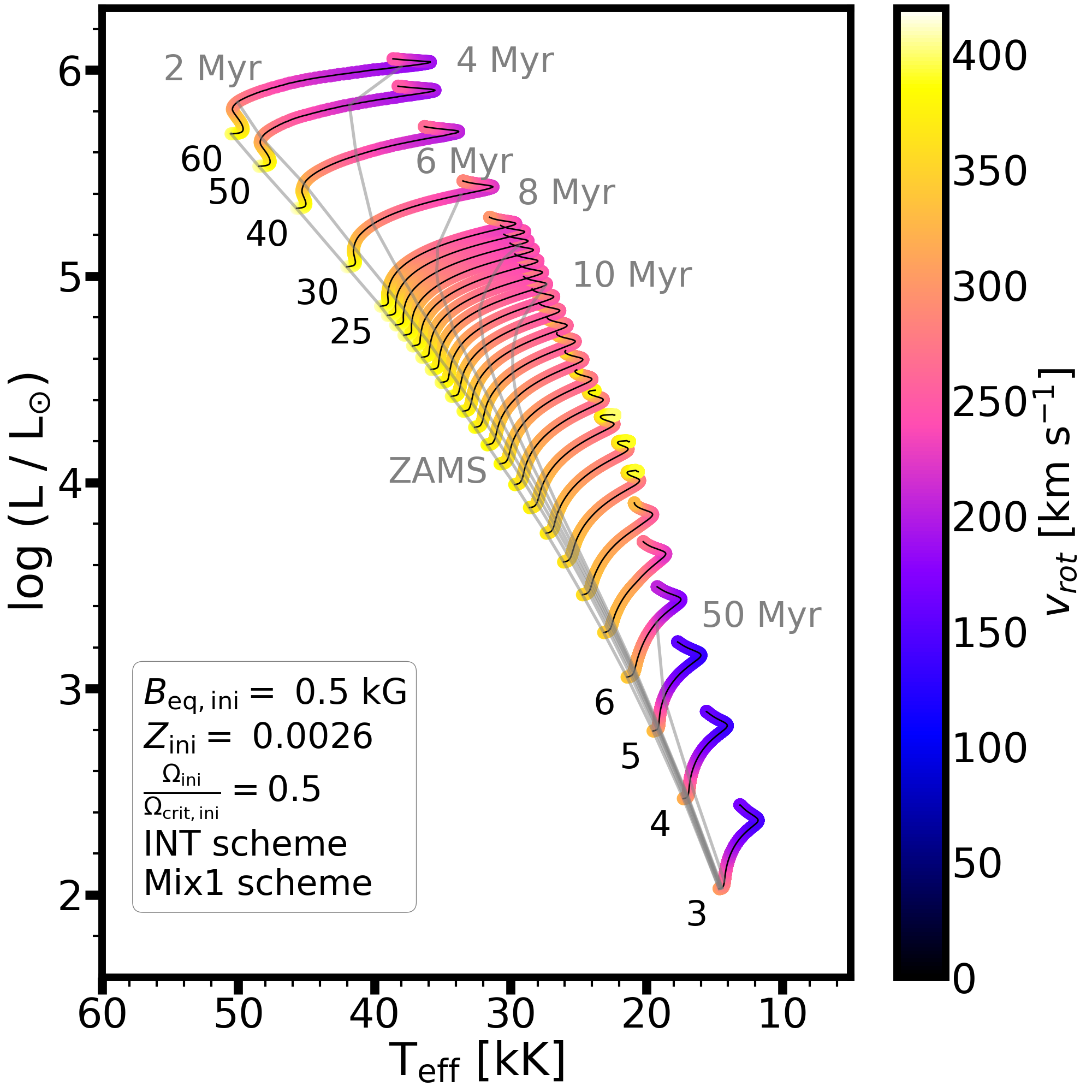}\includegraphics[width=6cm]{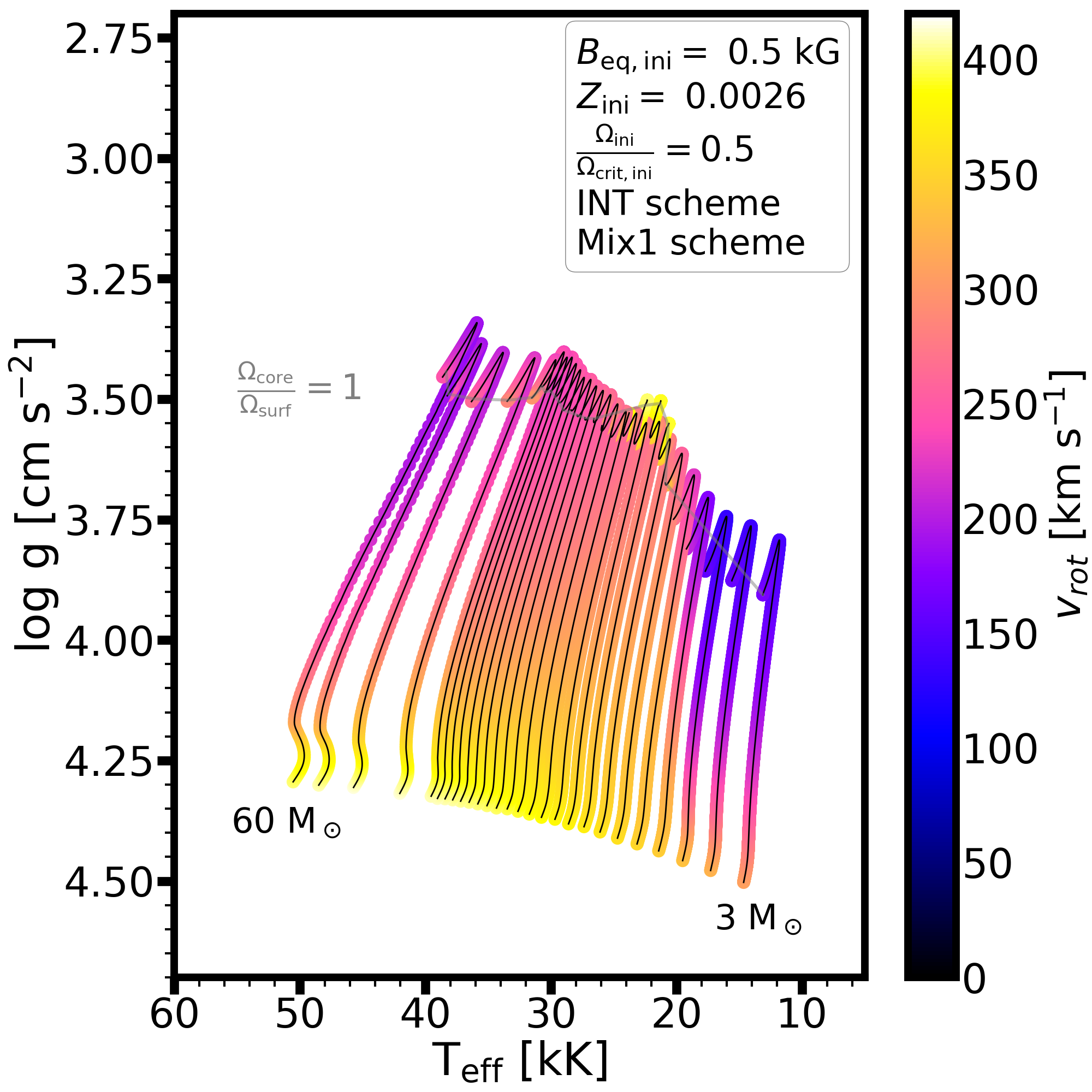}\includegraphics[width=6cm]{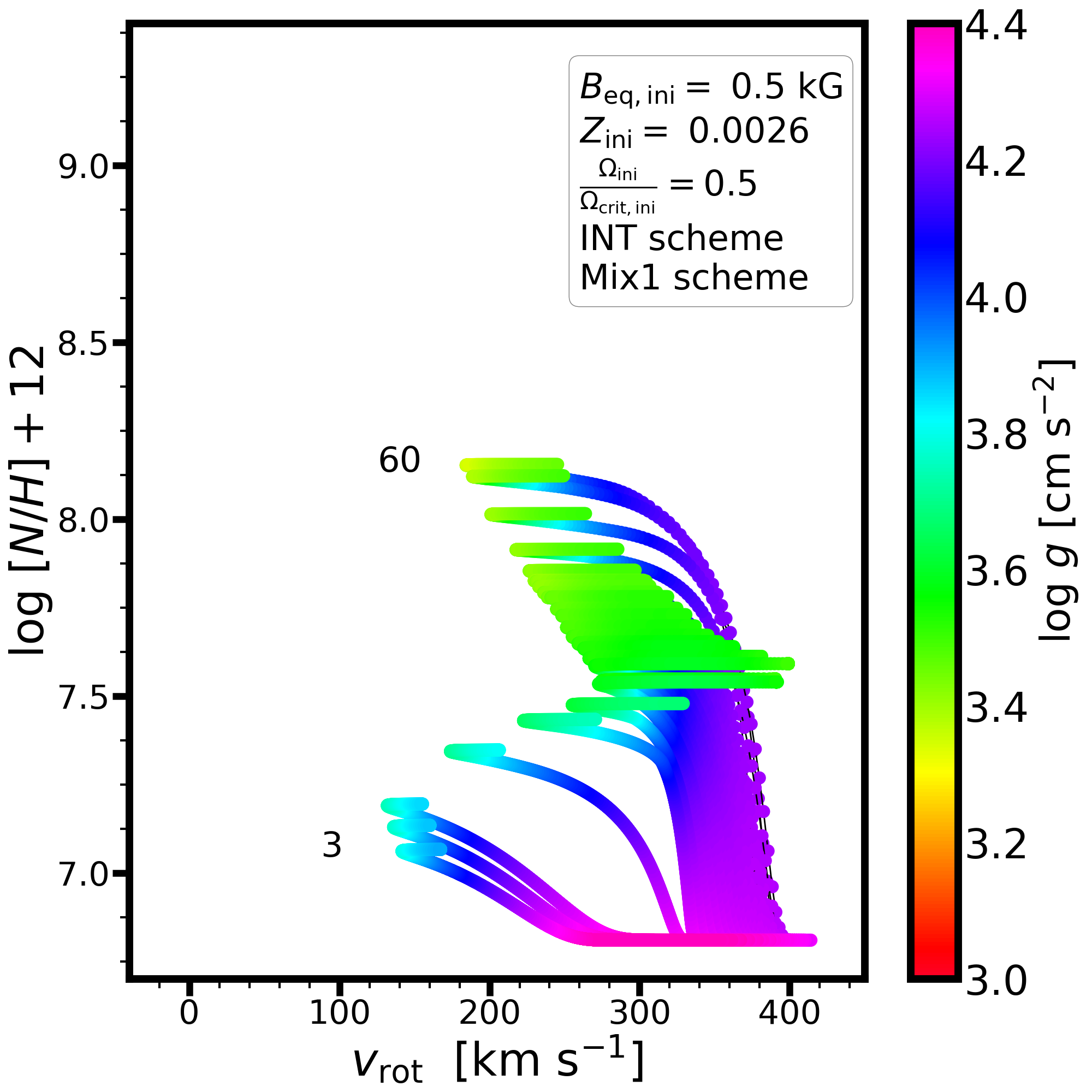}
\caption{Same as Figure \ref{fig:metale1} but for the NOMAG/Mix1 scheme (top) and INT/Mix1 scheme with an initial equatorial magnetic field strength of 0.5 kG (lower panel) at $Z = 0.0026$. The INT/Mix1 model with initial equatorial magnetic field strength of 3~kG is presented in Figure \ref{fig:metale1}. }\label{fig:smc_intmix1}
\end{figure*}
%
%
%
\begin{figure*}
\includegraphics[width=6cm]{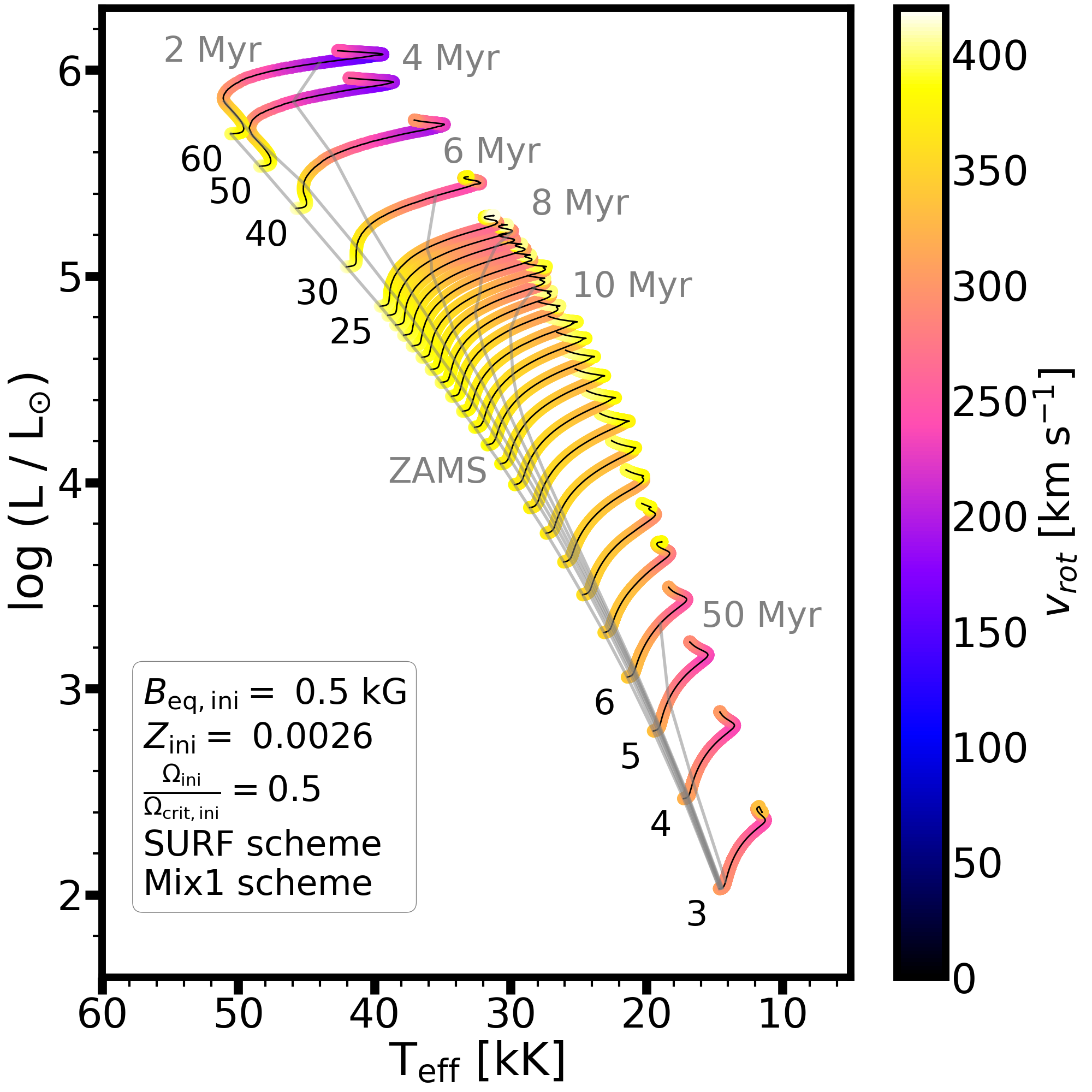}\includegraphics[width=6cm]{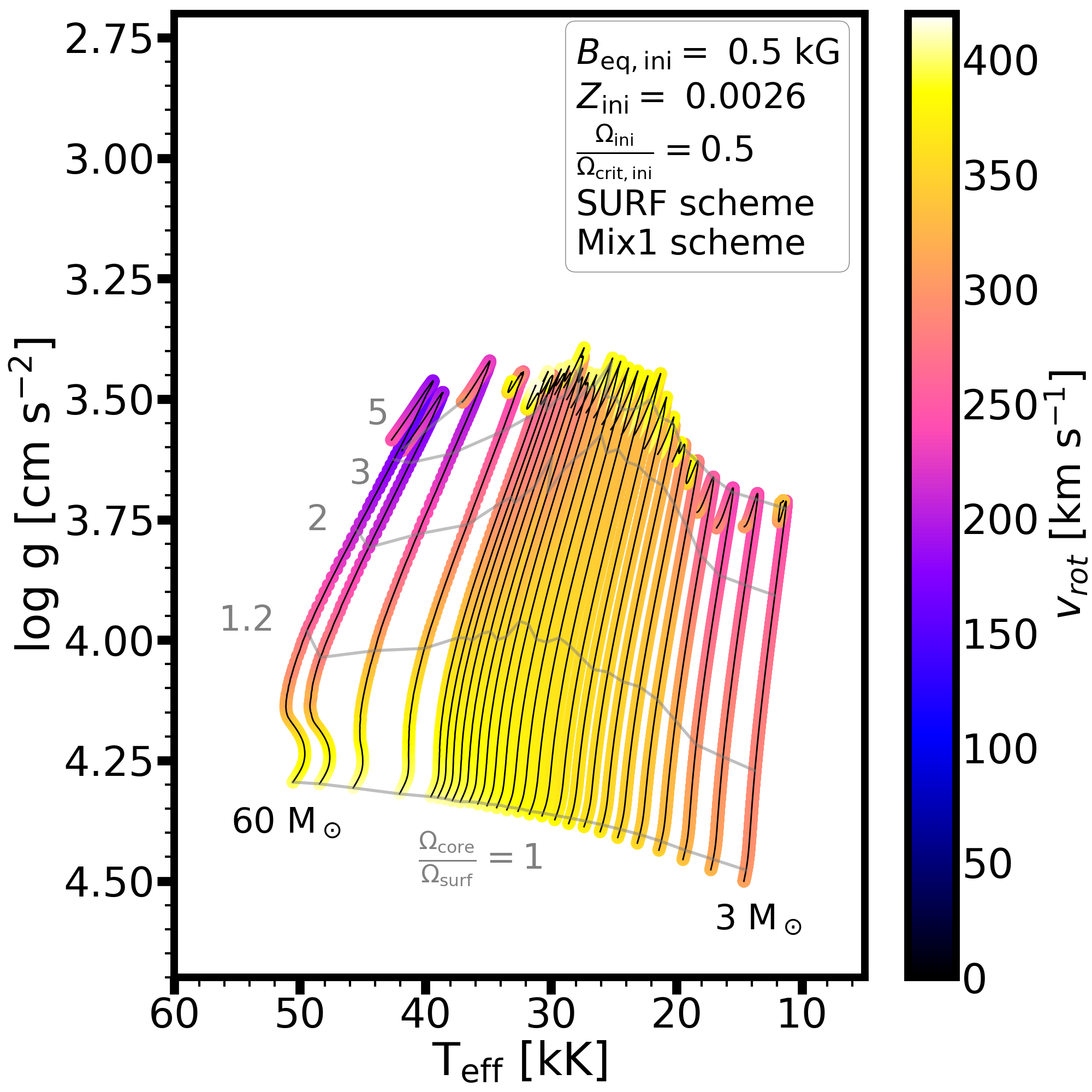}\includegraphics[width=6cm]{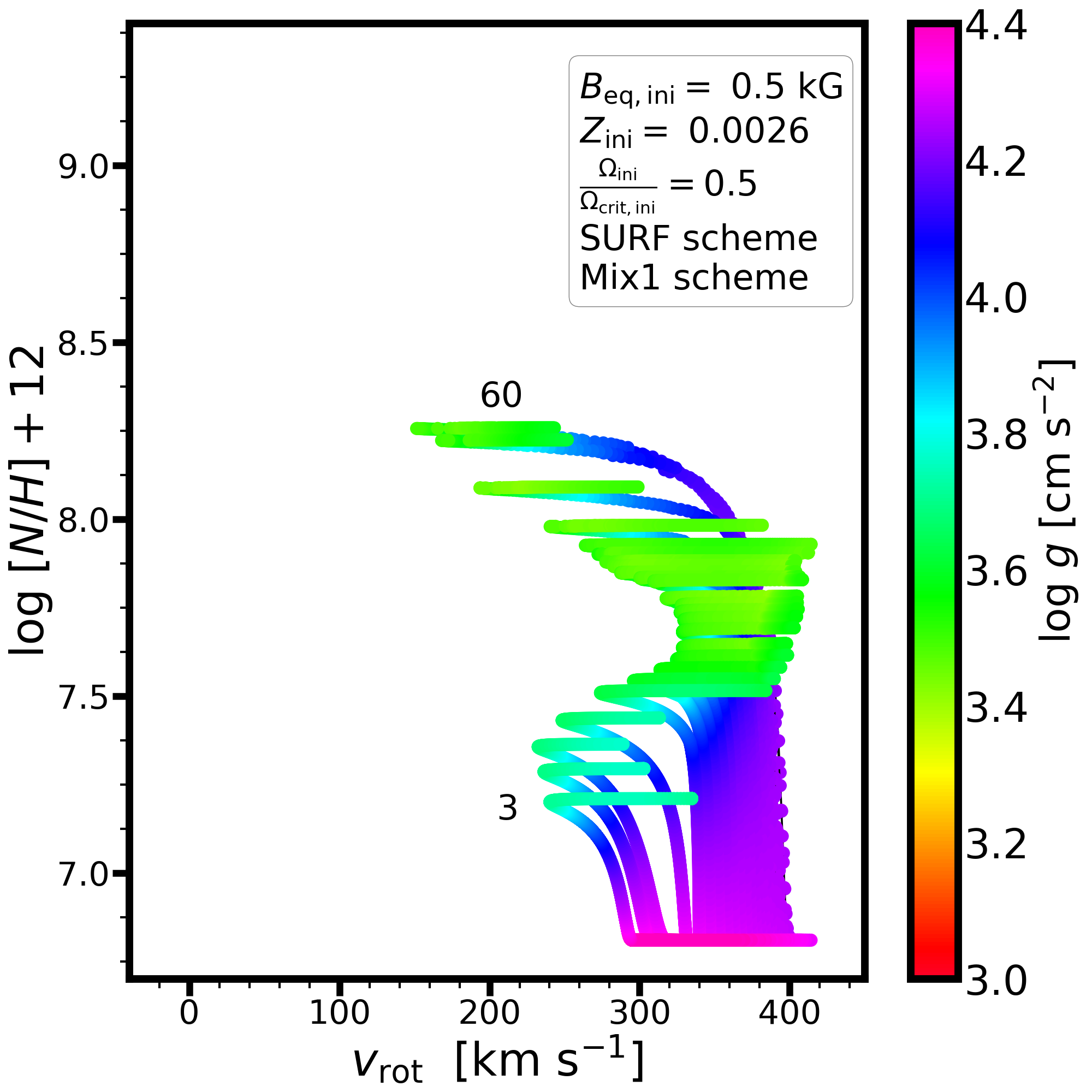}
\includegraphics[width=6cm]{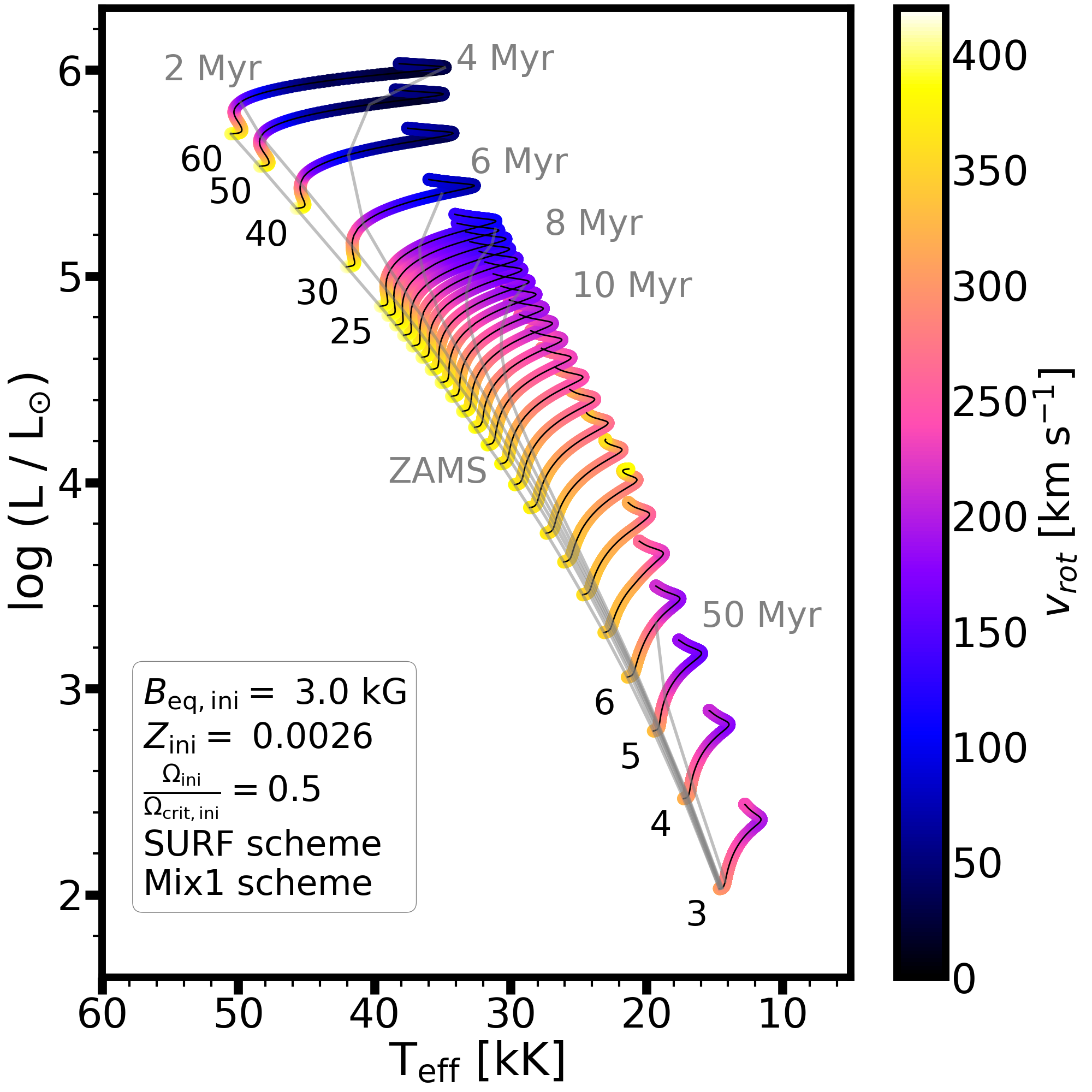}\includegraphics[width=6cm]{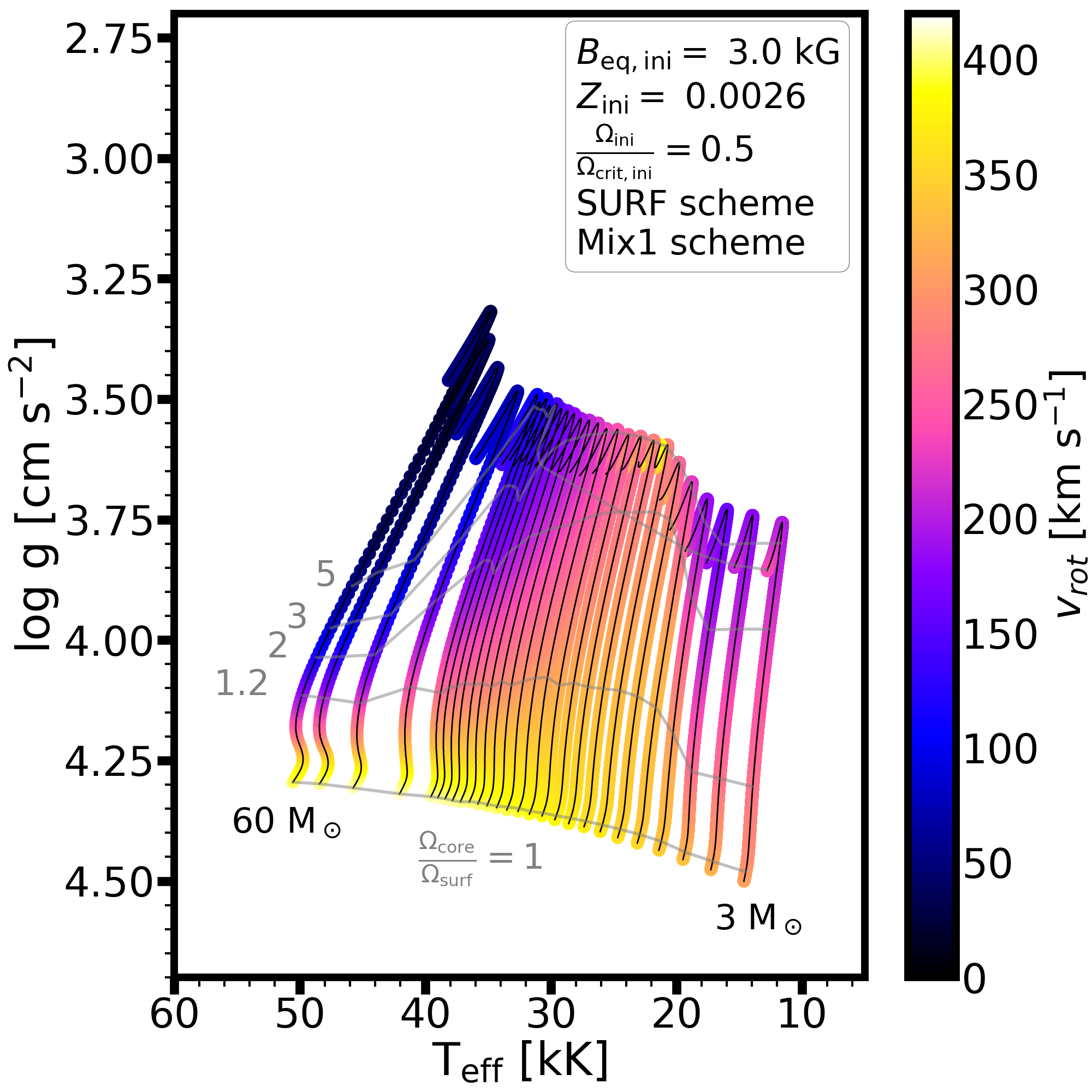}\includegraphics[width=6cm]{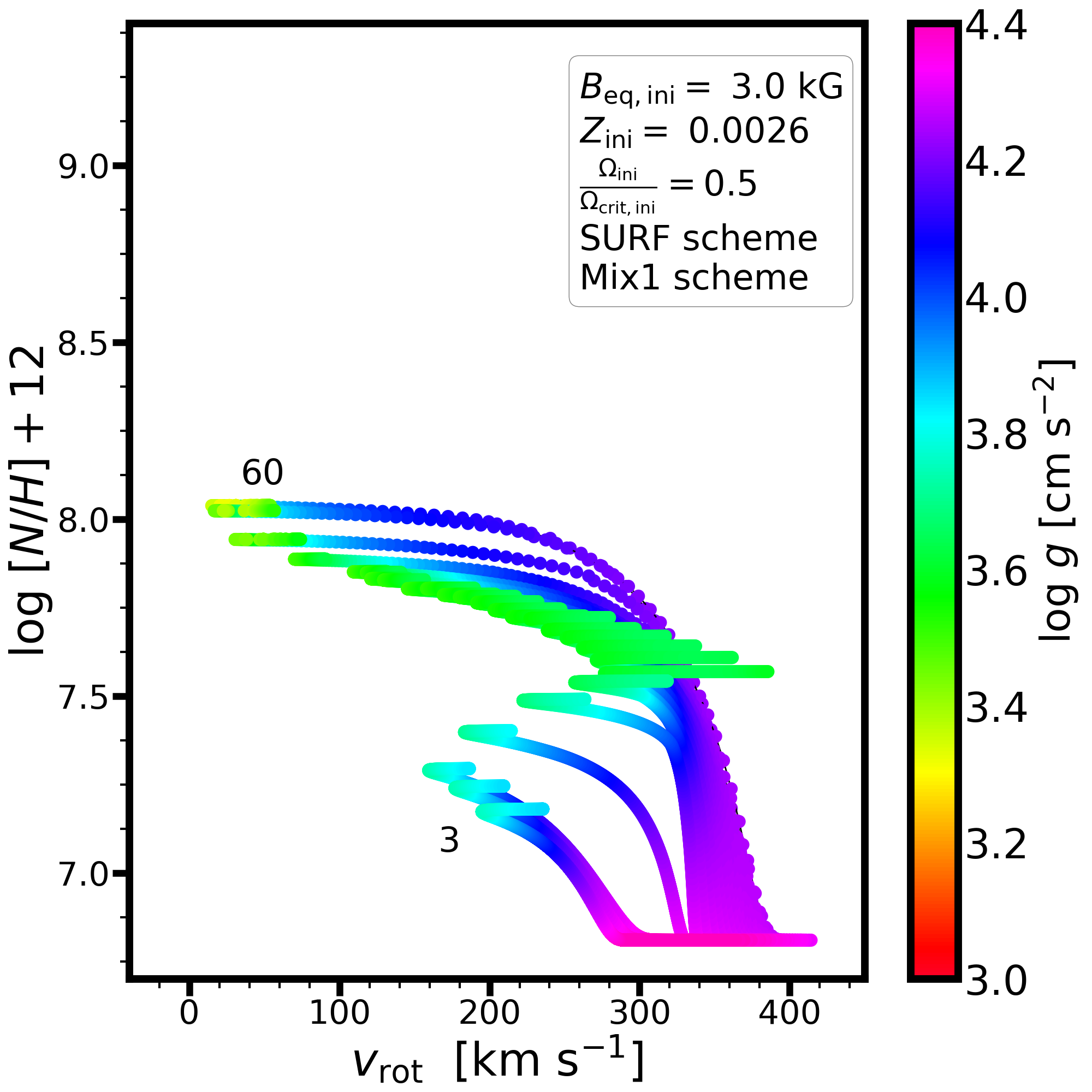}
\caption{Same as Figure \ref{fig:smc_intmix1} but for the SURF/Mix1 scheme. Top panels show models with an initial equatorial magnetic field strength of 0.5 kG, whereas the lower panels show models with 3 kG. The NOMAG/Mix1 model is presented in Figure \ref{fig:smc_intmix1}. }\label{fig:smc_surfmix1}
\end{figure*}
%
%
\begin{figure*}
\includegraphics[width=6cm]{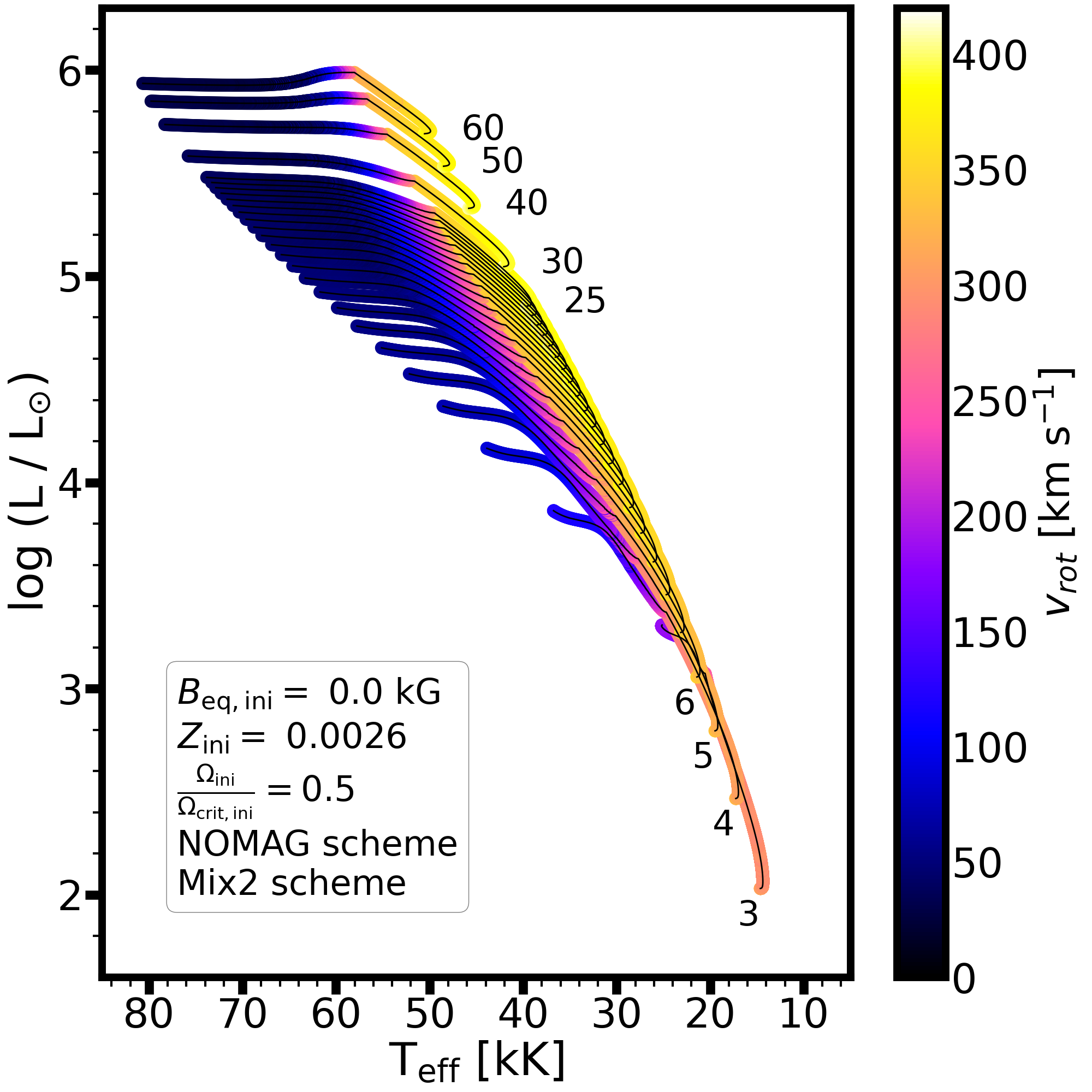}\includegraphics[width=6cm]{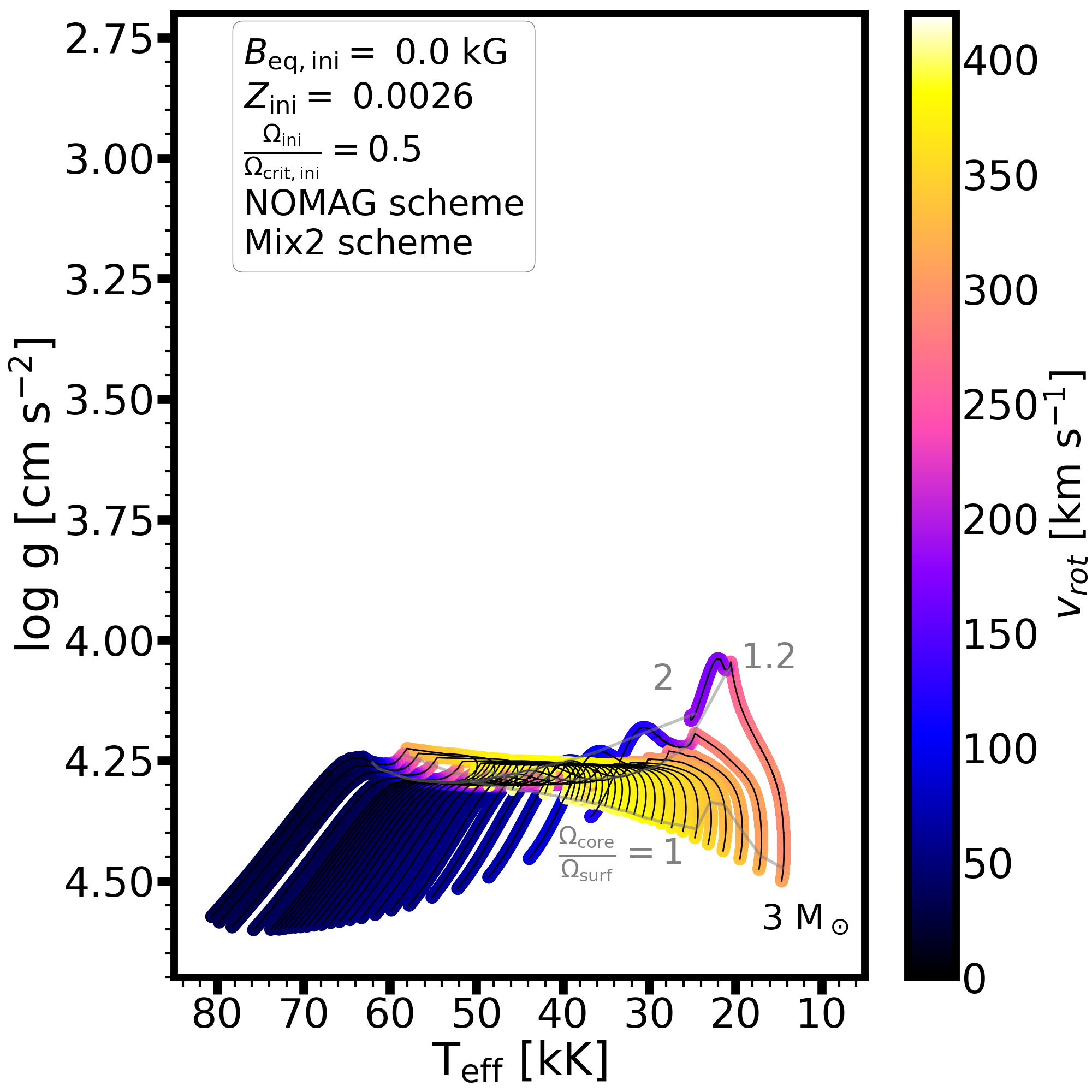}\includegraphics[width=6cm]{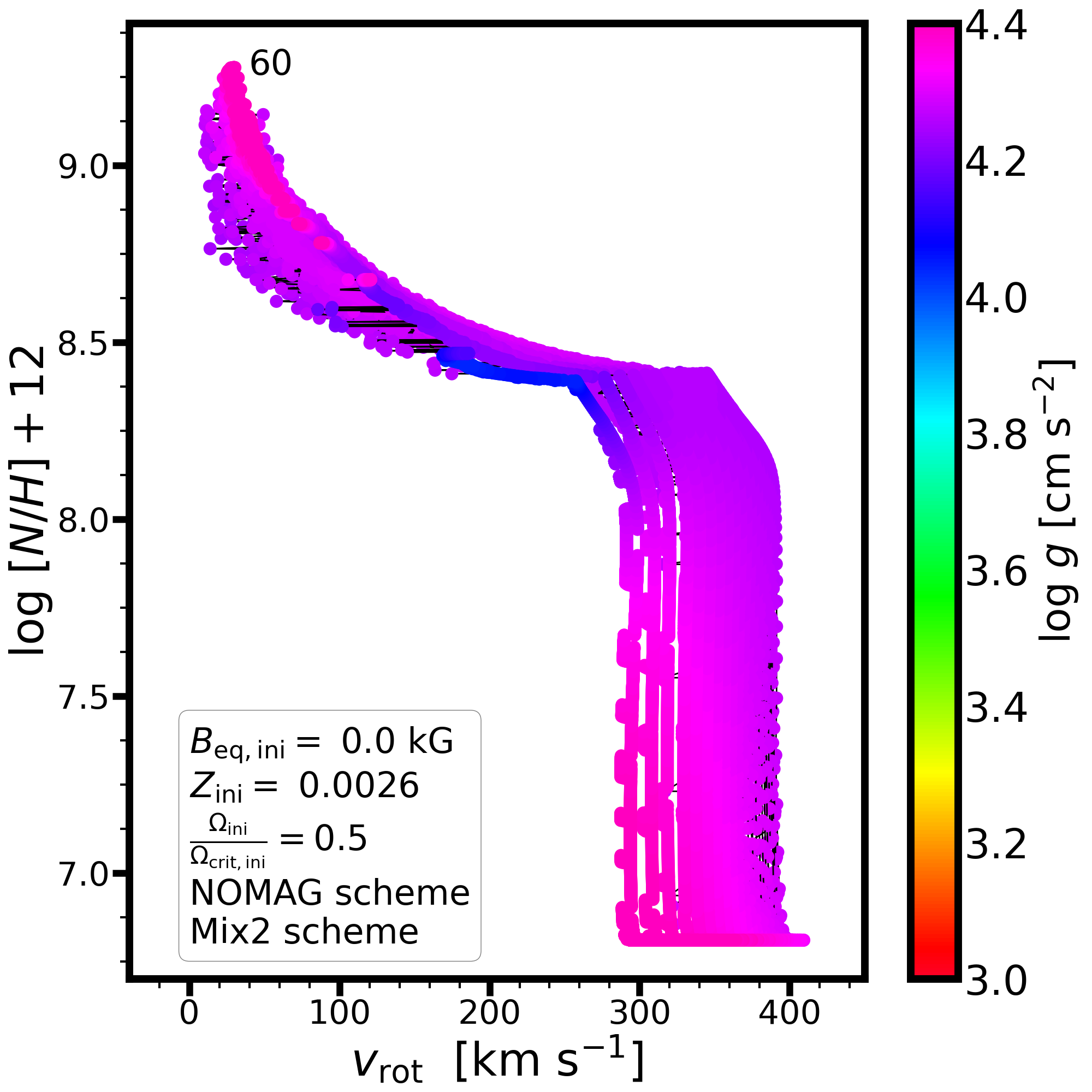} 
\includegraphics[width=6cm]{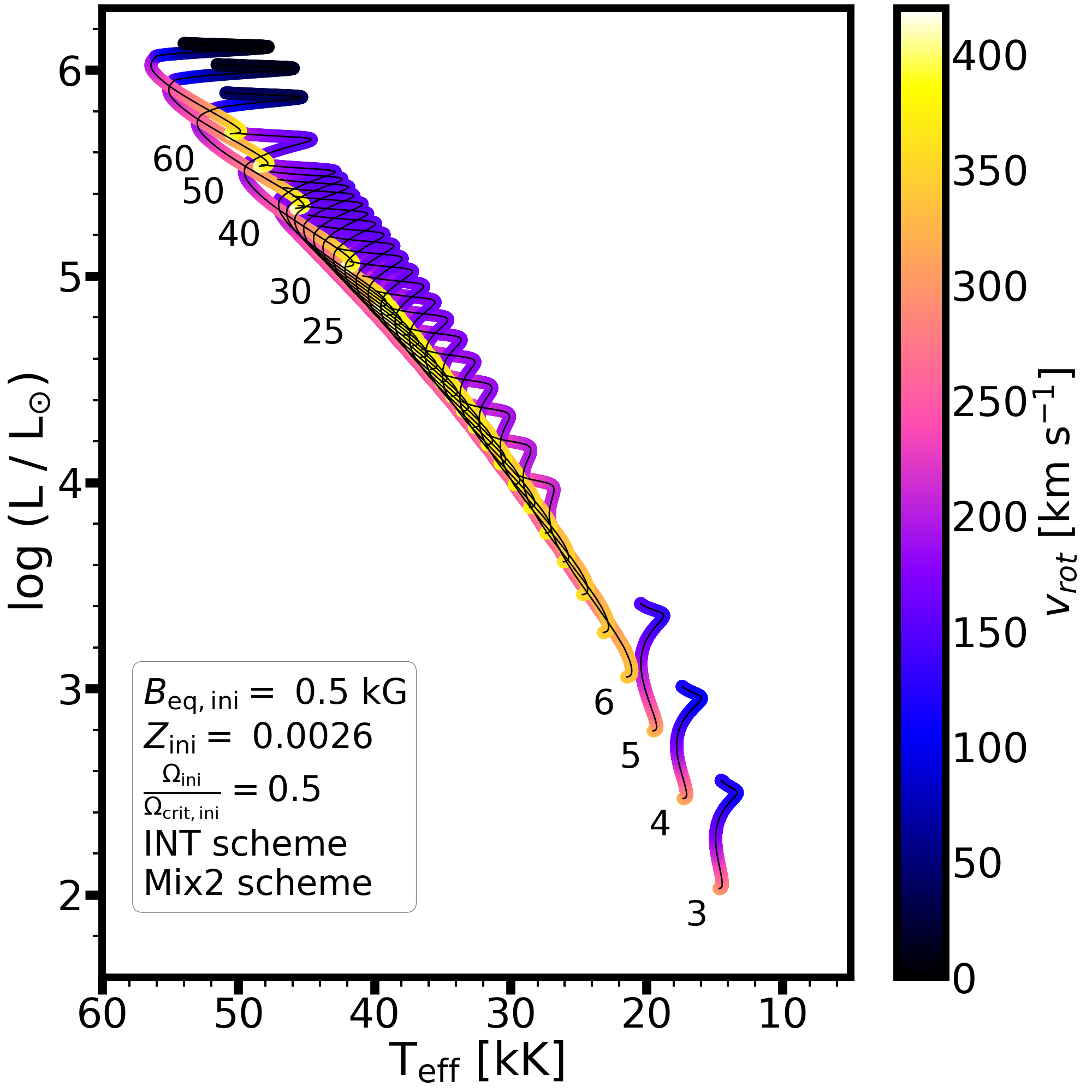}\includegraphics[width=6cm]{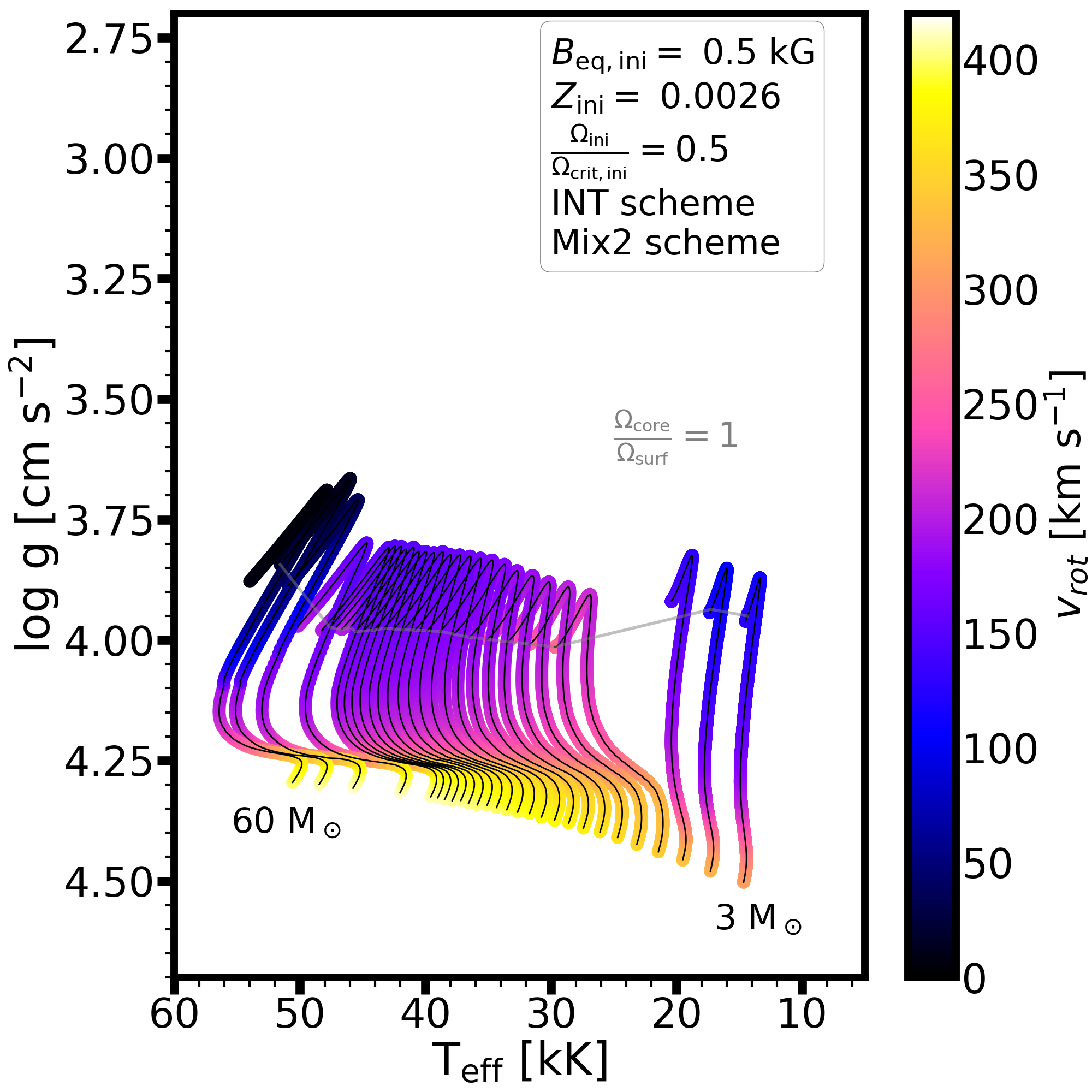}\includegraphics[width=6cm]{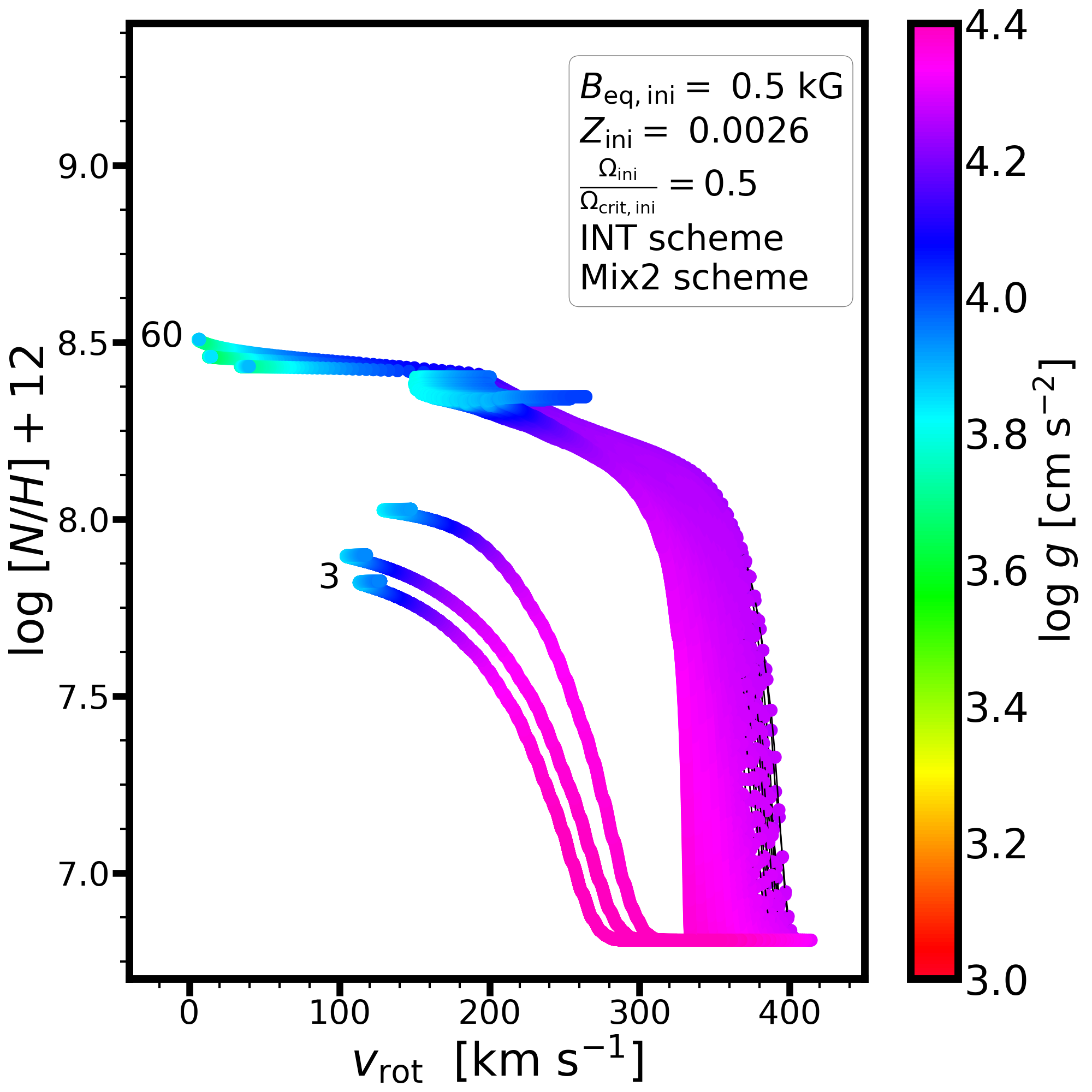}
\includegraphics[width=6cm]{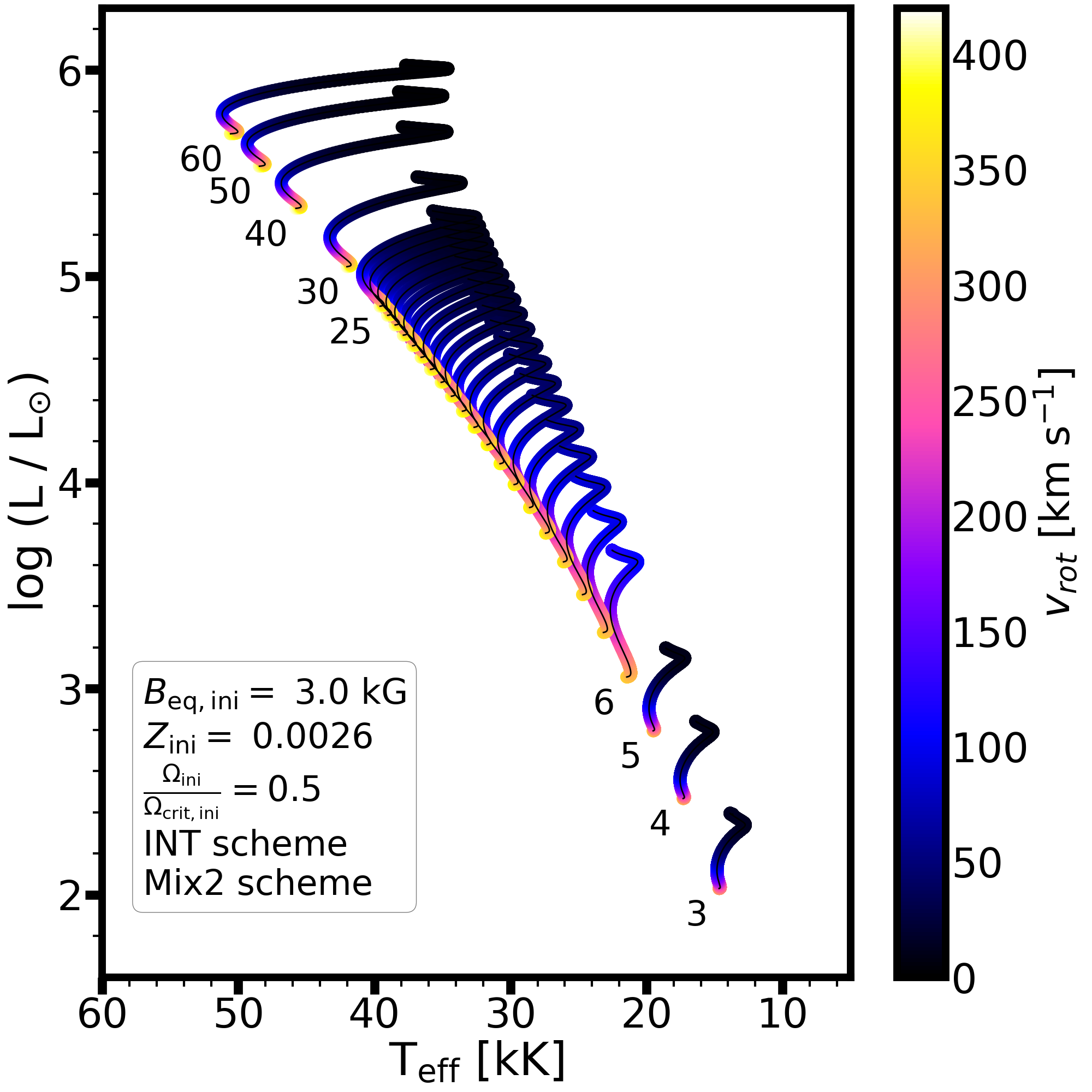}\includegraphics[width=6cm]{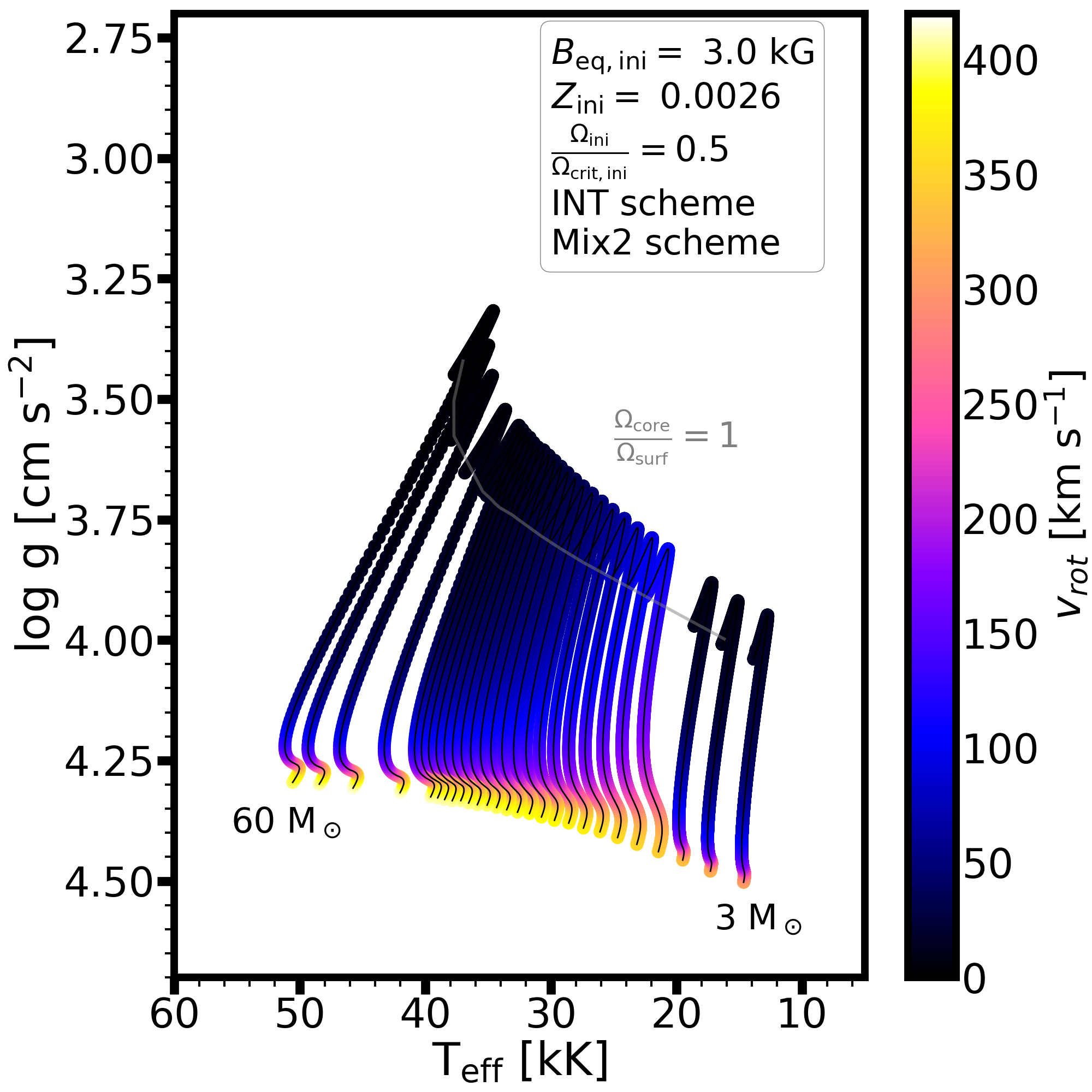}\includegraphics[width=6cm]{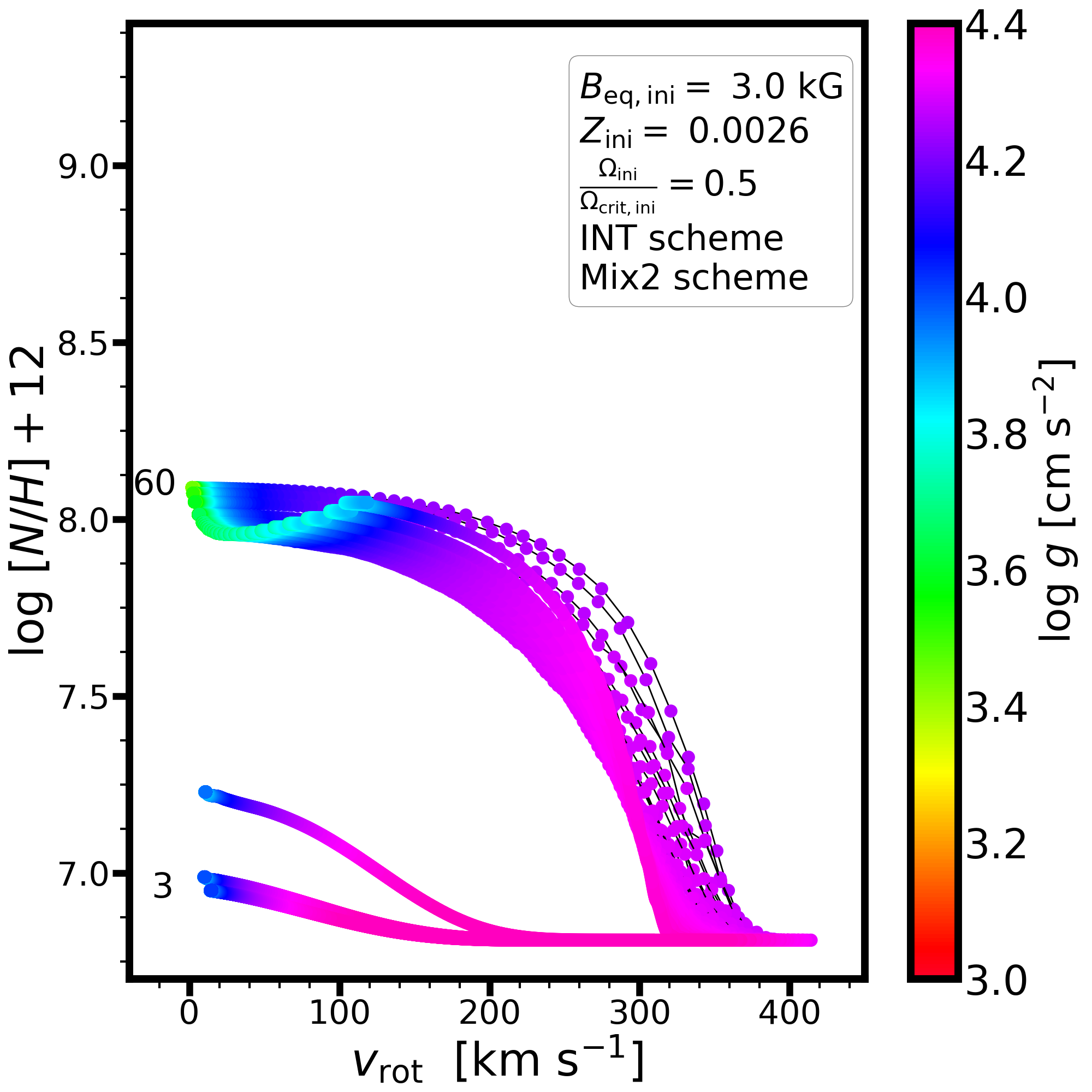}
\caption{Same as Figure \ref{fig:smc_intmix1} but for the NOMAG/Mix2 (top panels) and INT/Mix2 (middle and lower panels) schemes.}\label{fig:smc_intmix2}
\end{figure*}
%
%
\begin{figure*}
\includegraphics[width=6cm]{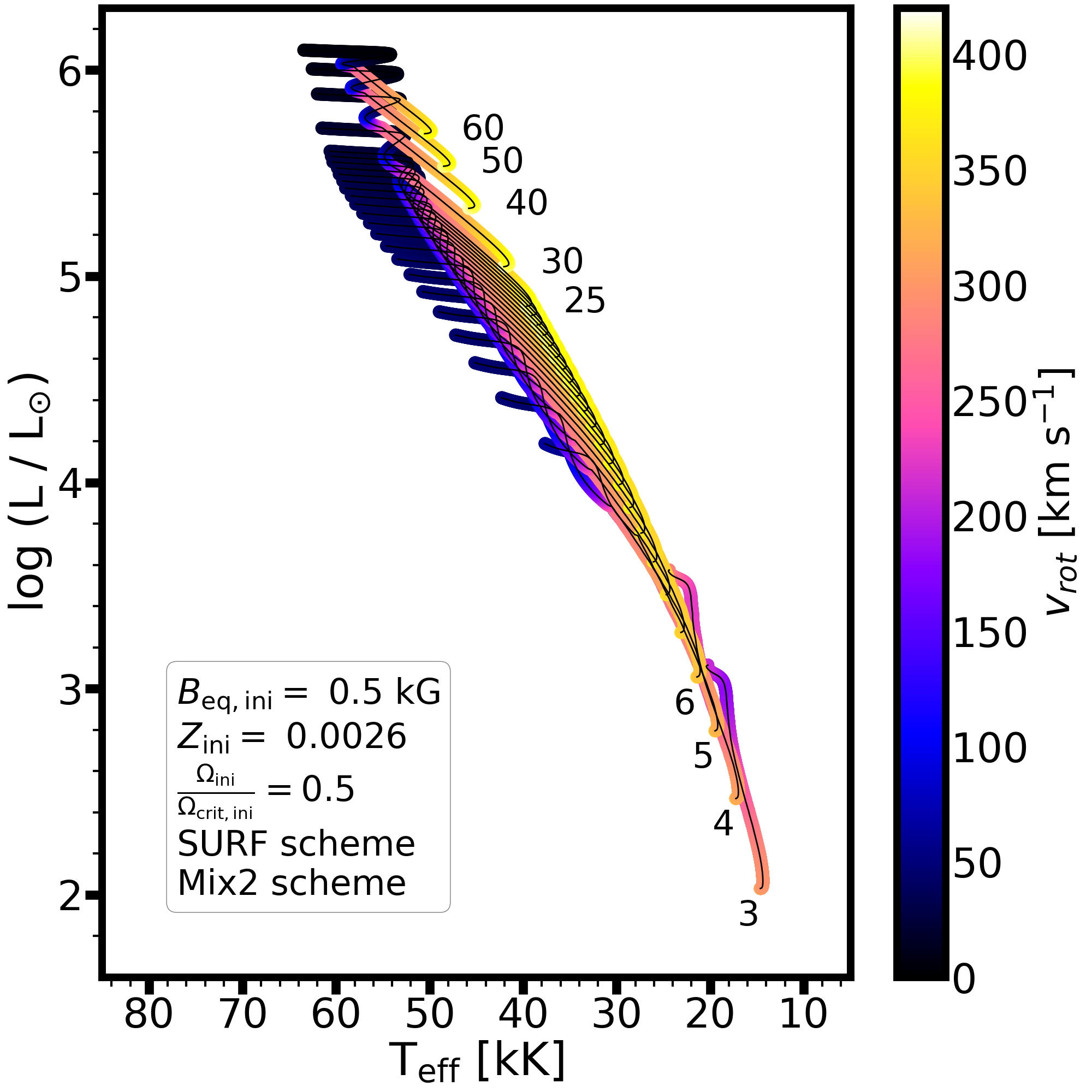}\includegraphics[width=6cm]{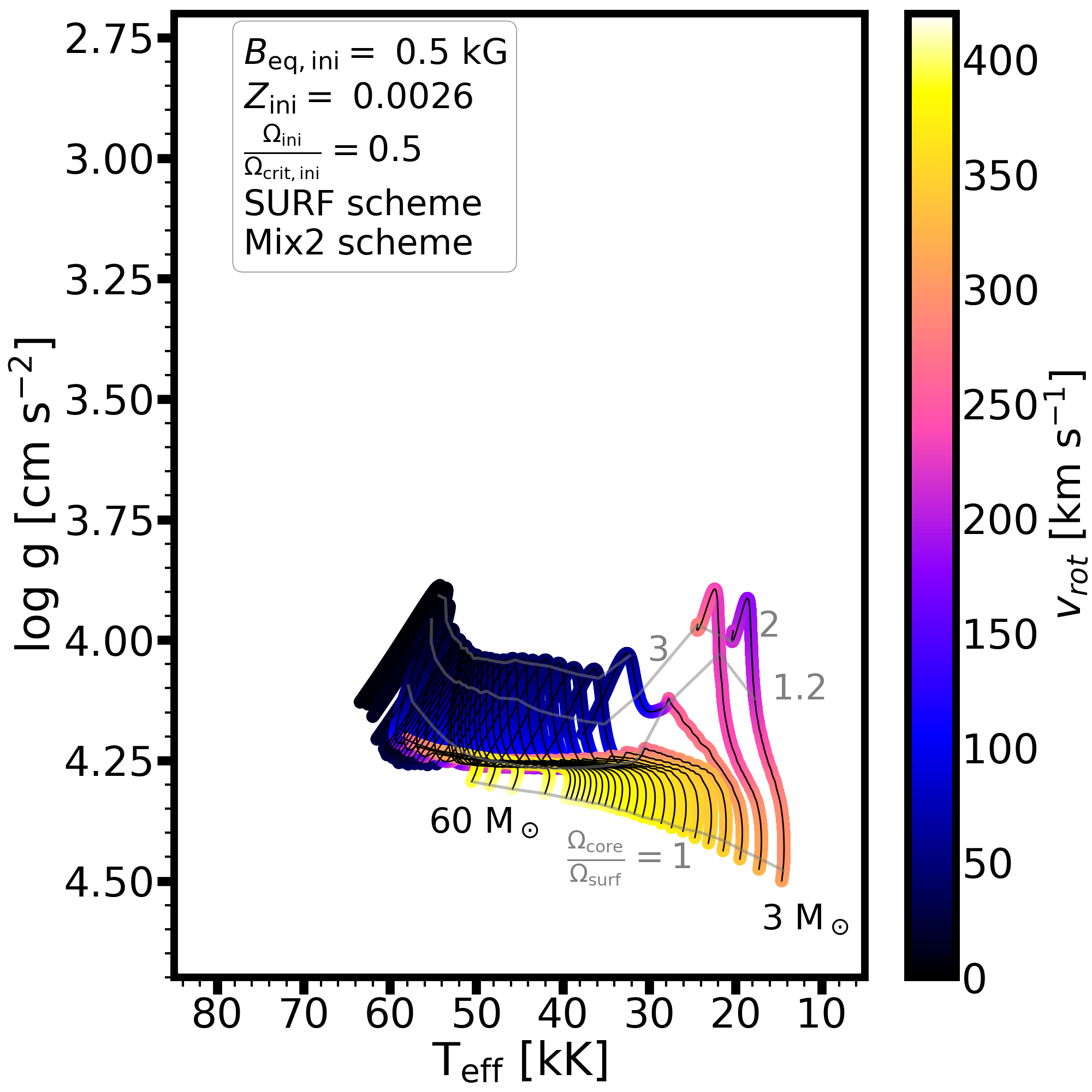}\includegraphics[width=6cm]{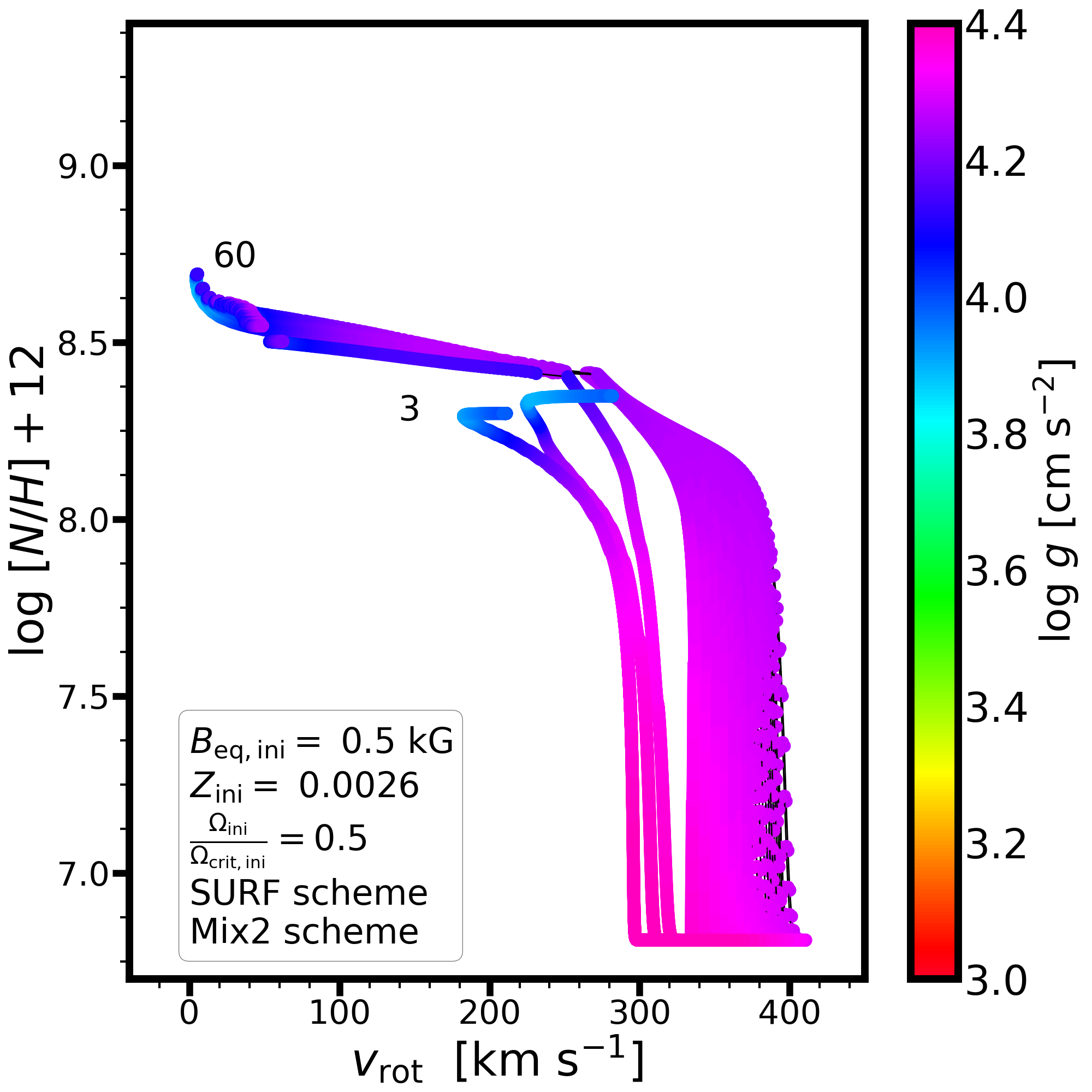}
\includegraphics[width=6cm]{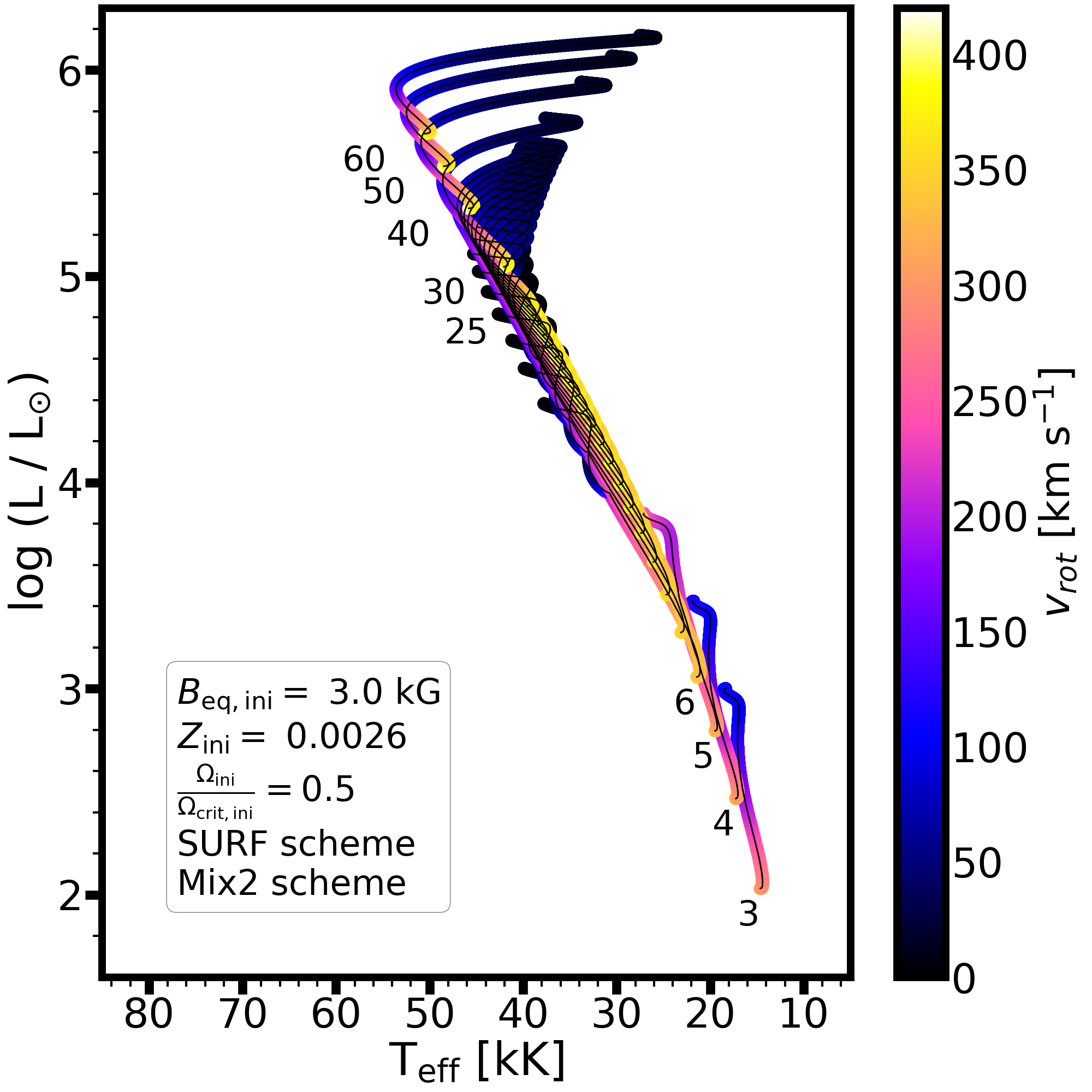}\includegraphics[width=6cm]{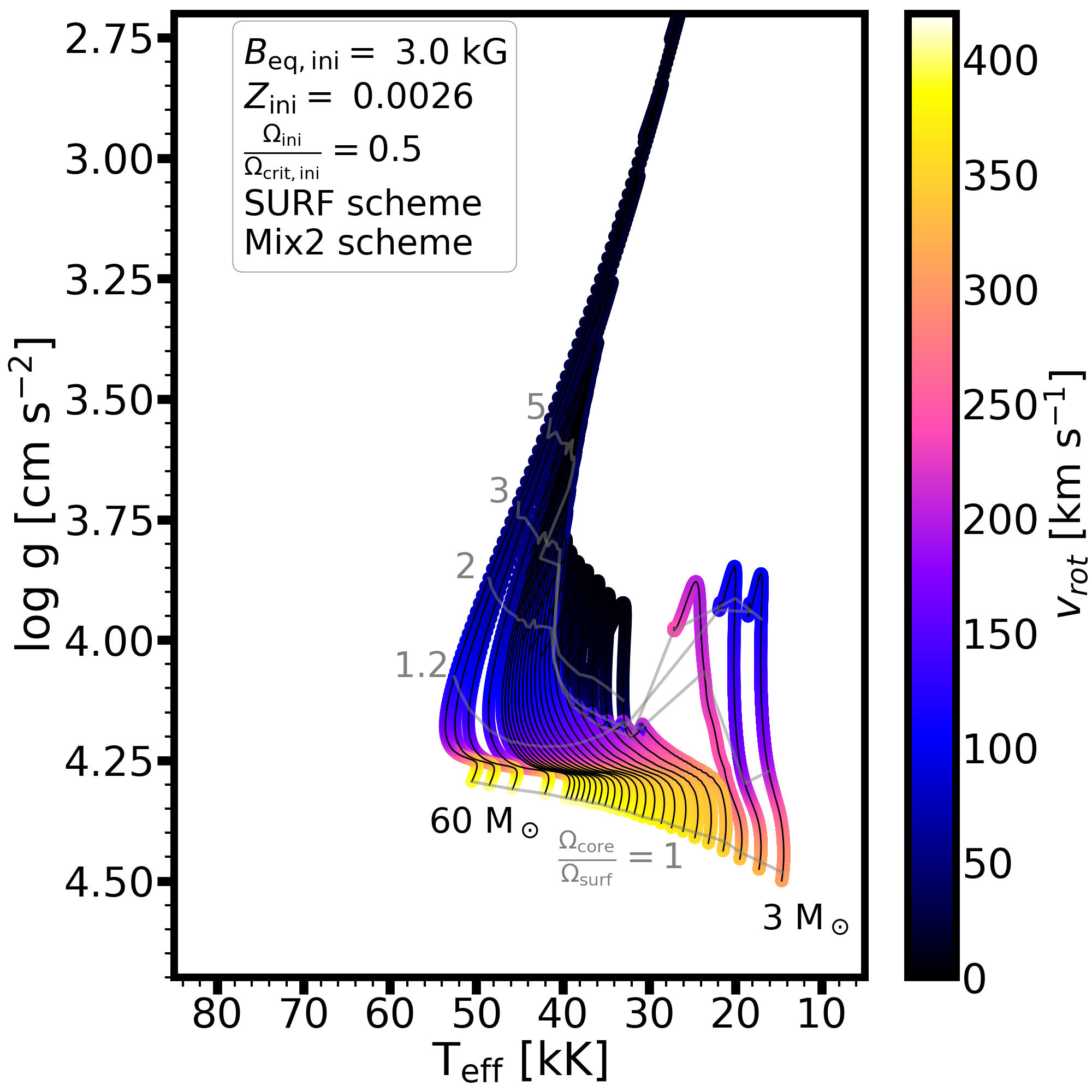}\includegraphics[width=6cm]{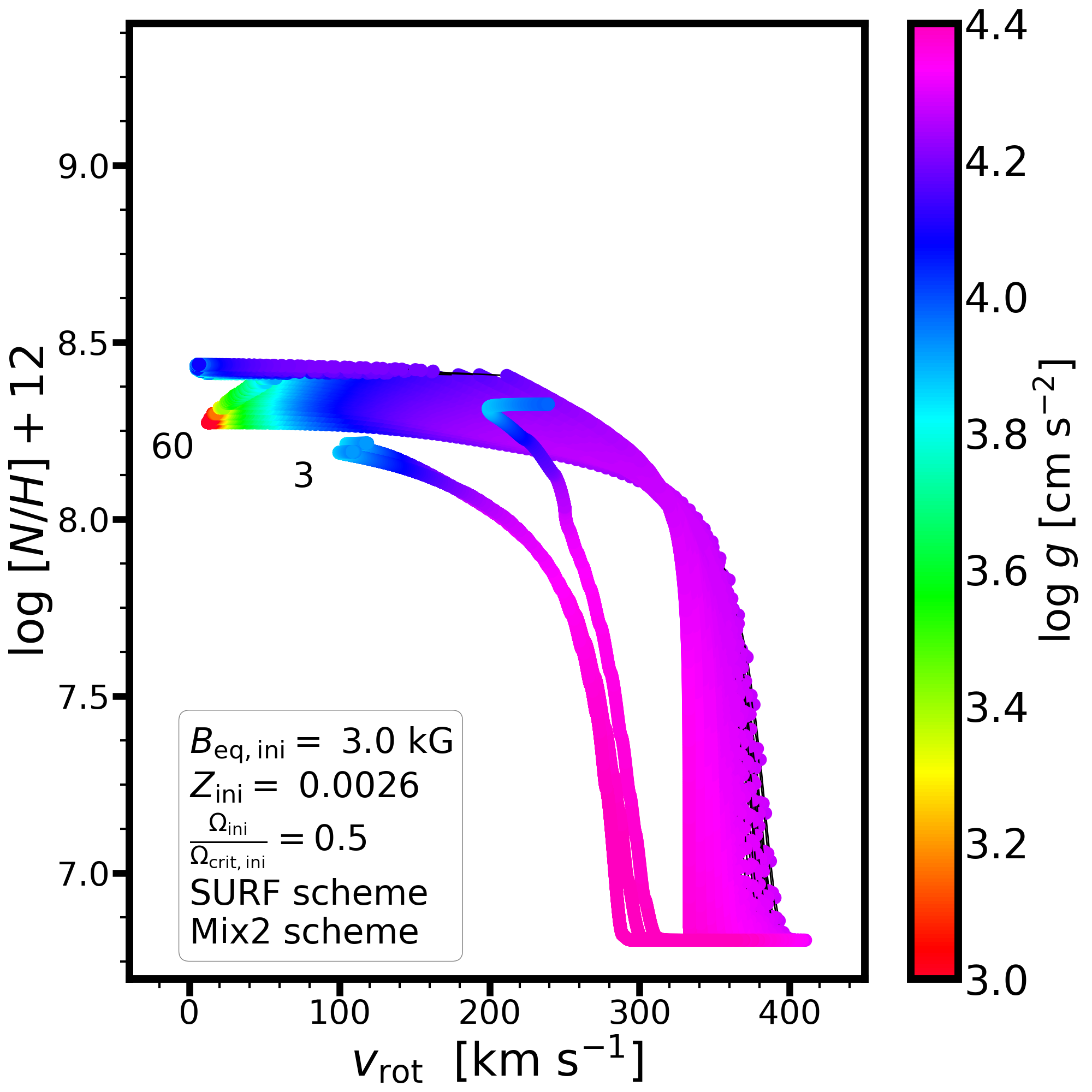}
\caption{Same as Figure \ref{fig:smc_intmix2} but for the SURF/Mix2 scheme. See top panels of Figure \ref{fig:smc_intmix2} for the NOMAG/Mix2 models.}\label{fig:smc_surfmix2}
\end{figure*}

\subsection{Impact of varying magnetic field strength at $Z = 0.0026$}

\subsubsection{INT/Mix1}

At SMC metallicity, the ZAMS position of the non-magnetic models is further shifted to higher effective temperatures compared to solar and LMC metallicity models (Figure \ref{fig:smc_intmix1}). For the 50 and 60~M$_\odot$ models, we note major differences in comparison to models at higher metallicity. The highest mass non-magnetic models at SMC metallicity are able to undergo a blueward evolutionary phase for some fraction of their main sequence. This results from the relatively rapid rotation of the models, the efficient core-envelope coupling (assumed via hydrodynamic processes in a diffusive scheme), and the effects of stellar winds. These models can enhance their surface $[N/H]$ ratio by about a factor of 50 compared to the initial baseline value. 

When a 0.5 kG initial equatorial magnetic field strength is considered, the effects of the spin-down are more pronounced than in higher metallicity models. We see that the blueward evolution of the most massive models is prevented. The impact on the nitrogen abundances is also notable as less enrichment is predicted compared to non-magnetic models. However, we should also note that in contrast to the most massive models in the INT/Mix1 scheme at higher metallicities, the rotational velocities of these models at SMC metallicity is actually less impacted. This is because stellar winds are less efficient at lower metallicity. Thus the combination of magnetic braking by a "weak" field (initially 0.5 kG) and weaker stellar winds allows for maintaining higher surface rotational velocities.

\subsubsection{SURF/Mix1}

In the SURF magnetic braking scheme we notice small differences compared to the INT scheme when keeping chemical mixing the same in the Mix1 scheme (Figure \ref{fig:smc_surfmix1}). For the same initial mass the SURF models tend to maintain a somewhat higher rotational velocity than the INT models. As expected, when increasing the value of the initial equatorial magnetic field strength from 0.5 to 3 kG, the spin-down is stronger, and consequently lower surface rotational velocities can be reached in conjunction with a reduction in the surface nitrogen enrichment. Nonetheless, the most massive models in this configuration with a 3 kG initial equatorial magnetic field strength are able to produce slowly-spinning, nitrogen-enriched Group 2 stars.

\subsubsection{INT/Mix2}

In the NOMAG/Mix2 at SMC metallicity, the most extended blueward evolution is produced, reaching 80~kK for the most massive stellar models (Figure \ref{fig:smc_intmix2}). The behaviour of these models is qualitatively similar to those of solar and LMC metallicity. The predicted increase in $[N/H]$ is first a factor of about 50 related to mixing nitrogen, and then further almost a factor of 10 related to losing hydrogen from the stellar surface. Due to the blueward evolution of the models the surface gravities are expected to remain high. 

In the INT/Mix2 case, a 0.5 and 3 kG initial equatorial magnetic field strength can prevent the extensive blueward evolution of the models. However, in contrast to some previous cases, the most massive models do not reach low surface gravities. Although for the non-magnetic case, all models are predicted to show a very large increase in their surface nitrogen abundance, the magnetic models show a more moderate enrichment. In this particular setup we also see that models less massive than initially 6~M$_\odot$ tend to produce much lower $[N/H]$ ratios, whereas models with higher initial masses are closely grouped together.

\subsubsection{SURF/Mix2}

In the SURF/Mix2 scheme at SMC metallicity the 0.5 kG initial magnetic field strength is not quite strong enough to prevent the quasi-chemically homogeneous evolution of the models (Figure \ref{fig:smc_surfmix2}). On the other hand, a 3~kG initial equatorial magnetic field strength leads to a more pronounced redward evolution of the most massive models after their spin-down. In this case, we again observe that the most massive models can reach low surface gravities along with low surface rotational velocities. In all cases, the predicted enrichment in $[N/H]$ remains over an order of magnitude compared to the baseline.

%
%
%
%
%
\begin{figure*}
\includegraphics[width=9cm]{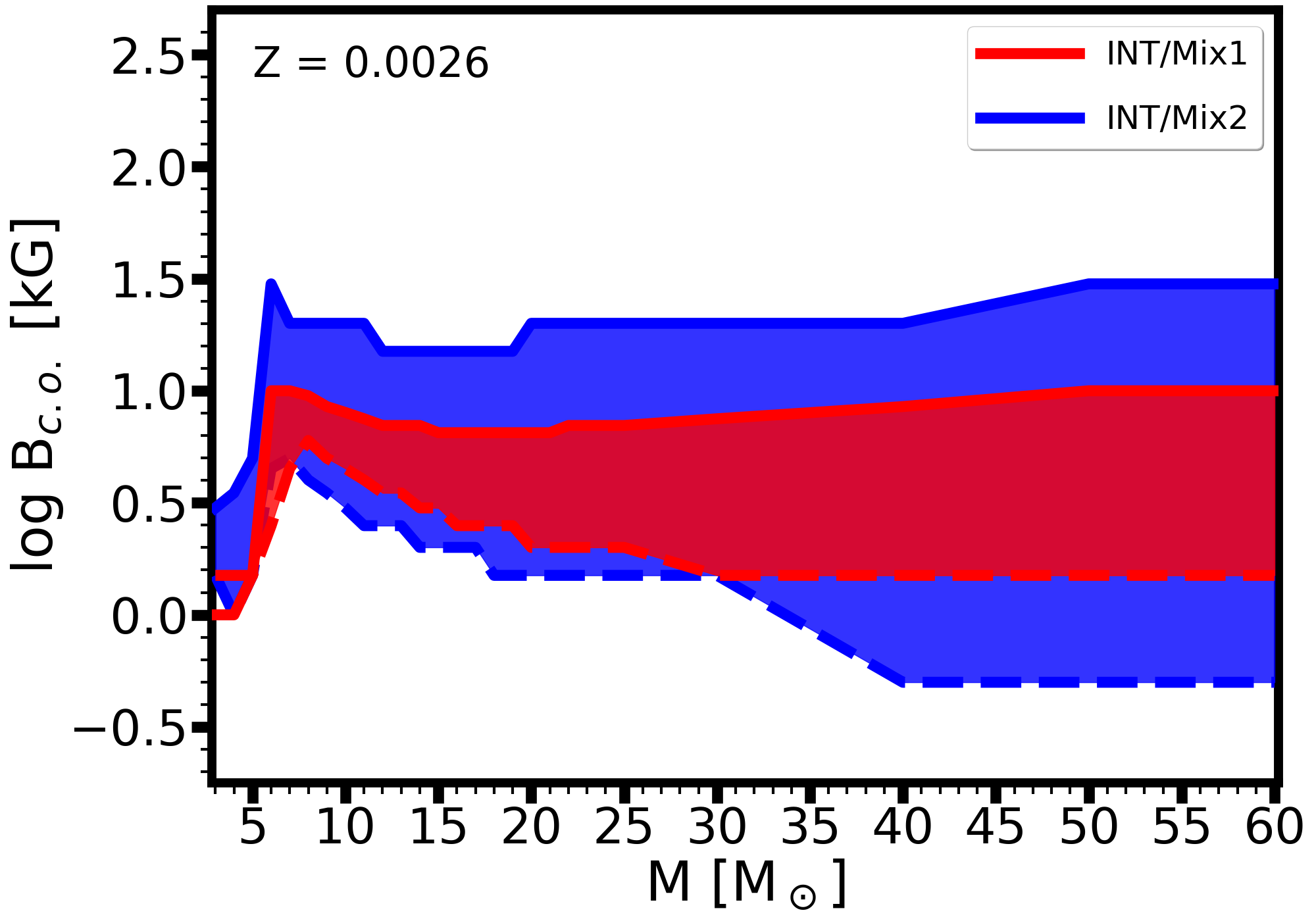}\includegraphics[width=9cm]{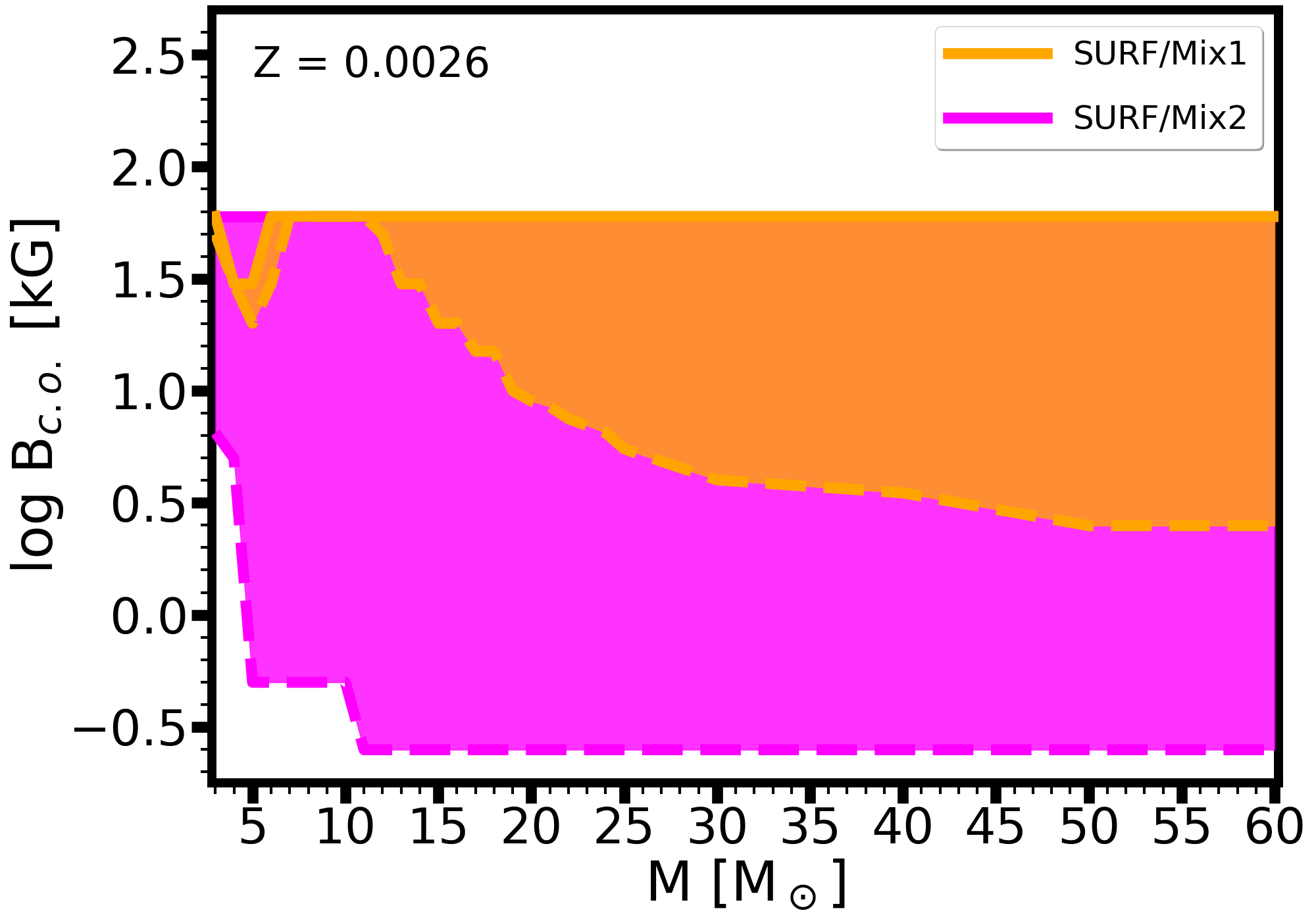}
\caption{Cutoff magnetic field strengths in the Small Magellanic Cloud to produce slowly-rotating nitrogen enriched stars as a function of mass. Left/right panels show the INT/SURF schemes. The models are considered for an initial rotation of $\Omega/\Omega_{\rm crit} = 0.5$ at SMC ($Z = 0.0026$) metallicity.  }\label{fig:cutoff2}
\end{figure*}
%
%
\begin{figure*}
\includegraphics[width=9cm]{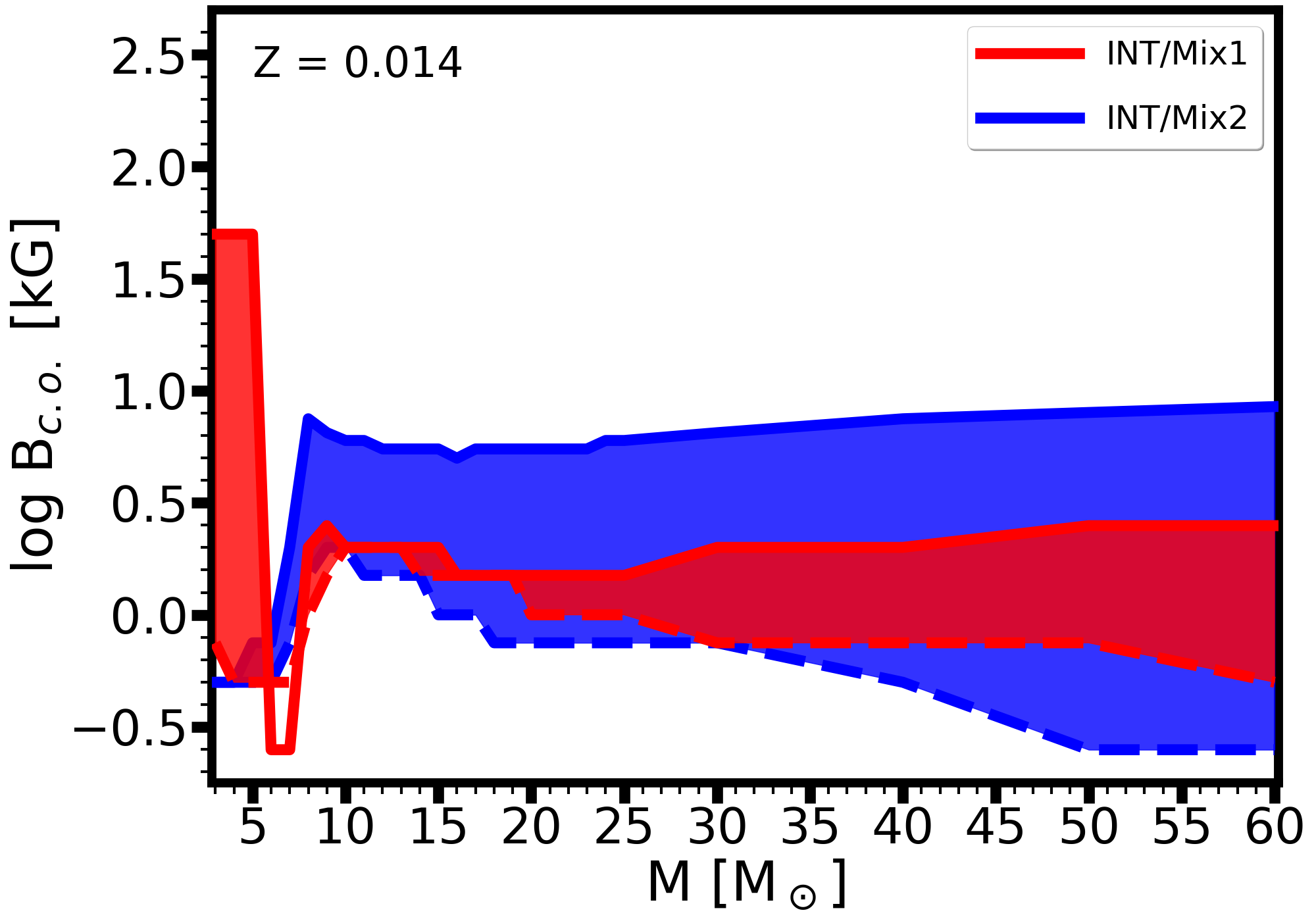}\includegraphics[width=9cm]{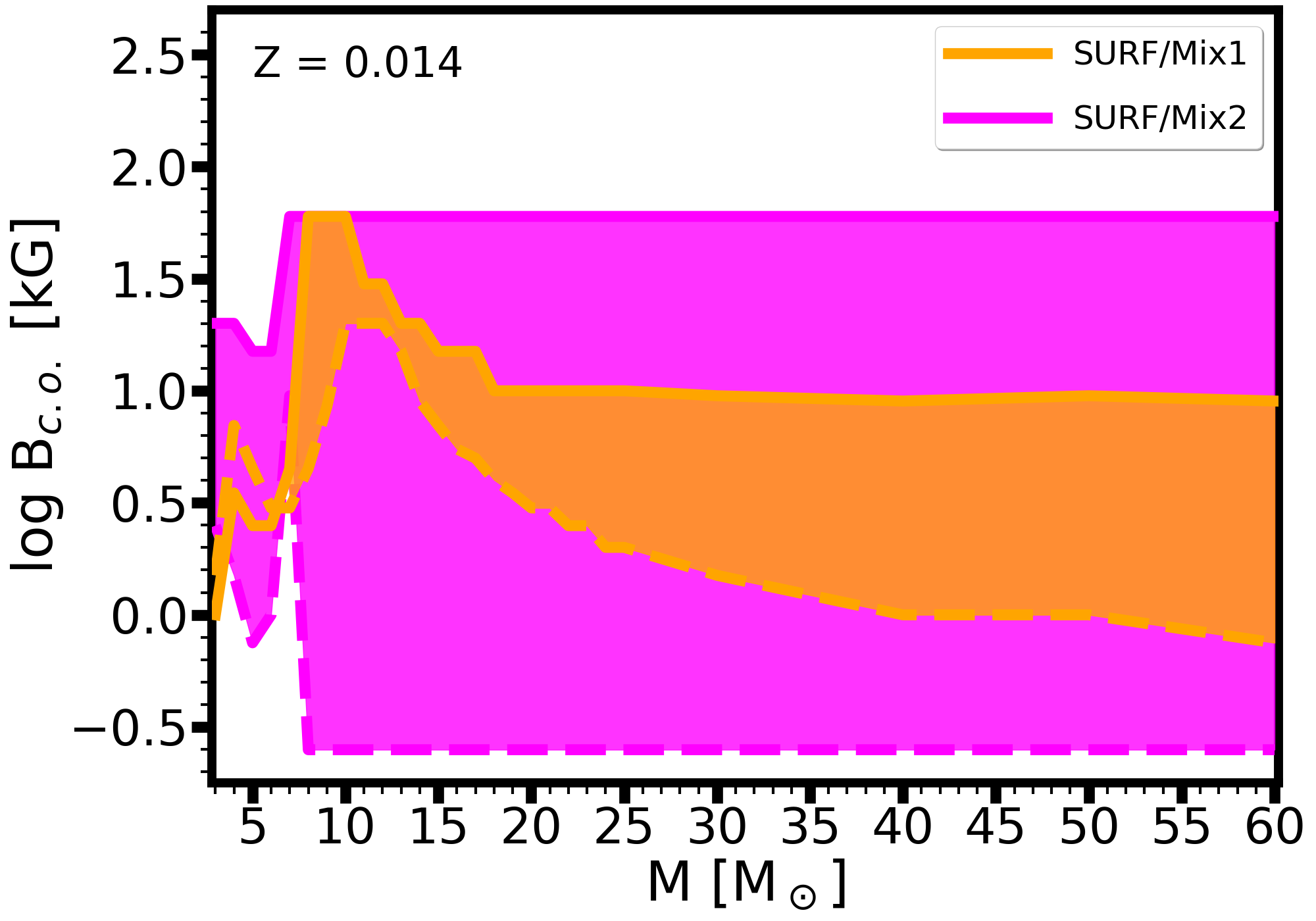}
\caption{Same as Figure \ref{fig:cutoff2} but for Solar metallicity models.}\label{fig:cutoff3}
\end{figure*}
\section{Cutoff magnetic field strength in the SMC and Solar neighbourhood}\label{sec:cutoff23}

As in Section~\ref{sec:cutoff}, we can quantify the cutoff magnetic field strengths in other metallicity environments needed to achieve slow-rotating, nitrogen-enriched stars. First, let us turn our attention to models at the SMC, shown in Figure~\ref{fig:cutoff2}. 
The four schemes predict essentially constant values for $B_{\rm max, N}$  for masses higher than about 10~M$_\odot$. The INT/Mix1 scheme is the most limiting scenario from the four schemes. The other schemes predict a wider range of magnetic field strengths that allow for producing Group 2 stars. Since nitrogen abundances and rotation rates are measurable in the SMC, the identified Group 2 stars \citep[][]{dufton2020} could be explained with a limited range of magnetic field strengths. Typically, the magnetic field strength decreases roughly by an order of magnitude on the main sequence (cf. Equation~\ref{eq:fieldevol}). Thus depending on the age and evolutionary state of observed Group 2 stars, their observable fields will be weaker than the ZAMS field strengths. 
A sort of minimum range of their putative magnetic field strength could evolve from the range of initially 250 G to 7.5 kG (equatorial strength) at the ZAMS to 25~G to 750~G at the TAMS if the INT/Mix1 scheme is considered. For massive star models, the SURF/Mix1 and SURF/Mix2 schemes allow for an upper limit of initially 50~kG, which we expect to weaken to a 5 kG equatorial field strength by the TAMS.

Figure~\ref{fig:cutoff3} shows models with an initial metallicity of $Z =~0.014$ (representative of the Solar neighbourhood). Consistently with previous findings, the INT/Mix1 is the most restrictive to produce Group~2 stars. In fact, in this case there are more forbidden regions than in models with lower metallicities. For example, the INT/Mix1 scheme essentially does not allow for producing Group~2 stars in the 6 - 23 M$_\odot$ range. 
In contrast, the SURF/Mix2 scheme allows for strong magnetic fields to brake the rotation and still produce nitrogen excess at the surface. Since high-resolution spectropolarimetric observations in the Galaxy have allowed for measuring the magnetic field strengths of massive stars, these model predictions allow for more direct comparison with observations than in the Magellanic Clouds. While the nitrogen abundance and (projected) rotational velocity are also measurable, a comprehensive study still needs to assess these three quantities jointly in comparison to magnetic stellar evolution models (however, see \citealt{morel2008,martins2012,martins2015,aerts2014}). Our single-star models provide testable predictions for these parameters, and such an investigation could help disentangle between the uncertain braking and mixing schemes. In particular, nitrogen-enriched slow rotators with strong magnetic fields favour efficient mixing.


\bsp	
\label{lastpage}
\end{document}